\documentclass[11pt,noindent]{article}
\usepackage{graphics,cite,amssymb,epsfig,float}
\usepackage[usenames,dvips]{color}
\usepackage{amsmath, mathbbol}
\usepackage{amsfonts,xfrac,slashed}
\usepackage{pstricks}
\usepackage{a4,graphicx}
\usepackage{pst-plot}
\usepackage{fancybox}
\usepackage{amsfonts}

\usepackage{ulem}
\usepackage{soul}
\usepackage{bm}

\usepackage{xcolor,color}
\definecolor{cblue}{RGB}{100,5,255}
\definecolor{cred}{RGB}{180,50,40} 
\definecolor{cgreen}{RGB}{40,255,40} 
\definecolor{corange}{RGB}{180,140,20} 
\definecolor{cmagenta}{RGB}{120,20,50} 
\definecolor{mathblue}{RGB}{100,100,250} 
     
\usepackage{array}
\usepackage{multirow}
\def\lsim{\;\raise0.3ex\hbox{$<$\kern-0.75em\raise-1.1ex\hbox{$\sim$}}\;}
\def\gsim{\;\raise0.3ex\hbox{$>$\kern-0.75em\raise-1.1ex\hbox{$\sim$}}\;}

\def\ben{\begin{enumerate}}  \def\een{\end{enumerate}}
\def\bit{\begin{itemize}}    \def\eit{\end{itemize}}
\def\beq{\begin{equation}}   \def\eeq{\end{equation}}
\def\ba{\begin{array}}       \def\ea{\end{array}}
\def\bea{\begin{eqnarray}}   \def\eea{\end{eqnarray}}

\newcolumntype{C}{ >{\centering\arraybackslash} m{4cm} }
\usepackage{hhline}

\begin{document}

\setcounter{footnote}{0}
\vspace*{-1.5cm}
\begin{flushright}
LPT Orsay 17-76 \\
%%PCCF RI 17-XX\\
TUM-HEP/1115/17\\
CP3-17-52\\

\vspace*{2mm}
%\today								
\end{flushright}
\begin{center}
\vspace*{1mm}

\vspace{1cm}
{\Large\bf 
Effective Majorana mass matrix from tau and pseudoscalar
\vspace*{1mm}\\ 
 meson lepton number violating decays
}

\vspace*{0.8cm}

{\bf Asmaa Abada$^{a}$, Valentina De Romeri$^{b}$, Michele Lucente$^{c}$,\\ 
Ana M. Teixeira$^{d}$ and Takashi Toma$^{e}$ }  
   
\vspace*{.5cm} 
$^{a}$Laboratoire de Physique Th\'eorique, CNRS, \\
Univ. Paris-Sud, Universit\'e Paris-Saclay, 91405 Orsay, France
\vspace*{.2cm} 

$^{b}$ AHEP Group, Instituto de F\'{\i}sica Corpuscular,
C.S.I.C./Universitat de Val\`encia,  \\
Calle Catedr\'atico Jos\'e Beltr\'an, 2 E-46980 Paterna, Spain
\vspace*{.2cm} 

$^{c}$ Centre for Cosmology, Particle Physics and Phenomenology - CP3, \\
Universit\'e Catholique de Louvain, Chemin du Cyclotron, 2 \\
1348 Louvain-la-Neuve, Belgium
\vspace*{.2cm} 

$^{d}$  Laboratoire de Physique de Clermont, CNRS/IN2P3 -- UMR 6533,\\ 
Campus des C\'ezeaux, 4 Av. Blaise Pascal, F-63178 Aubi\`ere Cedex, France
\vspace*{.2cm} 

$^{e}$ Physik-Department T30d, Technische Universit\"at M\"unchen, \\
James-Franck-Stra\ss e, D-85748 Garching, Germany
\end{center}

\vspace*{6mm}
\begin{abstract}
An  observation of any lepton number violating process will
undoubtedly point towards the existence of new physics and indirectly
to the clear Majorana nature of the exchanged fermion. 
In this work, we explore the potential of
a minimal extension of the Standard Model via heavy sterile fermions
with masses in the  $[ 0.1 - 10]$ GeV range
concerning an extensive array of ``neutrinoless'' meson and tau decay
processes. We assume that the Majorana neutrinos are produced
on-shell, and focus on three-body decays. 
We conduct an update on the bounds on the active-sterile
mixing elements, $|U_{\ell_\alpha 4} U_{\ell_\beta 4}|$, taking into
account the most recent experimental bounds (and constraints) and new
theoretical  
inputs, as well as the effects of a finite detector, imposing that the
heavy neutrino decay within the detector.  
This allows to establish up-to-date comprehensive
constraints on the sterile
fermion parameter space. Our results suggest that the branching
fractions of several decays are close to current sensitivities
(likely within reach of future facilities), some being already in
conflict  with current data (as is the case of $K^+ \to \ell_\alpha^+
\ell_\beta^+ \pi^-$, and $\tau^- \to \mu^+ \pi^- \pi^-$).
We use these processes  to extract constraints on all entries of 
an enlarged  definition of a $3\times 3$
``effective'' Majorana neutrino mass matrix $m_{\nu}^{\alpha \beta}$.
\end{abstract}
\vspace*{2mm}

\newpage
\section{Introduction}
In parallel to the direct searches being
currently carried at the LHC, the high-intensity frontier also offers
a rich laboratory to look for (very) rare processes which indirectly
manifest the presence of new physics.
Individual lepton flavours and total lepton number 
are strictly conserved quantities in the Standard
Model (SM).  By themselves, neutrino oscillations constitute evidence for
lepton flavour violation in the neutral lepton sector, and suggest the
need to consider extensions of the SM capable of accounting for the
necessarily massive neutrinos and for lepton mixing, as confirmed by
experimental
data~\cite{Tortola:2012te,Fogli:2012ua,GonzalezGarcia:2012sz,Forero:2014bxa,nufit,Gonzalez-Garcia:2014bfa,Esteban:2016qun,deSalas:2017kay}.   

The observation of a lepton number violation (LNV) process
would be a clear signal of new physics, in particular of the existence of
Majorana fermions. %, such as massive  neutrinos. 
Although neutrinoless double beta decay ($0\nu 2 \beta$) 
remains by excellence the observable associated with
the existence of Majorana neutrinos, many other processes 
reflecting $\Delta L=2$ are being actively searched for. 
At colliders, there are several possible signatures of LNV and/or
manifestations of the existence of Majorana fermions (see for
instance~\cite{Ali:2001gsa,Atre:2005eb,Atre:2009rg,Chrzaszcz:2013uz,Deppisch:2015qwa,Cai:2017mow} 
and references therein); in some cases these might even hint on 
the neutrino mass generation mechanism
(if interpreted in the light of such theoretical frameworks).  

Semileptonic meson decays are examples of
transitions which offer the possibility of studying LNV, for both
cases of three- or four-body final states, 
$M_1 \to  \ell^\pm \ell^\pm M_2$ and/or $M_1 \to \ell^\pm \ell^\pm
M_2 M_3$, $M_i$ denoting mesons. Likewise, there are several
semileptonic tau decays  
corresponding to $\Delta L=2$ transitions, $\tau^\pm  \to \ell^\mp M_1
M_2$, which can occur in the presence of Majorana neutral fermions
(such as neutrinos).  
However, whether such decays do have a non-negligible width 
strongly depends not only on the Majorana nature of the neutrinos,
but on their properties. If the processes are only mediated by the
three active light neutrinos, the corresponding widths will
be strongly suppressed by the smallness of their masses (as they are
proportional to $m_\nu^2$); 
should one consider extensions of the SM particle content
by heavier neutral fermion states, which have non-vanishing mixings with the
left-handed neutrinos of the SM, then the additional contribution to the 
widths will also be suppressed,
either by the heavy propagator's mass, or by tiny mixings.
The suppression can be nevertheless evaded if 
the mass of the heavy neutrino is such that it can be produced
on-shell in the decay of the heavy meson (or of the tau lepton). 
Due to the so-called ``resonant enhancement'', the LNV decay widths can
be strongly increased in this regime. 

In recent years, the r\^ole of heavy sterile fermions (sufficiently
light to be produced on-shell from the semileptonic LNV decays of
mesons or taus) has been addressed, both for three- and
four-body final 
states~\cite{Ali:2001gsa,Atre:2009rg,Helo:2010cw,Zhang:2010um,Cvetic:2010rw,Quintero:2011yh,Castro:2012gi,Castro:2012ma, Castro:2013jsn,Yuan:2013yba,Wang:2014lda,Cvetic:2014nla,Cvetic:2015naa,Mandal:2016hpr,Milanes:2016rzr,Quintero:2016iwi,Cvetic:2016fbv,Liu:2016oph,Asaka:2016rwd,Yuan:2017xdp,Cvetic:2017vwl, Mejia-Guisao:2017gqp,Yuan:2017uyq} 
and bounds were derived for the new
propagator's degrees of freedom. Just as valuable information on the
$ee$ entry of the effective Majorana mass matrix 
can be inferred from LNV $0\nu 2 \beta$ decays 
(see, e.g.~\cite{GomezCadenas:2011it}), bounds have been
established for other entries of 
$m_\nu^{\alpha \beta}$, using adapted expressions inspired from 
neutrinoless double beta
decays~\cite{Flanz:1999ah,Zuber:2000vy,Zuber:2000ca,Atre:2005eb, 
Rodejohann:2011mu,Liu:2016oph,Quintero:2016iwi}.  

B-factories (such as BaBar and Belle), together with the 
advent of the LHC (in its run 1 and 2), have allowed to establish 
increasingly stronger limits on LNV semileptonic 
decays~\cite{Patrignani:2016xqp,Amhis:2016xyh}. 
On the theoretical side, significant progress
has been made on the non-perturbative 
computation of several quantities - such as decay
constants - relevant for the LNV semileptonic transitions. 
Although these decays all imply the violation of individual lepton
flavours (by two units), some can lead to the appearance of same sign but 
distinct (i.e., different flavour) charged leptons in the final state, 
$M_1 \to  \ell^\pm_\alpha \ell^\pm_\beta M_2$ with $\alpha \neq
\beta$.  At the high-intensity frontier, 
a number of experiments dedicated to search for a signal of charged
lepton flavour violation (cLFV) - already collecting data, 
or due to operate in the near future - are
expected to significantly improve the bounds on several processes. 
Such is the case of $\mu \rightarrow  e \gamma$ for which the present
(future)  bound is $4.2\times 10^{-13}$~\cite{TheMEG:2016wtm} ($6\times
10^{-14}$~\cite{Baldini:2013ke}), of
$\mu \rightarrow e e e$ with a present (future) bound of $1.0 \times
10^{-12}$~\cite{Bellgardt:1987du} ($\sim10^{-16}$~\cite{Blondel:2013ia}), 
and of $\mu - e$  conversion in nuclei with a present bound of
$7\times10^{-13}$ for gold nuclei~\cite{Bertl:2006up} and an expected
sensitivity of $\sim 
10^{-17}$~\cite{Carey:2008zz,Cui:2009zz,Kuno:2013mha} for Aluminium targets. 
Likewise, valuable information can also be obtained from cLFV meson
decays, such as $M\to \ell_\alpha^+\ell_\beta^-$ and 
$M_1 \to M_2\ell_\alpha^+ \ell_\beta^-$,   
currently searched for both at B-factories and at the LHC~\cite{Patrignani:2016xqp}.

In view of the recent experimental and theoretical progress, and
in preparation for the expected improvement in the experimental fronts, in
this work we explore a variety of semileptonic meson and tau LNV
decays induced by the presence of sterile Majorana 
fermions, whose mass is such as to allow for the resonant enhancement
of the decay rates above referred to. Moreover, we propose a
generalised definition of the effective Majorana mass matrix
(encompassing the standard $m_\nu^{ee}$ associated with $0\nu2\beta$
decays), and infer constraints on {\it all} its entries from the
confrontation of a comprehensive set of LNV semileptonic decays to 
current experimental bounds.  

In the present study, we only consider processes leading to three-body
final states - in particular, when considering LNV meson decays, we
only address processes involving a
single meson in the final state;
furthermore, we focus on pseudoscalar meson decays. 
This implies that we will not address channels such as the decays of vector
mesons, nor the LNV four-body decays of $B$ mesons, 
which have recently been identified as promising experimental
probes, in particular at LHCb (see, 
e.g.~\cite{Chrzaszcz:2013uz,Castro:2013jsn,Yuan:2017xdp, Cvetic:2017vwl,Mejia-Guisao:2017gqp,Yuan:2017uyq}). 
Despite their experimental appeal, these decays are still subject to important  
theoretical uncertainties, such as poor control of the hadronic 
matrix elements (when vector mesons are involved), 
or interactions between final state hadrons, 
among others; we will not address them in the present study. 

Concerning the underlying theoretical framework, and as a 
first step, we do not consider a specific mechanism of
neutrino mass generation, but rather a minimal simple extension of the
SM via one Majorana sterile fermion, focusing on the 
$[0.1 - 10]$ GeV mass range. 
Any scenario of new physics involving sterile fermions - even such
minimal constructions - must comply with stringent phenomenological and
observational constraints arising from bounds derived from 
different experimental frontiers (high-intensity, high-energy
experiments, neutrino data, as well as from cosmology). All available
constraints will be taken into account in our
analysis. 

We conduct a systematic 
analysis of lepton number violating semileptonic meson decays, also
including lepton number violating tau-lepton decays. This allows to 
build upon and update earlier analysis, revising the bounds derived 
in~\cite{Flanz:1999ah,Zuber:2000vy,Zuber:2000ca,Atre:2005eb,
Atre:2009rg,Rodejohann:2011mu,Helo:2011yg,Liu:2016oph,Quintero:2016iwi,
Zamora-Saa:2016qlk,Zamora-Saa:2016ito}, 
and obtaining an up-to-date overview of
the sterile fermion parameter space (mass and combinations of the
active-sterile mixing angles, $U_{\ell_\alpha 4} U_{\ell_\beta 4}$).   
We explore the allowed parameter space in order to identify regimes 
for which the LNV branching ratios might be within experimental reach, 
for instance of NA62 
(for the light mesons), BES-III for charmed mesons and  
LHCb (and in the future, Belle II) for $B$ mesons and tau leptons. 
(Our study identified decays which already are in conflict with current
bounds, as is the case of $K^+ \to \ell_\alpha^+
\ell_\beta^+ \pi^-$, and $\tau^- \to \mu^+ \pi^- \pi^-$.) 
We also impose that the on-shell heavy fermion be
sufficiently short-lived as to decay within a finite size detector (we
will work for a ``benchmark'' value of 10 metres). 

As mentioned above, we further translate all the 
collected bounds into limits for the flavour conserving 
and flavour violating entries of the $3\times 3$ Majorana effective mass matrix,
$|m_{\nu}^{\alpha \beta}|$. Other than (strongly) improving existing
bounds, we propose bounds for the $|m_{\nu}^{\tau \tau}|$ entry. 

This work is organised as follows: in Section~\ref{sec:sterile:LNV.cLFV}
we update the constraints on
minimal SM extensions by a sterile neutral fermion arising from LNV 
(and cLFV) semileptonic three-body decays, reviewing their experimental
status, and we revisit the theoretical estimation of the corresponding
decay widths in the ``resonant enhancement'' regime;  once all 
phenomenological and observational constraints have been applied, we 
update the sterile fermion parameter space. We further 
address the reconstruction of the $3\times 3$
``effective'' Majorana mass matrix. Section~\ref{sec:Results} collects our
results (and predictions): LNV branching fractions for the
semileptonic decays of mesons ($B$, $D$, $K$) and $\tau$ leptons, 
as well as their impact for the effective Majorana mass entries. 
Further discussion and remarks are given in Section~\ref{sec:disc.concs}. 
Other than the detailed description of the minimal SM extension we
consider, Appendix~\ref{app:SM.sterile} summarises the 
relevant constraints applied in our work. 
Appendix~\ref{app:sec:widths} details the computation of 
the LNV meson and tau lepton decay widths, while in 
Appendix~\ref{app:sec:widths:nu4} we describe the calculation of the 
decay width of the exchanged Majorana state. 
Finally, the expressions for the cLFV meson decay widths are 
collected in Appendix~\ref{app:sec:widths:lfv}.

\section{An additional sterile fermion: impact for LNV and cLFV}
\label{sec:sterile:LNV.cLFV}

Extensions of the SM via 
the addition of Majorana sterile neutral fermions, with non-negligible
mixings with the light, active neutrinos, open the door to numerous
processes which violate total lepton number, charged lepton flavour (or
both), among other phenomena. While many complete models of neutrino
mass generation do include such states (for example right-handed
neutrinos in several seesaw realisations), considering 
a minimal scenario in which a single Majorana sterile neutrino is
added to the SM field 
content\footnote{The additional sterile state can also be
interpreted as encoding the effects of a larger number of states
possibly present in the underlying new physics model.}, without any 
assumption on the mechanism
of neutrino mass generation, proves to be a useful first step in
evaluating the potential contribution of these states to a wide array
of observables, including cLFV and LNV decays. 

The aim of
this section is to discuss the updated constraints on
minimal SM extensions by sterile fermions arising from LNV (and cLFV) 
semileptonic three-body decays. After a summary of the corresponding
experimental status, we discuss the theoretical computation of the
decay widths, and update the constraints on SM extensions
  with sterile Majorana fermions from the  
non-observation of LNV decays. We further address the reconstruction of
the effective $3\times 3$ Majorana mass matrix. In order to do so, we
thus rely on a simple ``toy model'' which is
built under the single hypothesis that
interaction and physical neutrino eigenstates are related via a
$4\times 4$ unitary mixing matrix, ${\bf U}_{ij}$ (see
Appendix~\ref{app:SM.sterile}).  Other than the
masses of the three light (mostly active) neutrinos, and their mixing
parameters, the simple 
``3+1 model'' is parametrised by the heavier (mostly sterile)
neutrino mass $m_4$, three active-sterile mixing angles, as well as
three new CP violating phases (two Dirac and one Majorana). 

In our discussion, and unless otherwise stated, the sterile state is
assumed to be produced (on-shell) from the semileptonic decays of either tau
leptons or mesons. 

\subsection{LNV and cLFV decays: experimental status}

As mentioned in the Introduction, 
searches for LNV and/or cLFV decays are (and have been) an important goal 
of numerous experimental collaborations.

Especially in a context in which LNV arises from the presence of
Majorana fermions, 
neutrinoless double beta decay is one of the most promising
observables: KamLAND-ZEN~\cite{Gando:2012zm,KamLAND-Zen:2016pfg},
GERDA~\cite{Agostini:2013mzu}  
and EXO-200~\cite{Auger:2012ar,Albert:2014awa,Albert:2017owj}, have
all set strong 
bounds on the $m_\nu^{ee}$ effective mass, 
to which the amplitude of $0\nu 2 \beta$  process is proportional
(see detailed expressions in Section~\ref{sec:effective.mass.th}).
The most recent and strongest constraint has been obtained by 
the KamLAND-ZEN collaboration~\cite{KamLAND-Zen:2016pfg}
\begin{equation}\label{eq:onu2beta:limit.KZ}
m_\nu^{ee}\, \lesssim \, 0.165 \text{ eV} \, \, (90\% \, \text{C.L.})\,;
\end{equation}
concerning future prospects, we summarise in
Table~\ref{tab:nulesssensitivities} the   
sensitivity of ongoing and planned $0\nu 2 \beta$ dedicated experiments.
In this work, we take a representative benchmark for the 
future sensitivity $|m_\nu^{ee}| \lesssim 0.01$ eV 
(for details concerning the theoretical uncertainties, see for 
instance~\cite{DellOro:2016tmg,Maneschg:2017mzu}). 

\begin{table}[h!]
\begin{center}
{\begin{tabular}{| l | l | c |}  \hline                       
Experiment & Ref. &  $ |m_{\nu}{ee} | $ (eV) \\
  \hline
 EXO-200 (4 yr) & \cite{Auger:2012ar,Albert:2014awa} & $0.075 - 0.2$  \\
nEXO (5 yr)  & \cite{Tosi:2014zza,Licciardi:2017oqg}& $0.012 - 0.029$  \\
nEXO (5 yr + 5 yr w/ Ba tagging) &
\cite{Tosi:2014zza} 
& $0.005 - 0.011$  \\
KamLAND-Zen (300~kg, 3 yr)& \cite{Gando:2012zm}  & $0.045 - 0.11$ \\
GERDA  phase II & \cite{Agostini:2013mzu,Agostini:2017iyd} & $0.09 - 0.29$ \\
CUORE (5 yr) & \cite{Gorla:2012gd,Aguirre:2014lua,Artusa:2014lgv} &
$0.051 - 0.133$ \\ 
SNO+  & \cite{Hartnell:2012qd,Lozza:2016ghw} & $0.07 - 0.14$ \\
SuperNEMO & \cite{Barabash:2011aa} & $0.05 - 0.15$ \\
NEXT & \cite{Granena:2009it,Gomez-Cadenas:2013lta}& $0.03 - 0.1$ \\
MAJORANA Demonstrator& \cite{Phillips:2011db,Wilkerson:2012ga} & $0.06 - 0.17$ \\
 \hline                       
\end{tabular}
}
\caption{Sensitivity of several $0\nu 2 \beta$ experiments.}
\label{tab:nulesssensitivities}
\end{center}
\end{table}

\bigskip
Semileptonic meson decays offer numerous channels to look for LNV;
final states comprising two same-sign leptons and one or two mesons, 
$M_1 \to  \ell^\pm \ell^\pm M_2$ and $M_1 \to \ell^\pm \ell^\pm
M_2 M_3$, have been experimentally searched for in recent years. In
Table~\ref{tab:LNV:meson:exp}  
we collect the bounds for the case of LNV 
three-body final states 
(including same and different 
charged leptons in the final states), further referring to the 
$m_\nu^{\alpha \beta}$ effective mass entries thus constrained.
The current bounds listed 
are expected to be improved by one or two orders of
magnitude with future experiments, such as LHCb in its
run 2~\cite{LHCb:2011dta} as well as NA62~\cite{na62}; 
for the channel $B^+\to\mu^+\mu^+\pi^-$, the
future experiments Belle~II\footnote{It is also worth mentioning that 
the  sensitivity of Belle~II to tau LFV decays is over 100 times greater than
its predecessor for the cleanest channels (as tau three-body decays), and 
about 10 times better for radiative tau decays (important irreducible
backgrounds precluding further improvements in 
sensitivity)~\cite{HerediadelaCruz:2016yqi}.} 
and FCC-ee are
expected to improve the current bounds (for $|U_{\mu4}|^2$) by about
two orders of magnitude~\cite{Asaka:2016rwd}.
Although not included in Table~\ref{tab:LNV:meson:exp}, 
the decay mode $B_c^-\to\tau^-\tau^-\pi^+$ can also be explored aiming
at constraining the
matrix element $m_\nu^{\tau\tau}$~\cite{Mandal:2016hpr} (and other
elements~\cite{Cvetic:2010rw, Milanes:2016rzr} as well).

\begin{table}[h!]
\begin{center}
\begin{tabular}{|c||c|c|c|}\hline
\multirow{2}{*}{LNV decay} & 
\multicolumn{3}{c|}{Current bound}\\\cline{2-4}
& 
$\ell_\alpha=e,~\ell_\beta=e$ & 
$\ell_\alpha=e,~\ell_\beta=\mu$ & 
$\ell_\alpha=\mu,~\ell_\beta=\mu$\\
\hhline{|=#=|=|=|}
$K^-\to \ell_\alpha^-{\ell_\beta}^-\pi^+$ & 
$6.4\times10^{-10}$~\cite{Patrignani:2016xqp} & 
$5.0\times10^{-10}$~\cite{Patrignani:2016xqp} & 
$1.1\times10^{-9}$~\cite{Patrignani:2016xqp}\\
$D^-\to\ell_\alpha^-{\ell_\beta}^-\pi^+$ & 
$1.1\times10^{-6}$~\cite{Patrignani:2016xqp} & 
$2.0\times10^{-6}$~\cite{Lees:2011hb} & 
$2.2\times10^{-8}$~\cite{Aaij:2013sua}\\
$D^-\to\ell_\alpha^-{\ell_\beta}^-K^+$ & 
$9.0\times10^{-7}$~\cite{Lees:2011hb} & 
$1.9\times10^{-6}$~\cite{Lees:2011hb} & 
$1.0\times10^{-5}$~\cite{Lees:2011hb}\\
$D^-\to\ell_\alpha^-{\ell_\beta}^-\rho^+$ &
-----------&
-----------& 
$5.6\times10^{-4}$~\cite{Patrignani:2016xqp}\\
$D^-\to \ell_\alpha^-{\ell_\beta}^-K^{*+}$ &
-----------&
-----------& 
$8.5\times10^{-4}$~\cite{Patrignani:2016xqp}\\
$D_s^-\to \ell_\alpha^-{\ell_\beta}^-\pi^+$ & 
$4.1\times10^{-6}$~\cite{Patrignani:2016xqp} & 
$8.4\times10^{-6}$~\cite{Lees:2011hb} & 
$1.2\times10^{-7}$~\cite{Aaij:2013sua}\\
$D_s^-\to \ell_\alpha^-{\ell_\beta}^-K^+$ & 
$5.2\times10^{-6}$~\cite{Lees:2011hb} & 
$6.1\times10^{-6}$~\cite{Lees:2011hb} & 
$1.3\times10^{-5}$~\cite{Lees:2011hb}\\
$D_s^-\to \ell_\alpha^-{\ell_\beta}^-K^{*+}$ & 
-----------&
-----------& 
$1.4\times10^{-3}$~\cite{Patrignani:2016xqp}\\
$B^-\to \ell_\alpha^-{\ell_\beta}^-\pi^+$ & 
$2.3\times10^{-8}$~\cite{BABAR:2012aa} & 
$1.5\times10^{-7}$~\cite{Lees:2013gdj} & 
$4.0\times10^{-9}$~\cite{Aaij:2014aba}\\
$B^-\to \ell_\alpha^-{\ell_\beta}^-K^+$ & 
$3.0\times10^{-8}$~\cite{BABAR:2012aa} & 
$1.6\times10^{-7}$~\cite{Lees:2013gdj} & 
$4.1\times10^{-8}$~\cite{Aaij:2011ex}\\
$B^-\to \ell_\alpha^-{\ell_\beta}^-\rho^+$ & 
$1.7\times10^{-7}$~\cite{Lees:2013gdj} & 
$4.7\times10^{-7}$~\cite{Lees:2013gdj} & 
$4.2\times10^{-7}$~\cite{Lees:2013gdj}\\
$B^-\to \ell_\alpha^-{\ell_\beta}^-D^+$ & 
$2.6\times10^{-6}$~\cite{Seon:2011ni} & 
$1.8\times10^{-6}$~\cite{Seon:2011ni} & 
$6.9\times10^{-7}$~\cite{Aaij:2012zr}\\
$B^-\to \ell_\alpha^-{\ell_\beta}^-D^{*+}$ & 
 -----------& 
 -----------& 
$2.4\times10^{-6}$~\cite{Patrignani:2016xqp}\\
$B^-\to \ell_\alpha^-{\ell_\beta}^-D^{+}_s$ & 
 -----------& 
 -----------& 
$5.8\times10^{-7}$~\cite{Patrignani:2016xqp}\\
$B^-\to \ell_\alpha^-{\ell_\beta}^-K^{*+}$ & 
$4.0\times10^{-7}$~\cite{Lees:2013gdj} & 
$3.0\times10^{-7}$~\cite{Lees:2013gdj} & 
$5.9\times10^{-7}$~\cite{Lees:2013gdj}\\
\hhline{|=#=|=|=|}
LNV matrix $m_\nu$ &
$m_{\nu}^{ee}$ & 
$m_{\nu}^{e\mu}$ & 
$m_{\nu}^{\mu\mu}$\\\hline
\end{tabular}
\caption{LNV meson decay processes. The current bounds for Kaon, $D$
 and $B$ meson decays were obtained by Belle~\cite{Seon:2011ni},
 BABAR~\cite{Lees:2013gdj, 
 BABAR:2012aa, Lees:2011hb} and LHCb~\cite{Aaij:2011ex, Aaij:2012zr,
 Aaij:2013sua, Aaij:2014aba}, and have been summarised 
in~\cite{Patrignani:2016xqp,Aoki:2016frl}. 
}\label{tab:LNV:meson:exp}
\end{center}
\end{table}

\bigskip
Important constraints, albeit indirect,  
on the effective mass entries $m_\nu^{\alpha \beta}$ can also be
inferred from lepton number conserving (but cLFV) semileptonic meson
decays, as the latter 
constrain individual entries (or combinations) of the lepton
mixing matrix,  $U_{\ell_\alpha i}U_{\ell_\beta i}$. 
While cLFV radiative and three-body decays
(such as $\ell_\alpha \to \ell_\beta \gamma$, $\ell_\alpha \to 3 \ell_\beta$, or
neutrinoless conversion in Nuclei) in general lead to severe constraints for
sterile masses typically
above the tenths of GeV, in the mass regimes
in which the heavy neutrino is produced on-shell from the above
discussed tau and/or meson decays,
constraints arising from cLFV meson decays can be important, and 
should thus be carefully evaluated and taken into account. 
In Tables~\ref{tab:LFV:meson:exp} and~\ref{tab:LFV:Bmeson:exp}
we summarise some of these bounds, which are of relevance to our
study. 

\begin{table}[h!]
\begin{center}
\begin{tabular}{|c||c|c|c|}\hline
\multirow{2}{*}{cLFV decay} & 
\multicolumn{3}{c|}{Current bound}\\\cline{2-4}
& 
$\ell_\alpha=e,~\ell_\beta=\mu$ & 
$\ell_\alpha=e,~\ell_\beta=\tau$ & 
$\ell_\alpha=\mu,~\ell_\beta=\tau$\\
%\hline\hline
\hhline{|=#=|=|=|}
$K^+\to \ell_\alpha^\pm{\ell_\beta}^\mp \pi^+$ & 
$5.2\times10^{-10} \,(1.3\times10^{-11})$ & 
-----------& 
-----------\\
$D^+\to\ell_\alpha^\pm{\ell_\beta}^\mp \pi^+$ & 
$2.9 (3.6)\times10^{-6}$ & 
----------- & 
-----------\\
$D^+\to\ell_\alpha^\pm{\ell_\beta}^\mp K^+$ & 
$1.2 (2.8)\times10^{-6}$ & 
----------- & 
-----------\\
$D_s^+\to \ell_\alpha^\pm{\ell_\beta}^\mp \pi^+$ & 
$1.2 (2.0)\times10^{-5}$ & 
-----------& 
-----------\\
$D_s^+\to \ell_\alpha^\pm{\ell_\beta}^\mp K^+$ & 
$14 (9.7)\times10^{-6}$ & 
-----------& 
-----------\\
$B^+\to \ell_\alpha^\pm{\ell_\beta}^\mp \pi^+$ & 
$0.17\times10^{-6}$ & 
$75\times10^{-6}$ & 
$72\times10^{-6}$\\
$B^+\to \ell_\alpha^\pm{\ell_\beta}^\mp K^+$ & 
$91\times10^{-6}$& 
$30\times10^{-6}$& 
$48\times10^{-6}$\\
$B^+\to \ell_\alpha^\pm{\ell_\beta}^\mp K^{*+}$ & 
$1.4\times10^{-6}$& 
-----------& 
-----------\\
$B^0\to \ell_\alpha^\pm{\ell_\beta}^\mp \pi^0$ & 
$0.14\times10^{-6}$& 
-----------& 
-----------\\
$B^0\to \ell_\alpha^\pm{\ell_\beta}^\mp K^0$ & 
$0.27\times10^{-6}$& 
-----------& 
-----------\\
$B^0\to \ell_\alpha^\pm{\ell_\beta}^\mp K^{*0}$ & 
$0.53\times10^{-6}$& 
-----------& 
-----------\\
\hhline{|=#=|=|=|}
\end{tabular}
\caption{cLFV meson decay processes relevant in constraining the LNV
  modes~\cite{Patrignani:2016xqp}. }\label{tab:LFV:meson:exp}
\end{center}
\end{table}

\begin{table}[h!]
\begin{center}
{
\begin{tabular}{|c||c|}\hline 
$B$ meson decay & Current bound \\
\hhline{|=#=|}
$B^+ \to e^+ \nu$ & $0.98\times10^{-6}$ \\
$B^+ \to \mu^+ \nu$ & $1.0\times10^{-6}$ \\
$\dagger$ $B^+ \to \tau^+ \nu$ & $ = (106 \pm 19) \times10^{-6}$ \\
\hline
$B^0 \to e^\pm \mu^\mp$ & $0.0028\times10^{-6}$ \\
$B^0 \to e^\pm \tau^\mp$ & $28\times10^{-6}$ \\
$B^0 \to \mu^\pm \tau^\mp$ & $22\times10^{-6}$ \\
\hline
\end{tabular}
\caption{Leptonic (flavour violating and flavour conserving) 
$B$-meson decay modes. The symbol $\dagger$ 
denotes a measurement rather than an upper bound.}\label{tab:LFV:Bmeson:exp}
}
\end{center}
\end{table}

\medskip
Due to its higher mass, the tau lepton can also decay
semileptonically and, in the presence of new physics sources 
several channels might signal $\Delta L=2$ transitions such as 
$\tau^\pm  \to \ell^\mp M_1 M_2$ (which is necessarily cLFV). 
The current bounds regarding $\tau$ decays are summarised in 
Table~\ref{tab:LNV.taudecay.exp}, and are used to infer constraints 
on the $m_\nu^{\ell\tau}$ effective mass entries. 
\begin{table}[h!]
\begin{center}
\begin{tabular}{|c||c|c|c|}\hline
\multirow{2}{*}{LNV decay} & 
\multicolumn{2}{|c|}{Current bound}\\\cline{2-3}
& $\ell=e$ & $\ell=\mu$\\
\hhline{|=#=|=|}
$\tau^-\to \ell^+\pi^-\pi^-$ & $2.0\times10^{-8}$ & $3.9\times10^{-8}$\\
$\tau^-\to \ell^+\pi^-K^-$   & $3.2\times10^{-8}$ & $4.8\times10^{-8}$\\
$\tau^-\to \ell^+K^-K^-$     & $3.3\times10^{-8}$ & $4.7\times10^{-8}$\\
\hhline{|=#=|=|}
LNV matrix $m_\nu$ & 
$m_{\nu}^{e\tau}$ & 
$m_{\nu}^{\mu\tau}$\\\hline
\end{tabular}
\caption{LNV $\tau$ decay processes. The upper
 bounds are from the Belle collaboration~\cite{Miyazaki:2012mx}.}
\label{tab:LNV.taudecay.exp}
\end{center}
\end{table}

\subsection{Meson and tau lepton decay widths}

\begin{figure}[t!]
\begin{center}
\epsfig{file=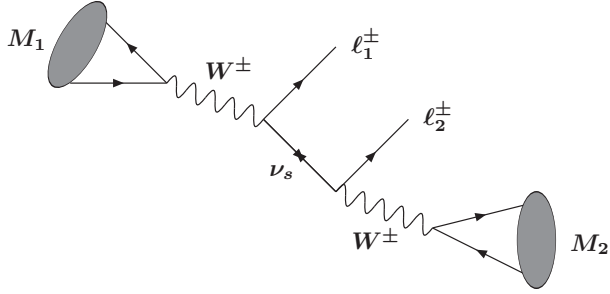, width=80mm}
\end{center}
\caption{Dominant contribution to the lepton number 
violating semileptonic meson decay, $M_1 \to
\ell^\pm_1 \ell^\pm_2 M_2$. Note that the $\ell_1^\pm \leftrightarrow \ell_2^\pm$
exchanged diagram also exists.}\label{fig:feynman:M1decay} 
\end{figure}

We now proceed to discuss and highlight relevant points leading to
the computation (theoretical derivation) of the LNV meson and tau
semileptonic decay widths. These are schematically depicted in 
Fig.~\ref{fig:feynman:M1decay} for the case of a semileptonic LNV meson
decay.
 
\subsubsection{Theoretical estimation}

As already mentioned, leading to the computation of the LNV
semileptonic decays, we have made several assumptions, which we
proceed to discuss. 
\begin{itemize}  
\item We consider semileptonic decay modes leading to three-body final
  states; moreover, we only consider the decays of pseudoscalar mesons 
  and do not address vector meson decays, as their  (non-perturbative)
  decay constants are plagued by larger theoretical
  uncertainties, and the resonances (and excitations) are not well
  determined;  
\item The only source of lepton number violation (and lepton flavour
  violation) at the origin of the distinct decays above mentioned 
  stems from the presence of (heavy) Majorana neutrinos; 
\item In order to avoid excessive suppression due to the propagation of
  a virtual heavy state, we focus on the resonant regimes in which,
  depending on the decaying particle and final state considered,
  sterile neutrinos with a mass within the kinematically allowed
  interval are produced on-shell, and propagate before decaying;
\item Via their interaction with the light (mostly active) neutrinos,
  the extra sterile fermions
  lead to a modification of the leptonic neutral and charged
  currents, as given by
  Eqs.~(\ref{eq:lagrangian:WGHZ}), (\ref{eq:modified.currents.G}) of
  Appendix~\ref{app:SM.sterile:formalism}. In turn, this means that many other
  observables are also affected, hence leading to extensive experimental
  and observational constraints on
  the sterile neutrino degrees of freedom; the latter are summarised
  in Appendix~\ref{app:constraints}.
\end{itemize}

The full expressions for the LNV decay widths of mesons and tau 
leptons\footnote{We have verified that our full analytical results
  agree with those of Refs.~\cite{Atre:2009rg, Helo:2010cw}.}
can be found in Appendix~\ref{app:sec:widths}; they can be however cast in the
following compact form, 
\begin{align}\label{eq:gamma.LNV.compact} 
\Gamma_\text{LNV}\, \propto \, 
G_F^4\,|V_{M{_1}}|^2\,|V_{M{_2}}|^2\,|U_{\ell_\alpha i}|^2\,|U_{\ell_\beta i}|^2\,
f_{M{_1}}^2\,f_{M{_2}}^2 \,\times\,
\Phi(m_i, M_1,M_2, m_\ell)\,,
\end{align}
in which $V_{M}$ denotes the CKM matrix element relevant for the quark
content of meson $M$, and $f_M$ is the meson's decay constant. Likewise, 
$U_{\ell i}$ is the leptonic mixing matrix element for the $W \ell
\nu$ interaction. The function $\Phi$ encodes the integrals over the
corresponding phase space, and depends on the masses and momenta
involved in the transitions. 
As mentioned before, the decay widths are computed under the
assumption of an on-shell neutrino; moreover, and as detailed in
Appendix~\ref{app:sec:widths}, one works in the narrow-width
approximation, which is valid and appropriate in the regimes here
discussed. 

In recent years, important progress has been made regarding meson
data, particularly in what concerns the computation of the decay
constants which are crucial in the estimation of $\Gamma_\text{LNV}$
(see Eq.~(\ref{eq:gamma.LNV.compact})). 
In Tables~\ref{tab:mpfp:pseudoscalar} - \ref{tab:mpfp:mesons.vector.2}  
we summarise the most recent data 
(the reported values are either experimentally determined, 
or non-perturbatively computed) for the mesons relevant to 
our analysis.

\begin{table}[h!]
\begin{center}
\begin{tabular}{|c||cccccccc|}\hline
(MeV)& $\pi^\pm$ & $K^\pm$ & $D^\pm$ & $D_s^\pm$ &
$B^\pm$&$\eta$&$\eta'$&$\eta_c$\\ 
\hhline{|=#========|}
$f_P$   & $130.2$ & $155.6$ & $211.9$ & $249.0$ &
$187.1$&$109.8$\cite{Duplancic:2015zna} &$88.4$
\cite{Duplancic:2015zna}&$387(7)$
\cite{Becirevic:2013bsa,McNeile:2012qf}\\ 
$M_P$   & $139.6$ & $493.7$ & $1869.4$ & $1968.5$ &
$5279$&$547.9$ & $957.8$&$2983.4$\\
\hline 
\end{tabular}
\caption{Decay constants and masses for pseudoscalar mesons: 
$\pi^\pm$, $K^\pm$, $D^\pm$, 
$D_s^\pm$, $B^\pm$, as well as 
$\eta$, $\eta'$ and $\eta_c$.}\label{tab:mpfp:pseudoscalar}
\end{center}
\end{table}

\begin{table}[h!]
\begin{center}
\begin{tabular}{|c||ccc|}\hline
(MeV) & $B$ & $B_s$ & $B_c$ \\ 
\hhline{|=#===|}
$f_P$ & $186(4)$& $224(5)$& 
$427(6)$ \\
$M_P$ & $5276.6$ & $5366.8$& 
$6275.1$
\\
\hline 
\end{tabular}
\caption{Decay constants and masses for $B$ mesons ($f_B$ is an
  average for $B^0$ and $B^\pm$), $B_s$, $B_c$~\cite{Aoki:2016frl}.  
}\label{tab:mpfp:Bmesons}
\end{center}
\end{table}

\begin{table}[h!]
\begin{center}
\begin{tabular}{|c||ccccc|}\hline
(MeV)  
& $B^{*}$ & $B_s^{*}$&$B_c^{*}$&$\eta_b$&$\Upsilon$
\\ 
\hhline{|=#=====|}
$f_V$ & $175(6)$\cite{Colquhoun:2015oha}  & $213$\cite{Colquhoun:2015oha}
 &$422$\cite{Colquhoun:2015oha} &$667(6)$
 \cite{McNeile:2012qf}&$689(5)$ \cite{McNeile:2012qf}
\\
$M_V$ & $5324.65$\cite{Aoki:2016frl} &
 $5415.4$\cite{Aoki:2016frl} &$6330(2)$\cite{Gregory:2009hq} &
 $9399$\cite{Aoki:2016frl}&$9460.3$\cite{Aoki:2016frl}
\\
\hline 
\end{tabular}
\caption{Decay constants and masses for the vectorial $B$
  mesons~\cite{Rosner:2015wva}.}\label{tab:mpfp:Bmesons.vector} 
\end{center}
\end{table}

\hspace*{-1cm}\begin{table}[h!]
\begin{center}
{
\begin{tabular}{|c||c|c|c|c|c|}\hline
 (MeV)
&$\rho^{\pm}$
& $\rho^0$ 
& $K^{*\pm}$ 
& $D^{*\pm}$ 
& $D_s^{*\pm}$ \\
\hhline{|=#=====|}
$f_V$
& 210(4)~\cite{Straub:2015ica}   
& $213(5)$~\cite{Straub:2015ica}
& $204(7)$~\cite{Straub:2015ica} 
& $252(22)$~\cite{Lucha:2014xla} 
&$305(26)$~\cite{Lucha:2014xla} 
\\
$M_V$\cite{Patrignani:2016xqp}
& $775.26(25)$  
&$775.26(25)$  
& $891.66(26)$ 
&$2010.26(05)$ 
&$2112.1(4)$ 
\\
\hline 
\end{tabular}
\caption{Decay constants and masses for 
$\rho$, $K^*$ and $D^*_{(s)}$ vector meson states 
(the values for the decay constants of $K^{*0},\bar K^{*0}$
  ($D^{*0},\bar D^{*0}$) are the same as those of $K^{*\pm}$
  ($D^{*\pm}$) 
since they are related via isospin symmetry). 
}\label{tab:mpfp:mesons.vector.1} 
}
\end{center}
\end{table}

\hspace*{-1cm}\begin{table}[h!]
\begin{center}
{
\begin{tabular}{|c||c|c|c|}\hline
(MeV)
& $\omega$ 
& $\phi$
&$J/\Psi$\\
\hhline{|=#===|}
$f_V$
&$197(8)$~\cite{Straub:2015ica} 
&$233(4)$~\cite{Straub:2015ica} 
&$418(9)$\cite{Becirevic:2013bsa}
\\
$M_V$\cite{Patrignani:2016xqp}   
&$782.65(12)$ 
&$1019.460(2)$
&$3096.900(6)$
\\
\hline 
\end{tabular}
\caption{Decay constants and masses for $\omega$, $\phi$ and 
$J/\psi$ vector mesons.
}\label{tab:mpfp:mesons.vector.2}
}
\end{center}
\end{table}

\noindent
Although in our study we do not address the decays of vector mesons, 
the corresponding decay constants and masses are collected 
in Tables \ref{tab:mpfp:Bmesons.vector},
\ref{tab:mpfp:mesons.vector.1} and \ref{tab:mpfp:mesons.vector.2}, 
as such inputs are relevant regarding the computation of 
the (on-shell) heavy 
neutrino decay width.

\subsubsection{On-shell sterile neutrinos and finite detector effects}
\label{sect:finitedectector}

As previously mentioned, LNV semileptonic meson
and tau decays can be strongly enhanced when the exchanged
Majorana fermion is on-shell. This is equivalent to having the
production of a real state, which propagates and has a well-defined
width. 
Depending on its mixing with the leptonic doublets, a
sterile state with a mass in the range relevant for our study (i.e. 
$100 \text{ MeV} \lesssim m_{\nu_s}\lesssim 10 \text{ GeV}$) can have
several possible decay channels. In addition to those which are
directly related with the LNV final states, $\nu_s \to \ell
M$, with $M$ denoting a meson, it can also have 
(simple) two-body decays, $\nu_s \to \nu_i \ell$, $i=1,2, 3$.  
In the present study, we carry a full computation of the heavy
neutrino decay width, including {\it all} possible channels which
are open for a given mass range. We collect the relevant expressions in
Appendix~\ref{app:sec:widths:nu4}.

\begin{figure}[t!]
\begin{center}
\epsfig{file=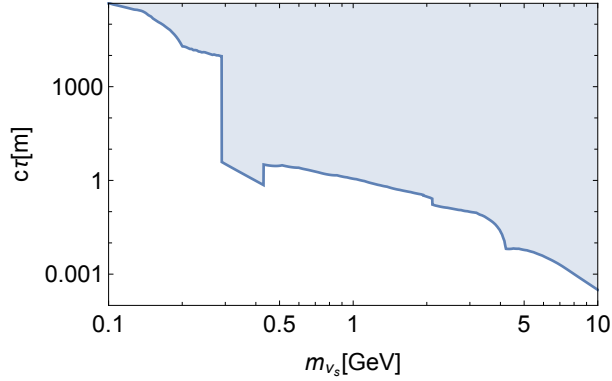, width=80mm}
\caption{Expected lower bounds on the length of flight 
($L^\text{flight}_{\nu_s} =\gamma_s \beta_s
\tau_s c$, in metres) as a function of the heavy sterile neutrino mass
(in GeV), assuming the maximal allowed mixings with the light active
states. 
}\label{fig:length.flight} 
\end{center}
\end{figure}

It is important to stress that the observability of the different
processes referred to in the previous section is not only related to
the expected number of events, but also to whether or not the final
states can be indeed observed - if sufficiently 
long-lived, the LNV decays of the heavy neutrino can occur 
{\it outside} the detector. Thus, in our analysis we will
impose that the distance travelled
prior to the decay does not exceed the typical size of a detector.
The actual distance travelled by the neutrino depends on its lifetime
and on its velocity $\beta_s$, $L^\text{flight}_{\nu_s} = \gamma_s
\beta_s \tau_s c$. For relatively heavy neutrinos (with a mass typically
above 500~MeV), $\tau_s$ can be sufficiently small to ensure that the
length of flight does not exceed a few metres (or even less). On the
other hand, for lighter states, the mixings tend to be very small (due
to the important experimental constraints, see
Appendix.~\ref{app:constraints}), and such states might decay 
well outside the detector. In the latter case, and independently of the
actual decay widths, the final states would not be observed.  
In Fig.~\ref{fig:length.flight} we display the heavy neutrino length of
flight, obtained for the maximal allowed mixings to the active
states. This corresponds to a lower bound on
$L^\text{flight}_{\nu_s}$, 
as the lifetime will be longer for smaller mixings. 
For instance, a heavy neutrino with a mass around 150~MeV
 produced with an energy of $\simeq 
200~\text{MeV}$, already has $\gamma_s \beta_s \sim 0.88$, 
and will certainly escape detection.

In order to have realistic results concerning the constraints arising
from the LNV bounds, the effect of a real (finite) detector volume must then
be taken into account upon computation of the LNV branching ratios:
the probability that the sterile neutrino length of flight is smaller 
than the detector size, $L_{\nu_s}^\text{flight} \lesssim L^\text{det.}$, is
given by
\begin{equation}\label{eq:finite.detector.prob}
P_{\nu_s}^\text{det.}\, =
\, 1-\mathrm{Exp}\left(-L^\text{det.}\, \Gamma_{\nu_s} \, 
\frac{m_{\nu_s}}{p_{\nu_s}}\right),
\end{equation}
where $p_{\nu_s}$, $m_{\nu_s}$ and 
$\Gamma_{\nu_s}$ are the momentum, mass and
decay width of the heavy (mostly) sterile neutrino,  respectively. 
The momentum $p_{\nu_s}$ depends on the energy of the mother particle ($M_1$ or
$\tau$) produced by experiments; as done in Ref.~\cite{Helo:2010cw},
here we simply consider that the mother particle is produced at rest or
with a comparatively small momentum. Should that be the case, 
$p_{\nu_s}$ is then given by
$p_{\nu_s}=\lambda^{1/2}(m_A^2,m_{\nu_s}^2,m_B^2)/(2m_A)$ where 
$m_A=m_{M_1}$ and $m_B=m_{\ell_\alpha},m_{\ell_\beta}$ respectively for the LNV
meson decay, or $m_A=m_\tau$ and $m_B=m_{M_1},m_{M_2}$ for the LNV
$\tau$ decay; $m_{\nu_s}$ is the mass of the sterile state
${\nu_s}$ and the kinematical function
$\lambda(m_A^2,m_{\nu_s}^2,m_B^2)$ is defined in
Eq.~(\ref{lambda.function}). 

The theoretically computed branching fraction of each LNV decay
channel is then corrected by the within-detector decay probability
(cf. Eq.~(\ref{eq:finite.detector.prob})).

In what follows - and as already stated in the beginning of this
section - we will consider a simple extension of the SM by one sterile 
Majorana fermion, which mixes with the light (mostly active)
neutrinos. 
We will thus subsequently denote the corresponding parameters as 
$m_{\nu_s} = m_4$,  $p_{\nu_s}= p_4$, $\Gamma_{\nu_s} = \Gamma_4$, 
$\tau_s = \tau_4$ and $L_{\nu_s}^\text{flight}=L_{\nu_4}^\text{flight}$.

\subsection{Updated constraints from LNV decays}\label{sec:updatedLNV}

\begin{figure}[h!]
\begin{center}
\begin{tabular}{cc}
\epsfig{file=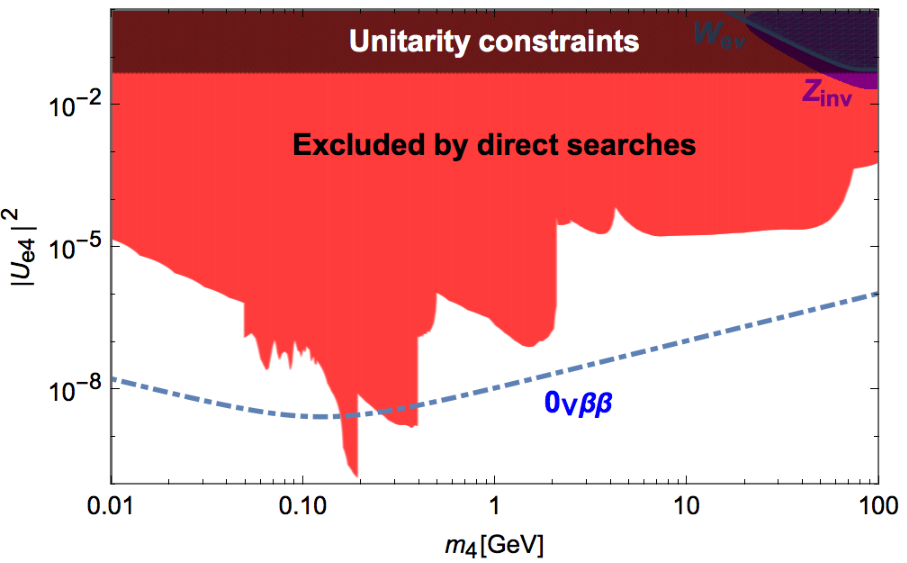, width=73mm}  \hspace*{4mm}&\hspace*{2mm}
\epsfig{file=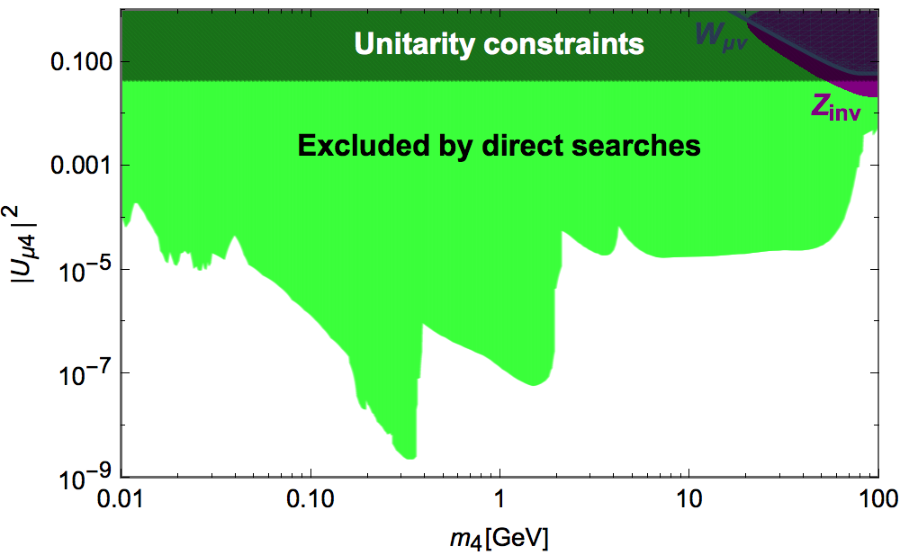, width=73mm}  \\
\epsfig{file=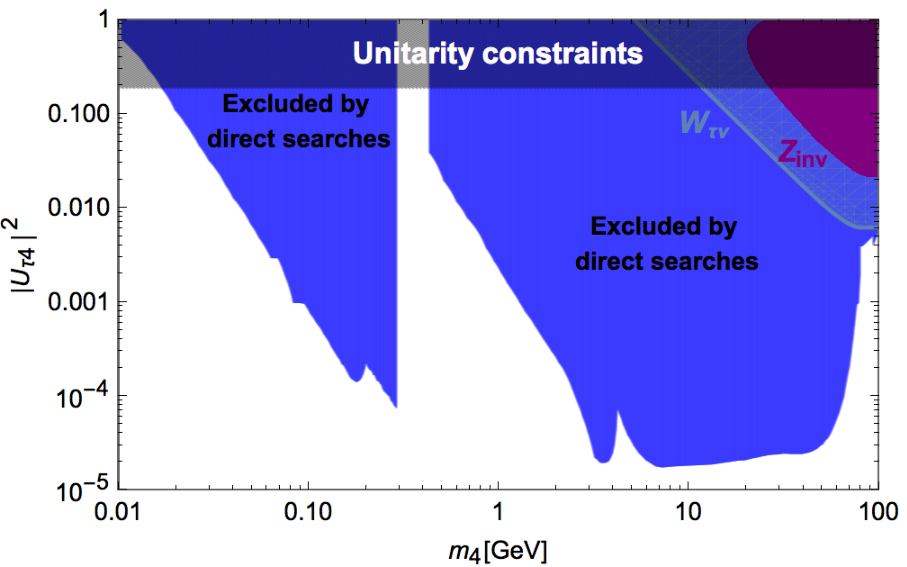, width=73mm}  \hspace*{4mm}&\hspace*{2mm} 
\epsfig{file=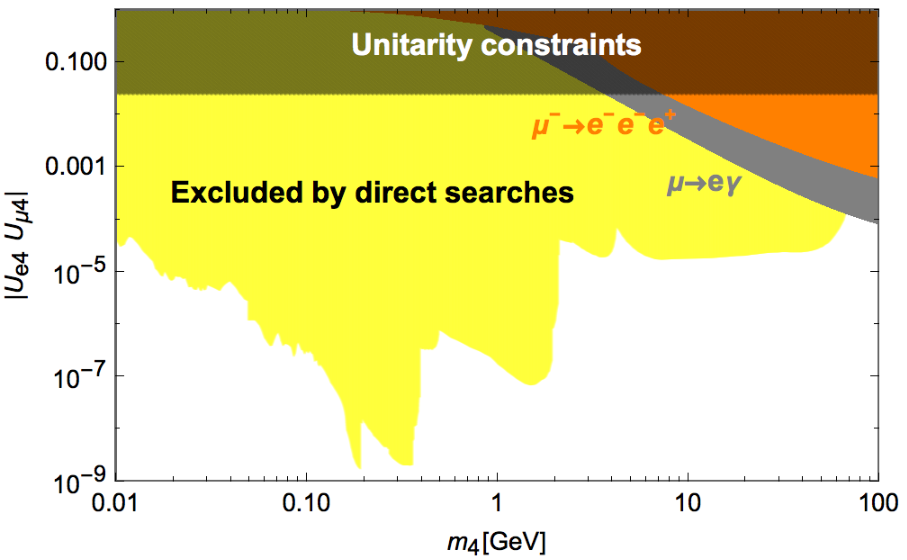, width=73mm}  \\
\epsfig{file=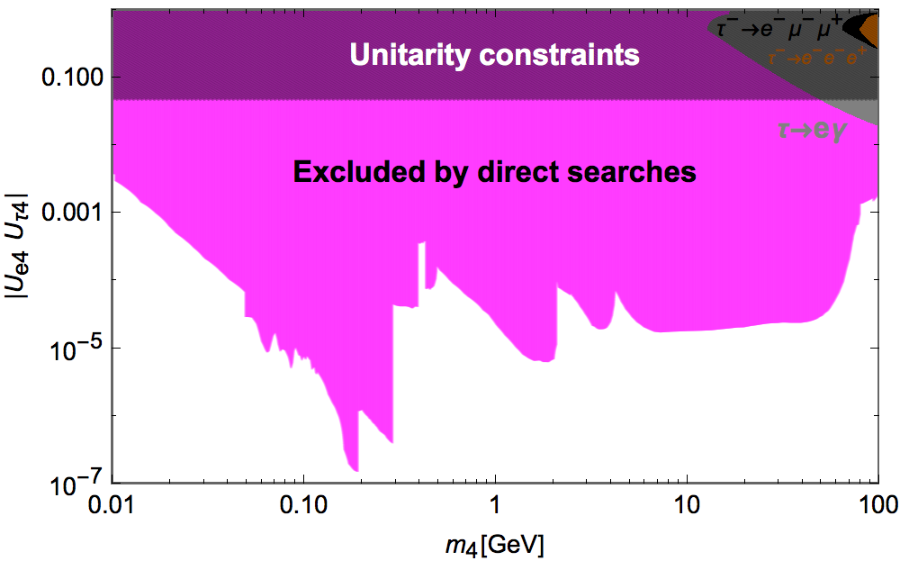, width=73mm} \hspace*{4mm}&\hspace*{2mm}
\epsfig{file=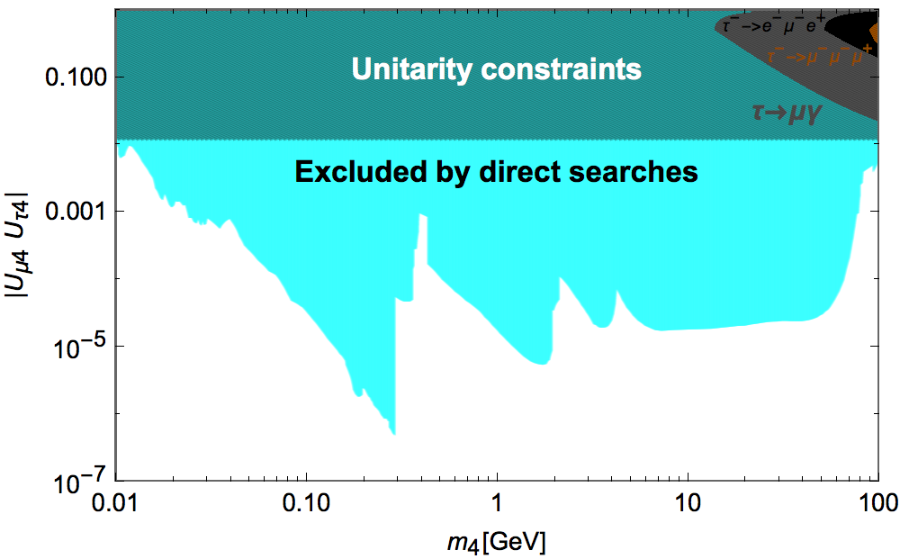, width=73mm} \\
\epsfig{file=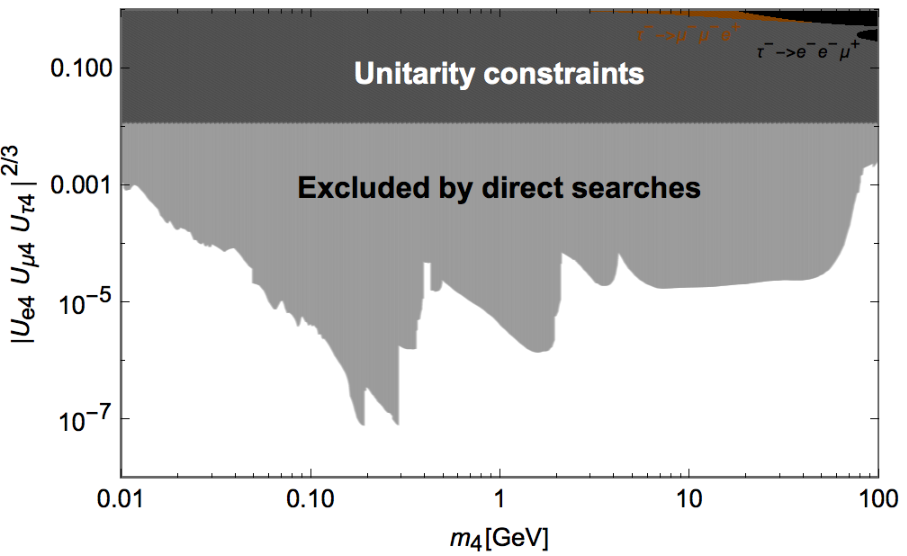, width=73mm} \hspace*{4mm}&\hspace*{2mm}
\end{tabular}
\caption{Updated constraints on the sterile neutrino parameter space, as
  spanned by various combinations of $U_{\ell_\alpha 4}$ mixing elements   
  and $m_4$ (in GeV): from left to right, top to bottom, 
  $|U_{e 4}|^2$, $|U_{\mu 4}|^2$, 
  $|U_{\tau 4}|^2$, $|U_{e 4} U_{\mu 4}|$, $|U_{e 4} U_{\tau 4}|$, 
  $|U_{\mu 4} U_{\tau 4}|$ and $|U_{e 4} U_{\mu 4} U_{\tau 4}|^{2/3}$.  
  Solid surfaces denote excluded regimes due
  to violation of at least one experimental or observational bound.
}\label{fig:ML.U-m4constraints}
\end{center}
\end{figure}

Due to the significant contributions of a wide array of experiments,
the sterile neutrino parameter space (as spanned by its 
degrees of freedom, $m_4$ and 
$U_{\ell_\alpha 4}$, $\ell_\alpha=e,\mu,\tau$) 
is already subject to extensive constraints. 
Before revisiting those arising from LNV, we summarise in
Fig.~\ref{fig:ML.U-m4constraints} the current status of the sterile neutrino
parameter space (the bounds here imposed are 
enumerated and described in Appendix~\ref{app:constraints}).  
The different panels displaying the 
bounds on the active-sterile mixing angles (as a function
of the mass of the heavy, mostly sterile neutrino) arise from direct searches,
cLFV bounds, invisible $Z$ decays, $W\to \ell \nu$  decays,
among others~\cite{Ue_exclusions,
  Umu_exclusions,Utau_exclusions,UeUmu_exclusions,UeUmuUtau_exclusions,
  Das:2016hof,Das:2017zjc,Das:2017rsu}.  
Although not included in these plots, it is important to
emphasise that leptonic observables such as $R_\tau$, $R_\pi$ and $R_K$ 
(or their corresponding deviations from the SM predictions, 
$\Delta r_K$, $\Delta r_\pi$ and $\Delta r_\tau$) are
extremely constraining, and will be taken into account in the
subsequent numerical analyses.

Leading to the exclusion regions of Fig.~\ref{fig:ML.U-m4constraints},
the limits on a given mixing element $U_{\ell 4}$ are obtained by setting the
other mixings with the sterile state to zero (for instance, the
constraints on $U_{\mu 4}$ are inferred assuming $U_{e 4}=U_{\tau
  4}=0$); for the exclusions on combined mixing elements  
$U_{\ell 4} U_{\ell' 4}$, with $\ell\ne\ell'$, one assumes the latter to be equal,
and the remaining one is set to zero (for instance, leading to the constraints on 
$U_{e 4} U_{\mu 4}$, one has $U_{e 4}= U_{\mu 4}$ and $U_{\tau 4}=0$); 
finally, for the final panel regarding bounds on the triple product of
couplings\footnote{The final panel summarises the exclusion regimes on the
$|U_{e4} U_{\mu 4} U_{\tau 4}|^{2/3}$ vs. $m_4$ plane: 
the  constraints leading to the exclusion zone arise from box diagram 
contributions to 3-body tau decays, which do actually depend on   
powers of $|U_{e4}^2 U_{\mu 4} U_{\tau 4}|$ 
and/or $|U_{e4} U_{\mu 4}^2 U_{\tau 4}|$; 
for  simplicity, and to keep a similar scaling of this panel compared 
to the others, we have summarised the dependency via the quantity
$|U_{e4} U_{\mu 4} U_{\tau 4}|^{2/3}$, 
without entailing any loss of physical content.} 
($|U_{e 4} U_{\mu 4} U_{\tau 4}|$), all entries are taken to
be equal.
The dot-dashed line on the first panel ($|U_{e 4}|^2$
vs. $m_4$), corresponds to neutrinoless double beta decay searches, and
is obtained by requiring that, by itself, the contribution of 
the sterile neutrino does not exceed 0.165~eV (we assume that
no cancellations occur between
the heavy and the light neutrino contributions); 
this should be interpreted as a representative bound, 
since it can be strengthened (relaxed) in the case of constructive
(destructive) interference with the active neutrino contributions. 
The discontinuity in the exclusion for $|U_{\tau 4}|^2$
vs. $m_4$ (for the interval $m_4 \in [0.3, 0.5]\text{ GeV}$)
arises from the lack of experimental bounds applicable to
that particular mass regime.
For completeness, the sterile mass regime constrained by the results
of Fig.~\ref{fig:ML.U-m4constraints} extends to masses heavier than the mass
interval characteristic of the meson and tau LNV decays here
addressed (below 10~GeV); this is done in order to explicitly obtain a
comprehensive picture of all available bounds - including those from 
cLFV decays (relevant for the combined mixing
elements), which only become significantly constraining for sterile mass
close to the electroweak scale (as manifest in the $|U_{e 4} U_{\mu 4}|$
vs. $m_4$ panel, for example). 
Bounds from invisible (leptonic) decays of $Z$ ($W$)
bosons are also only relevant for a heavy sterile mass regime, and
are typically superseded by the constraints arising from direct searches
- and in the case of $|U_{e 4}|^2$, from neutrinoless double beta decay.
Finally, the unitarity bound arising from non-standard neutrino
interactions with matter~\cite{Antusch:2008tz} or from non-standard
oscillation schemes, constrains the deviation from unitarity 
of the left-handed lepton mixing matrix 
($\tilde U_\text{PMNS}$), which leads to the exclusion of the large
mixing regimes (for all the sterile mass regimes considered). 
In deriving these bounds we used the recent
results~\cite{Blennow:2016jkn} (see also~\cite{Antusch:2014woa}). 

The different panels of Fig.~\ref{fig:ML.U-m4constraints} thus summarise
the most recent and up-to-date available constraints on the parameter
space of the SM extended by one additional massive Majorana fermion
with a mass between 10~MeV and 100~GeV.  

\bigskip
The bounds and exclusion regions on the sterile neutrino degrees of
freedom, as identified in the different panels of
Fig.~\ref{fig:ML.U-m4constraints}, must now be combined with 
the bounds arising from LNV decays; together, they 
will have a strong impact on the maximal currently allowed values of
the LNV semileptonic tau and meson decays, and most importantly, on
the values of the entries of the $3\times 3$ Majorana effective mass matrix. 

We thus begin by discussing the 
updated constraints on the relevant combination of leptonic  
mixing matrix elements, stemming from the most recent experimental
bounds. 
We recall that resonant production of sterile neutrinos (which is
crucial to enhance the LNV meson and tau decays, 
$M_1\to \ell_\alpha \nu_4^*\to\ell_\alpha \ell_\beta M_2$
and $\tau\to M_1\nu_4^*\to
M_1M_2\overline{\ell}$) implies that the sterile state is on-shell. 
Should the decay width of the sterile neutrino be sufficiently small, 
the decay may occur after a length of flight larger than a realistic
detector size, thus rendering the LNV decay processes invisible.
Whenever relevant, the derived bounds take into account the 
requirement that the heavy neutrino decays within
a finite detector (i.e. imposing that its length of flight does not
exceed a nominal value $L_{\nu_4}^\text{flight}=10$~m), as described in
Section~\ref{sect:finitedectector}.  
The final results are then compared to the experimental limits 
listed in Tables~\ref{tab:LNV:meson:exp} and~\ref{tab:LNV.taudecay.exp}.

\begin{figure}[h!]
\begin{center}
\begin{tabular}{cc}
\epsfig{file=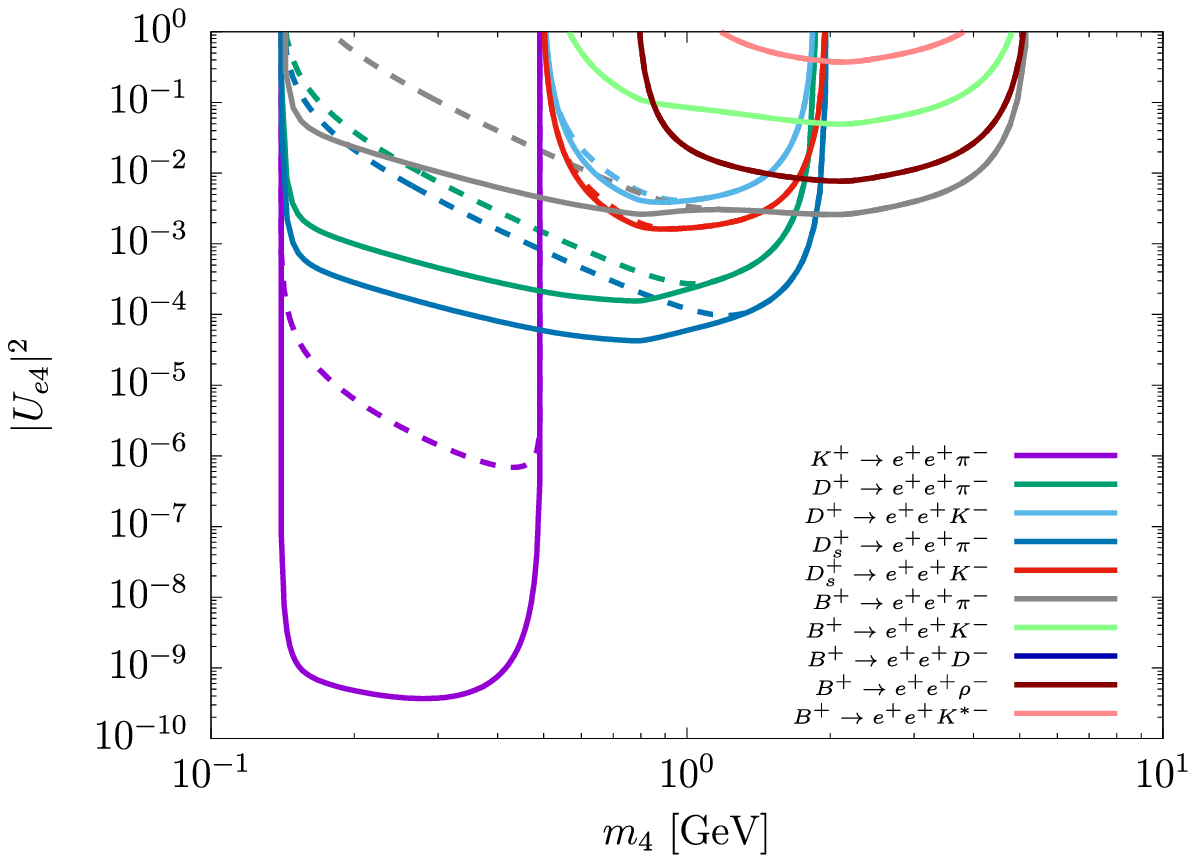, width=73mm} &
\epsfig{file=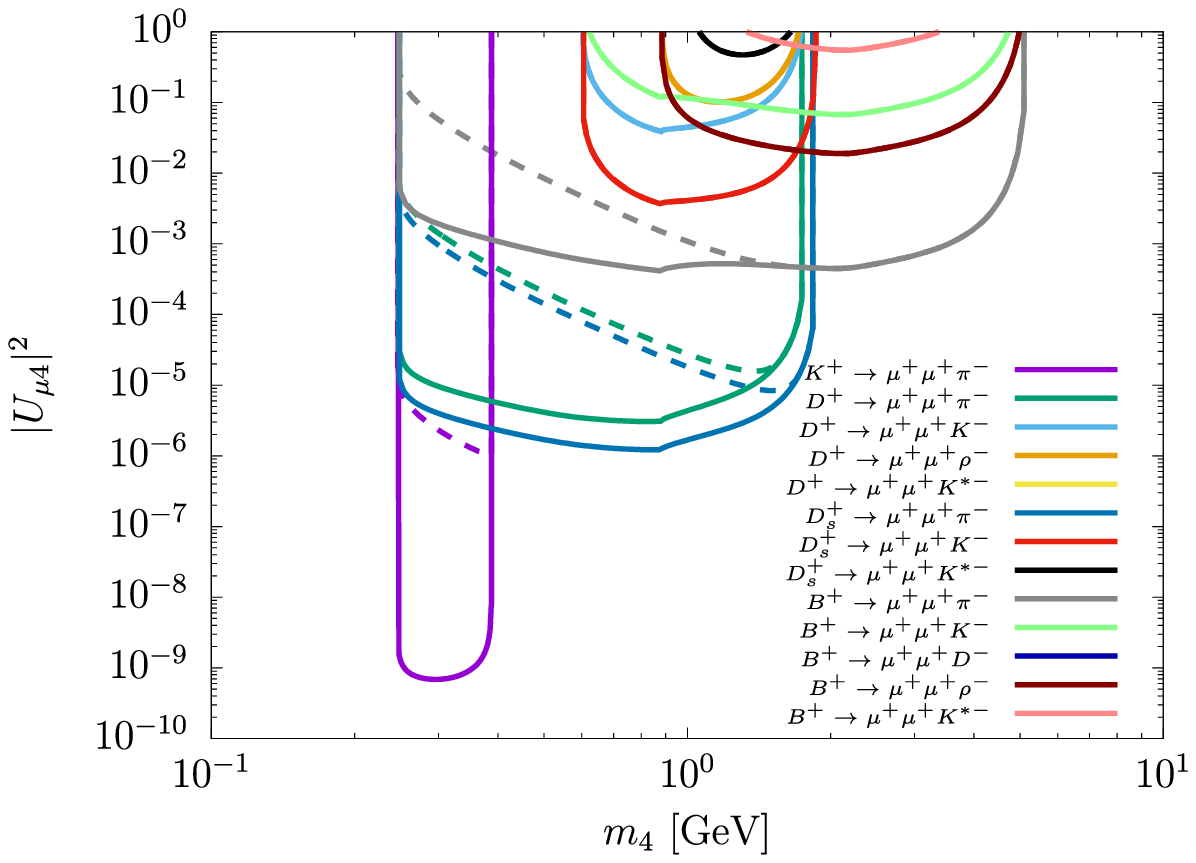, width=73mm} \\
\epsfig{file=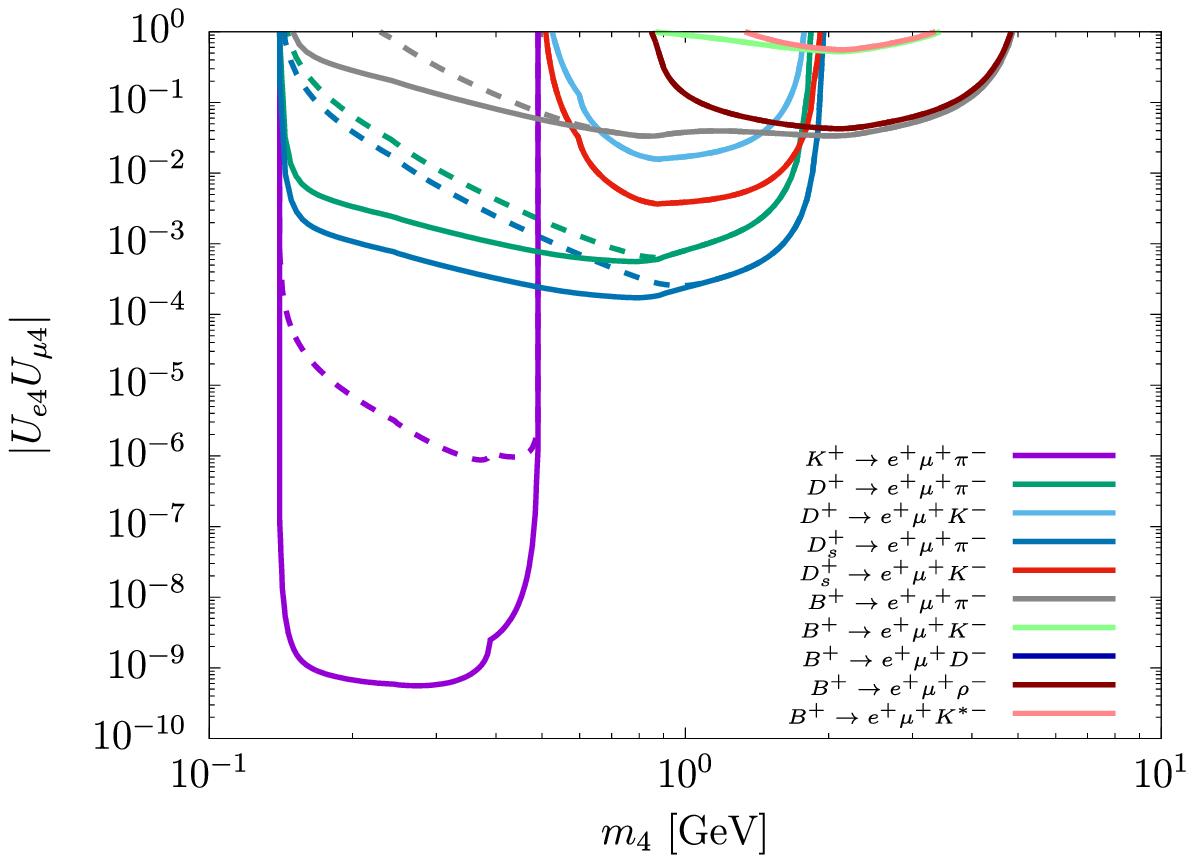, width=73mm} & 
\end{tabular}
\caption{Updated constraints on the relevant combination of leptonic
  mixing matrix elements ($|U_{\ell_\alpha 4} U_{\ell_\beta 4}|$) arising
  from LNV pseudoscalar meson decays,  
as a function of the heavy sterile neutrino mass (GeV). Same
assumptions on $U_{\ell_\alpha 4}$ as leading to
Fig.~\ref{fig:ML.U-m4constraints}.
Dashed lines denote the bounds derived under the requirement 
$L_{\nu_4}^\text{flight}\lesssim10$~m. 
}\label{fig:updated.UU.m4.M.2}
\end{center}
\end{figure}

\begin{figure}[h!]
\begin{center}
\begin{tabular}{cc}
\epsfig{file=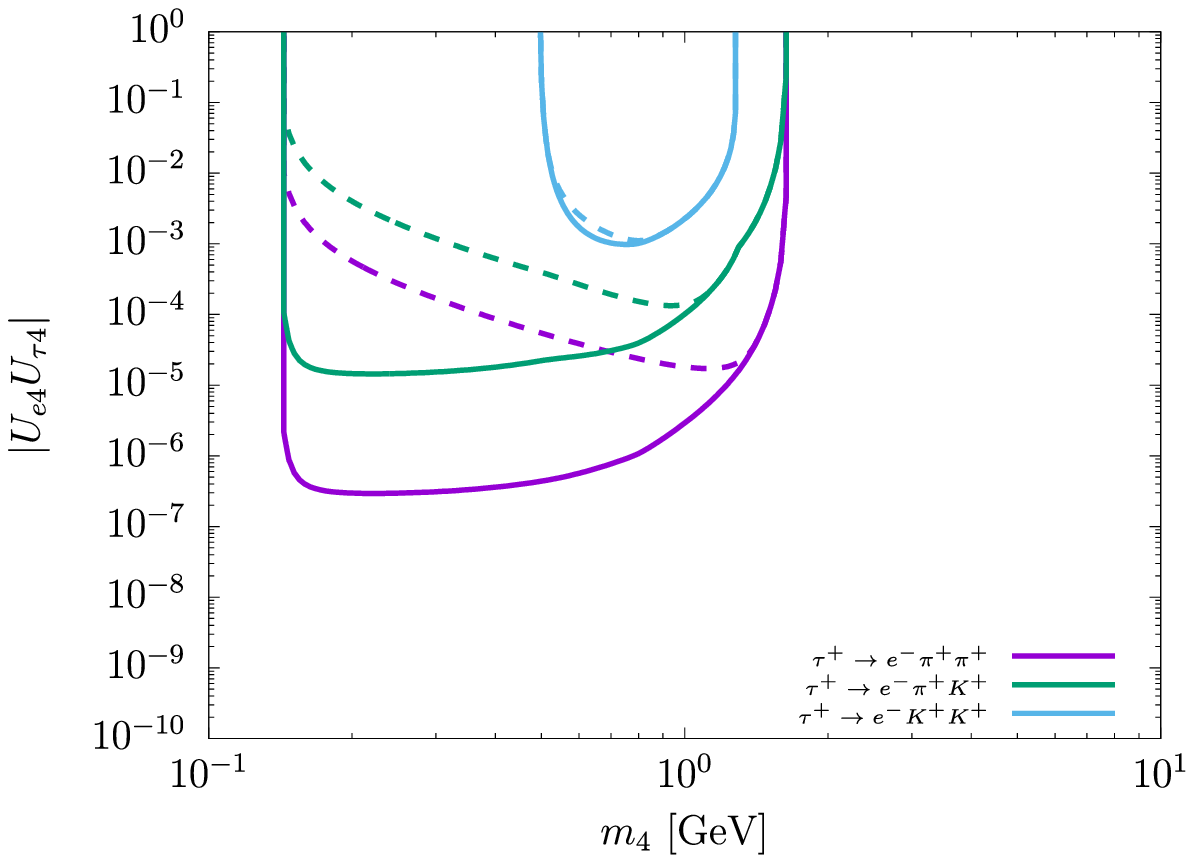, width=73mm} &
\epsfig{file=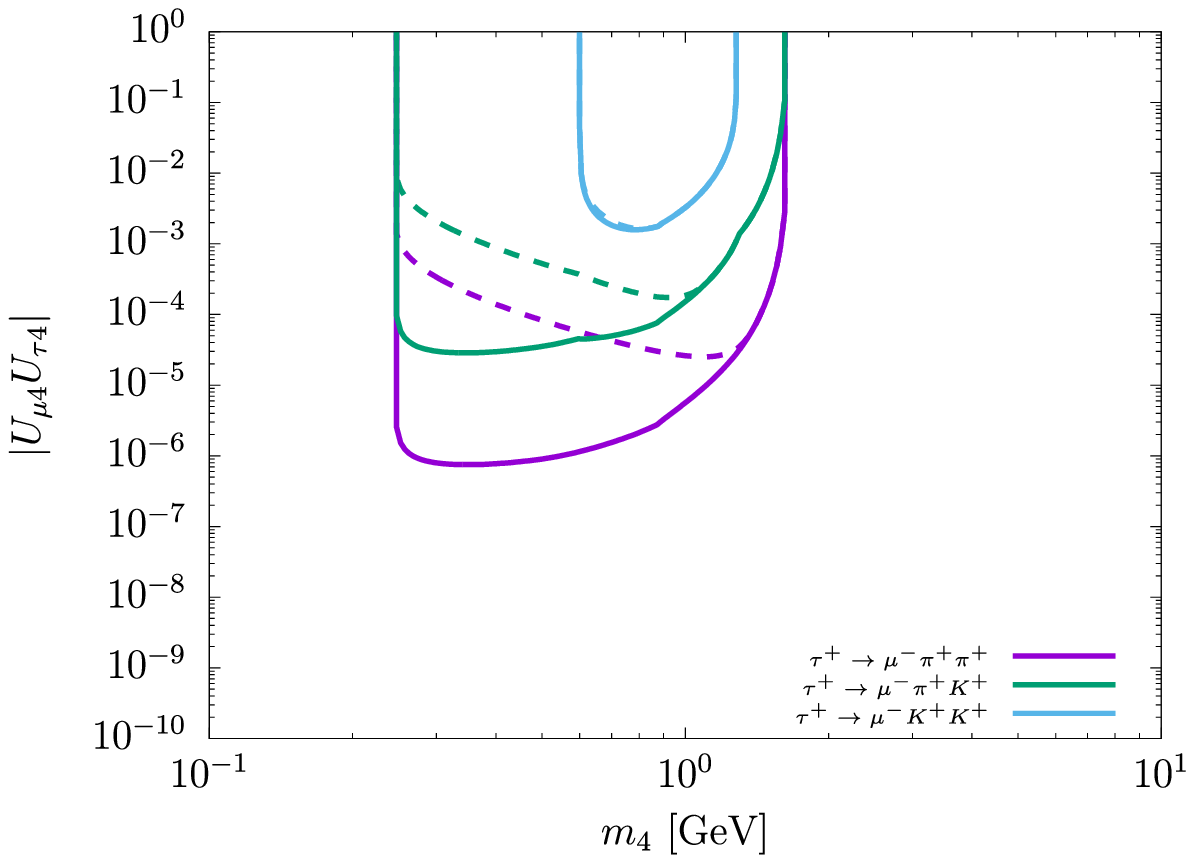, width=73mm} 
\end{tabular}
\caption{Updated constraints on the relevant combination of leptonic
  mixing matrix elements ($|U_{\ell_\alpha 4} U_{\tau 4}|$) arising
  from LNV semileptonic tau decays,   
as a function of the heavy sterile neutrino mass (GeV). Same
assumptions on $U_{\ell_\alpha 4}$ as leading to 
Fig.~\ref{fig:ML.U-m4constraints}. 
Dashed lines denote the bounds derived under the requirement 
$L_{\nu_4}^\text{flight}\lesssim 10$~m.}\label{fig:updated.UU.m4.tau.2}
\end{center}
\end{figure}

Working under the same assumptions as those leading to
Fig.~\ref{fig:ML.U-m4constraints}, we thus display in 
Figs.~\ref{fig:updated.UU.m4.M.2} and~\ref{fig:updated.UU.m4.tau.2}
the bounds on active-sterile mixing angles, as a function
of the (mostly) sterile heavy neutrino mass, arising from
LNV meson and tau decays. When present, dashed lines denote the
bounds derived taking into account the requirement of having the 
heavy neutrino decaying within 10~m (see discussion above). 
These constraints will be subsequently used in our numerical analysis.

As expected, given the current experimental bounds, the most stringent
LNV constraints on the active-sterile mixing angles arise from
semileptonic kaon decays ($K^+ \to \ell_\alpha^+ \ell_\beta^+ \pi^-$), leading
to constraints on (combinations of) mixings of $\mathcal{O}(10^{-9})$. Even
when corrected to account for a finite detector size 
(within-detector decay), semileptonic kaon decays - especially leading
to a final state containing at least one electron - are still the most
stringent ones.

In recent years, similar analyses have led to the derivation of
increasingly stronger bounds on the sterile fermion parameter space. 
Although very recent works already include updated experimental bounds
in their results, our study considers the most recent data for {\it all}
the LNV modes addressed, leading to leptons of same or different
flavour in the final states. 

Lepton number violating semileptonic tau decays have a strong impact on
constraining combinations of mixings involving the $\tau$ lepton, as
can be inferred from a direct comparison of Fig.~\ref{fig:ML.U-m4constraints} 
with the corresponding panels of Fig.~\ref{fig:updated.UU.m4.tau.2}.
It is interesting to notice that (as will be
discussed in greater detail in Section~\ref{sec:results:LNVtau}) LNV
tau decays lead to constraints on $\mu-\tau$ mixing elements that can
be as stringent as $|U_{\mu 4} U_{\tau4}| \lesssim 10^{-5}$ for
sterile mass close to 1~GeV - for conservative limits on the size 
of the detector. Such bounds already appear to supersede those displayed in the
corresponding panel of Fig.~\ref{fig:ML.U-m4constraints}.  

The results collected in the above plots reflect two distinct regimes
for the sterile neutrino lifetime. We have also investigated more
extreme limits (for example $L_{\nu_4}^\text{flight}\lesssim 1$~m, as
will be the case at Belle II), and our findings suggest a loss in
sensitivity amounting to a factor 2 to 3 in the different
combinations $|U_{\ell_\alpha 4} U_{\ell_\beta 4}|$.

\begin{figure}[h!]
\begin{center}
\begin{tabular}{cc}
\epsfig{file=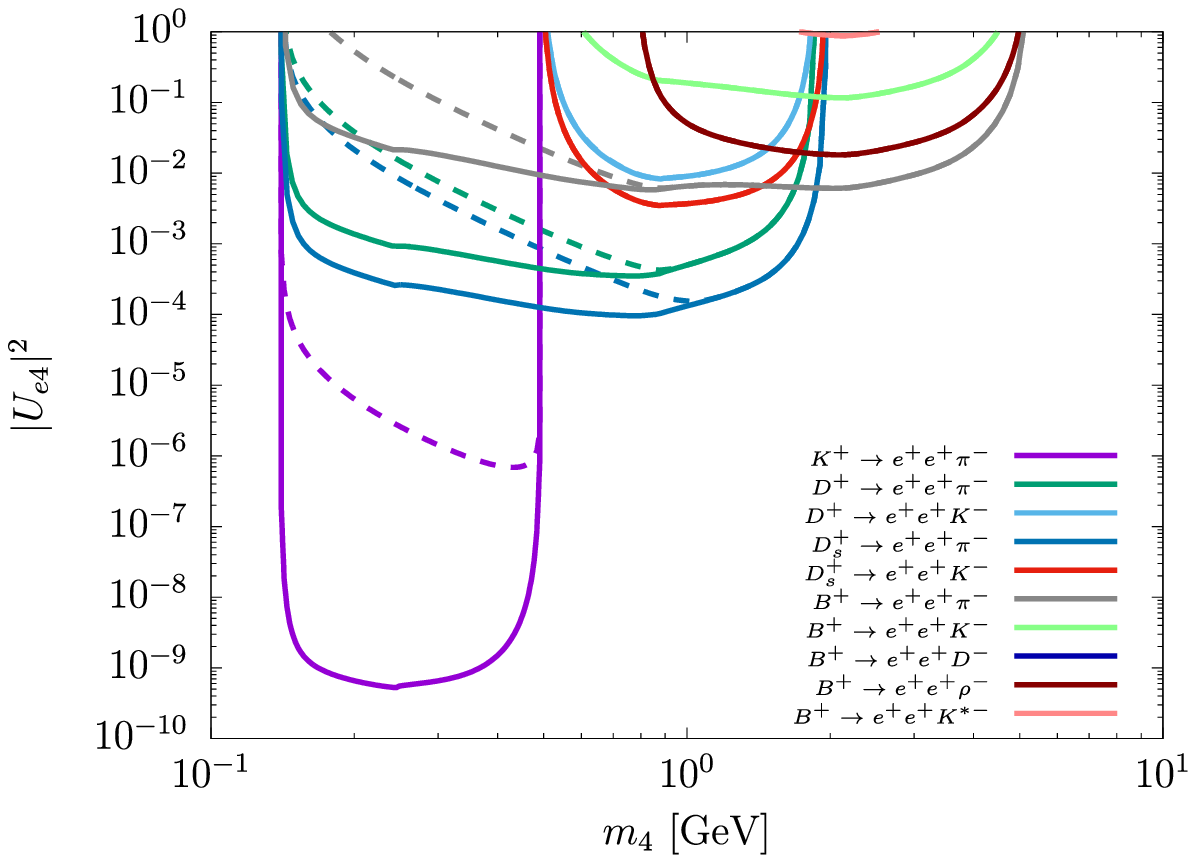, width=73mm} &
\epsfig{file=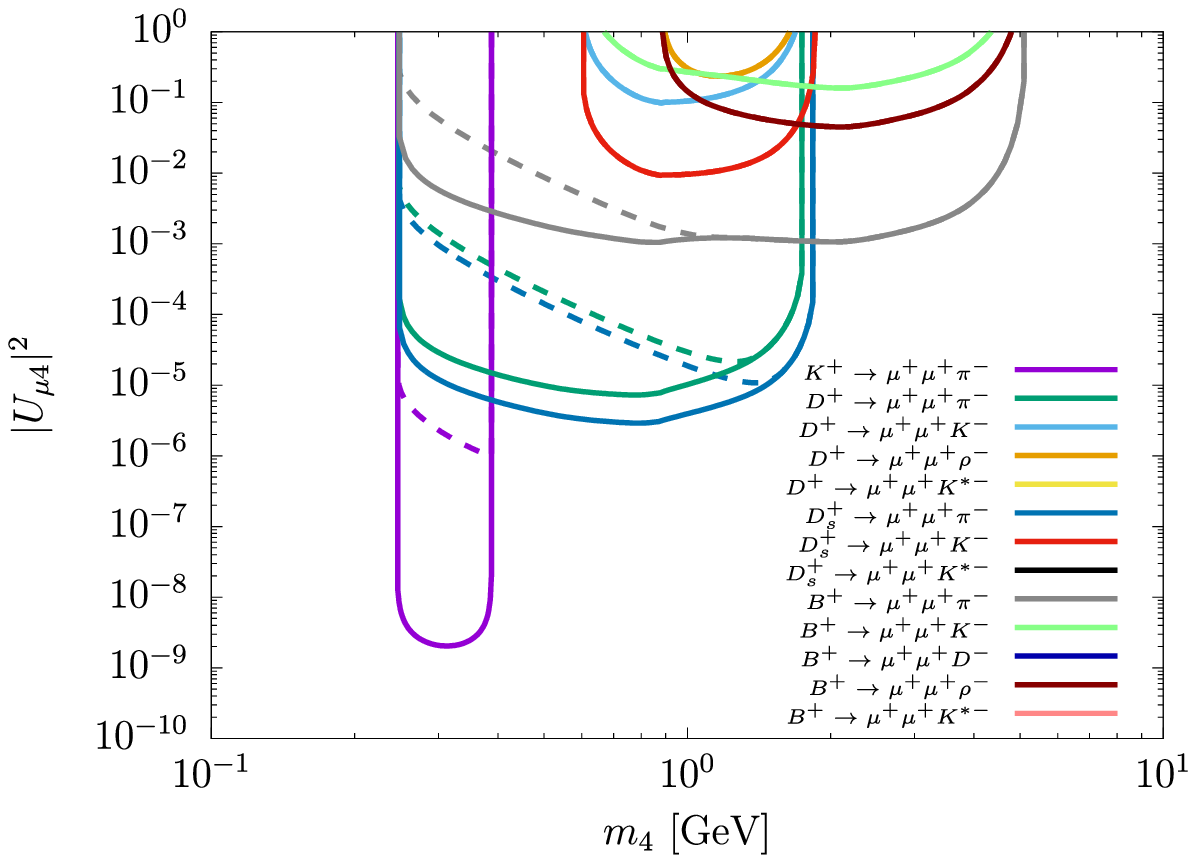, width=73mm} \\
\epsfig{file=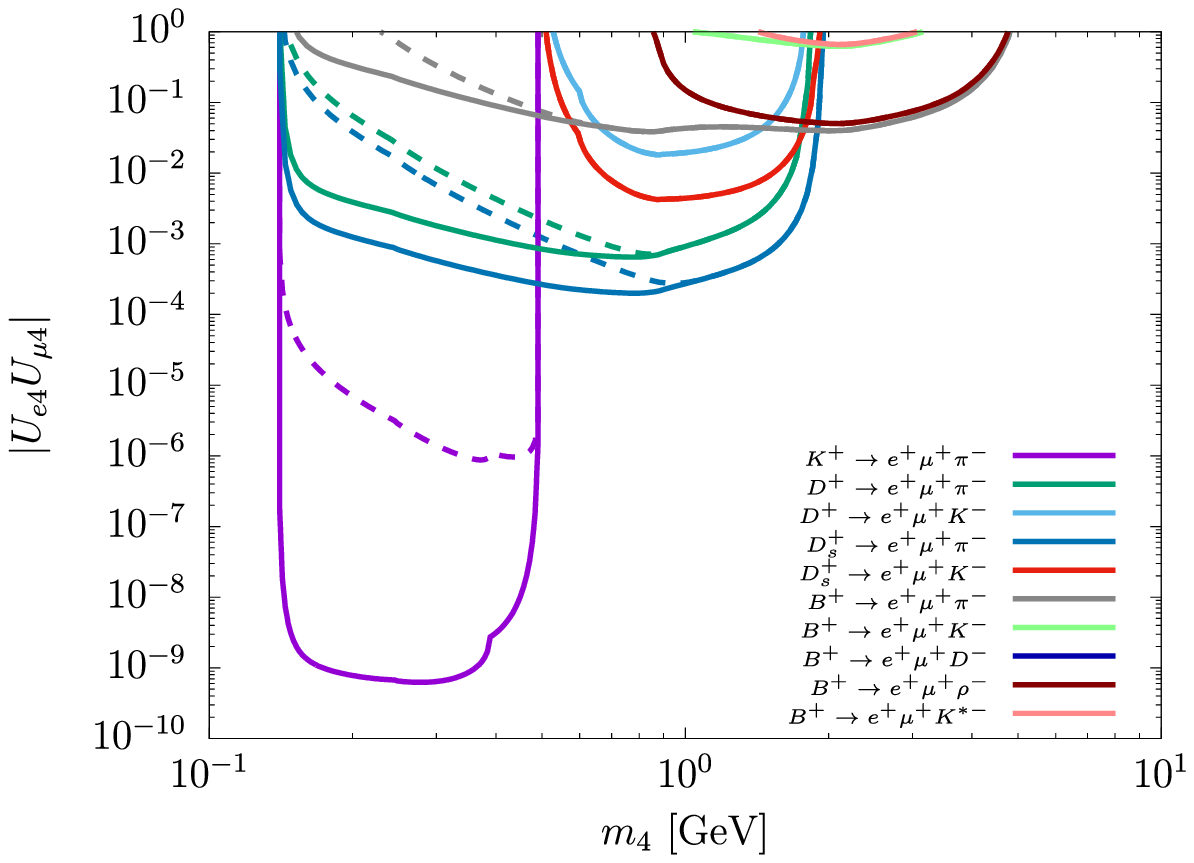, width=73mm} & 
\end{tabular}
\caption{Updated constraints on the relevant combination of leptonic
  mixing matrix elements ($|U_{\ell_\alpha 4} U_{\ell_\beta 4}|$) arising
  from LNV pseudoscalar meson decays,  
as a function of the heavy sterile neutrino mass (GeV). Leading to the
above plots, all active-sterile mixing elements were taken to be equal
(i.e., $|U_{e 4}| = |U_{\mu 4}| = |U_{\tau 4}|$). Line and colour code
as in Fig.~\ref{fig:updated.UU.m4.M.2}.
}\label{fig:updated.UU.m4.M.1}
\end{center}
\end{figure}

\begin{figure}[h!]
\begin{center}
\begin{tabular}{cc}
\epsfig{file=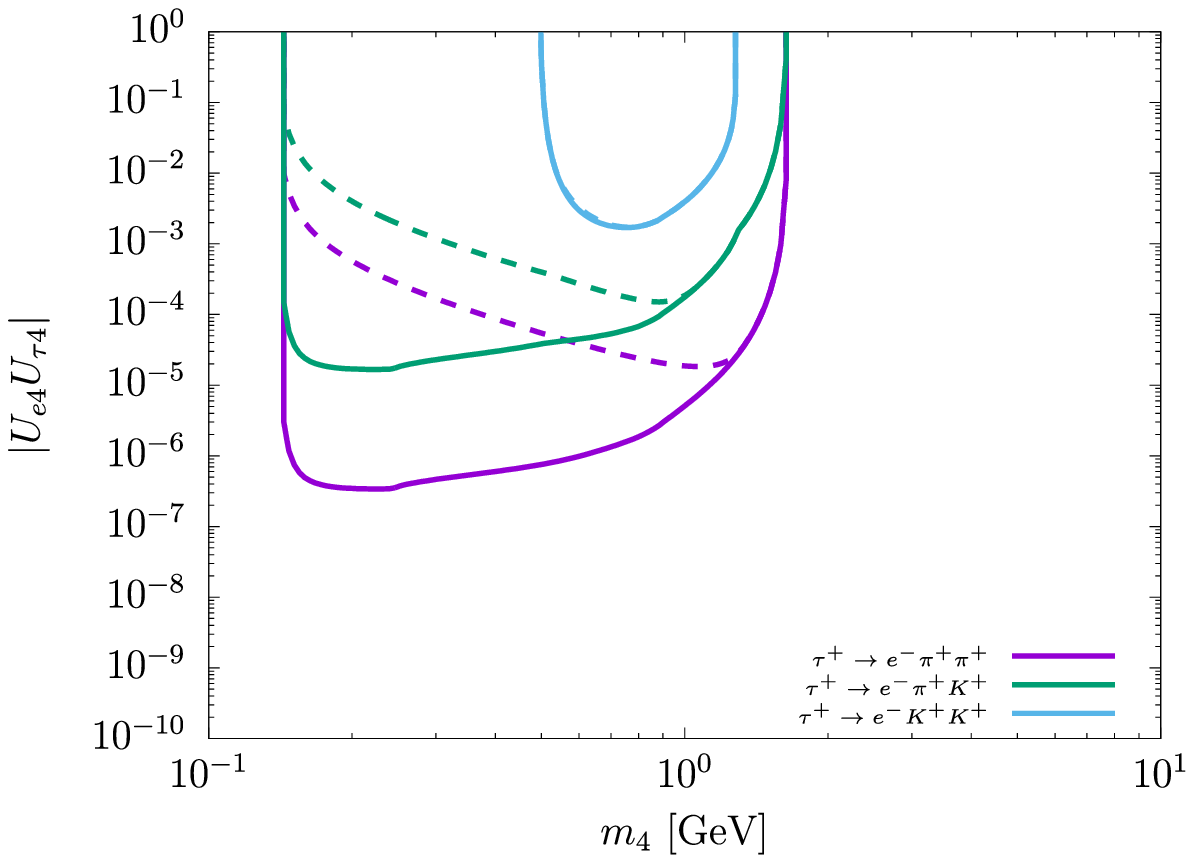, width=73mm} &
\epsfig{file=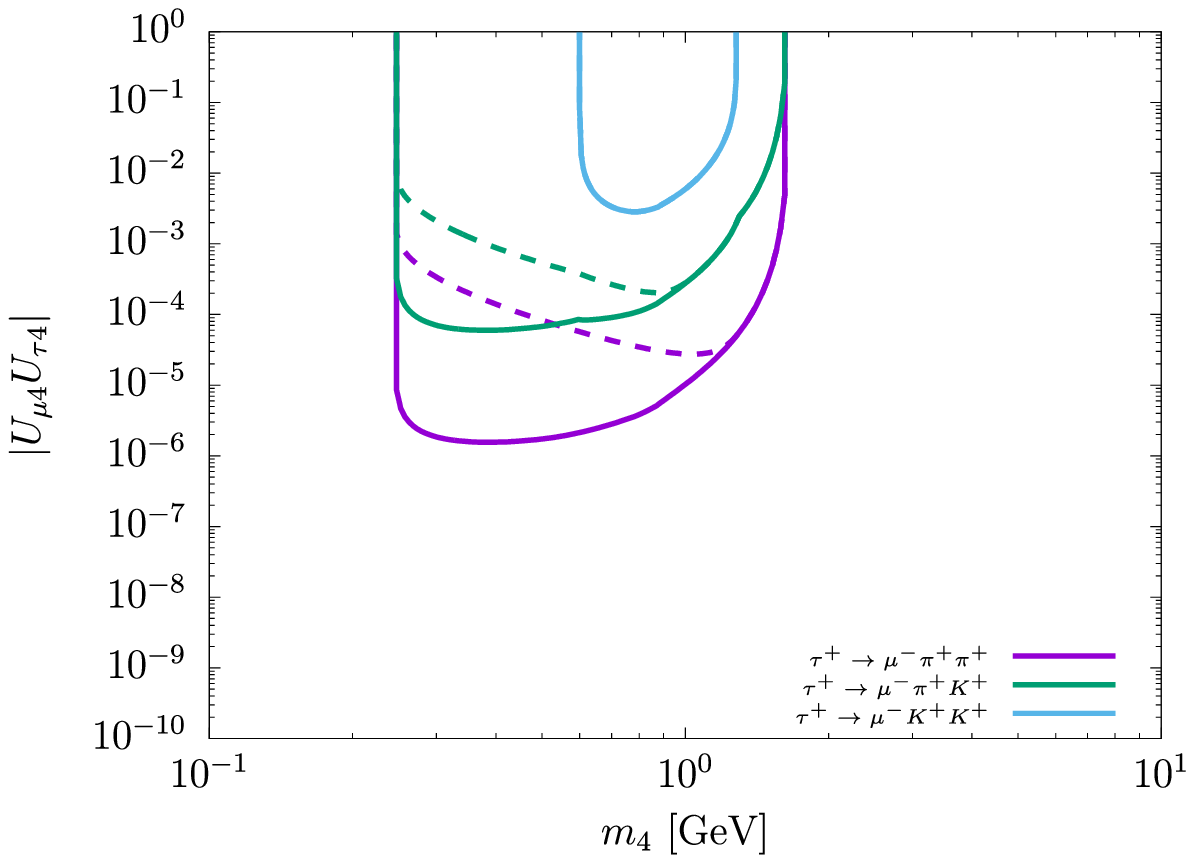, width=73mm} 
\end{tabular}
\caption{Updated constraints on the relevant combination of leptonic
  mixing matrix elements ($|U_{\ell_\alpha 4} U_{\tau 4}|$) arising
  from LNV semileptonic tau decays,   
as a function of the heavy sterile neutrino mass (GeV). Leading to the
above plots, all active-sterile mixing elements were taken to be equal
(i.e., $|U_{e 4}| = |U_{\mu 4}| = |U_{\tau 4}|$). Line and colour code
as in Fig.~\ref{fig:updated.UU.m4.tau.2}
}\label{fig:updated.UU.m4.tau.1}
\end{center}
\end{figure}

\bigskip
In order to compare our results with previous analyses carried in the
literature, it is convenient to recast the current 
experimental bounds on LNV decays using a most minimal assumption for
the active-sterile mixings, in particular that of degenerate mixing
angles. Working under the hypothesis of $|U_{e 4}| = |U_{\mu 4}| =
|U_{\tau 4}|$, the bounds on active-sterile mixing angles, as a function
of the (mostly) sterile heavy neutrino mass, are displayed in 
Figs.~\ref{fig:updated.UU.m4.M.1}
and~\ref{fig:updated.UU.m4.tau.1}. 

We have thus compared our results with
those obtained by other groups; as an example let us mention that our
findings qualitatively agree with those
of~\cite{Atre:2009rg,Helo:2010cw,Yuan:2017uyq}. 
The general shape of the derived curves
is in good agreement\footnote{Although the analytical expressions here
derived do agree with those reported on~\cite{Atre:2009rg}, there are
minor discrepancies in what concerns the 
shape of the lower sides of the exclusion
curve, possibly arising due to different underlying assumptions leading to
the numerical results.}, but our results lead to stronger exclusions, as
they were obtained for the most recent experimental data. Moreover, we also
emphasise that we do compute the full contributions to the on-shell
heavy neutrino decay width.

Four-body semileptonic LNV meson decays\footnote{For LNV semileptonic
  4-body decays of the tau lepton, see~\cite{Yuan:2017xdp}.} 
have also been explored in recent
years as complementary means to constrain the active-sterile neutrino
mixings in the kinematically allowed heavy neutrino mass intervals. 
In~\cite{Castro:2013jsn}, the decays $B^- \to D^0\pi^+\mu^- \mu^-$ 
(and $D^0 \to \pi^- \pi^- \mu^- \mu^-, K^- \pi^- \mu^- \mu^-$) were first used
to derive bounds on the mass of the on-shell Majorana neutrino. 
The LNV decays $B \to D^{(*)} \mu^\pm \mu^\pm \pi^\mp$ (as well as 
$B \to \mu^\pm \mu^\pm \pi^\mp$) were addressed 
in~\cite{Cvetic:2016fbv,Cvetic:2017vwl}; the future
sensitivity of LHCb might allow to improve upon the constraints on
$U_{\mu 4}$ for the relevant mass intervals.
Very recent studies~\cite{Mejia-Guisao:2017gqp} 
further suggested that the 4-body decay modes of 
$B^0_s \to P^- \mu^+ \mu^+ \pi^-$ ($P=K, D_s$) could lead to a small
amelioration of the limits inferred by LHCb from the LNV three-body decay 
$B^- \to \mu^- \mu^- \pi^+$. The comparison of our results with 
the bounds derived for LNV 
4-body $B^0$ decays in~\cite{Yuan:2017uyq} - 
which appeared during the final stages of this work - suggest that
while qualitatively akin in their constraining power, 
certain 3 body processes, such as 
$K \to e \mu \pi$ or $D_s \to \mu \mu \pi$, are in fact more stringent.

Before proceeding to the following section we nevertheless
re-emphasise that in general we will not work under the
assumption of degenerate mixings.
We also recall that we only consider three-body LNV final states.

\subsection{Reconstructing the effective Majorana mass 
$\bm{m_{\nu}^{\alpha \beta}}$}\label{sec:effective.mass.th}

The general amplitude for a LNV process mediated by Majorana (active
or sterile) neutrinos typically depends on the following elements:
\begin{equation}  
\sum_{j=1}^n \frac{ U_{\alpha j} \, m_j\, U_{\beta j}}{q^2 - m_j^2+
i\ m_j\,\Gamma_j}=
\frac{1}{q^2} \sum_{j=1}^n \frac{ U_{\alpha j} \, 
m_j\, U_{\beta j}}{1 - m_j^2/q^2+i\ m_j\,\Gamma_j/q^2}\equiv 
\frac{1}{q^2} m_\nu^{\alpha \beta}(q^2),
\end{equation}
where $q^2$ is the momentum of the neutrinos mediating the process 
and $\Gamma_j$ is their corresponding decay width. 
This general expression can be simplified in limiting cases (depending
on the properties of the exchanged Majorana particle): 
for the active neutrinos
$\Gamma_i \simeq 0$, $m_i^2/q^2 \ll 1$ and one recovers the usual
definition of the $0\nu2 \beta$ decay effective mass $m_\nu^{ee}$;
if $q^2 < 0$ the (virtual) neutrinos are not on-shell, and $\Gamma_i$
can be neglected for the heavy states as well. Notice that in the
presence of heavy (almost sterile) Majorana neutrinos, the
$0\nu 2\beta$ effective mass is correctly parametrised by
$m_\nu^{ee}\left( -(125 \text{ MeV})^2\right)$~\cite{Blennow:2010th},
in which $q^2 = -(125 \text{ MeV})^2$ is an average of the virtual
momenta in different decaying nuclei.

Under the hypothesis that Majorana neutrinos are at the origin of all 
lepton number violating processes, 
the LNV meson semileptonic decays $M_1 \to  \ell_\alpha^\pm
\ell^\pm_\beta M_2$ 
allow to infer the corresponding  
contributions to the ($\alpha,\beta$) entry of the ``effective neutrino
mass matrix'', $m_\nu^{\alpha \beta}(q^2)$. 
The amplitude of the LNV decay processes under study (see
Appendix~\ref{app:sec:widths}) includes the following terms  
\begin{align}\label{eq:effective.mass.def.1}
\sum_{i=1}^3  
\frac{ U_{\ell_\alpha i} \, m_i\, U_{\ell_\beta i}}{q^2 - m_i^2}\, +\, 
\frac{U_{\ell_\alpha 4}\, m_4\, U_{\ell_\beta 4}}{q^2 - m_4^2 + im_4\,\Gamma_4} \,,
\end{align}
in which $q^2 >0 $ denotes the momentum transfer, and $\Gamma_4$ the width
of the additional sterile fermion with mass $m_4$. 
For the specific LNV decay $M_1 \to  \ell_\alpha^\pm
\ell^\pm_\beta M_2$, the transfer momentum is of order  
$q^2 \approx p^2_{12} = {m_{M_1}^2 - m_{M_2}^2}$ (for tau decays 
$p_{12}^2 \to {m_{\tau}^2 - \text{max}(m_{M_{1,2}})^2}$). 
The above equation thus becomes
\begin{align}\label{eq:effective.mass.def.2}
\text{Eq.}~(\ref{eq:effective.mass.def.1})
\approx  \left( \frac{1}{p_{12}^2}
\sum_{i=1}^4 \frac{U_{\ell_\alpha i}\, m_i\, U_{\ell_\beta i}}{1 -
  m_i^2/p_{12}^2\, + i\, m_i \Gamma_i/p_{12}^2} 
\right )\,,
\end{align}
allowing to cast a generic definition of the neutrino effective
mass matrix as (following the approach of~\cite{Atre:2005eb})
\begin{align}\label{eq:effective.mass.def.3}
m_\nu^{\alpha \beta} (p_{12}^2) \, \equiv \, 
\left |\sum_{i=1}^4 \frac{U_{\ell_\alpha i}\, m_i\, U_{\ell_\beta i}}{1 -
  m_i^2/p_{12}^2\, + i\, m_i \Gamma_i/p_{12}^2} 
\right |\,.
\end{align}

The combination of lepton matrix elements entering in the computation
of the effective mass is heavily constrained - not only
from the LNV bounds, but from a plethora of other processes which, 
as mentioned above, includes cLFV decays, direct and indirect searches, 
as well as cosmology (see Section~\ref{sec:updatedLNV} and
Appendix~\ref{app:constraints}). 
Although Eq.~(\ref{eq:effective.mass.def.3}) does offer a means to
infer constraints on the effective mass matrix, it is important to
stress that the latter depend on the momentum transfer ($q^2$,
implicitly entering $p^2_{12}$), and thus no absolute bounds can be
obtained\footnote{Notice that in general this is indeed the case. Although in the usual ($0\nu 2\beta$) definition $m_\nu^{ee}$ is apparently 
independent of $q^2$, this is a consequence of only having
contributions of active (light) neutrinos (for which $m_i^2 \ll
\left|q^2\right|$). 
In the presence of heavy (virtual) neutrinos, the momentum dependence is restored (although in a milder way with respect to the case of resonant enhancements considered in this work).}.

\section{Results in minimal frameworks}\label{sec:Results}
We proceed to report the results obtained for the distinct types of 
LNV decays we have considered, emphasising on how the latter 
can be 
further constrained by considering the improved sensitivity to other
modes also violating lepton flavour, and
finally discussing their implication towards the constraining of the
effective mass matrix. 
As mentioned in Section~\ref{sec:sterile:LNV.cLFV}, 
we work in the framework of a minimal toy model
(a ``3+1'' SM extension),  
described in Appendix~\ref{sec:3+1}. Here we just recall that the simple 
``3+1 model'' is parametrised by the heavier (mostly sterile)
neutrino mass $m_4$, three active-sterile mixing angles as well as
three new CP violating phases (two Dirac and one Majorana). 
In the numerical analyses (which will in general assume a 
normal ordering for the light neutrino spectra) we will consider a range
for the mass of the additional heavy
state so that it can be produced as an on-shell state from the tau and
meson decays, 
\begin{equation}\label{eq:effective:m4range}
0.1 \text{ GeV }\lesssim \, m_4 \, \lesssim 10 \text{ GeV},
\end{equation}
and we sample the active-sterile mixing angles from the 
interval $[0, 2 \pi]$ (likewise for the various CP violating phases).

\subsection{LNV semileptonic decays}\label{sec:Bdecays}
We begin our presentation of the numerical results by investigating
the contributions of the on-shell mostly sterile state concerning
several lepton number violating processes. 
In particular, we investigate {\it all} kinematically viable 3-body decay
modes of charged $B$, $B_c$, $D_s$, $D$ and $K$ mesons, 
as well as the 3-body LNV decay modes of the tau lepton. We consider both the semileptonic decays into pseudoscalar and vector meson final states. 

Leading to the plots presented in this section - with the exception of
those of Fig.~\ref{fig:LNV.cLFV:correl} - 
in order to optimally enhance the LNV rates, one considers the
maximally allowed active-sterile mixing regimes for the lepton
flavours involved in a given decay, as obtained from the
study whose results are summarised in
Fig.~\ref{fig:ML.U-m4constraints}. This allows to maximise 
the couplings responsible for the process, while setting to zero
the mixings involving the flavour(s) absent from the decay (mother particle or
products), which in turn minimises the neutrino decay width. 
In some mass regimes this mixing pattern leads to a neutrino lifetime
$\tau_4$ such that $\tau_4 c > 10 $ m, rendering the process difficult
to observe in realistic detectors. In these cases we provide an
additional prediction for the expected branching fractions, obtained
by considering non-zero mixings also with the flavours not involved in
the process, in order to comply with the condition $\tau_4 c = 10$~m. 
The presentation of our results (see e.g. Fig.~\ref{fig:res:LNV:Bdecays} and
similar ones in the present section) includes both predictions for
the maximal allowed LNV branching ratios, encoding via a full
line the more general result and using a dashed one whenever the condition
$\tau_4 c < 10 $ m is taken into account. In the mass regimes where
the two lines merge (or if no dashed line is present) the above
condition is automatically fulfilled. 

An important constraint in the mass range $m_4 \in \left[ 0.14 - 0.49
  \right]$ GeV stems from the strong bound on cLFV kaon decays, BR$(K^+
\rightarrow \pi^+ \mu^+ e^-) < 1.3 \times
10^{-11}$~\cite{Patrignani:2016xqp}, which can be mediated by
on-shell neutrinos in this mass regime; by limiting the product
$U_{e4} U_{\mu 4}$ and/or the sterile neutrino width, this bound
indirectly constrains the widths of the 
LNV channels involving both an electron and
a muon as final states. When our default mixing pattern violates the
experimental bound on $K^+ \rightarrow \pi^+ \mu^+ e^-$ we consider
non-vanishing $U_{\tau 4}$, thus allowing to comply with the latter bound. 
We have verified that the existing bounds on the remaining cLFV meson
decay channels are not competitive with the constraints presented in
Fig.~\ref{fig:ML.U-m4constraints}.

\subsubsection{$\bm{B}$ meson decays}
\begin{figure}[h!]
\begin{center}
\begin{tabular}{cc}
\epsfig{file=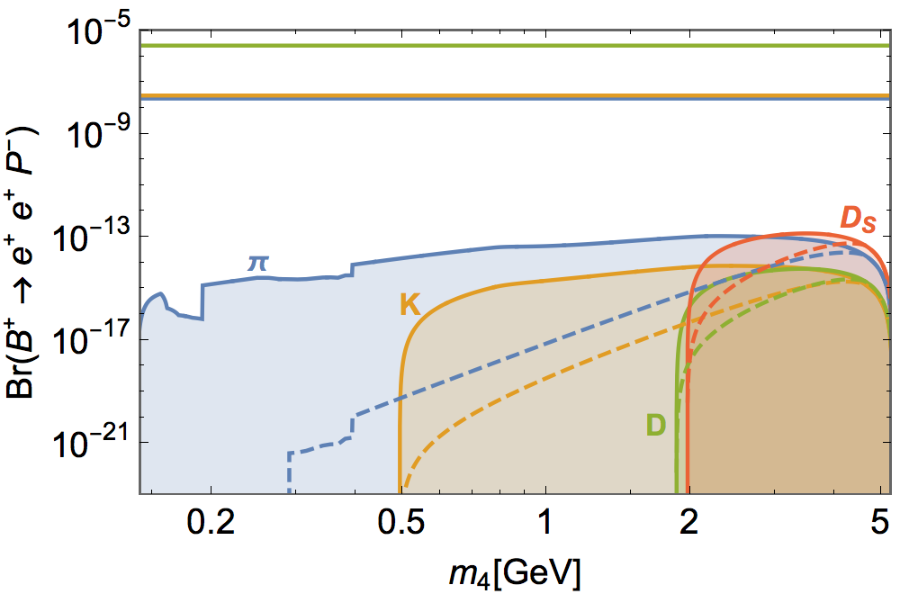,
width=73mm, angle =0} 
\hspace*{2mm}&\hspace*{2mm}
\epsfig{file=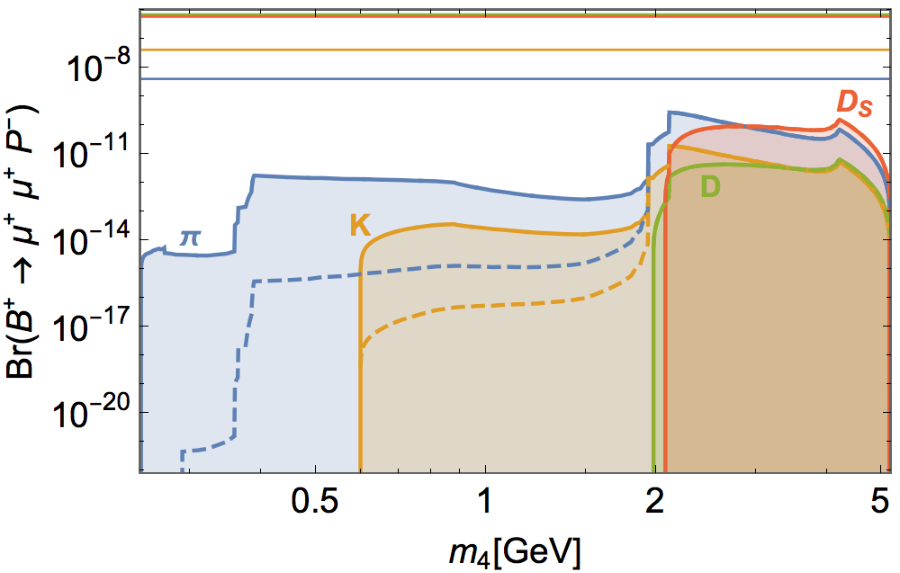,
width=73mm, angle =0}
\\
\epsfig{file=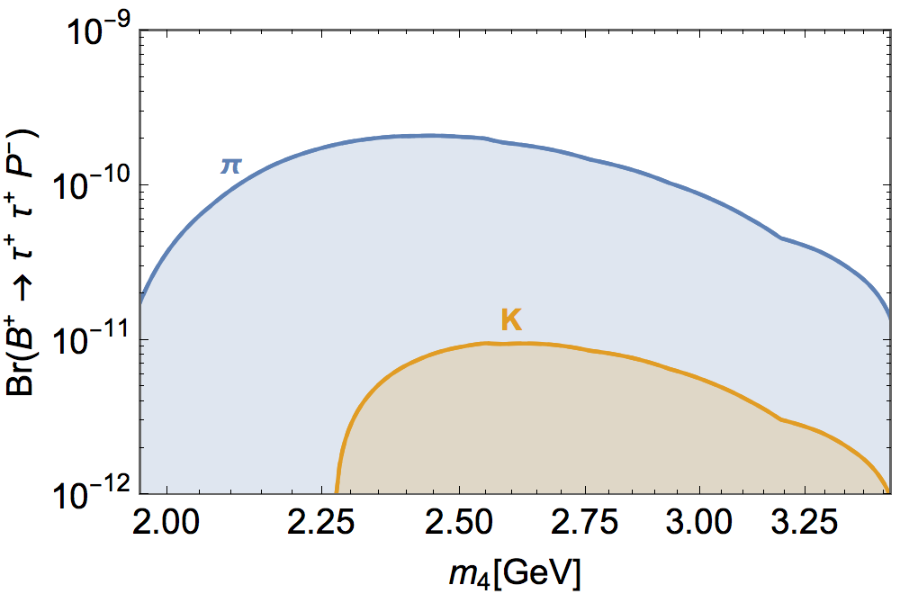,
width=73mm, angle =0}
\hspace*{2mm}&\hspace*{2mm}
\epsfig{file=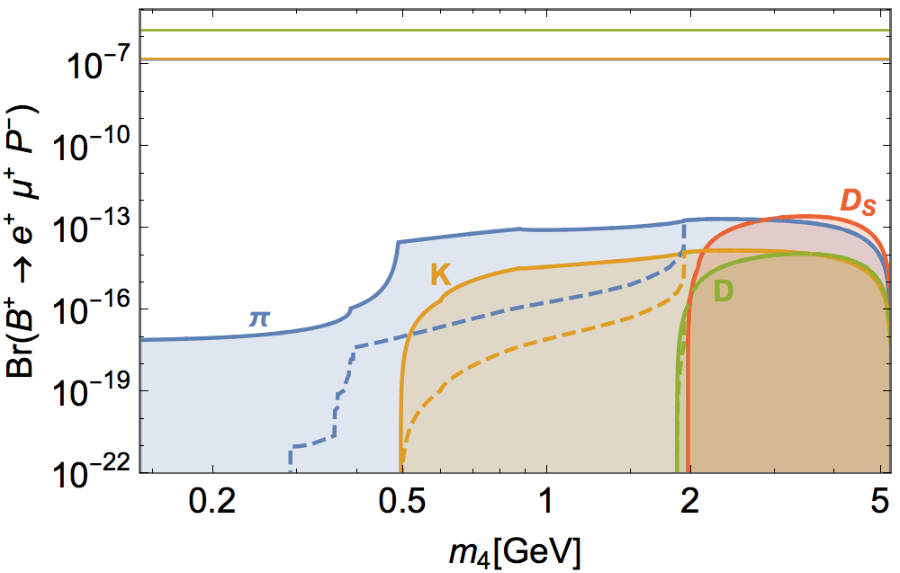,
width=73mm, angle =0}
\\
\epsfig{file=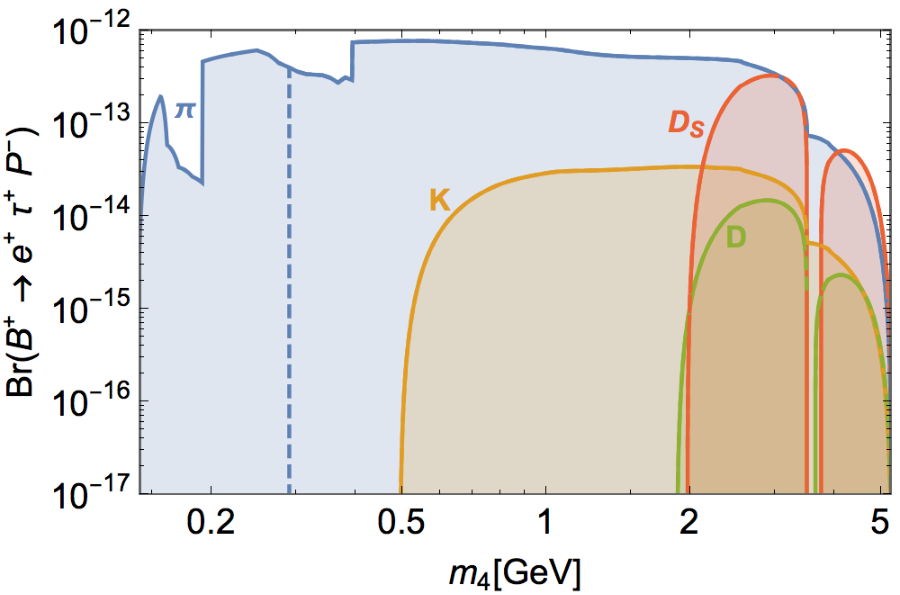,
width=73mm, angle =0}
\hspace*{2mm}&\hspace*{2mm}
\epsfig{file=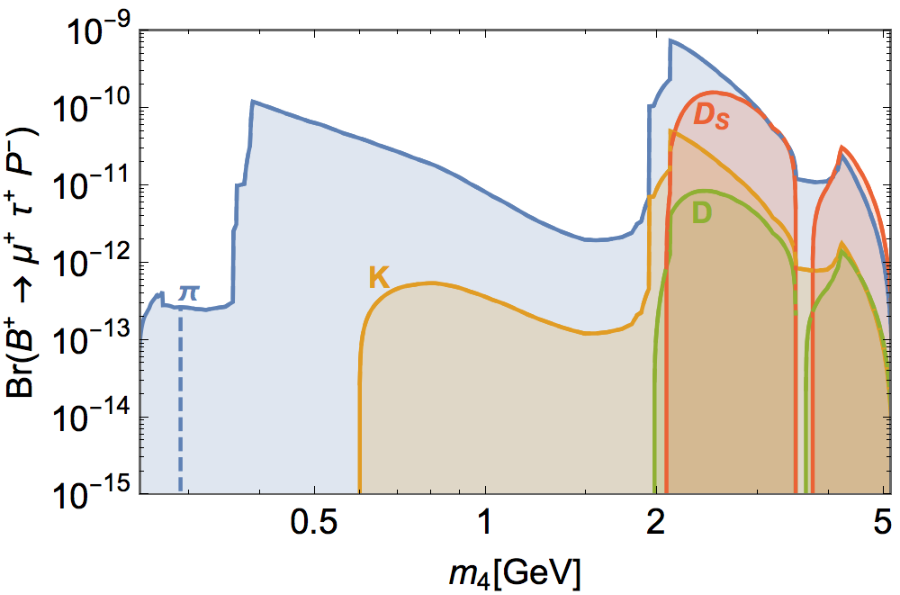,
width=73mm, angle =0}
\end{tabular}
\end{center}
\caption{Branching ratios of LNV decay modes of $B^+$
  mesons (into pseudoscalar final states, plus leptons of same or
  different flavour), 
  as a function of the (mostly sterile) heavy neutrino mass,
  $m_4$. Pale blue, yellow, red and green curves (surfaces) respectively denote
  the maximal (allowed) values of the BR($B^+ \to \ell_\alpha^+ \ell_\beta^+ P^-$),
  with $P^-= \pi^-$, $K^-$, $D_s^-$ and $D^-$; coloured dashed curves denote 
  the corresponding maximal values of the BRs once
  $c \tau_4 \leq 10$~m is imposed. The coloured horizontal
  lines correspond to the present experimental bounds, when available
  (cf. Table~\ref{tab:LNV:meson:exp}). From top to bottom, left to
  right, the final state lepton content is $e^+ e^+$, $\mu^+ \mu^+$,
  $\tau^+ \tau^+$, $e^+ \mu^+$, $e^+ \tau^+$ and $\mu^+ \tau^+$. 
}\label{fig:res:LNV:Bdecays}
\end{figure}

\begin{figure}[h!]
\begin{center}
\begin{tabular}{cc}
\epsfig{file=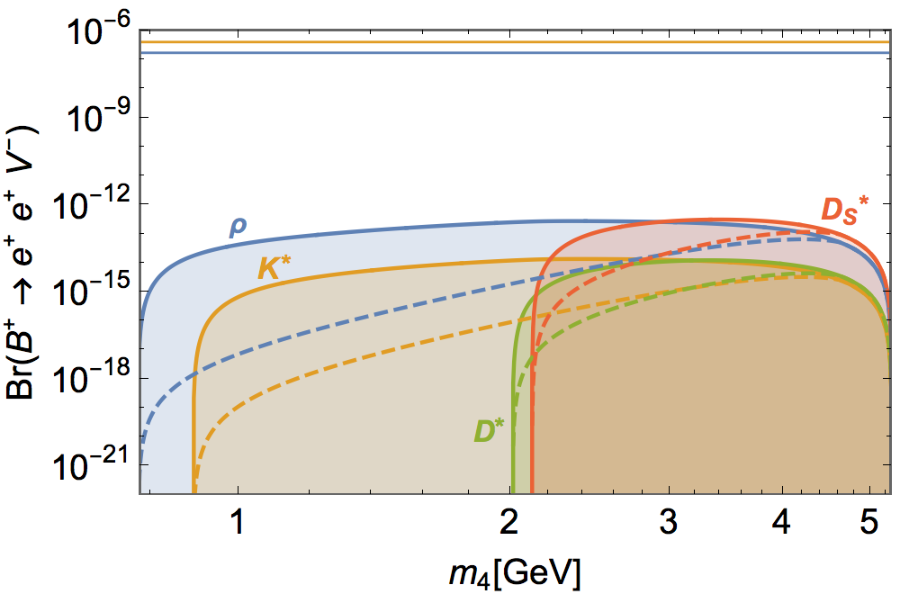,
width=73mm, angle =0} 
\hspace*{2mm}&\hspace*{2mm}
\epsfig{file=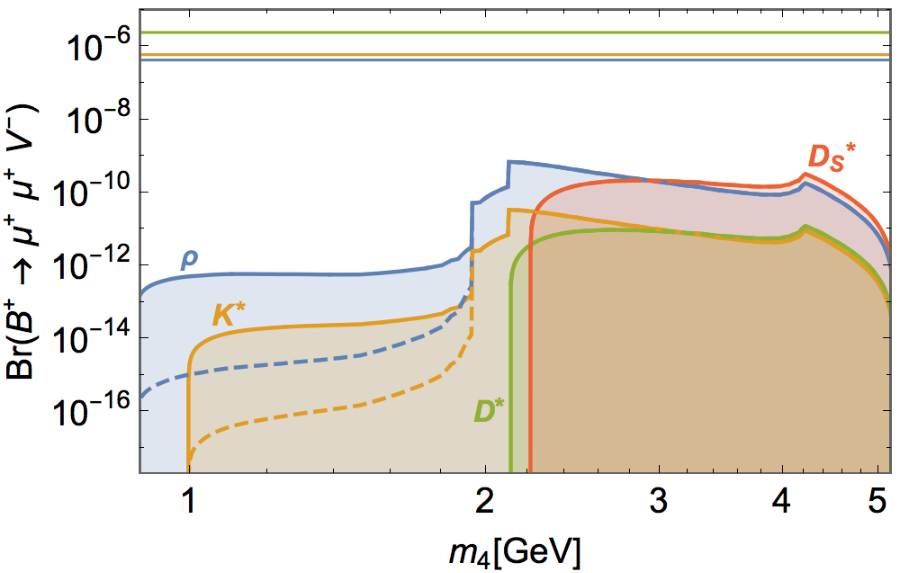,
width=73mm, angle =0}
\\
\epsfig{file=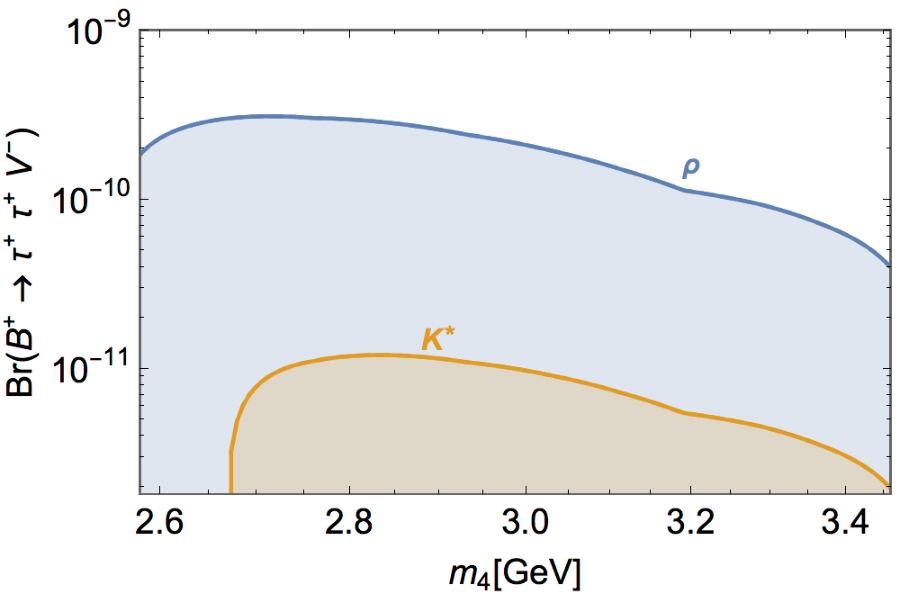,
width=73mm, angle =0}
\hspace*{2mm}&\hspace*{2mm}
\epsfig{file=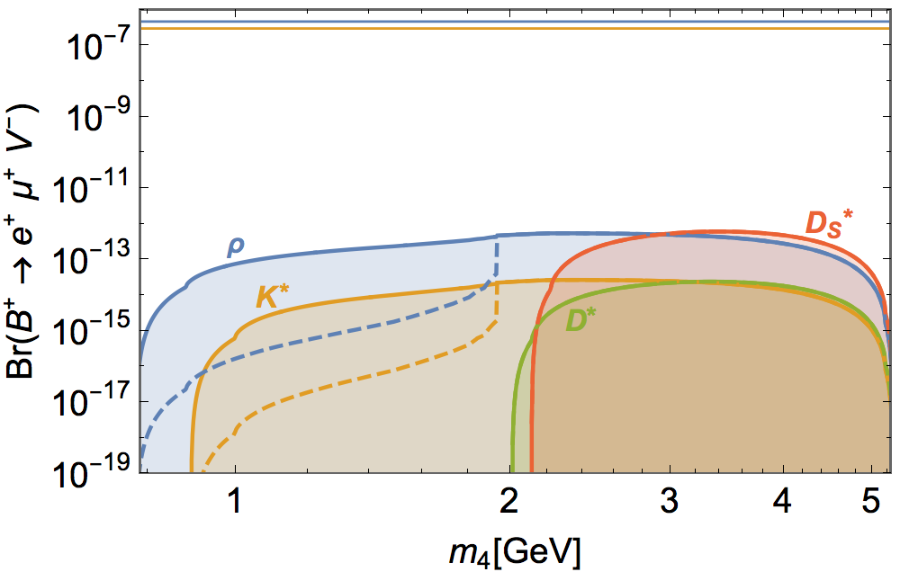,
width=73mm, angle =0}
\\
\epsfig{file=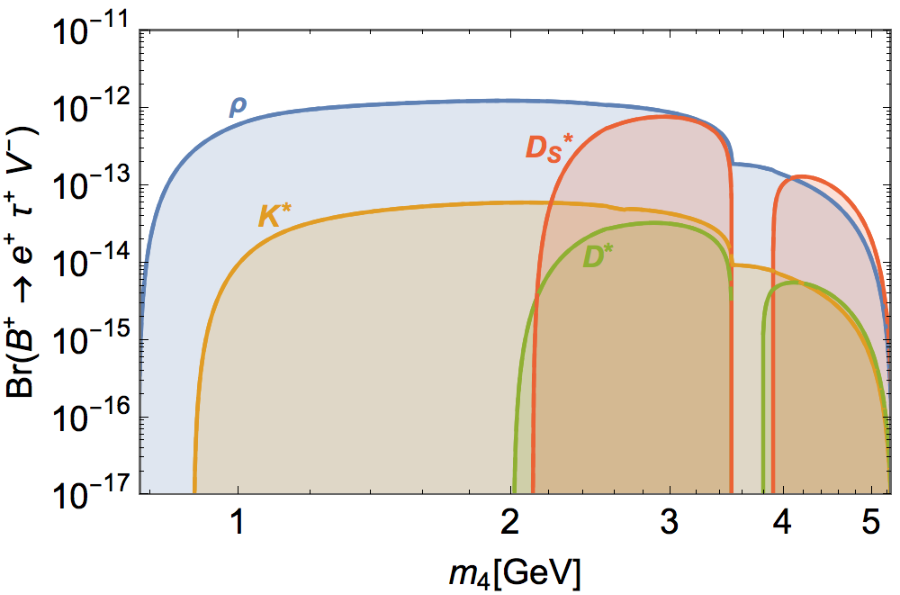,
width=73mm, angle =0}
\hspace*{2mm}&\hspace*{2mm}
\epsfig{file=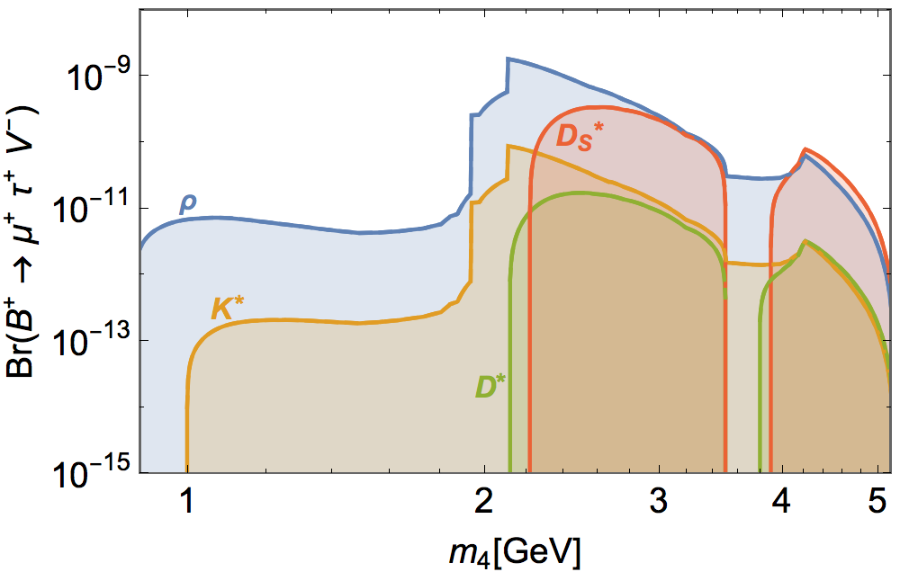,
width=73mm, angle =0}
\end{tabular}
\end{center}
\caption{Branching ratios of LNV decay modes of $B^+$
  mesons (into vector final states, plus leptons of same or different flavour), 
  as a function of the (mostly sterile) heavy neutrino mass,
  $m_4$. 
  Pale blue, yellow, red and green curves (surfaces) respectively denote
  the maximal (allowed) values of the BR($B^+ \to \ell_\alpha^+ \ell_\beta^+ V^-$),
  with $V^-= \rho$, $K^*$, $D_s^*$ and $D^*$; coloured dashed curves denote 
  the corresponding maximal values of the BRs once
  $c \tau_4 \leq 10$~m is imposed. The coloured horizontal
  lines correspond to the present experimental bounds, when available
  (cf. Table~\ref{tab:LNV:meson:exp}). 
}\label{fig:res:LNV:BdecaysV}
\end{figure}

The allowed branching fractions for the decays of the 
pseudoscalar $B$ meson into pairs of leptons of same or different flavour
and a meson are summarised in the panels of  
Figs.~\ref{fig:res:LNV:Bdecays} and~\ref{fig:res:LNV:BdecaysV}, 
for the case of pseudoscalar and vector meson final
states. 
With the exception of the final states
including a pair of tau leptons (in which only light mesons as $K$ and
$\pi$ can be produced), all other decay modes include pions, kaons,
as well as $D$ and $D_s$ pseudoscalar mesons. For the case of vector mesons,
$\rho$, $K^*$, $D^*$ and $D_s^*$ are kinematically allowed (with the
exception of the $\tau \tau$ mode, for which only $\rho$ and $K^*$
are kinematically allowed). Whenever the regime would allow for
long-lived Majorana mediators, we further recompute the maximal
branching fractions imposing that the on-shell neutrino 
lifetime should comply with $c \tau_4 \leq 10~\text{m}$ (dashed lines).

\hspace*{-8mm}

A common feature\footnote{Notice that for sterile mass regimes close
  to 0.1~GeV, the contour of the line associated with final states
  comprising a pion strongly reflects the excluded regions due to 
  laboratory constraints (direct searches), as seen in 
  Fig.~\ref{fig:ML.U-m4constraints}. }
 to several of the curves displayed in
Figs.~\ref{fig:res:LNV:Bdecays} and~\ref{fig:res:LNV:BdecaysV} (as
well as to others presented in this section) is the appearance of two
regimes in the decay widths: while in some cases this is perceived as
a double plateau, or a ``bump'' within a curve, in others it is
evidenced as a clear discontinuity for given intervals of
$m_4$. To understand this, recall that the total decay
amplitude stems from two contributions - the one directly displayed in
Fig.~\ref{fig:feynman:M1decay}, 
and the one corresponding to the exchange of the two lepton flavours 
as final states (either arising from the production or from the 
decay of the heavy
neutrino). Under the assumption that one cannot distinguish the
lepton's production vertex, the condition of having an on-shell heavy
neutrino mediating, for instance, the process $M_1 \to \ell_\alpha
\ell_\beta M_2$ (i.e. $m_{M_1} > m_4 + m_{\ell_\alpha}$ and 
$m_4 > m_{\ell_\beta}+ m_{M_2}$), must be 
verified for both channels. 
Although in general the regions fulfilling the above
conditions tend to overlap under the exchange $\alpha \leftrightarrow \beta$ 
(nevertheless giving rise to the
double-plateau behaviour), this is not always the case, especially when 
$m_{\ell_\alpha} \gg m_{\ell_\beta}$. An example can be found in $B^- \to e^+ \tau^+
D_s^-$ decay, where the on-shell mass regimes, [1.97, 3.50]~GeV
and [3.75, 5.28]~GeV do not overlap. 
The contribution to the
amplitude of the two regimes can also be very different - one regime
might dominate over the other by as much as one order of magnitude, 
giving rise to ``kinks'' at the sterile neutrino mass values where 
the dominant contribution becomes kinematically forbidden.

Whether or not these processes are likely to be experimentally
observed in the near future - for example at the LHC run 2, be it at
LHCb, CMS, or ATLAS - calls upon a more dedicated discussion, taking
into account not only the $\Delta L=2$ decay widths, but the $B$-meson
production prospects, and detector capabilities for such final
states. 
A na\"ive approximation suggests the following estimated number of
events, for example regarding $B^+ \to \ell_\alpha^+ \ell_\beta^+ M^-$:
\begin{align}
N_{B\to \ell_\alpha \ell_\beta M} \, \approx \, \mathcal{L}^\text{int}\, \times \, 
\sigma^\text{prod}(pp \to B^+ + X) \, \times \, 
\text{BR}(B^+ \to \ell_\alpha^+ \ell_\beta^+ M^-) \, \times \, \mathcal{D}\, ,
\end{align}
in which, and for a given experiment, $\mathcal{L}^\text{int}$ denotes
the integrated luminosity, $\sigma^\text{prod}$ is
the $B$ meson production cross section, 
and $\mathcal{D} = \varepsilon_D \times \rho_N$ corresponds to a factor which
parametrises the overall detection efficiency for a given process
($\varepsilon_D$), times the acceptance factor of the detector ($\rho_N$). 

For the LHC operating at $\sqrt{s} =13$~TeV,
the $B$ meson production cross section has been recently reported 
by LHCb~\cite{Aaij:2017qml} 
to be $\sigma^\text{prod}_\text{LHCb}(pp \to
B^+ + X) = (86.6 \pm 0.5 \pm 5.4 \pm 3.4)~\mu\text{b}$, for 
$ \mathcal{L}^\text{int}_\text{LHCb}= 0.3~\text{fb}^{-1}$. Assuming 
that $ \mathcal{D}_\text{LHCb}$ should not be far from the percent level (i.e., 
$\sim 1\%$), one finds for the expected number of events
\begin{equation}
N^\text{LHCb}_{B\to \ell_\alpha \ell_\beta M} \, \approx \, 0.25 \, 
\times \frac{\text{BR}(B^+ \to \ell_\alpha^+ \ell_\beta^+ M^-)}{10^{-9}}\,.
\end{equation}
Should the integrated luminosity augment to $10~\text{fb}^{-1}$ (or
possibly to $50~\text{fb}^{-1}$ at the end of LHC run 3), then 
$\Delta L=2$ processes with branching fractions in the ballpark of 
$\mathcal{O}(10^{-8} - 10^{-10})$ could lead to a non-negligible
number of events\footnote{In~\cite{Mejia-Guisao:2017gqp}, and working for
  similar operating benchmarks (luminosity and detector performance),
  it was found that a significant sensitivity could also be expected
  for LNV 4-body decays, in particular for $B_s^0 \to K^- \pi^- \mu^+
  \mu^+$ and $B_s^0 \to D_s^- \pi^- \mu^+ \mu^+$.}.  
In particular, and given that the current bounds on $B^+ \to \mu^+ \mu^+ \pi^-$
are already $\mathcal{O}(10^{-9})$
(cf. Table~\ref{tab:LNV:meson:exp}), a future improvement of the
experimental sensitivity\footnote{In~\cite{Cvetic:2017vwl}, the
  number of $B$ mesons expected to be produced at the LHCb upgrade and
  at Belle II were reported to be $4.8 \times 10^{12}$ and 
  $5 \times 10^{10}$, respectively. } 
should be sensitive to $\Delta L=2$
transitions as induced by the SM minimally extended by a sterile fermion:
as seen from Fig.~\ref{fig:res:LNV:Bdecays} maximal
branching fractions for the latter decay 
lie around $\mathcal{O}(10^{-10})$, for sterile mass
between 2 and 3~GeV. 

Naturally, this is also subject to having a
conservative estimation of the detector size: assuming a smaller
effective detector length might lead to different constraints on the
sterile fermion parameter space, and thus modify the above
discussion. For instance, for the case of a 1~m detector (as will be
the case at Belle II) - which was already qualitatively discussed in
what concerns the bounds on the combinations of the active-sterile
mixing elements (see Section~\ref{sec:updatedLNV}) - 
we have verified that the LNV
branching rates would be reduced by between one and two orders of
magnitude, depending on the actual regime of $m_4$. Heavy mass
regimes, as visible from the panels of Fig.~\ref{fig:res:LNV:Bdecays},
would not be sensitive to finite (realistic) detector effects.

\hspace*{-8mm}

\begin{figure}[h!]
\begin{center}
\begin{tabular}{cc}
\epsfig{file=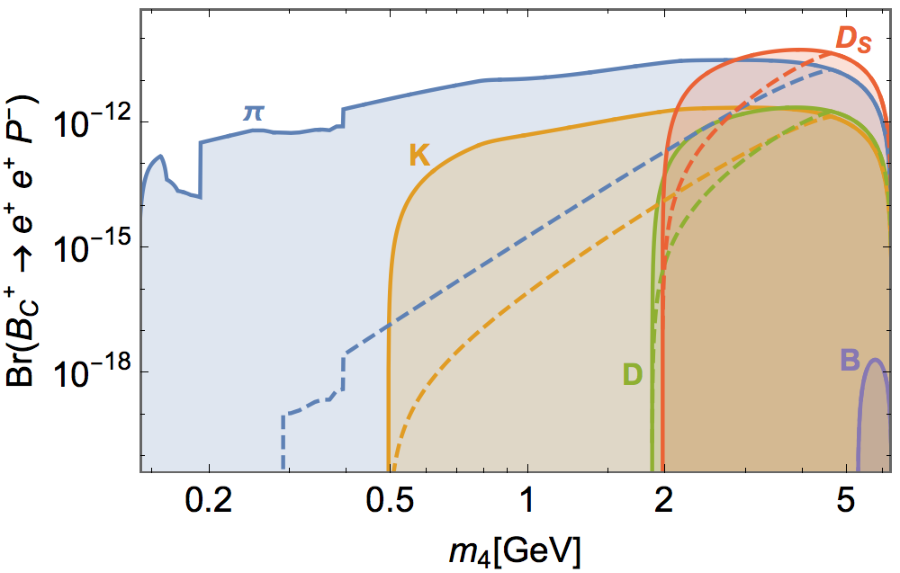,
width=73mm, angle =0} 
\hspace*{2mm}&\hspace*{2mm}
\epsfig{file=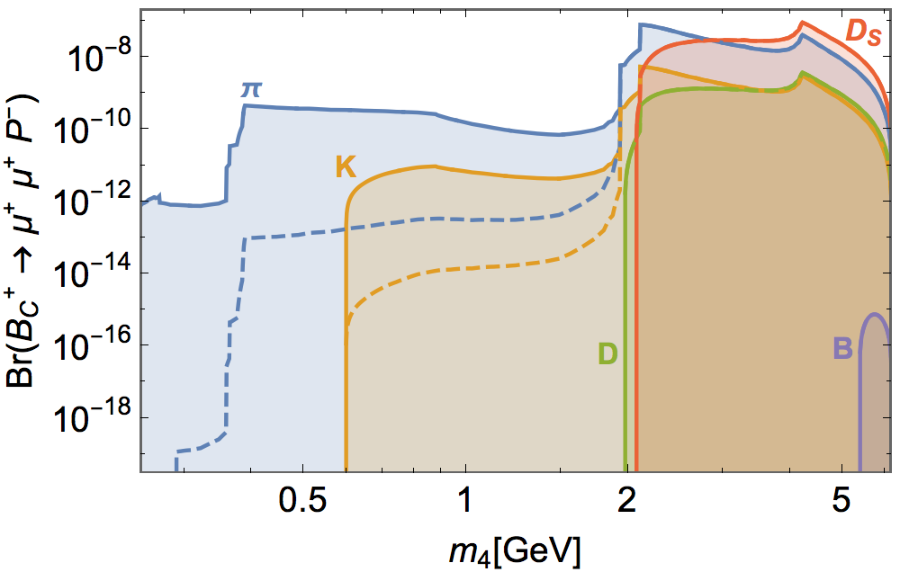,
width=73mm, angle =0}
\\
\epsfig{file=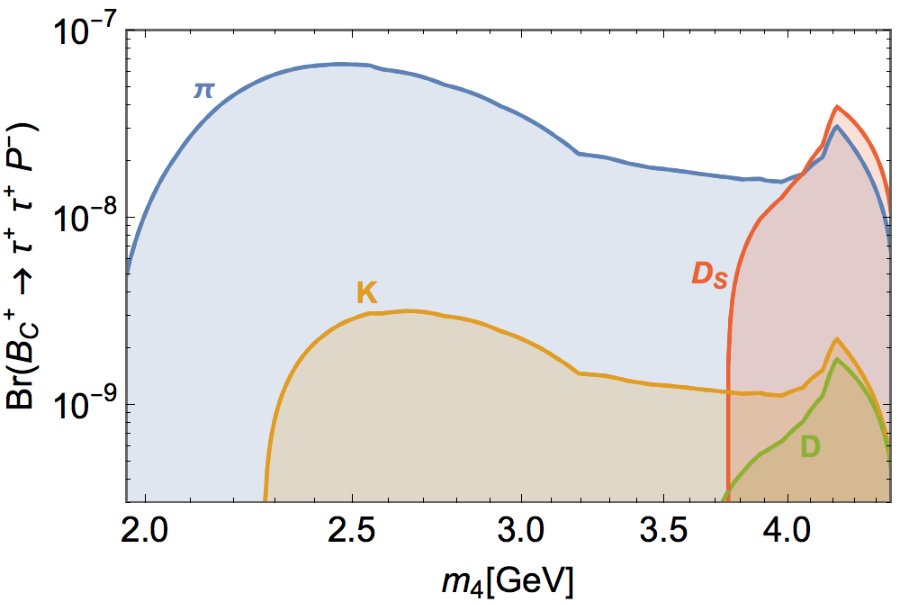,
width=73mm, angle =0}
\hspace*{2mm}&\hspace*{2mm}
\epsfig{file=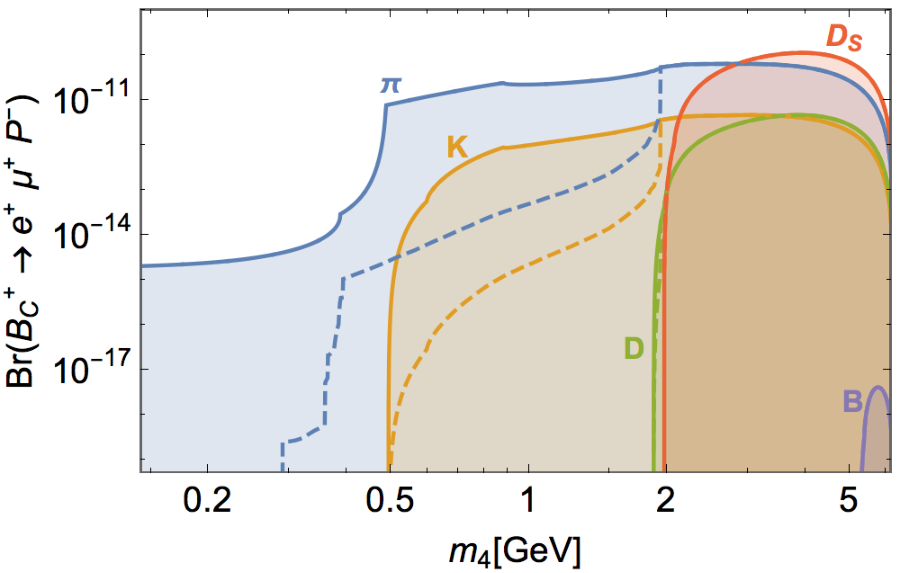,
width=73mm, angle =0}
\\
\epsfig{file=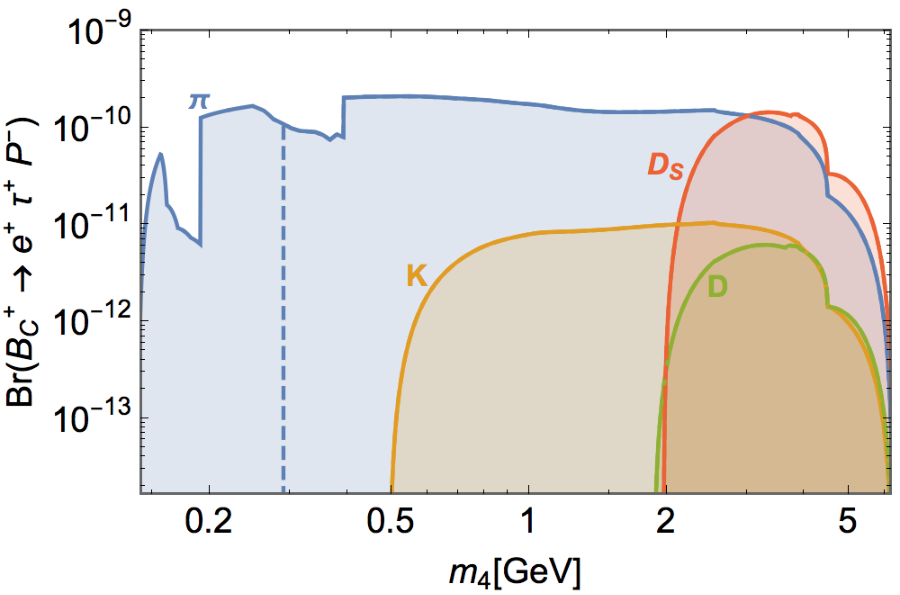,
width=73mm, angle =0}
\hspace*{2mm}&\hspace*{2mm}
\epsfig{file=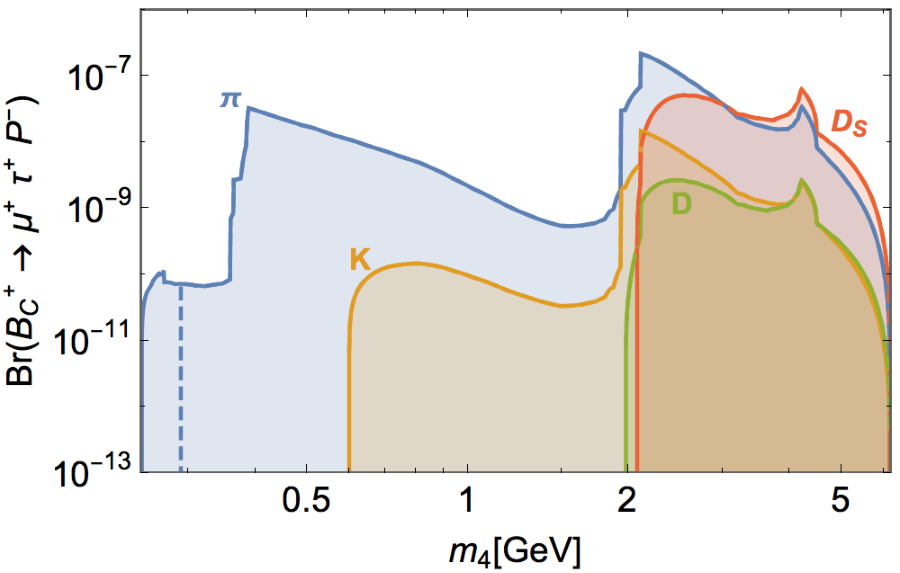,
width=73mm, angle =0}
\end{tabular}
\end{center}
\caption{Branching ratios of LNV decay modes of $B_c^+$
  mesons (into pseudoscalar final states, plus leptons of same or
  different flavour), 
  as a function of the (mostly sterile) heavy neutrino mass,
  $m_4$. Pale blue, yellow, red, green and blue curves (surfaces) 
  respectively denote
  the maximal (allowed) values of the BR($B_c^+ \to \ell_\alpha^+ \ell_\beta^+ P^-$),
  with $M^-= \pi^-$, $K^-$, $D_s^-$, $D^-$ and $B^-$; coloured dashed
  curves denote  
  the corresponding maximal values of the BRs once
  $c\tau_4 \leq 10$~m is imposed. 
}\label{fig:res:LNV:Bcdecays}
\end{figure}

\begin{figure}[h!]
\begin{center}
\begin{tabular}{cc}
\epsfig{file=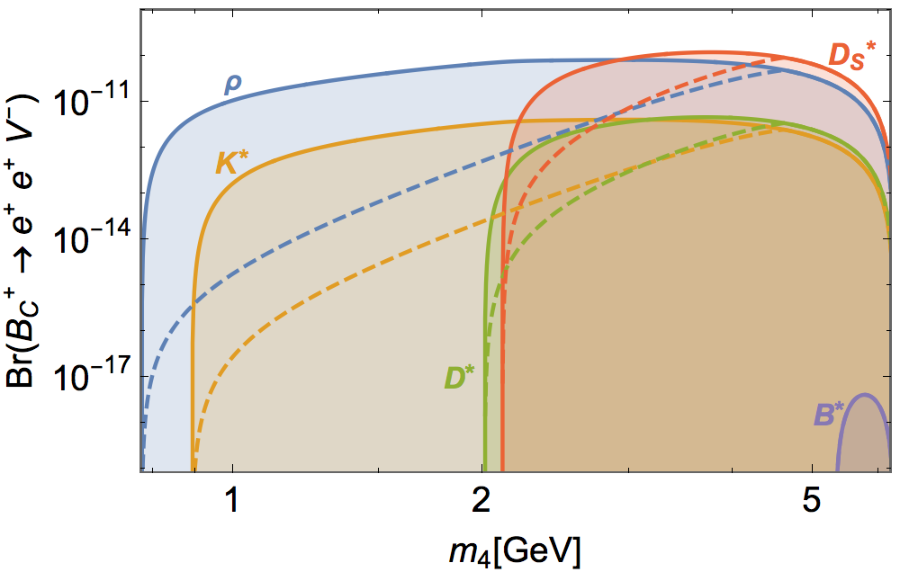,
width=73mm, angle =0} 
\hspace*{2mm}&\hspace*{2mm}
\epsfig{file=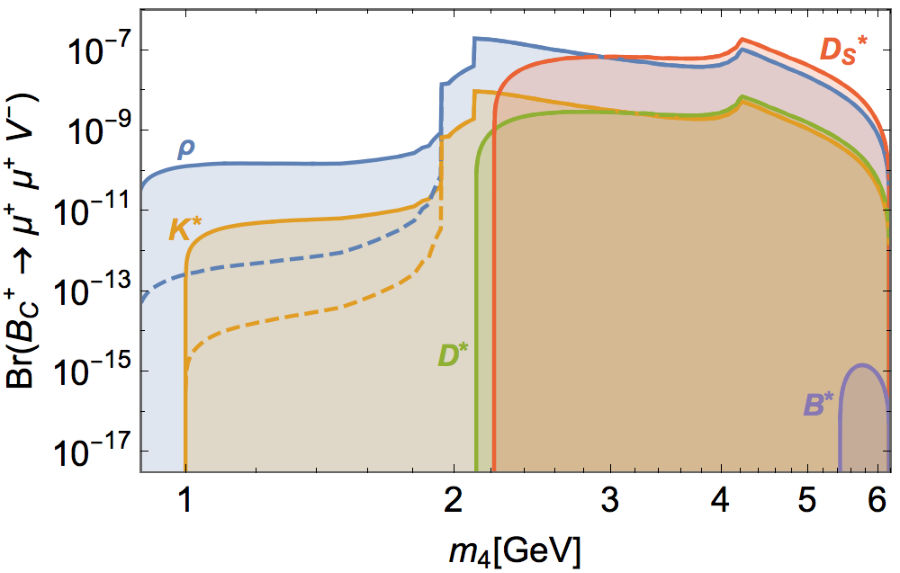,
width=73mm, angle =0}
\\
\epsfig{file=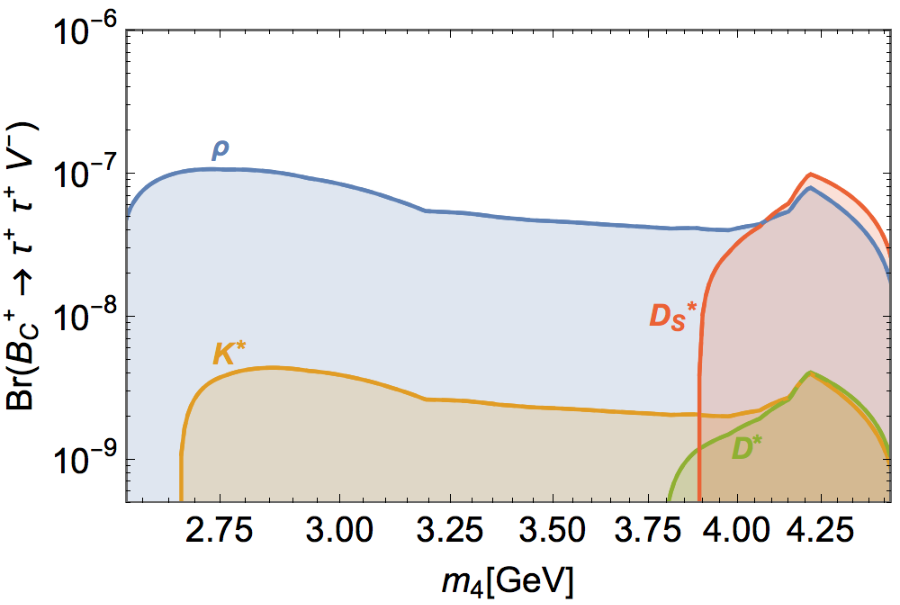,
width=73mm, angle =0}
\hspace*{2mm}&\hspace*{2mm}
\epsfig{file=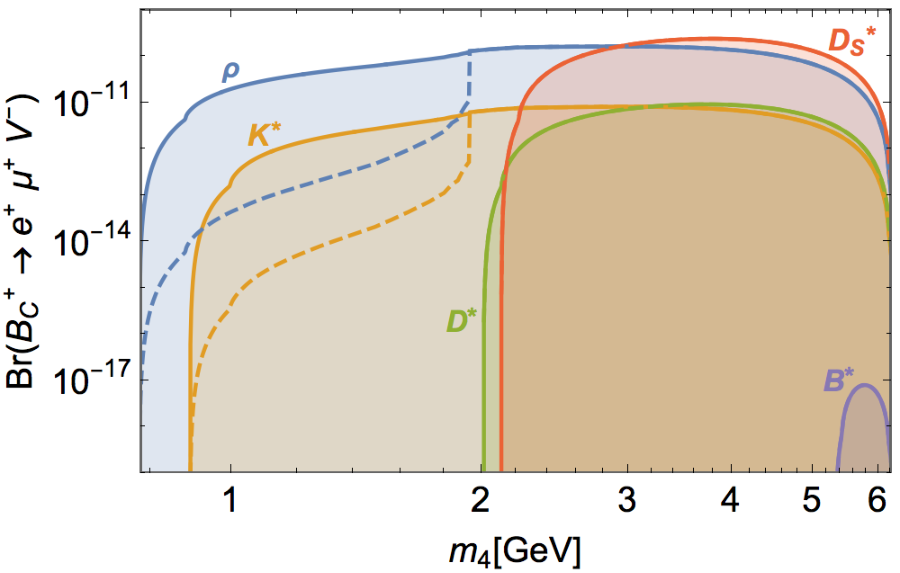,
width=73mm, angle =0}
\\
\epsfig{file=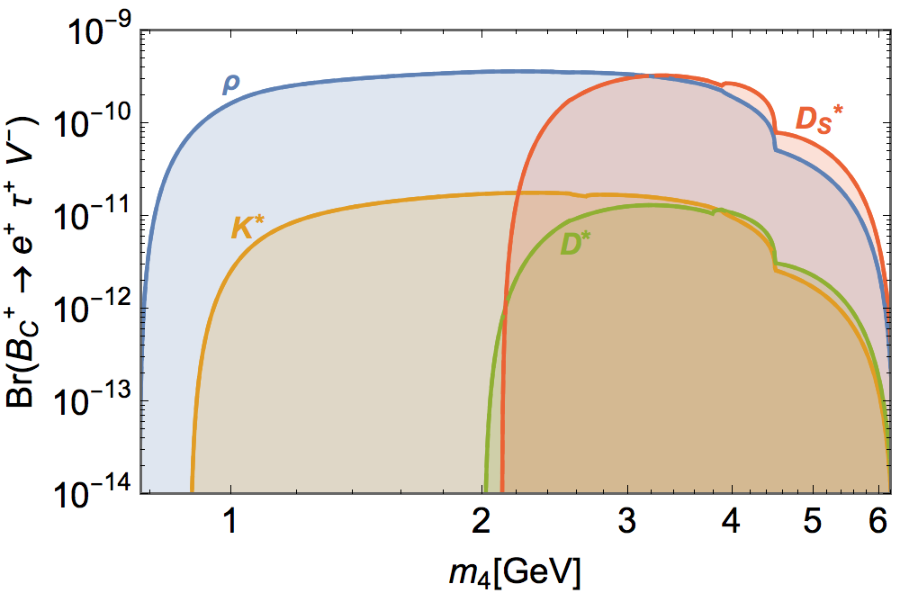,
width=73mm, angle =0}
\hspace*{2mm}&\hspace*{2mm}
\epsfig{file=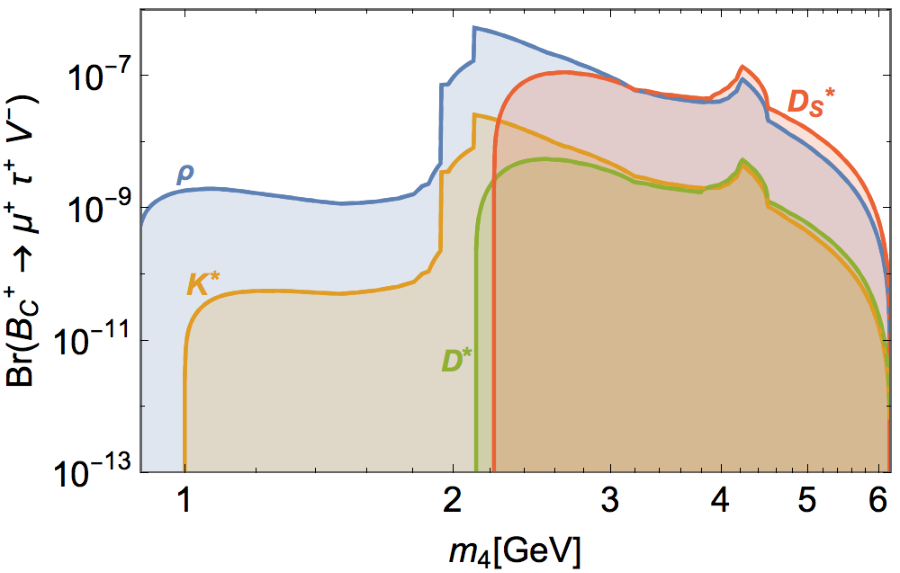,
width=73mm, angle =0}
\end{tabular}
\end{center}
\caption{Branching ratios of LNV decay modes of $B_c^+$
  mesons (into vector final states, plus leptons of same or different flavour), 
  as a function of the (mostly sterile) heavy neutrino mass,
  $m_4$. 
  Pale blue, yellow, red, green and blue curves (surfaces) 
  respectively denote
  the maximal (allowed) values of the BR($B_c^+ \to \ell_\alpha^+ \ell_\beta^+ V^-$),
  with $V^-= \rho$, $K^*$, $D_s^*$, $D^*$ and $B^*$; coloured dashed
  curves denote  
  the corresponding maximal values of the BRs once
  $c \tau_4 \leq 10$~m is imposed.  
}\label{fig:res:LNV:BcdecaysV}
\end{figure}

In all analogy, the panels of Fig.~\ref{fig:res:LNV:Bcdecays}
display the maximal (allowed) values of the LNV decays 
$B_c^+ \to \ell_\alpha^+ \ell_\beta^+ M^-$ (final state pseudoscalar meson). 
With the exception of final states
including at least one tau lepton (for which the final $B^-$ is
kinematically inaccessible) all other decay modes include pions, kaons,
$D$ and $D_s$, as well as $B$ mesons. In the absence of experimental
bounds (as no data is available concerning these very rare $B_c$
decays), one cannot comment upon the experimental impact of the LNV
decays; however, our analysis suggests that the maximal associated branching
fractions are larger than those obtained for $B^+$ mesons, by as much
as two orders of magnitude. For example, for a sterile state with mass 
$m_4 \approx 2.2$~GeV, BR($B_c^+ \to \mu^+\mu^+ \pi^-$)$|_\text{max}$
$\sim \mathcal{O}(10^{-7})$, while   
BR($B^+ \to \mu^+\mu^+ \pi^-$)$|_\text{max}$ $\sim \mathcal{O}(10^{-9})$.
Figure~\ref{fig:res:LNV:BcdecaysV} summarises the same information,
but this time for final states including one vector meson. 

\hspace*{-8mm}

\hspace*{-8mm}

\begin{figure}[h!]
\begin{center}
\begin{tabular}{ccc}
\epsfig{file=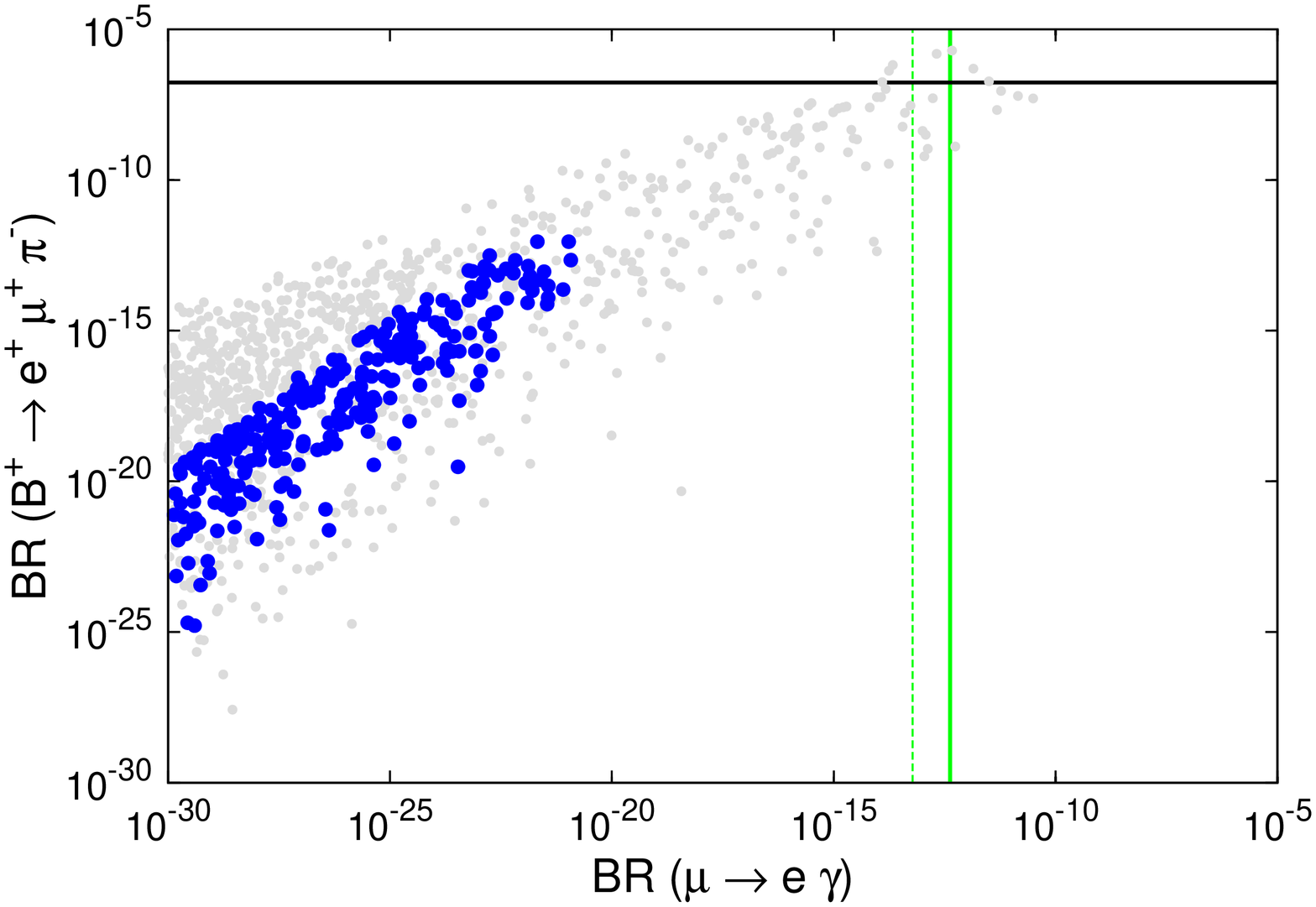,
width=58mm, angle =270}
 & 
\epsfig{file=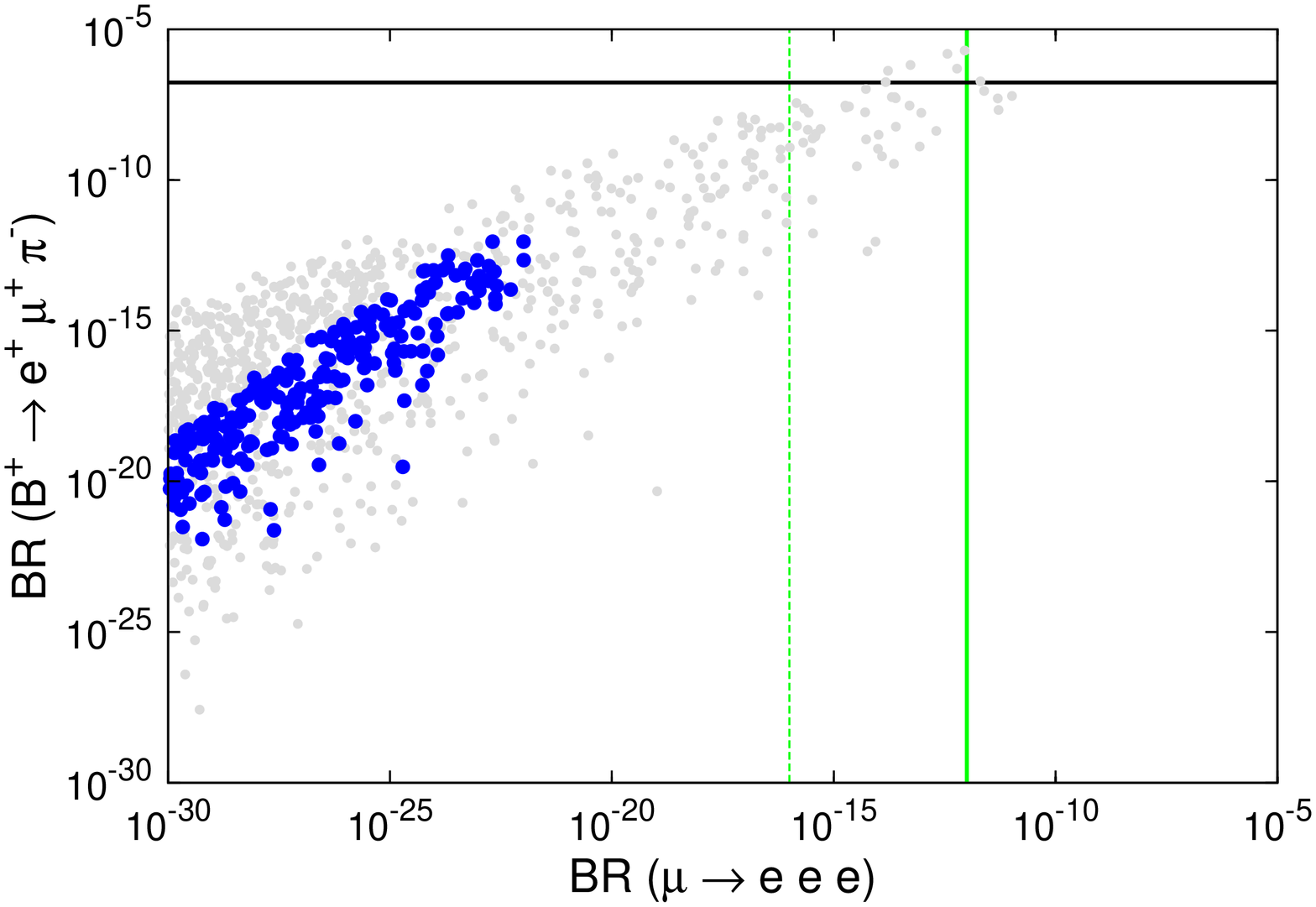,
width=58mm, angle =270} \\
\epsfig{file=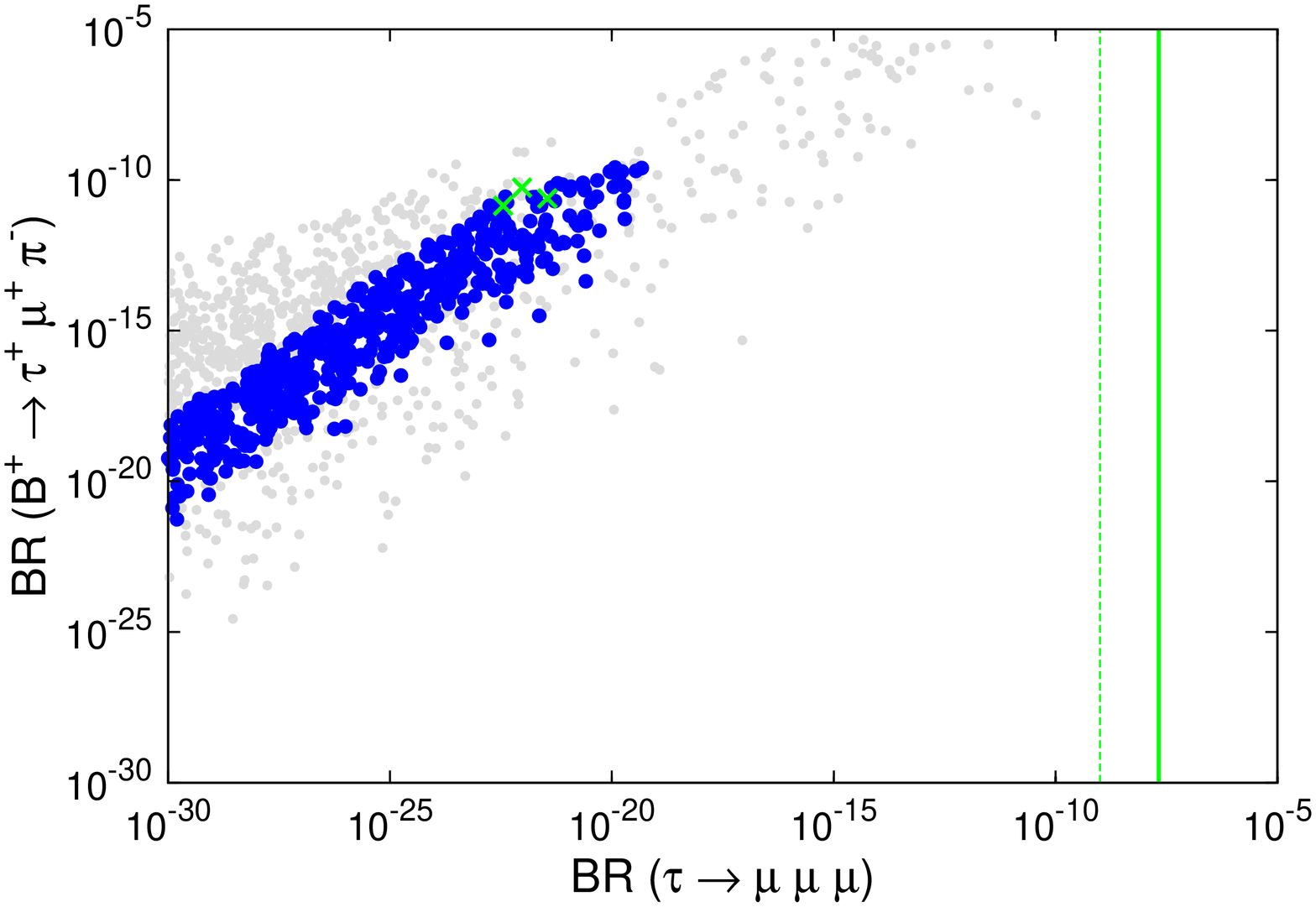,
width=58mm, angle =270}
 & 
 \epsfig{file=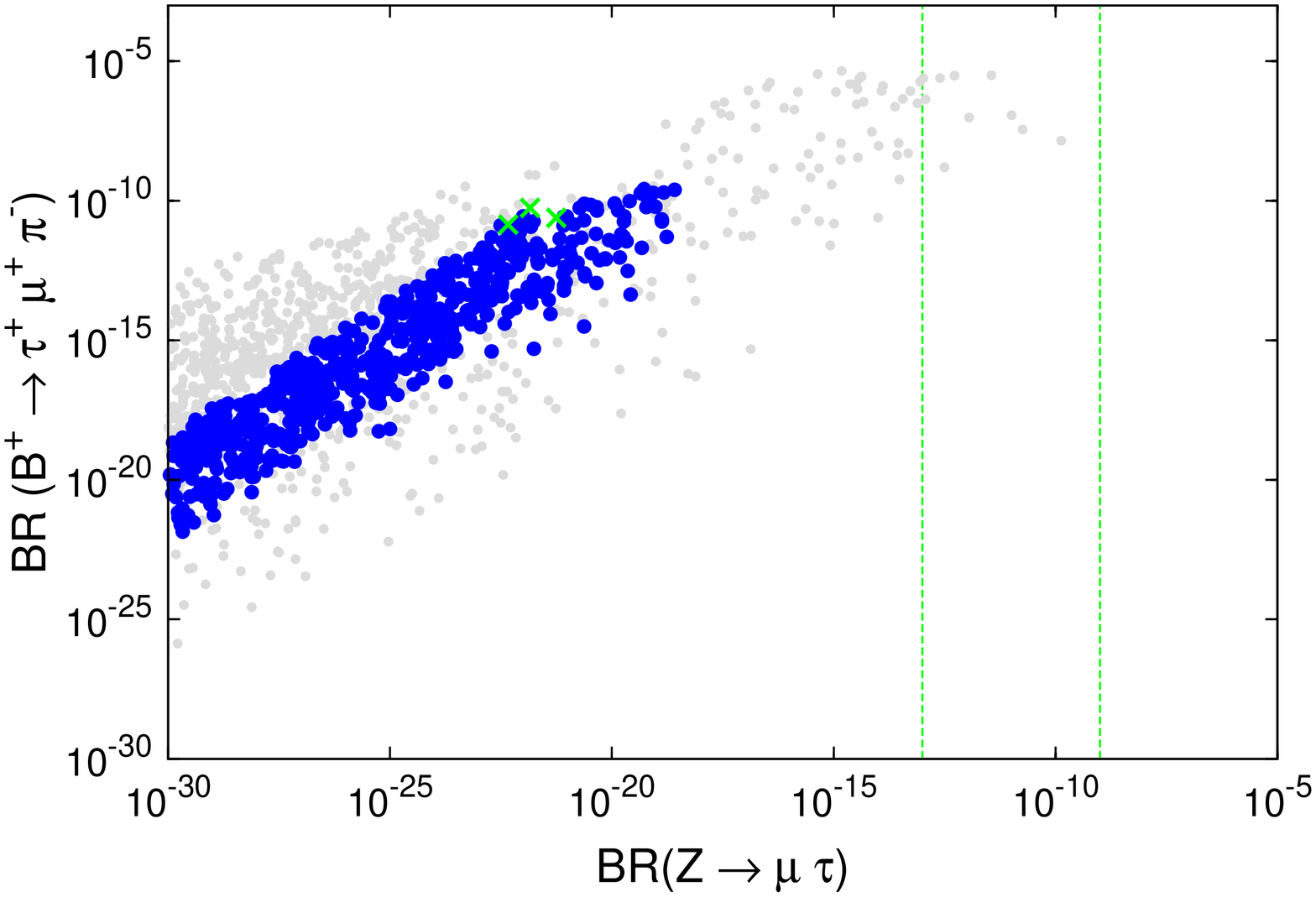,
 width=58mm, angle =270}
\end{tabular}
\end{center}
\caption{Correlation between leptonic cLFV decays 
and LNV $B$-meson decays (into leptons of same or different flavour). 
In each panel the vertical full (dashed) lines 
denote the current cLFV bounds (future expected sensitivity) of
the corresponding leptonic decay. Grey points are excluded due to
violation of at least one experimental or observational constraint 
(and by requiring $c \tau_4 \leq 10$~m), 
while the green crosses (present in the lower panels) denote points 
which are excluded due to excessive contributions to LNV decays.
}\label{fig:LNV.cLFV:correl}
\end{figure}

Before addressing other decays, and for completeness, we consider
the question of whether or not current cLFV bounds directly 
constrain the maximal allowed values of the LNV decay rates (leading
to final states with leptons of same or different flavour).
Recall that the most important contributions to
cLFV observables typically emerge for sterile states with masses well
above the GeV - in particular for states heavier than the electroweak
scale~\cite{Alonso:2012ji,Abada:2014kba,Abada:2014cca,Abada:2015oba,Abada:2016vzu};  
hence, and as can be verified
from the panels of Fig.~\ref{fig:LNV.cLFV:correl} (we chose
semileptonic $B$ decays into 
a pseudoscalar meson final state to illustrate this point), despite the
expected correlation between the similar ``flavour-content'' transitions, 
cLFV bounds (current as well as the expected future sensitivities), do
not directly constrain the LNV modes. We notice that leading to the
results of Fig.~\ref{fig:LNV.cLFV:correl}, a more comprehensive survey
of the parameter space has been conducted: the active-sterile mixings 
(both angles and phases) are
randomly sampled (without any underlying hypothesis) from the interval 
$[0, 2 \pi]$; likewise, the mass of the heavy, mostly sterile state,
is taken from a scan of $0.1~\text{ GeV} \lesssim m_4 \lesssim
10~\text{ GeV}$. 

It is relevant to notice that following our study of
the impact of the sterile fermion on LNV semileptonic decays, a small
subset of the points (denoted by green crosses in
Fig.~\ref{fig:LNV.cLFV:correl}) is excluded due to being associated
with excessive LNV decays, already in conflict with current bounds. 
This reinforces the expected experimental prospects of these
observables.

\hspace*{-8mm}

\subsubsection{$\bm{D}$ meson decays}
\begin{figure}[h!]
\begin{center}
\begin{tabular}{cc}
\epsfig{file=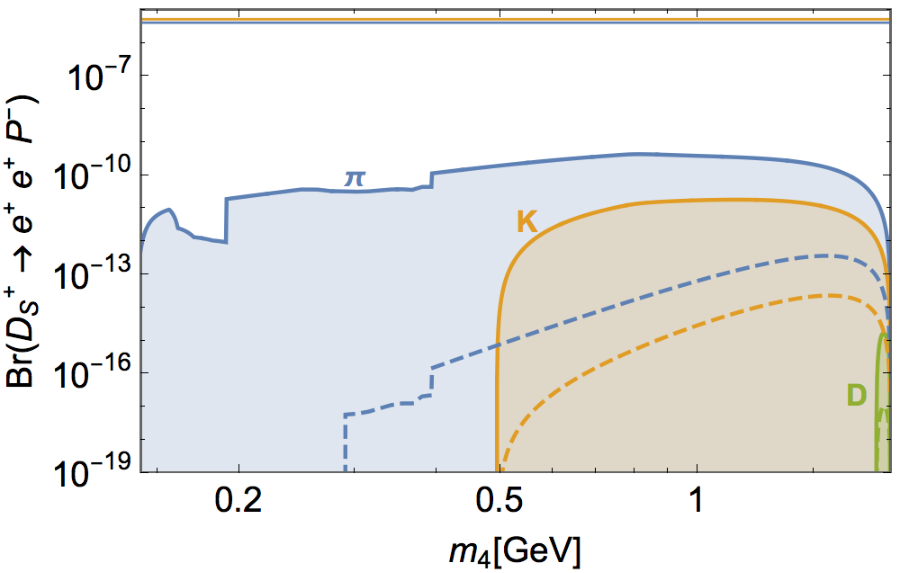,
width=73mm, angle =0} 
\hspace*{2mm}&\hspace*{2mm}
\epsfig{file=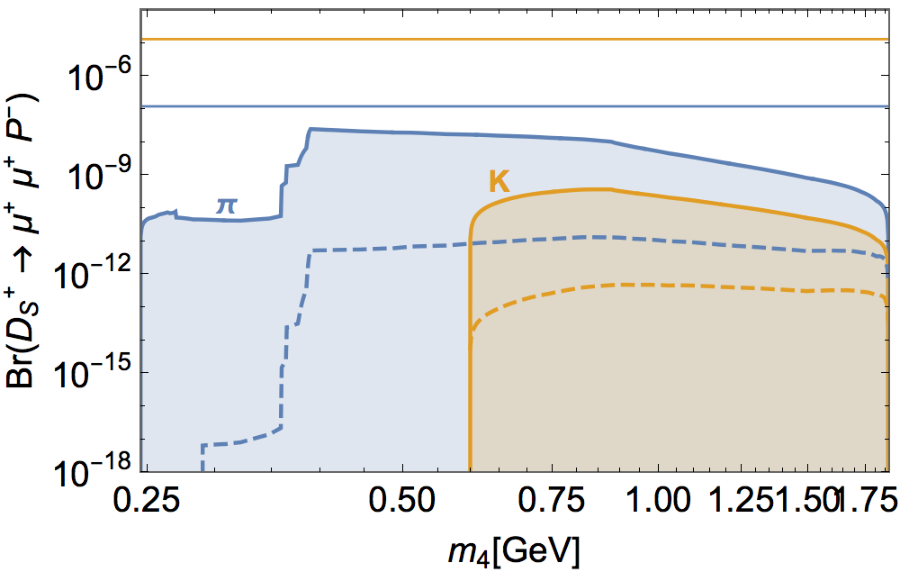,
width=73mm, angle =0}
\\
\epsfig{file=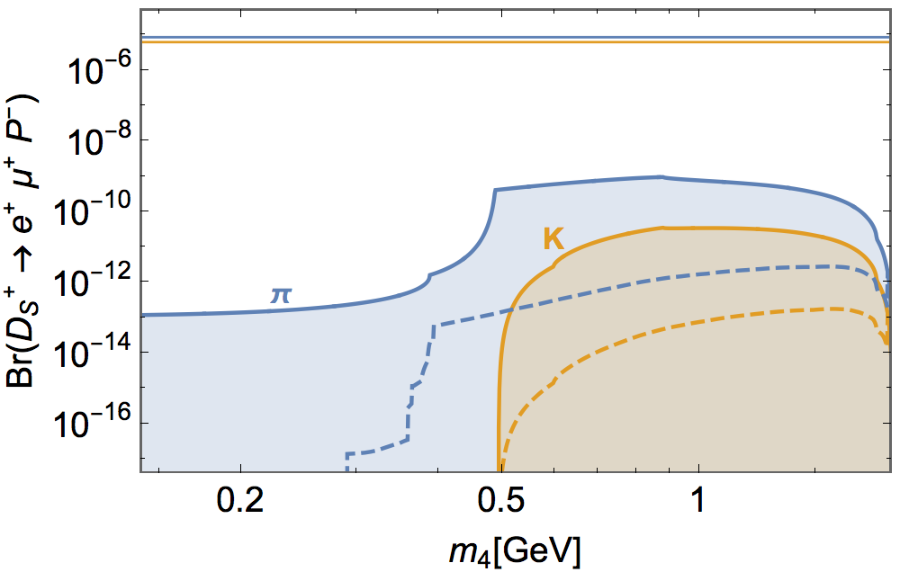,
width=73mm, angle =0} \hspace*{2mm}&\hspace*{2mm}
\epsfig{file=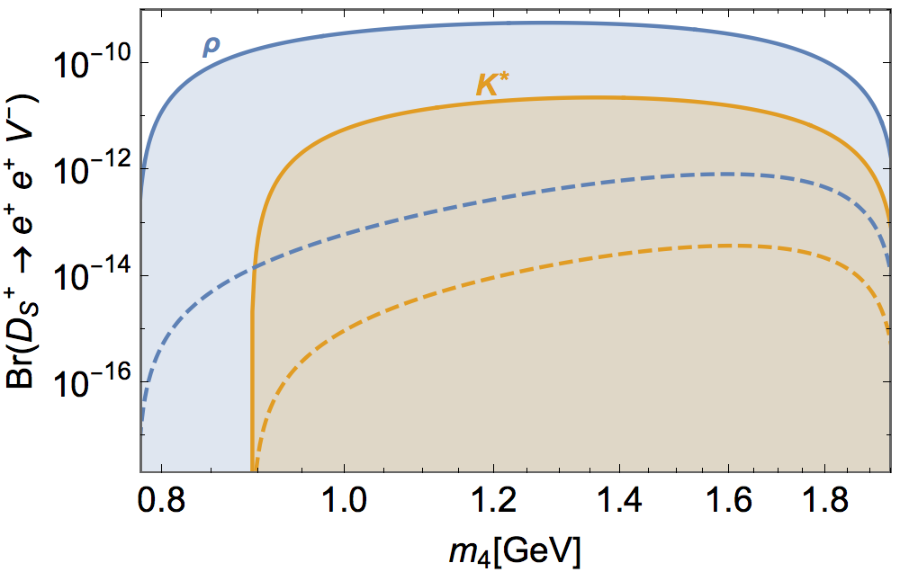,
width=73mm, angle =0}
\\
\epsfig{file=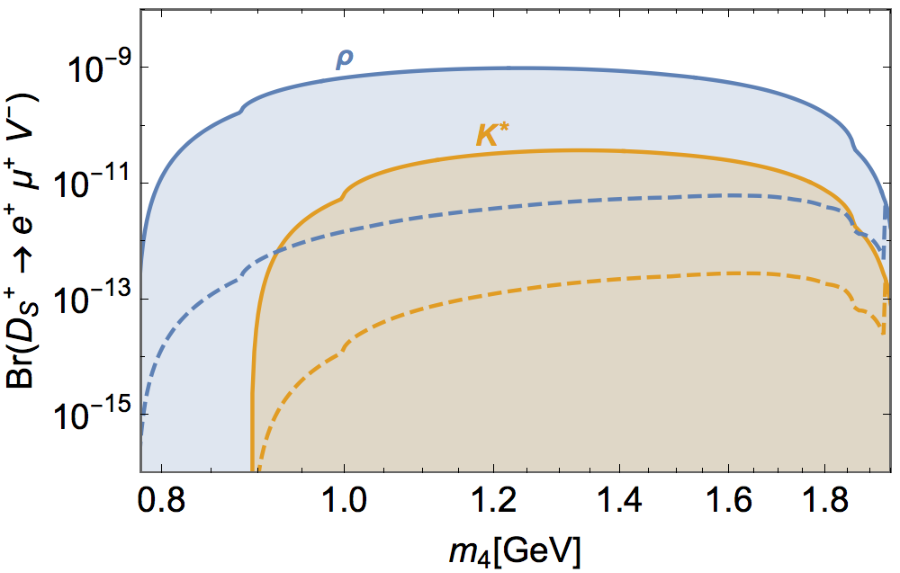,
width=73mm, angle =0} \hspace*{2mm}&\hspace*{2mm}
\epsfig{file=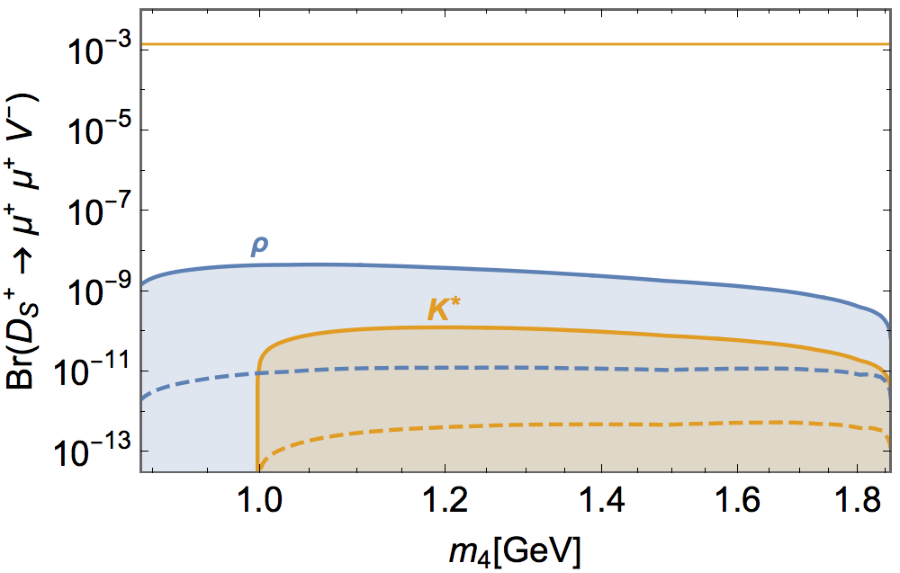,
width=73mm, angle =0}
\\
\epsfig{file=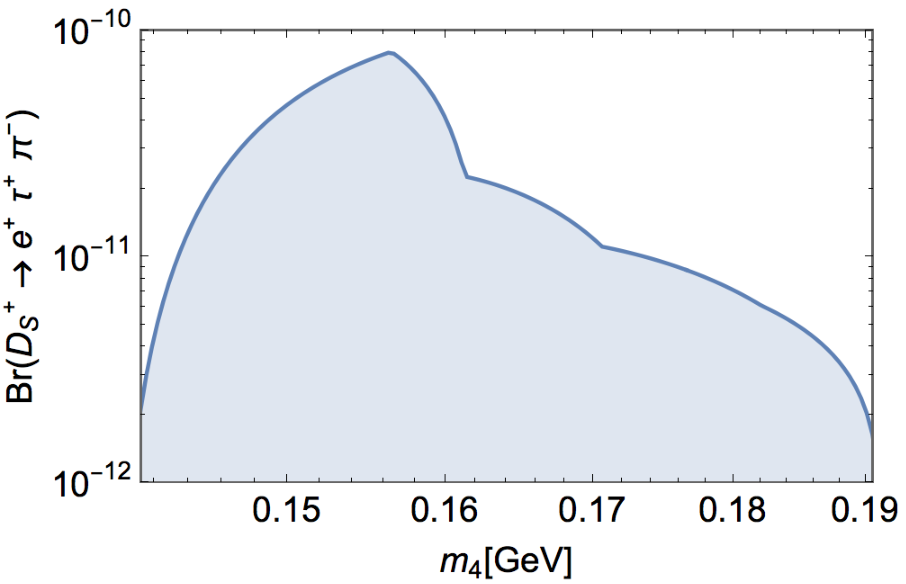,
width=73mm, angle =0}
\end{tabular}
\end{center}
\caption{Branching ratios of LNV decay modes of $D_s^+$
  mesons (into pseudoscalar and vector final states, plus 
  leptons of same or different flavour), 
  as a function of the (mostly sterile) heavy neutrino mass,
  $m_4$. Line and colour code as in Figs.~\ref{fig:res:LNV:Bdecays}
  and~\ref{fig:res:LNV:BdecaysV}.
}\label{fig:res:LNV:Dsdecays}
\end{figure}

We continue our discussion of LNV semileptonic meson decays by
investigating the prospects for $D_s$ and $D$ decays. 
For the latter decays, only
pions and kaons ($\rho$ and $K^*$) are candidates to 
the kinematically allowed final state pseudoscalar (vector) mesons. 
The results are summarised in the panels of
Fig.~\ref{fig:res:LNV:Dsdecays}. It is worth noticing that the maximal
allowed rates for $D_s^+ \to \mu^+\mu^+ \pi^-$ are very close to the
current experimental bound; in the near future, this observable can
thus play an important r\^ole in further constraining the
sterile neutrino degrees of freedom. 

\hspace*{-8mm}

\begin{figure}[h!]
\begin{center}
\begin{tabular}{cc}
\epsfig{file=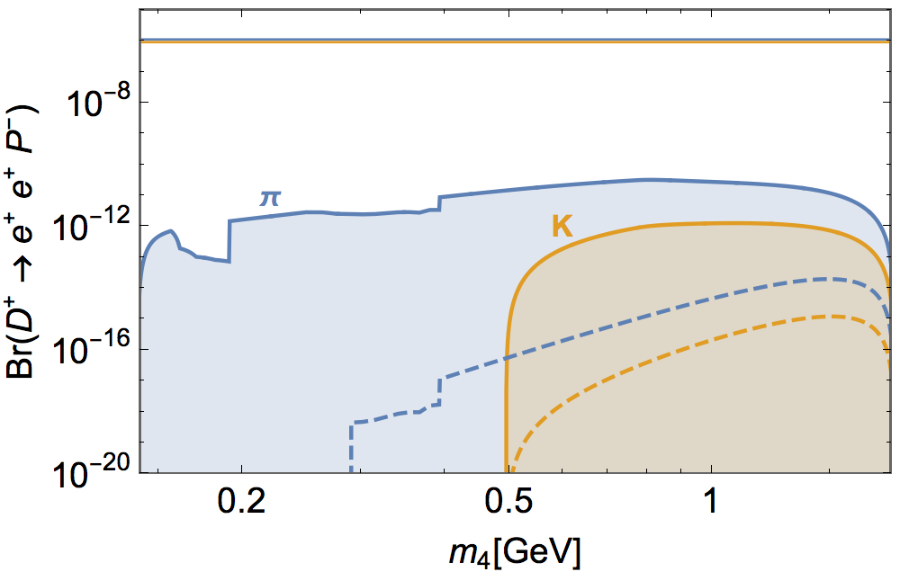,
width=73mm, angle =0} 
\hspace*{2mm}&\hspace*{2mm}
\epsfig{file=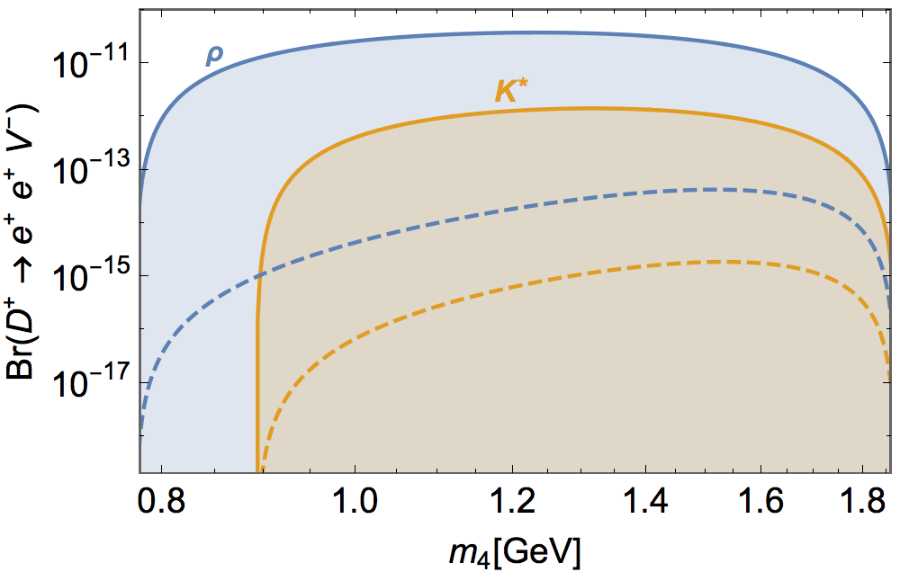,
width=73mm, angle =0} 
\\
\epsfig{file=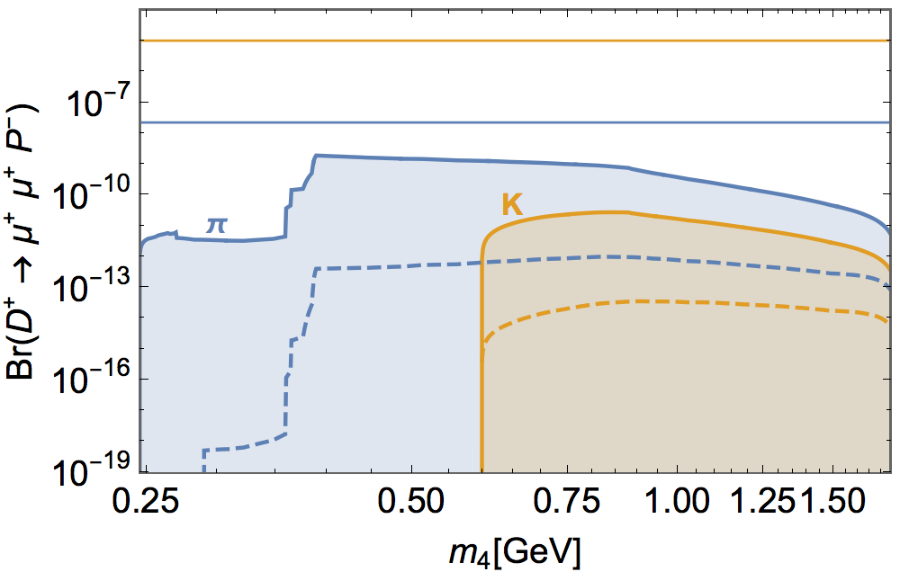,
width=73mm, angle =0}
\hspace*{2mm}&\hspace*{2mm}
\epsfig{file=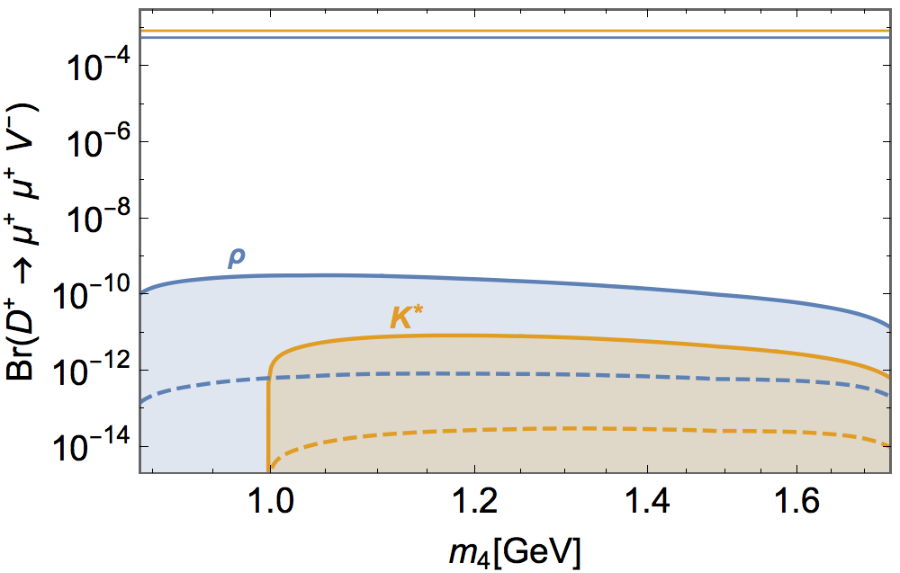,
width=73mm, angle =0}
\\
\epsfig{file=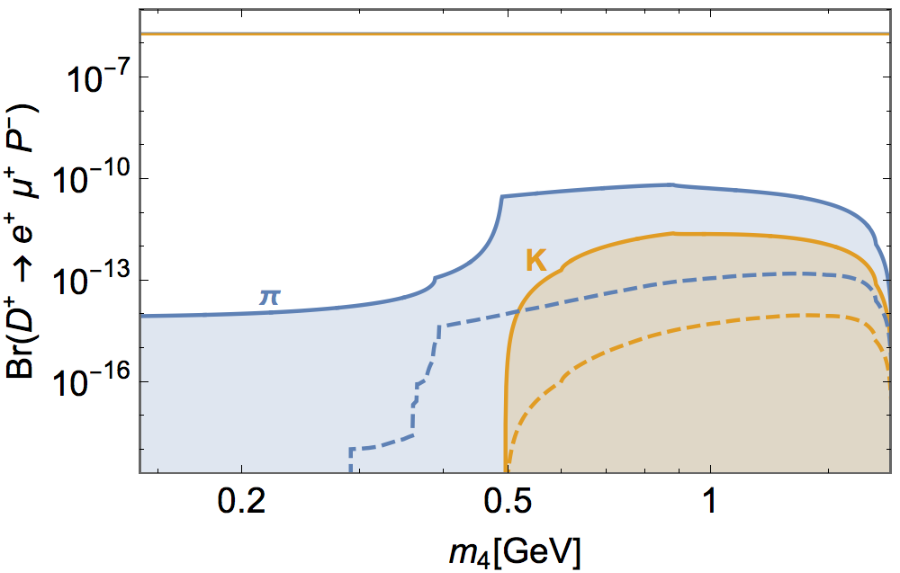,
width=73mm, angle =0}
\hspace*{2mm}&\hspace*{2mm}
\epsfig{file=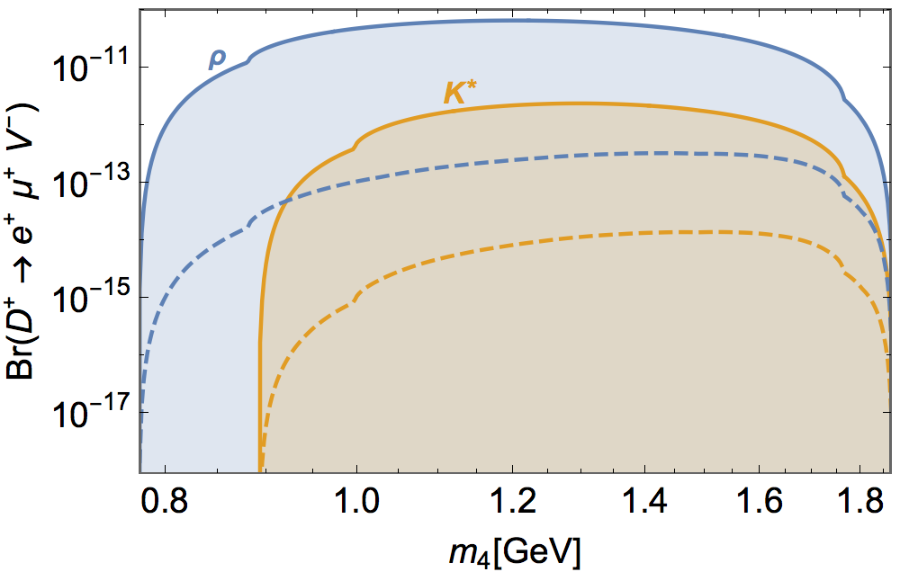,
width=73mm, angle =0} \hspace*{2mm}
\end{tabular}
\end{center}
\caption{Branching ratios of LNV decay modes of $D$
  mesons into pseudoscalar (left panels) and vector (right panels) final states, plus leptons of same or different flavour, 
  as a function of the (mostly sterile) heavy neutrino mass,
  $m_4$. Line and colour code as in Fig.~\ref{fig:res:LNV:BdecaysV}.
}\label{fig:res:LNV:Ddecays}
\end{figure}

For completeness, we also display in Fig.~\ref{fig:res:LNV:Ddecays} 
the LNV decays of $D$ mesons into pseudoscalar and
vector meson final states (accompanied by leptons of same or different
flavour), $D \to \ell_\alpha \ell_\beta V$, with $V=\rho, K^*$.

\hspace*{-8mm}

\subsubsection{$\bm{K}$ meson decays} 
\begin{figure}[h!]
\begin{center}
\epsfig{file=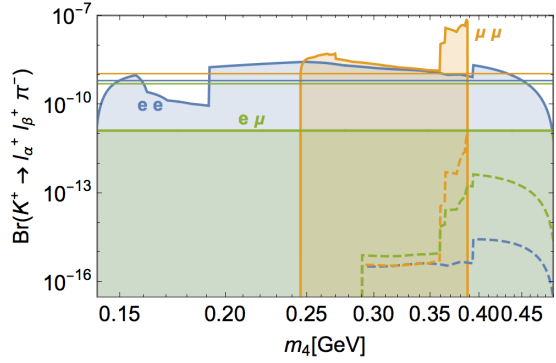,
width=73mm, angle =0}
\end{center}
\caption{Branching ratios of LNV decay modes of $K$
  mesons, $K^+ \to \ell_\alpha^+ \ell_\beta^+ \pi^-$,
  as a function of the (mostly sterile) heavy neutrino mass,
  $m_4$. 
  Pale blue, yellow and green curves (surfaces) respectively denote
  the maximal (allowed) values of the BR($K^+ \to \ell_\alpha^+
  \ell_\beta^+ \pi^-$), 
  with $( \ell_\alpha^+ \ell_\beta^+)= (e^+ e^+, \mu^+ \mu^+, e^+ \mu^+)$;
  coloured dashed curves denote  the corresponding maximal values of the BRs once
  $c\tau_4 \leq 10$~m is imposed.
  The coloured horizontal lines correspond to the present experimental bounds
  (cf. Table~\ref{tab:LNV:meson:exp}). 
  }\label{fig:res:LNV:Kdecays}
\end{figure}

We conclude by presenting in Fig.~\ref{fig:res:LNV:Kdecays} the
prospects for the LNV decays of the $K$ meson into a pion and a pair
of leptons: for the $ee$ and  $\mu \mu$  final states the maximal
expected branching ratios exceed the current experimental bounds over
the majority of the considered sterile neutrino mass range 
(furthermore, the $e\mu$ channel is
strongly constrained by the corresponding cLFV bound), although
realistic observable rates are strongly suppressed by the upper bound
on the sterile neutrino lifetime. Kaons are nevertheless an
excellent framework to probe new physics via contributions to 
LNV decays, since the already
existing stringent bounds reported in Table~\ref{tab:LNV:meson:exp}
are likely to be improved in the near future by the NA62 collaboration, 
with an expected sensitivity of $\mathcal{O}(10^{-11})$ for the
  $ee$ and $e\mu$ channels, and $\mathcal{O}(10^{-12})$ for the
$\mu\mu$ channel\footnote{E. Minucci, private communication.}: 
should any hint of LNV be manifest in
kaon decays, the results here presented  
can be useful in disentangling between the
sterile neutrino hypothesis or other different mechanisms at its origin. 

\hspace*{-8mm}

\subsubsection{Tau-lepton decays}\label{sec:results:LNVtau}
Due to its large mass, a tau lepton can also decay semileptonically; the
kinematically allowed channels (always for the case of a primary
production of an on-shell mostly sterile heavy neutrino) comprise
several $\Delta L=2$ final states, including electrons and muons, 
as well as light kaons and pions.   

\begin{figure}[t!]
\begin{center}
\begin{tabular}{cc}
\epsfig{file=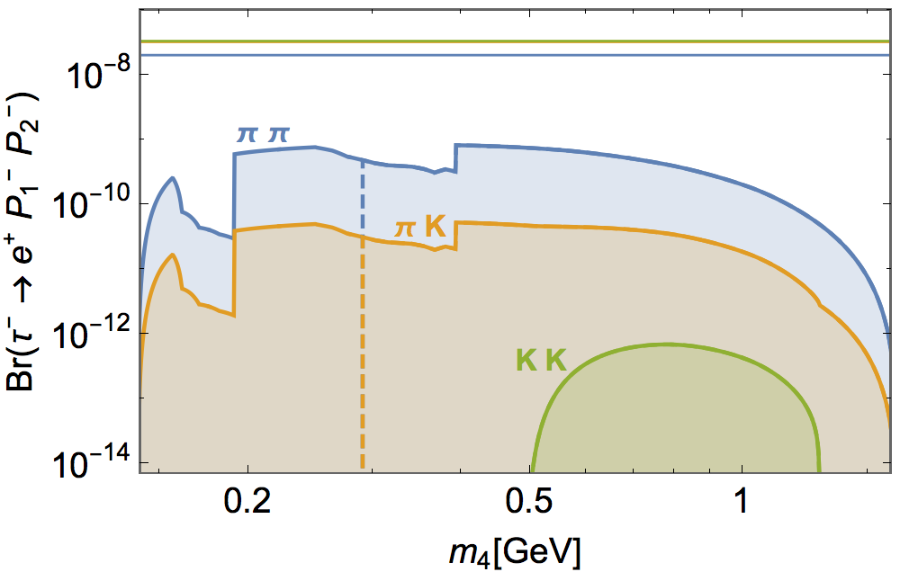,
width=73mm, angle =0} 
\hspace*{2mm}&\hspace*{2mm}
\epsfig{file=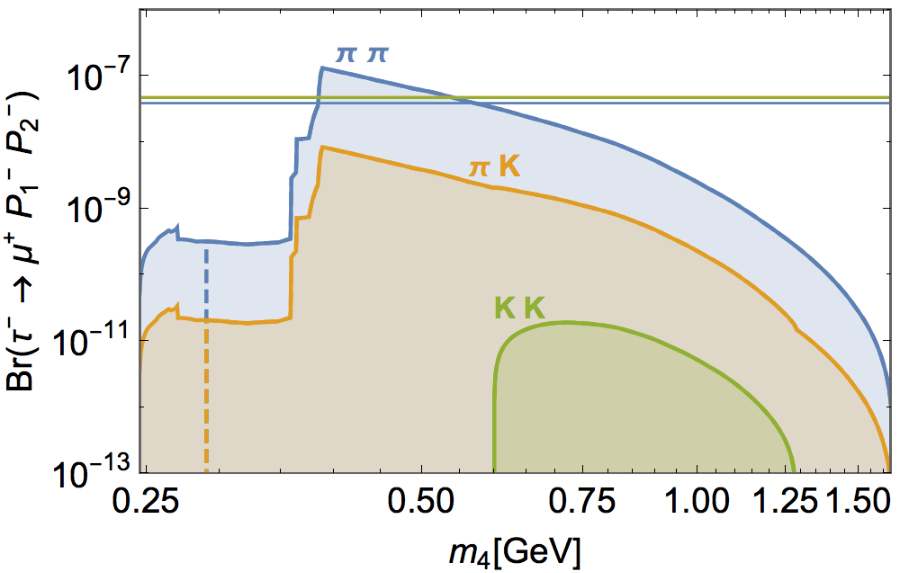,
width=73mm, angle =0}
\end{tabular}
\end{center}
\caption{Branching ratios of LNV decay modes of tau
  leptons, $\tau^- \to \ell_\alpha^+ P_1^- P_2^-$,
  as a function of the (mostly sterile) heavy neutrino mass,
  $m_4$. 
  Pale blue, yellow and green curves (surfaces) respectively denote
  the maximal (allowed) values of the BR($\tau^- \to \ell_\alpha^+ M_1^- M_2^-$),
  with $(M_1^- M_2^-)= (\pi^-\pi^-, \pi^-K^-, K^-K^-)$; coloured
  dashed curves denote  
  the corresponding maximal values of the BRs once
  $c \tau_4 \leq 10$~m is imposed. The coloured horizontal
  lines correspond to the present experimental bounds
  (cf. Table~\ref{tab:LNV.taudecay.exp}). 
}\label{fig:res:LNV:taudecays}
\end{figure}

The two panels of Fig.~\ref{fig:res:LNV:taudecays} display the 
branching ratios for final states with either an electron or a muon. 
It is interesting to notice that for the decay mode 
$\tau^- \to \mu^+ \pi^- \pi^-$, a heavy neutrino with a mass between 
0.4~GeV and 0.55~GeV can be at the origin of decay widths already 
in conflict with experimental observation. Lepton number violation, in
association with the $\tau-\mu$ sector, thus emerges as a new
constraint that must be taken into account in current analyses. 
Given the sizeable branching fractions for both 
$\mu^+ \pi^-\pi^-$ and $\mu^+ \pi^-K^-$ final states, any improvement
in the experimental sensitivity (at LHCb, or Belle II) will render the
LNV and cLFV observables a source of stringent constraints for models
with additional Majorana neutrinos with masses below 1~GeV. 
Conversely, these channels might also play a discovery r\^ole 
for new physics sources of LNV. 

\hspace*{-8mm}

\subsection{Impact for the effective mass matrix: constraining 
$\bm{m_\nu^{\alpha \beta}}$}

One of the most important implications of the study of semileptonic
LNV decays is the extraction of bounds on the relevant entries of the
 $3\times 3$ effective  neutrino Majorana mass matrix, $m_\nu^{\alpha \beta}$. 
With the exception of the decays leading to $\Delta L_e =2$, in which
case the most constraining bound on $m_\nu^{ee}$ stems from
neutrinoless double beta decays, the results we present emphasise
to which extent LNV semileptonic decays allow to
constrain the effective neutrino mass matrix.

As mentioned in the Introduction, past analyses have contributed to the
derivation of upper bounds for certain entries (for 
instance in~\cite{Flanz:1999ah,Zuber:2000vy,Zuber:2000ca,Atre:2005eb, 
Rodejohann:2011mu,Liu:2016oph,Quintero:2016iwi}); 
it is important to notice that contrary to the neutrinoless double
beta decay process, the dependence on the exchanged momentum 
(see Eq.~(\ref{eq:effective.mass.def.3}),
Section~\ref{sec:effective.mass.th}) - and hence on the masses of 
the different states (fermions and mesons) - does not favour a
generalisation of the bounds. In this section, we thus discuss  
the expected sensitivity to the entries of the effective
neutrino mass matrix, as inferred from the {\it distinct} three-body 
LNV semileptonic decays here addressed. 

We begin by discussing the allowed
values of the effective neutrino mass as a function of the mass of the
heavy (mostly sterile) fermion as obtained from specific 
semileptonic decays. In particular, we choose as illustrative
examples the predictions for the six independent entries of 
$m_\nu^{\alpha \beta}$ obtained from the LNV decays 
$B \to \ell_\alpha \ell_\beta \pi$ (allowing for a large interval of
sterile neutrino masses). These are complemented by the constraints for 
$m_\nu^{\mu \mu}$ extracted from 
$D \to \mu \mu \pi$ LNV decays,  and $m_\nu^{\tau \mu}$, which is directly
inferred from semileptonic  tau LNV decays ($\tau \to \mu \pi \pi$).

Depending on the heavy neutrino length of flight (which is in turn a
consequence of its mass and mixings - see
Section~\ref{sect:finitedectector}), the LNV decays might occur
outside a detector of finite size; leading to the results of all
subsequent figures, we systematically exclude regimes which would be
associated with $L^\text{flight}_{\nu_4} \gtrsim 10$~m (denoted also by
grey points).

\begin{figure}[h!]
\begin{center}
\begin{tabular}{cc}
\epsfig{file=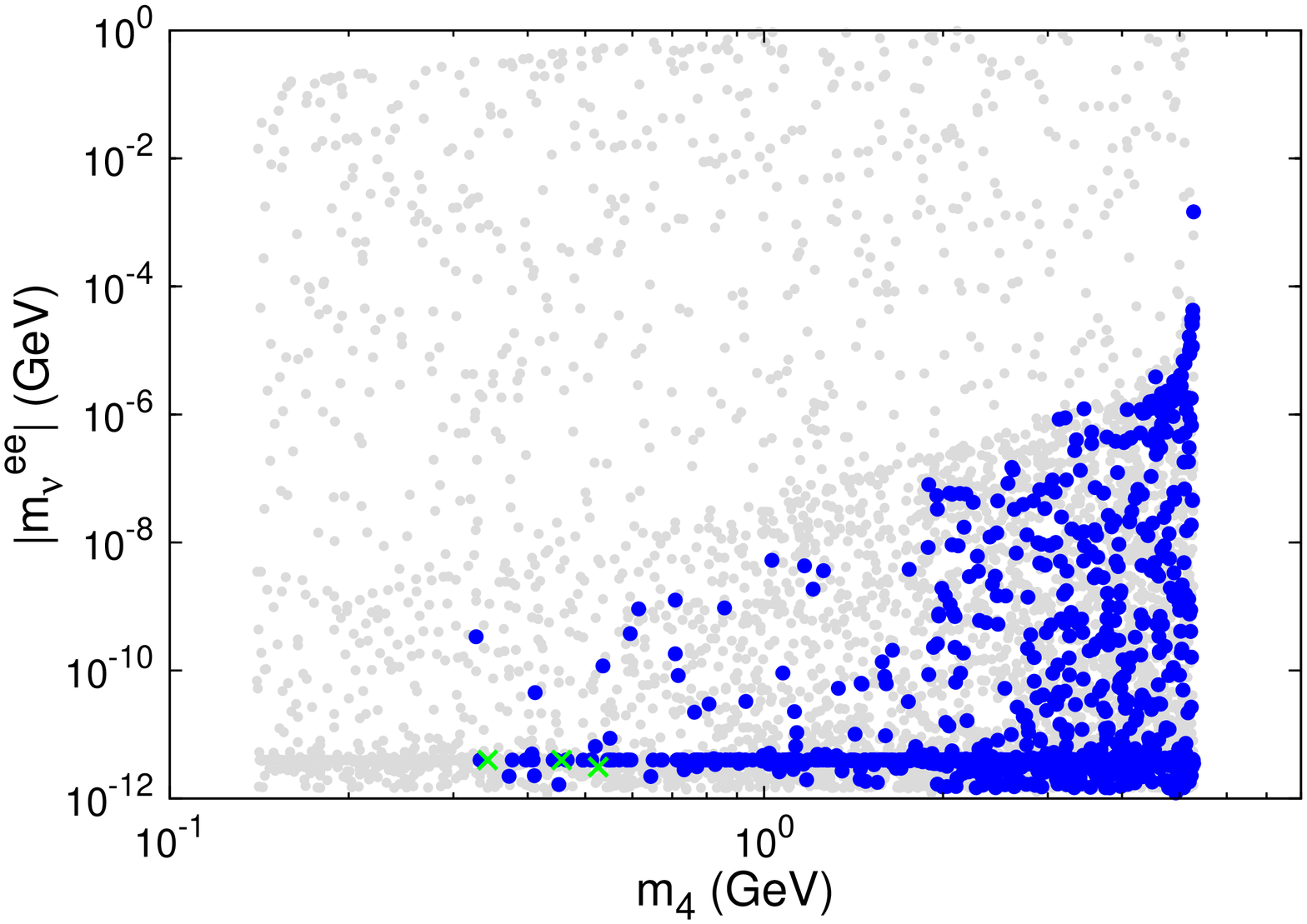,
width=53mm, angle =270}
\hspace*{2mm}&\hspace*{2mm}
\epsfig{file=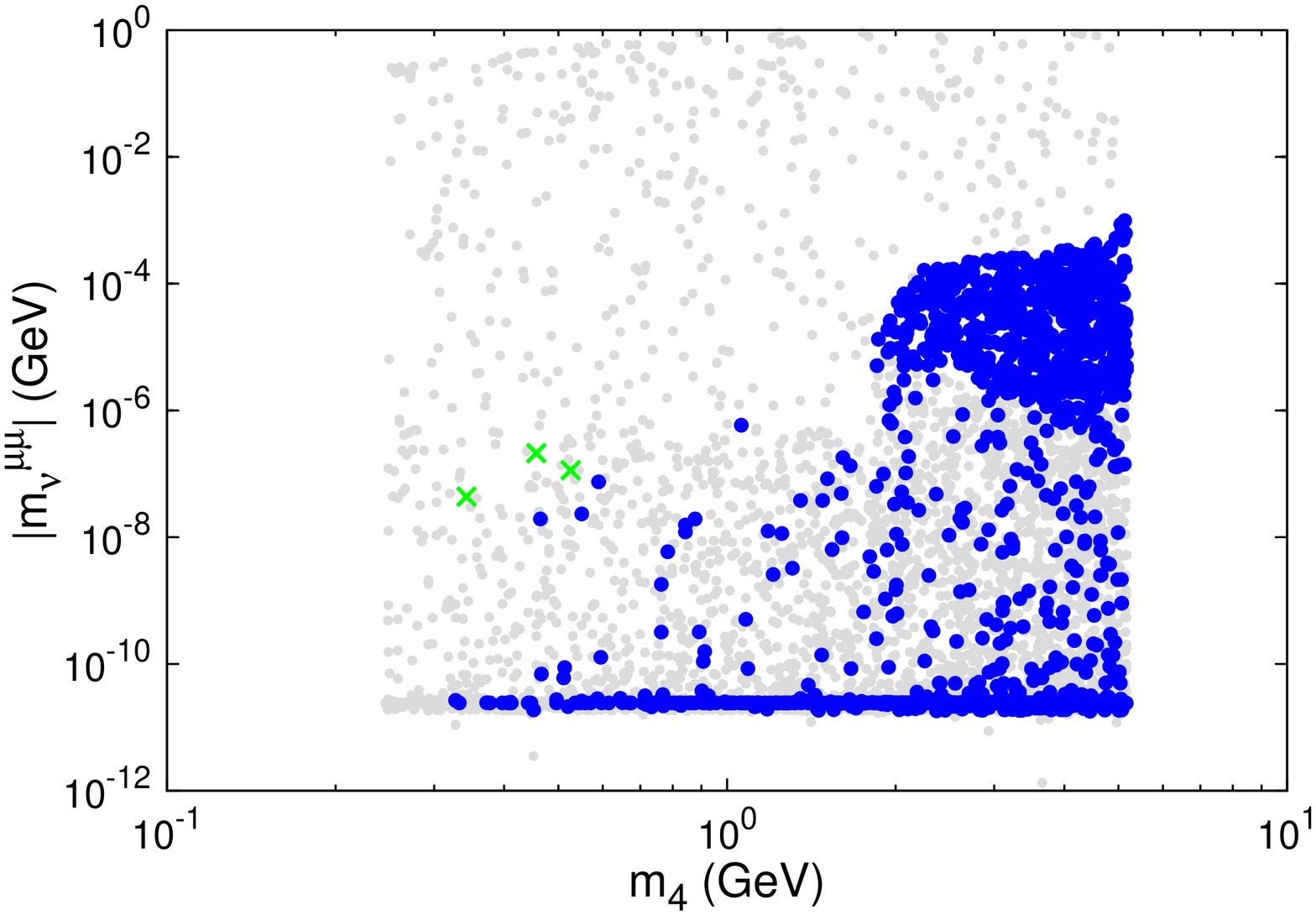,
width=53mm, angle =270}
\\
\epsfig{file=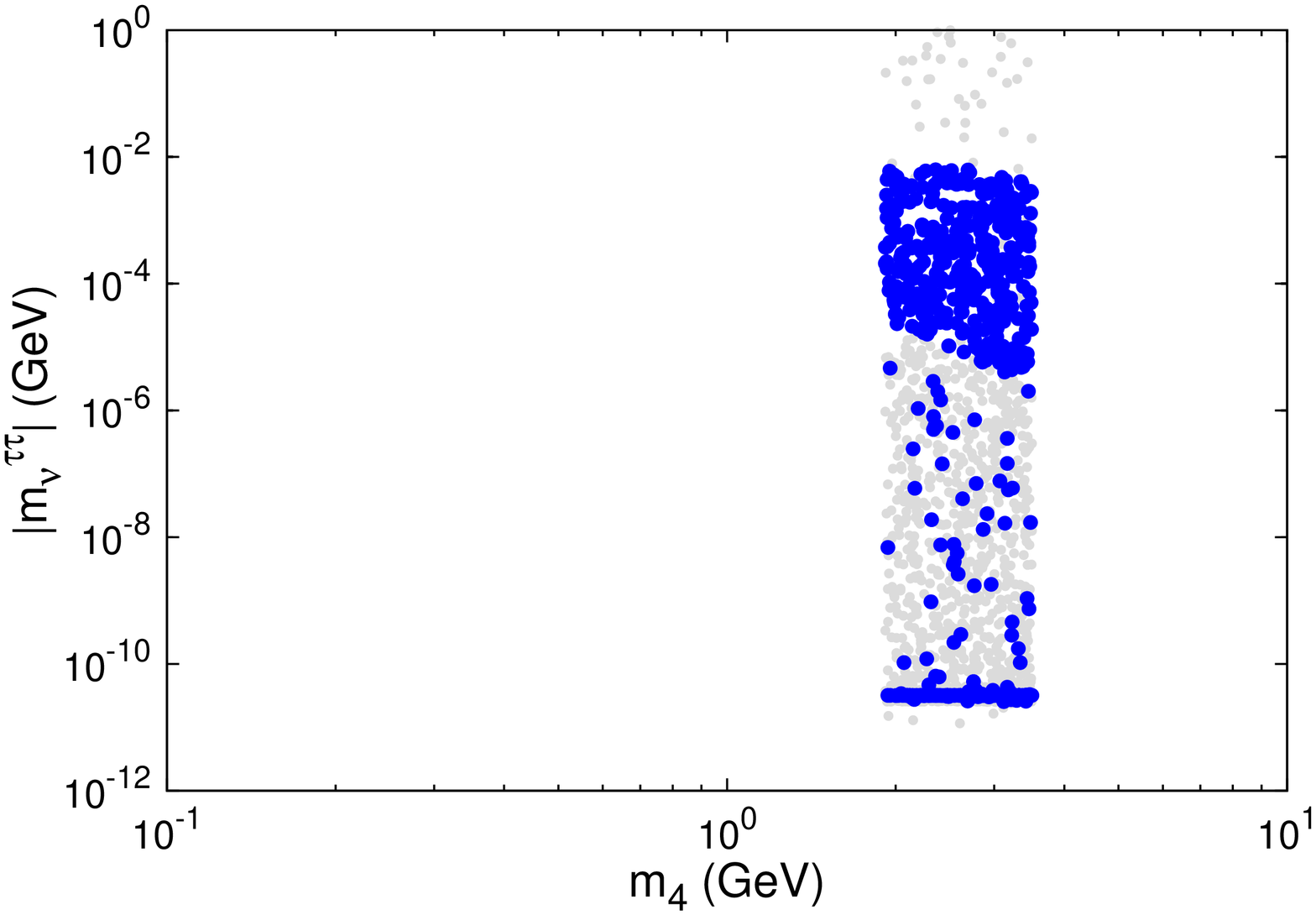,
width=53mm, angle =270}
\hspace*{2mm}&\hspace*{2mm}
\epsfig{file=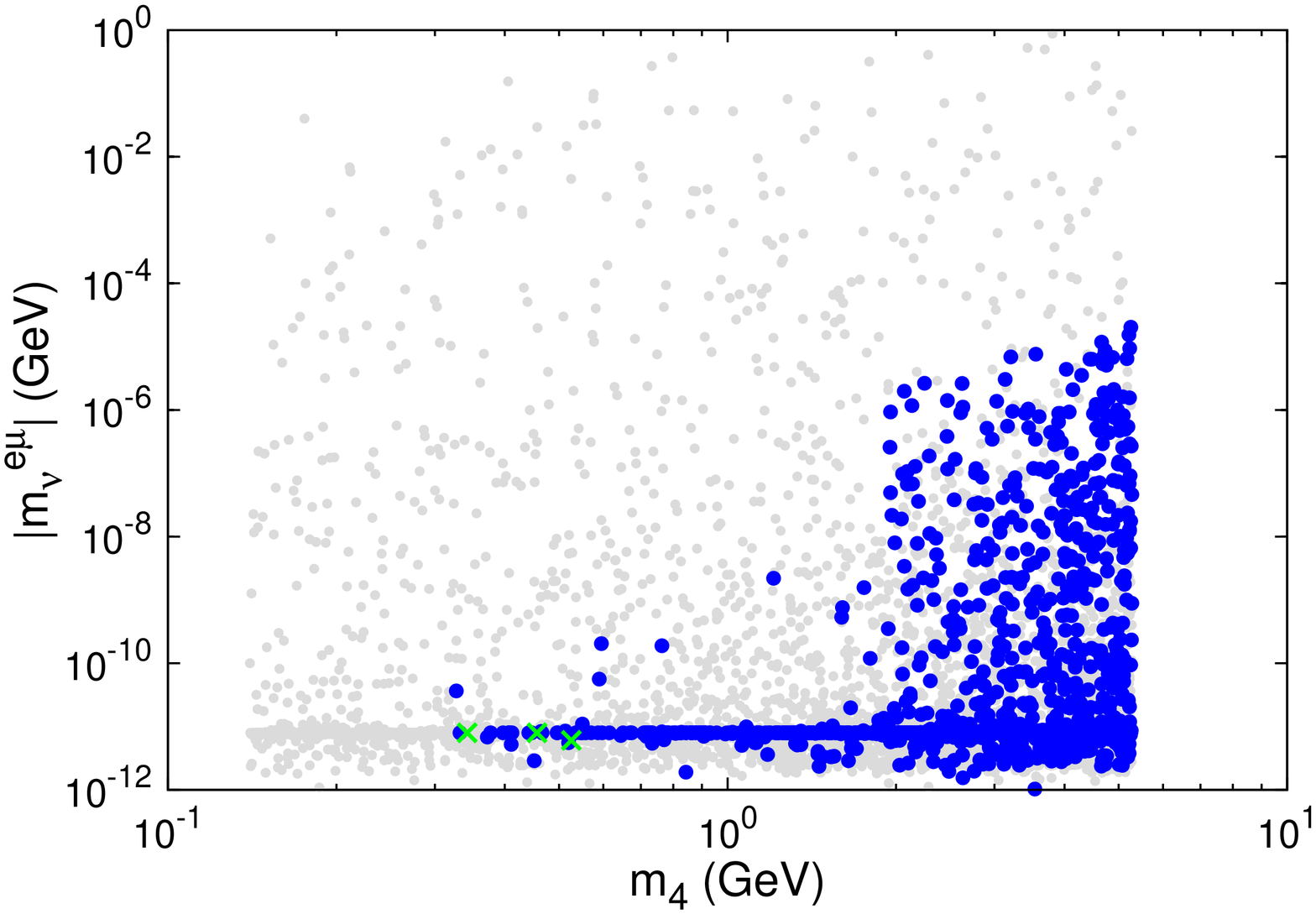,
width=53mm, angle =270}
\\
\epsfig{file=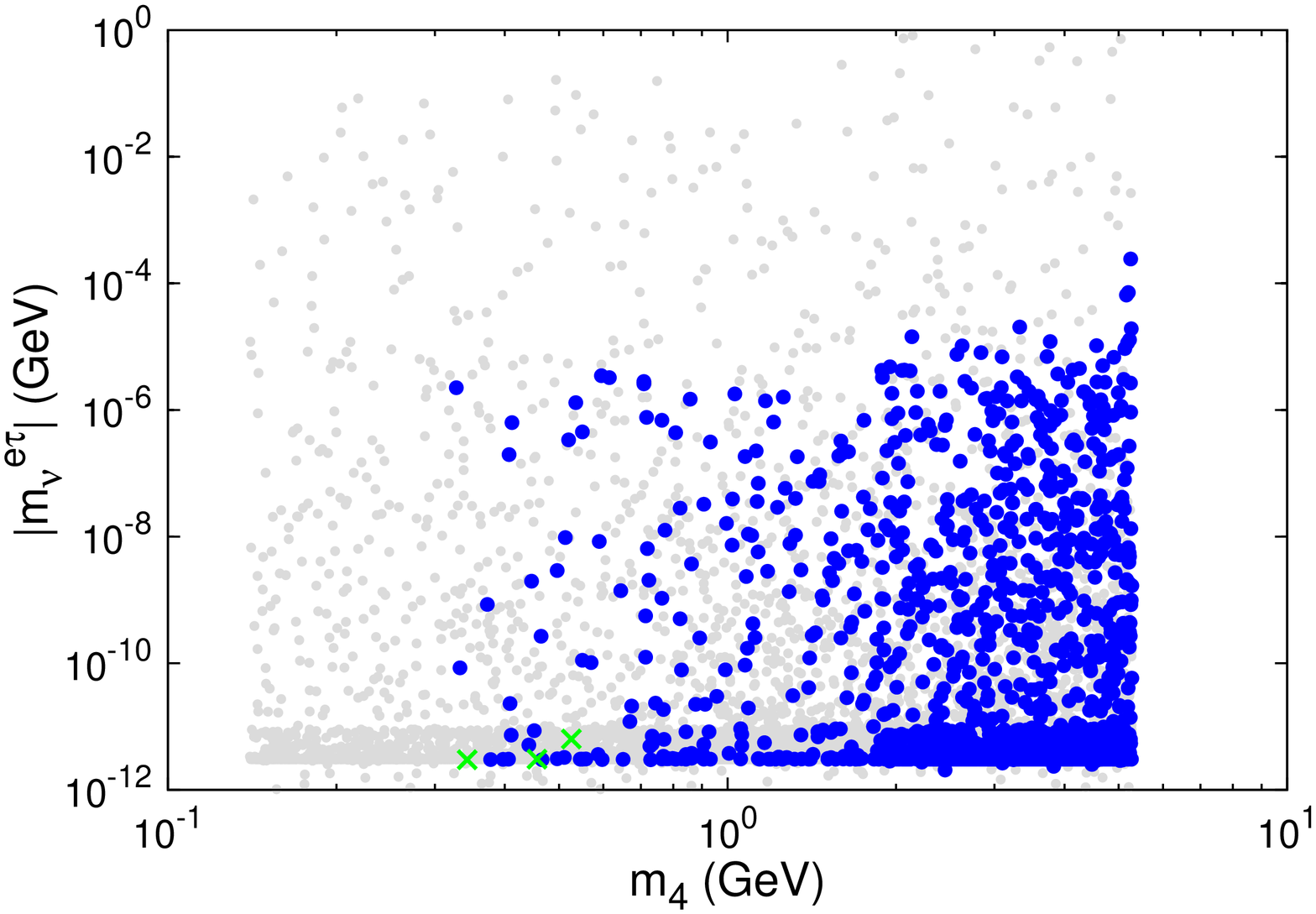,
width=53mm, angle =270}
\hspace*{2mm}&\hspace*{2mm}
\epsfig{file=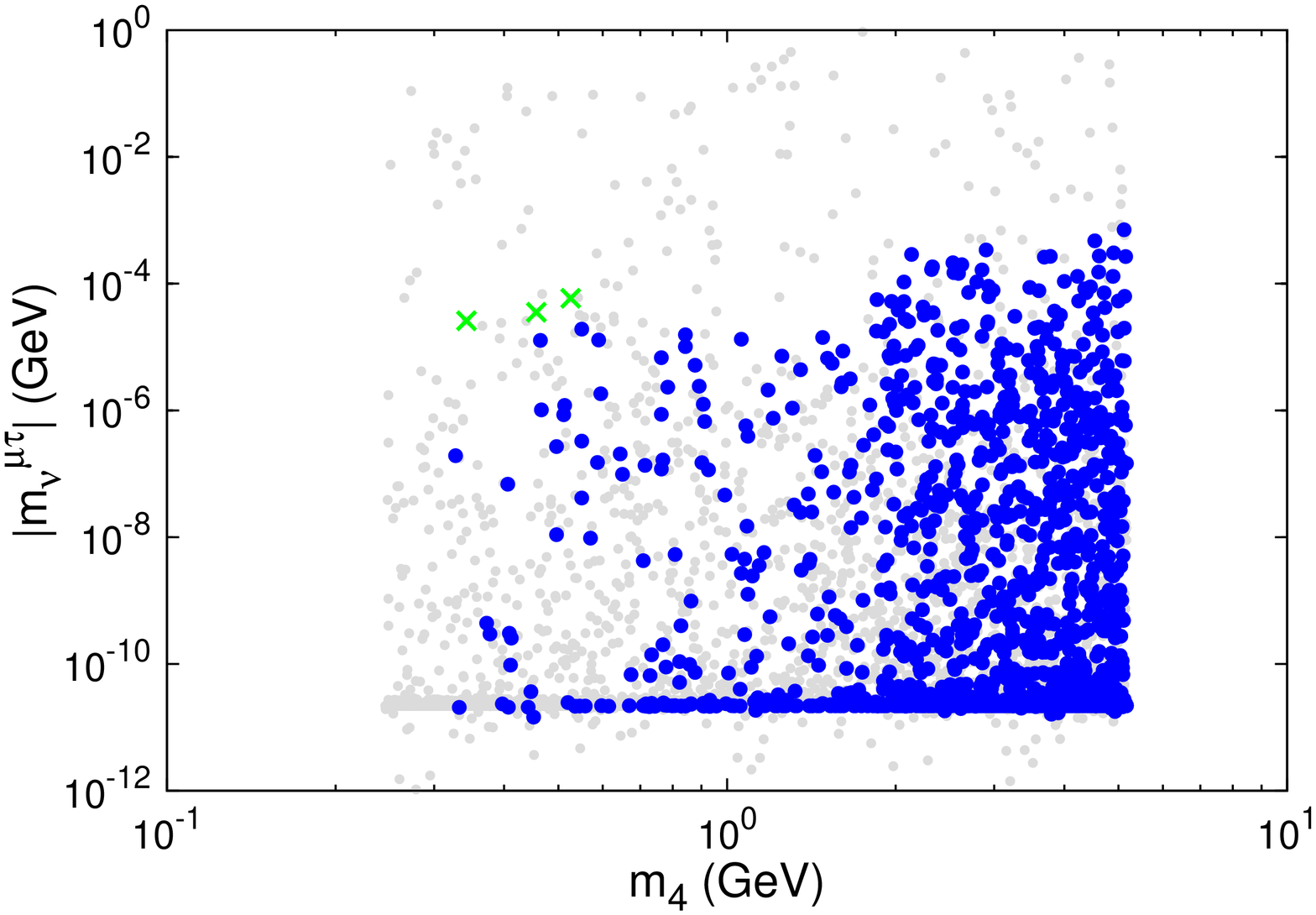,
width=53mm, angle =270}
\end{tabular}
\end{center}
\caption{Predictions for the effective mass $m_\nu^{\alpha \beta}$ 
as a function of the heavy (mostly sterile) fermion mass, $m_4$,
derived from the corresponding $B \to \ell_\alpha \ell_\beta \pi$ 
 LNV decay modes. 
Grey points are excluded due to the
violation of at least one experimental or observational constraint (or
to having $L^\text{flight}_{\nu_4} \gtrsim 10$~m), 
while the green crosses denote points 
which are excluded due to excessive contributions to LNV decays.
}\label{fig:res:m_eff.m4.Bpi}
\end{figure}

In Fig.~\ref{fig:res:m_eff.m4.Bpi}, the different panels display the
values of $|m_\nu^{\alpha \beta}|$ for the $m_4$ intervals kinematically
allowed. (Leading to this figure, and to all subsequent ones, the
underlying scan is similar to the one described for 
Fig.~\ref{fig:LNV.cLFV:correl} in Section~\ref{sec:Bdecays}.)
Two regimes can be identified for the experimentally allowed
points (blue): the contributions arising from the light (active) 
neutrinos, associated with the saturation (mostly independent
of $m_4$) observed for the smallest values of $|m_\nu^{\alpha
  \beta}|$; the contributions from the heavy additional state, which
span over several orders of magnitude (in GeV). 
The direct and indirect bounds on the branching ratios 
(as discussed in the previous
subsection), typically translate into upper
bounds of around $10^{-3}$ to $10^{-4}$~GeV for the 
entries of the effective mass
matrix, with the exception of the $\tau\tau$ entry.
This already implies an improvement with respect to the results 
of~\cite{Flanz:1999ah,Zuber:2000vy,Atre:2005eb,GomezCadenas:2011it}. 
In particular, our bounds ameliorate those 
of~\cite{Atre:2005eb,GomezCadenas:2011it} by 2 up to 7  orders of
magnitude for the $\mu \mu$, $e \tau$ and $\mu \tau$ 
entries\footnote{Further bounds on LNV $\mu-e$ transitions,
  and hence on $|m_\nu^{e\mu}|$ can
be obtained from searches for lepton number violating neutrinoless
$\mu-e$ conversion in Nuclei ($\mu^- - e^+$, N); these were reported
to be $\mathcal{O}(10^{-2}\text{ GeV})$ by~\cite{GomezCadenas:2011it}.}.

\begin{figure}[h!]
\begin{center}
\begin{tabular}{cc}
\epsfig{file=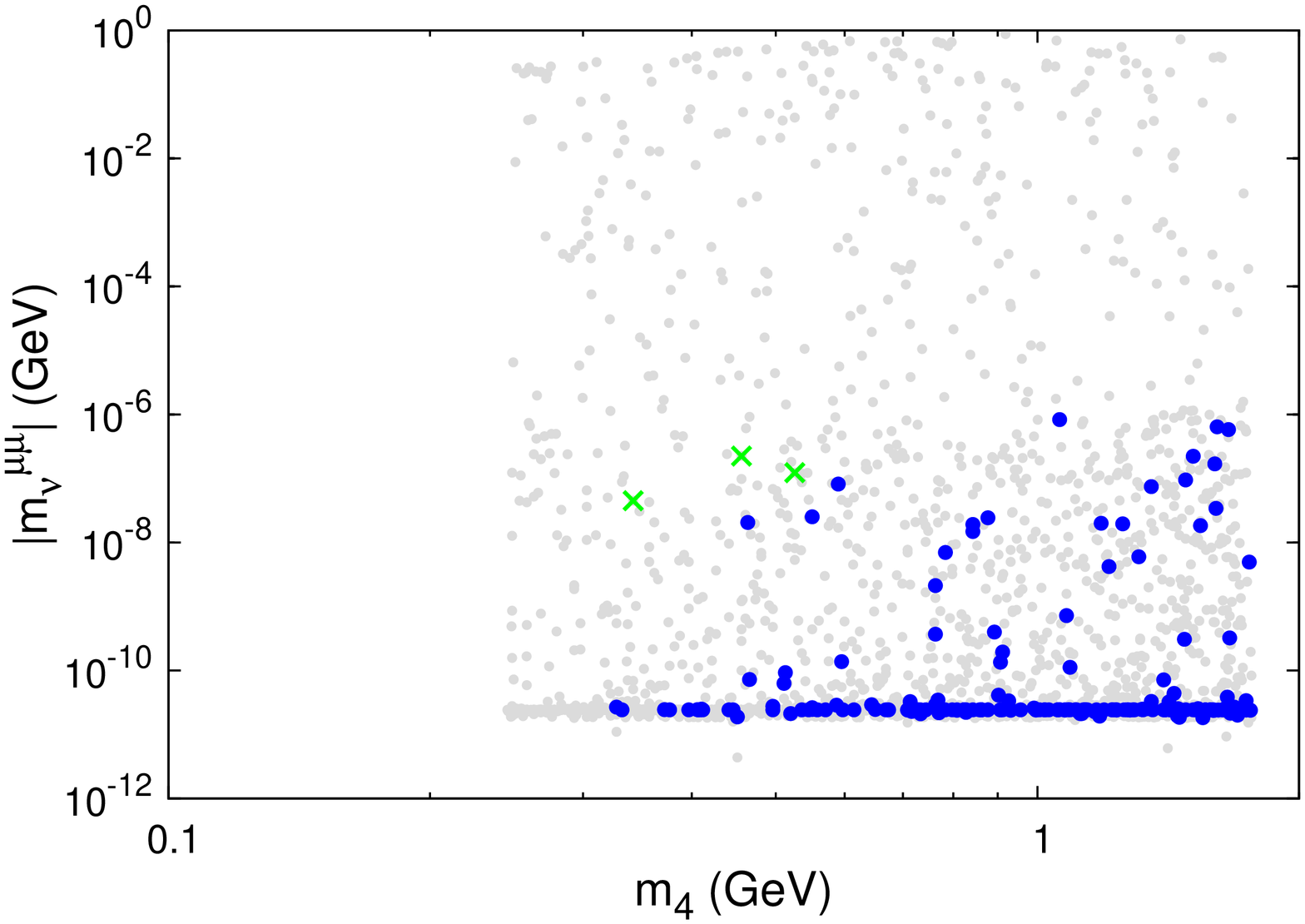,
width=53mm, angle =270}
\hspace*{2mm}&\hspace*{2mm}
\epsfig{file=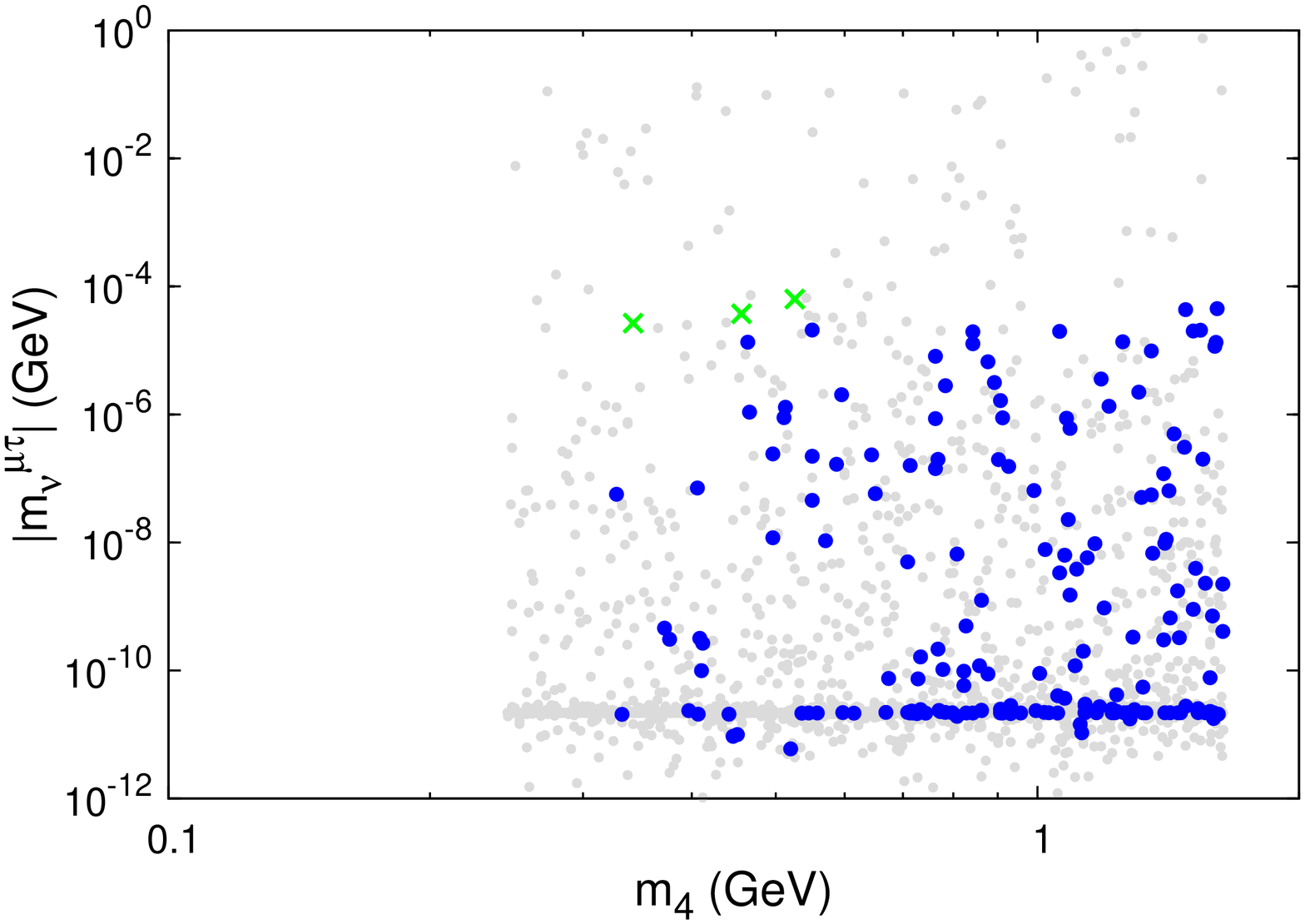,
width=53mm, angle =270}
\end{tabular}
\end{center}
\caption{Predictions for the effective mass $m_\nu^{\alpha \beta}$ 
as a function of the heavy (mostly sterile) fermion mass, $m_4$,
derived from the LNV $D \to \mu \mu \pi$ and $\tau \to \mu \pi \pi$
decay modes. Colour code as in Fig.~\ref{fig:res:m_eff.m4.Bpi}.
}\label{fig:res:m_eff.m4.D.tau}
\end{figure}

Other than being dependent on the sterile neutrino mass and mixings,
notice that  
these bounds are intrinsically related to the actual
LNV process. This is manifest when comparing the $\mu \mu$ panel of 
Fig.~\ref{fig:res:m_eff.m4.Bpi} with the left panel of 
Fig.~\ref{fig:res:m_eff.m4.D.tau}, in which the bounds on 
$|m_\nu^{\mu \mu}|$ are displayed - but now arising
from decays of $D$ mesons, $D \to \mu \mu \pi$: the bounds are
considerably stronger, leading in general to 
$|m_\nu^{\mu \mu}| \lesssim 10^{-6}$~GeV. Likewise, the bounds on 
$|m_\nu^{\mu \tau}|$ obtained from LNV $B$ decays can be compared with
those obtained from semileptonic tau decays (right panel of 
Fig.~\ref{fig:res:m_eff.m4.D.tau}): in this case both channels lead to
similar constraints. 
 
A final remark concerns the last entry, $m_\nu^{\tau \tau}$, for
which the available bounds are typically much less stringent.
Although a same-sign tau pair in the final state is not always
kinematically accessible, there are nevertheless decays, such as 
$B \to \tau \tau \pi$, which allow to constrain $m_\nu^{\tau \tau}$. 
As can be seen from the relevant panel of
Fig.~\ref{fig:res:m_eff.m4.Bpi}, one can already achieve important
constraints, $|m_\nu^{\tau \tau}| \lesssim 10^{-2}$~GeV reflecting a
significant improvement with respect to former results (bounds on 
$|m_\nu^{\tau \tau}| \lesssim 10^{4}$~GeV were suggested from an
analysis of LNV conducted for searches at HERA~\cite{Flanz:1999ah}).

\hspace*{-8mm}

\hspace*{-8mm}

\begin{figure}[h!]
\begin{center}
\begin{tabular}{cc}
\epsfig{file=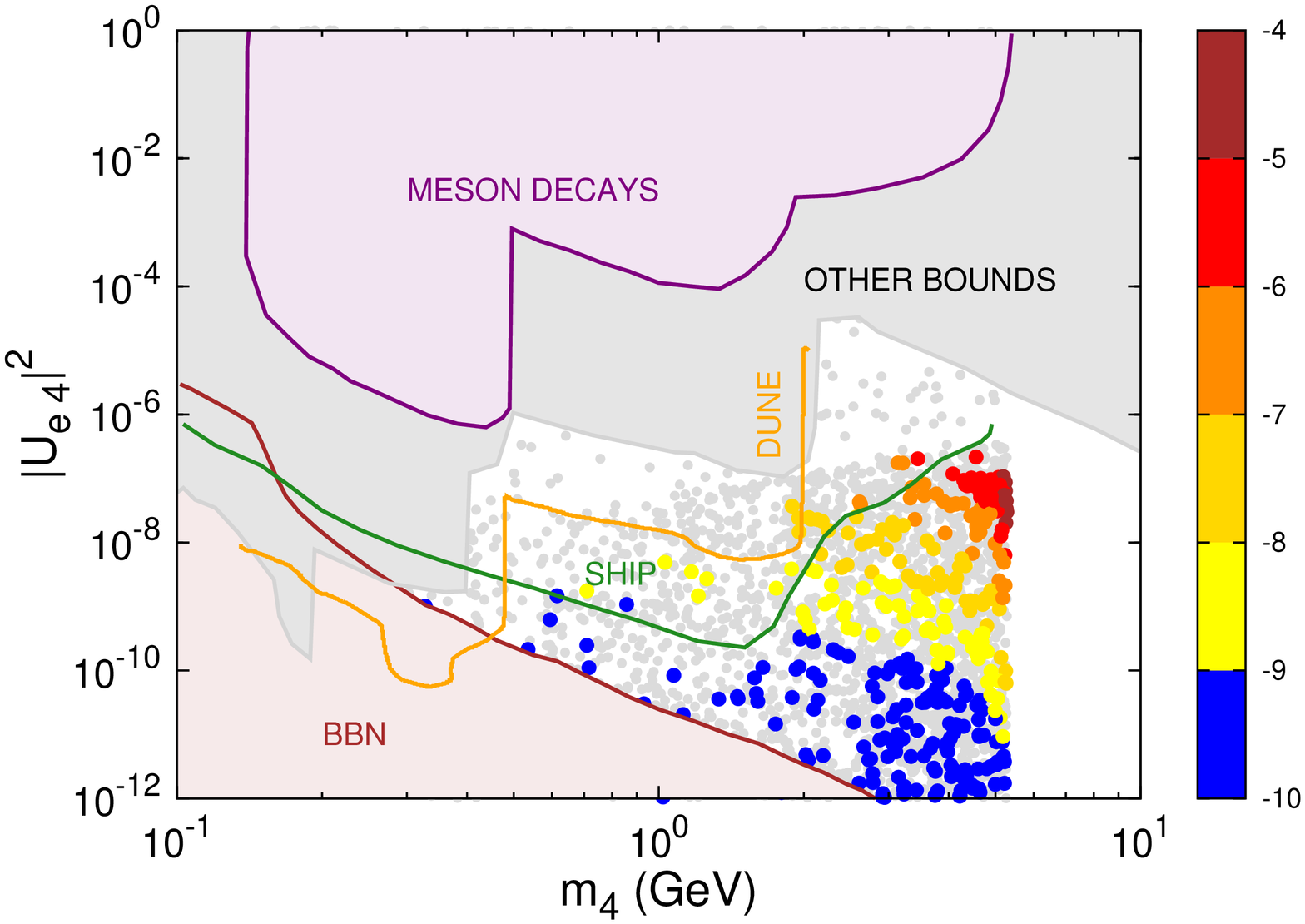,
width=53mm, angle =270}
&
\epsfig{file=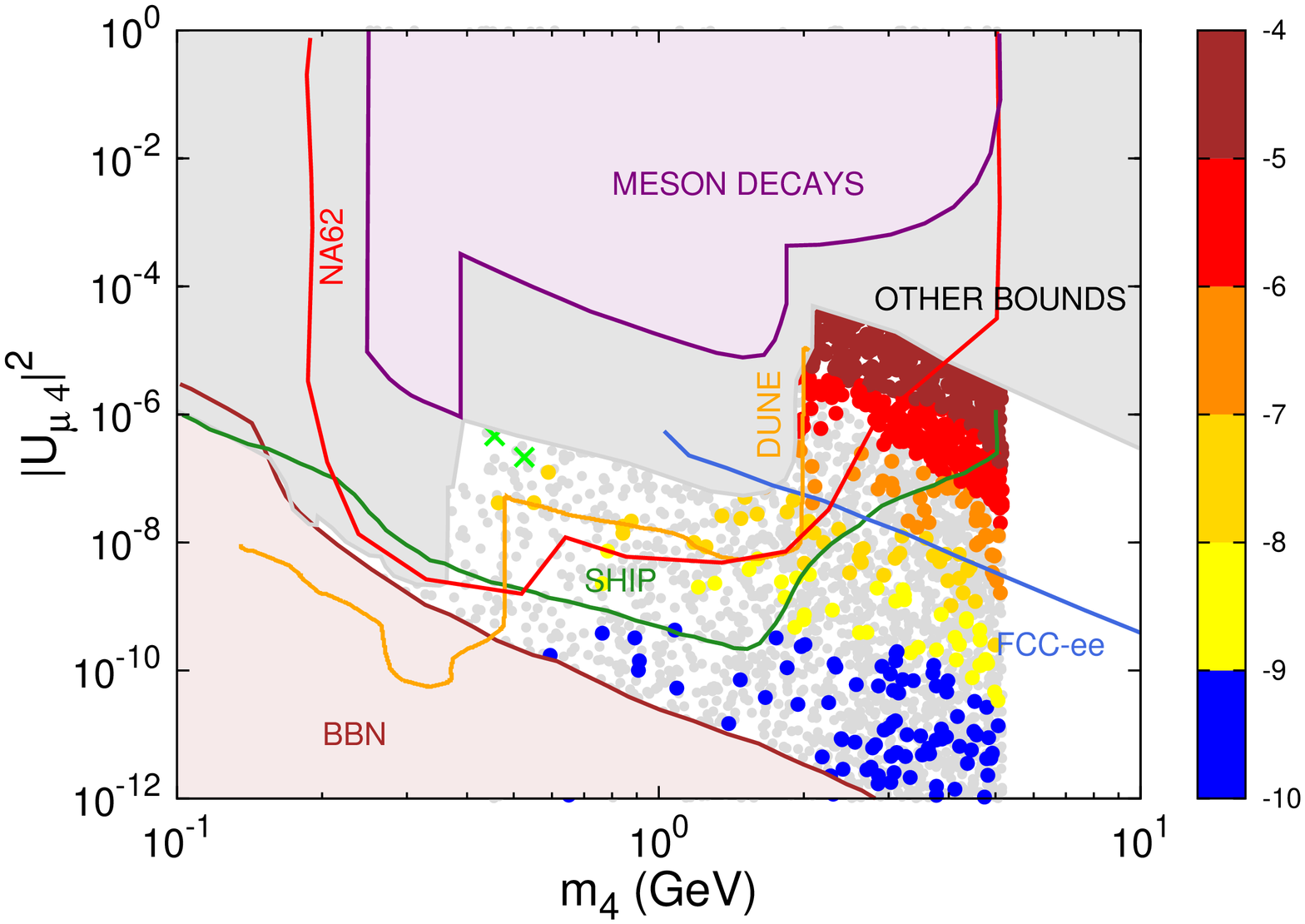,
width=53mm, angle =270}
\\
\epsfig{file=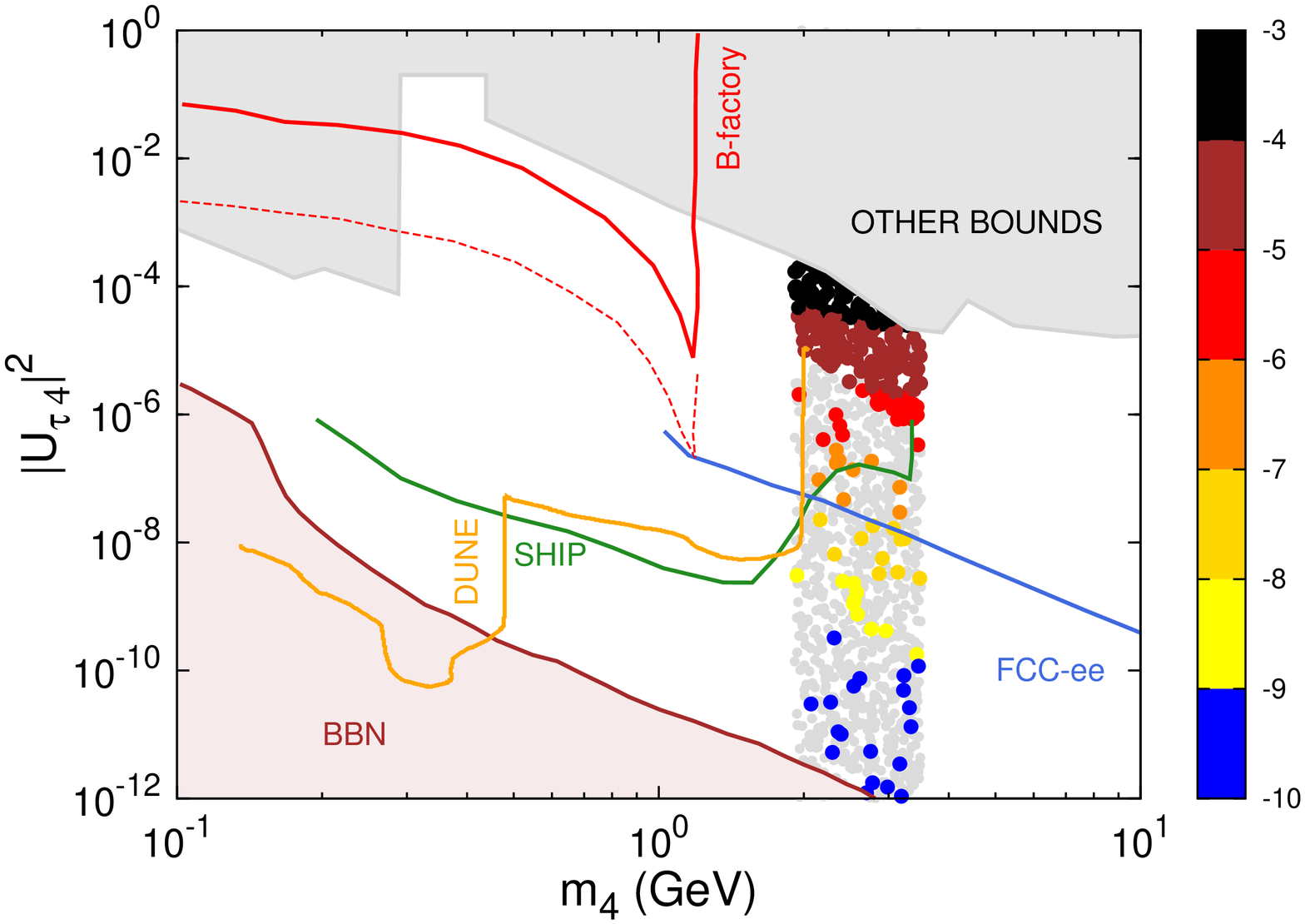,
width=53mm, angle =270}
&
\epsfig{file=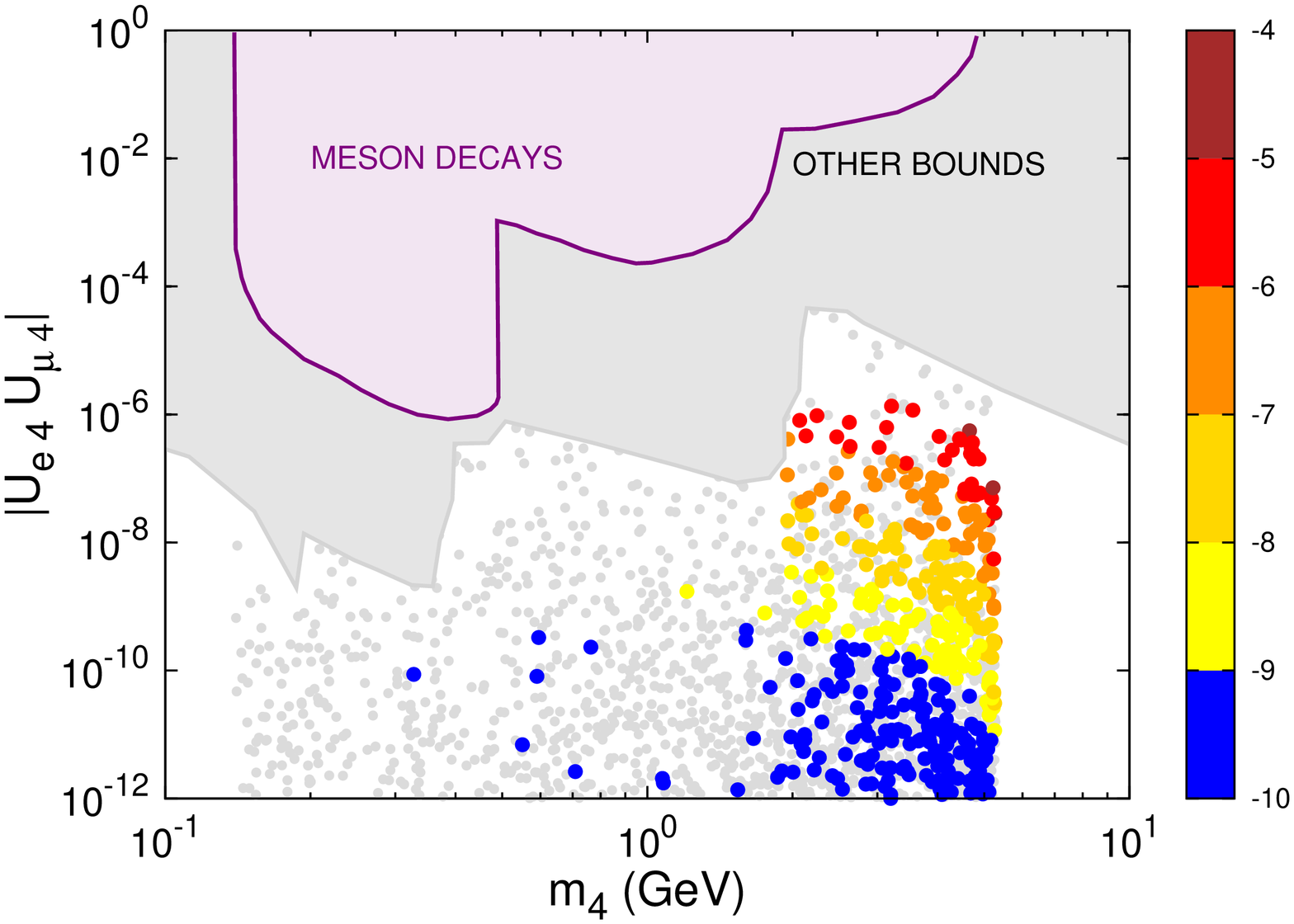,
width=53mm, angle =270}
\\
\epsfig{file=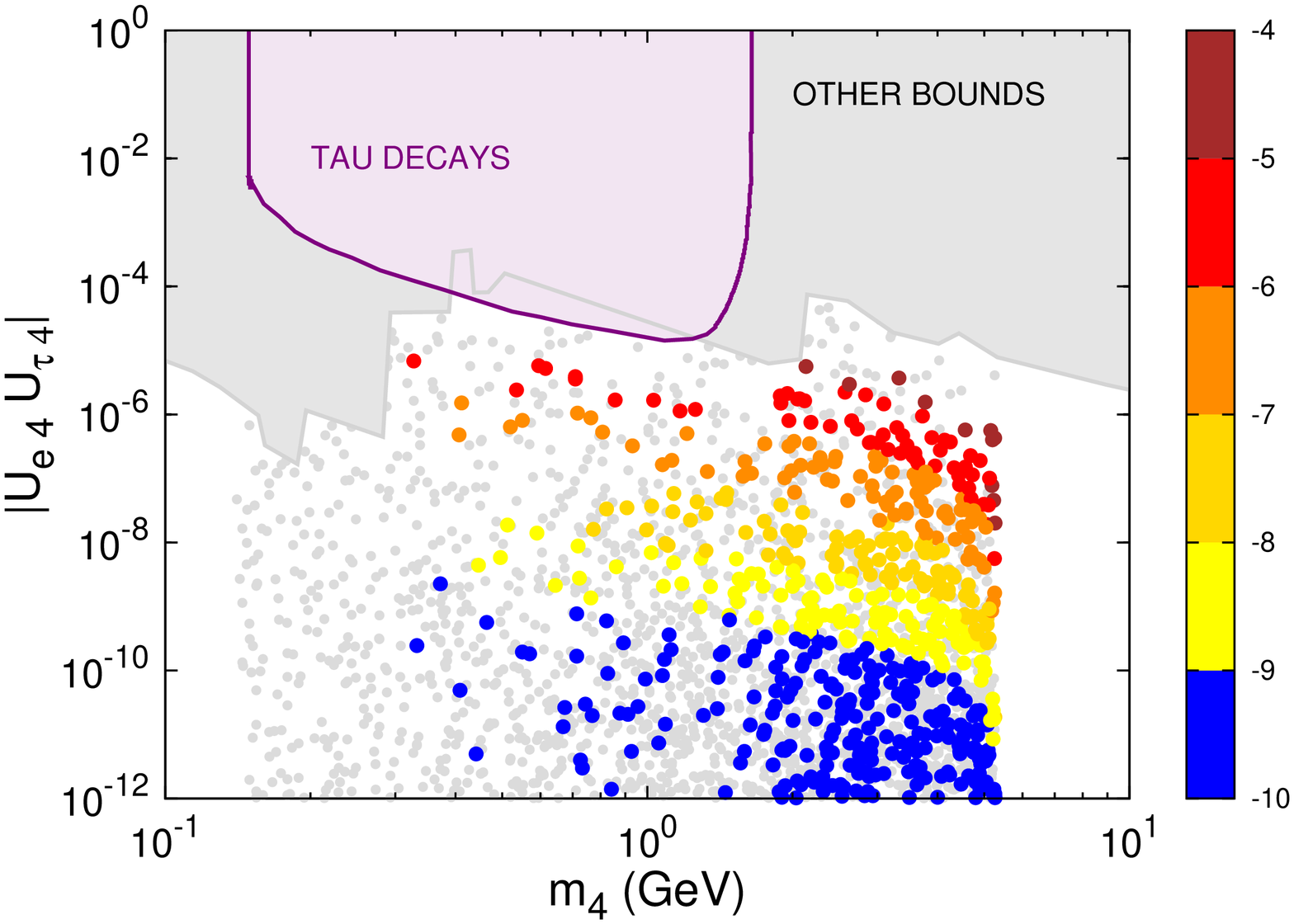,
width=53mm, angle =270}
&
\epsfig{file=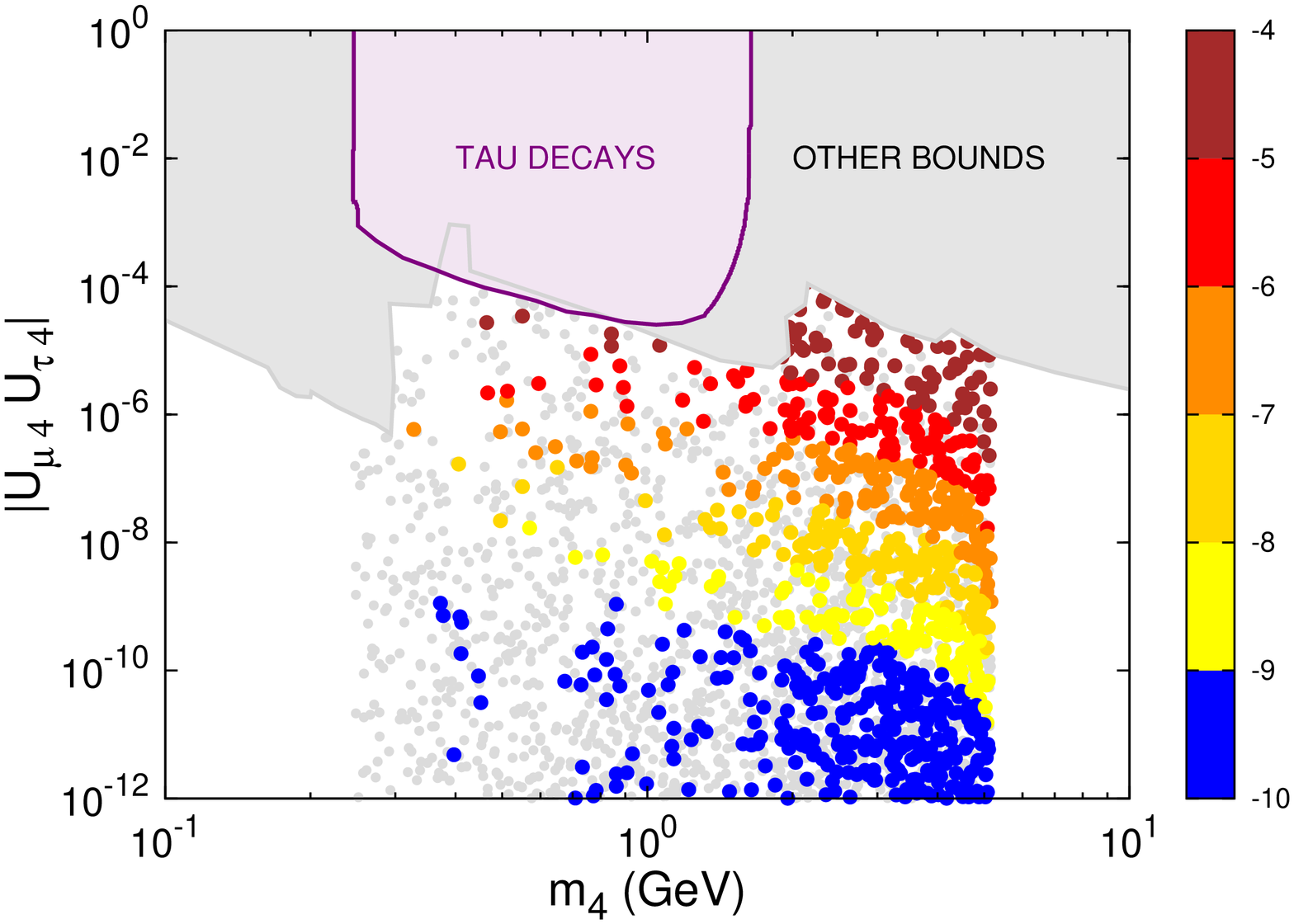,
width=53mm, angle =270}
\end{tabular}
\end{center}
\caption{Predictions for the effective mass $m_\nu^{\alpha \beta}$ 
in the sterile fermion parameter space $(|U_{\ell_\alpha 4} U_{\ell_\beta
  4}|,m_4)$, as derived 
from the allowed values for the LNV decays $B \to \ell_\alpha \ell_\beta \pi$.
The colour scheme of the 
in-laid points denotes regimes for the effective mass, 
 $\log(m_\nu^{\alpha \beta})$, in GeV; grey points denote exclusions
due to the violation of experimental/observational
bounds, as well as 
having an excessively long-lived sterile state.  
Solid surfaces (grey and pink) correspond to excluded regimes, and
the solid lines denote the sensitivity of future experiments. Green crosses 
(present in the upper right panel) denote points 
which are excluded due to excessive contributions to LNV decays.
}\label{fig:res:m_eff.Bpi}
\end{figure}

Complementary information (and an overview of the predictions for 
$m_\nu^{\alpha \beta}$ over the sterile neutrino parameter
space) is displayed in Fig.~\ref{fig:res:m_eff.Bpi}: 
coloured isosurfaces for 
$\log(m_\nu^{\alpha \beta})$ are identified in the parameter space
generated by the heavy neutrino mass and the relevant combination of
matrix elements,  $|U_{\ell_\alpha 4} U_{\ell_\beta 4}|$. Solid surfaces 
reflect excluded regimes due to the violation of experimental/observational
bounds, already summarised in Fig.~\ref{fig:ML.U-m4constraints}, 
or due to leading to $L^\text{flight}_{\nu_4} \gtrsim 10$~m; 
the coloured lines denote the sensitivity reach of several
future experiments (high energy and high intensity).
 
\begin{figure}[h!]
\begin{center}
\begin{tabular}{cc}
\epsfig{file=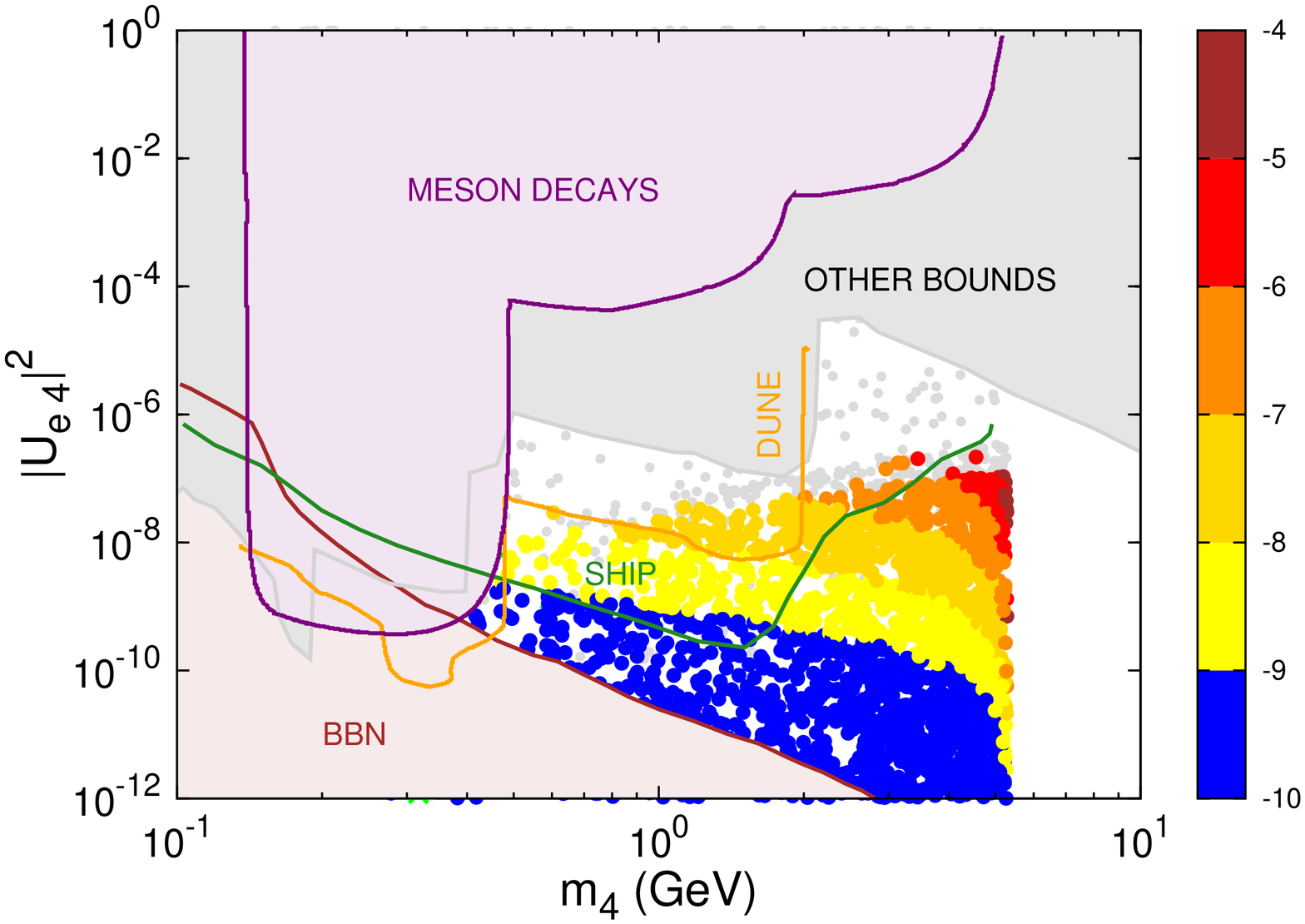,
width=53mm, angle =270}
&
\epsfig{file=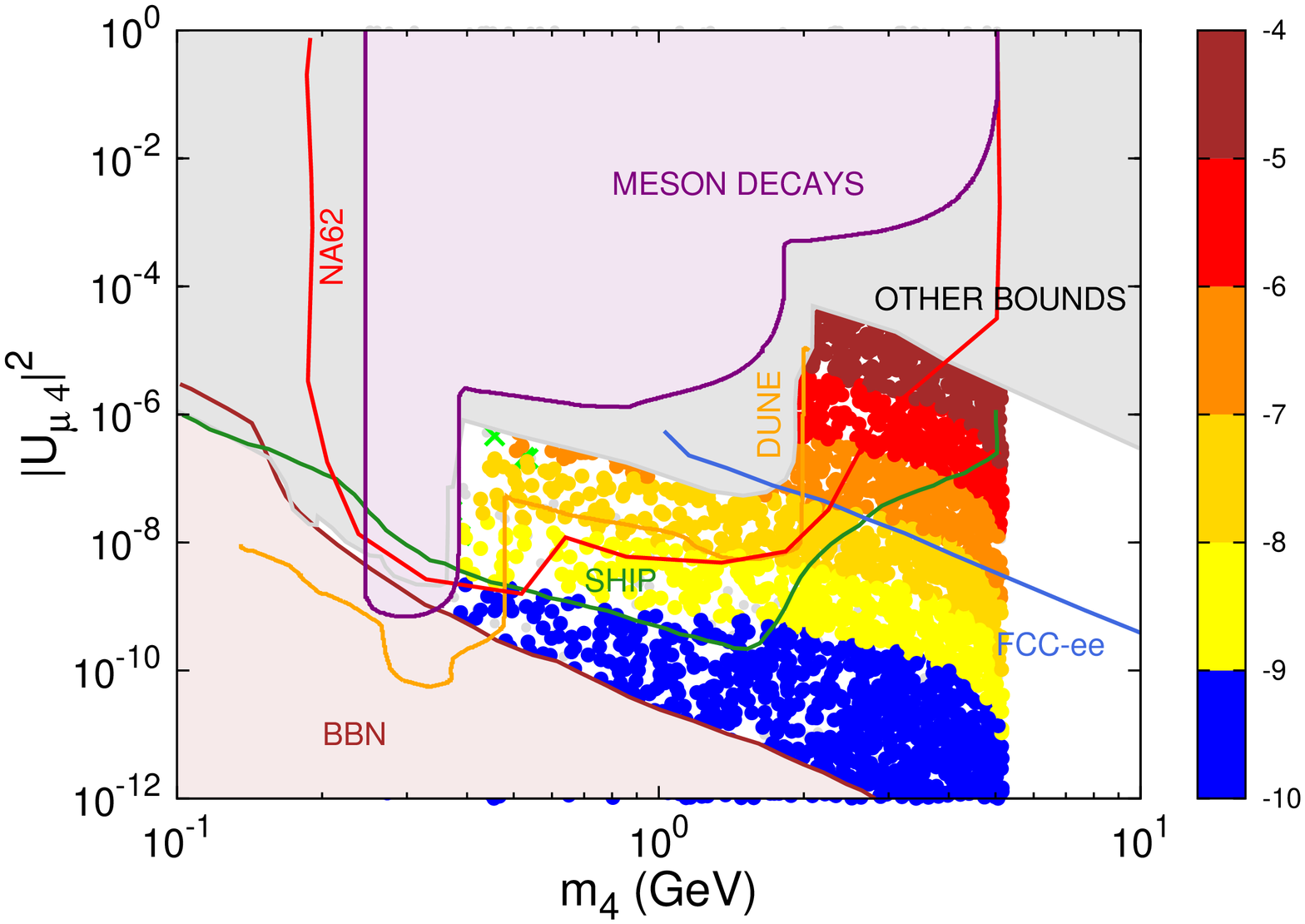,
width=53mm, angle =270}
\\
\epsfig{file=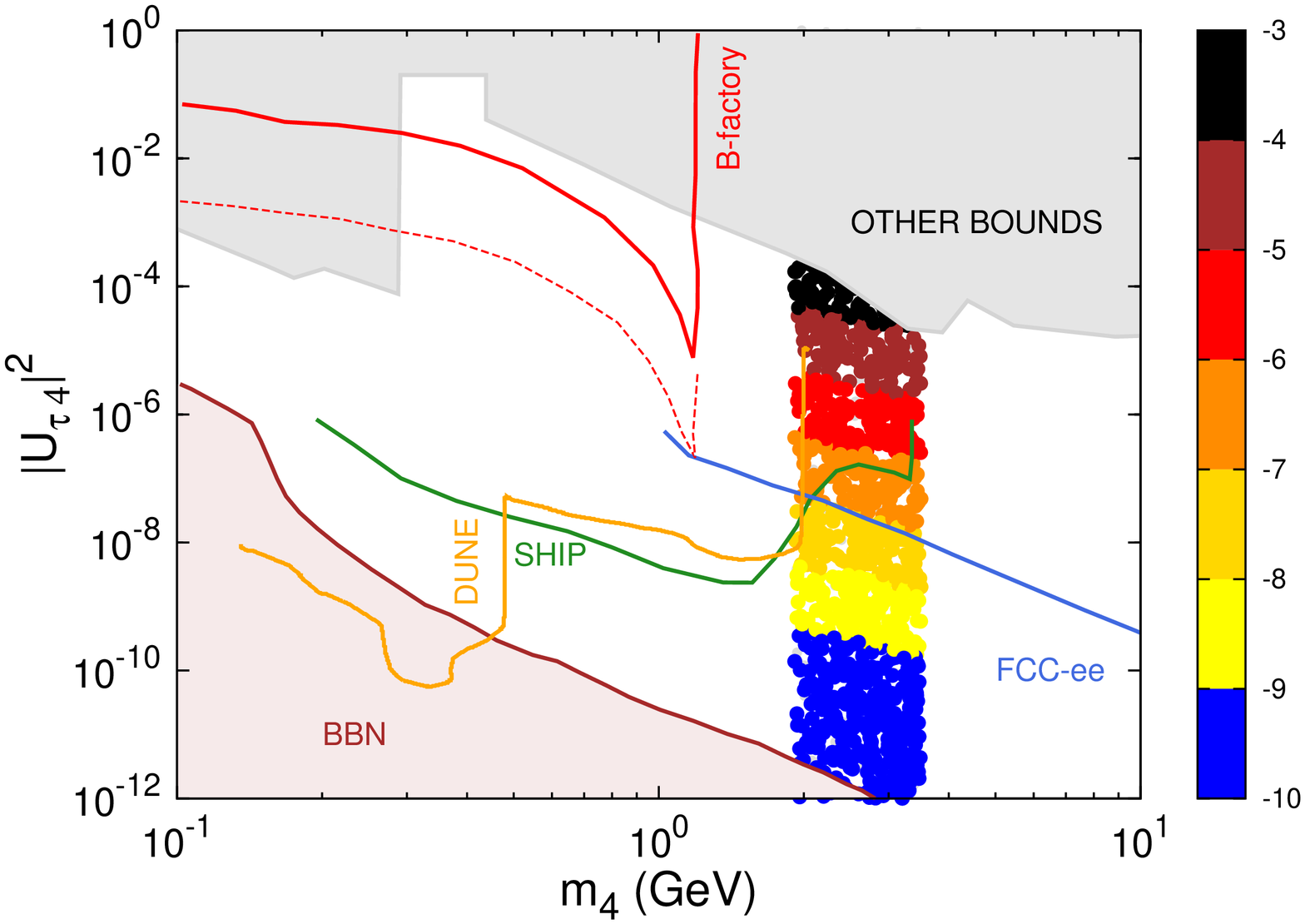,
width=53mm, angle =270}
&
\epsfig{file=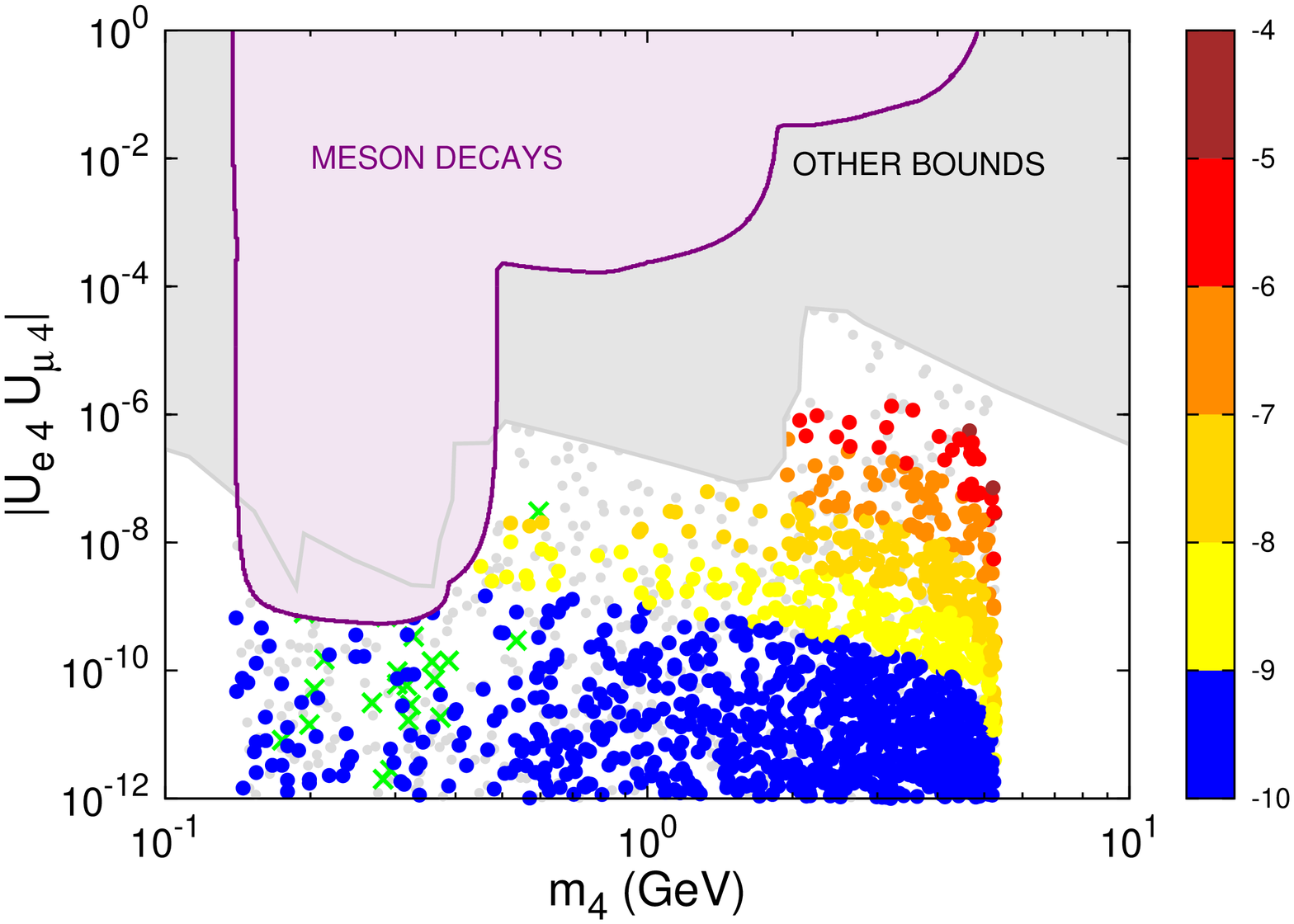,
width=53mm, angle =270}
\\
\epsfig{file=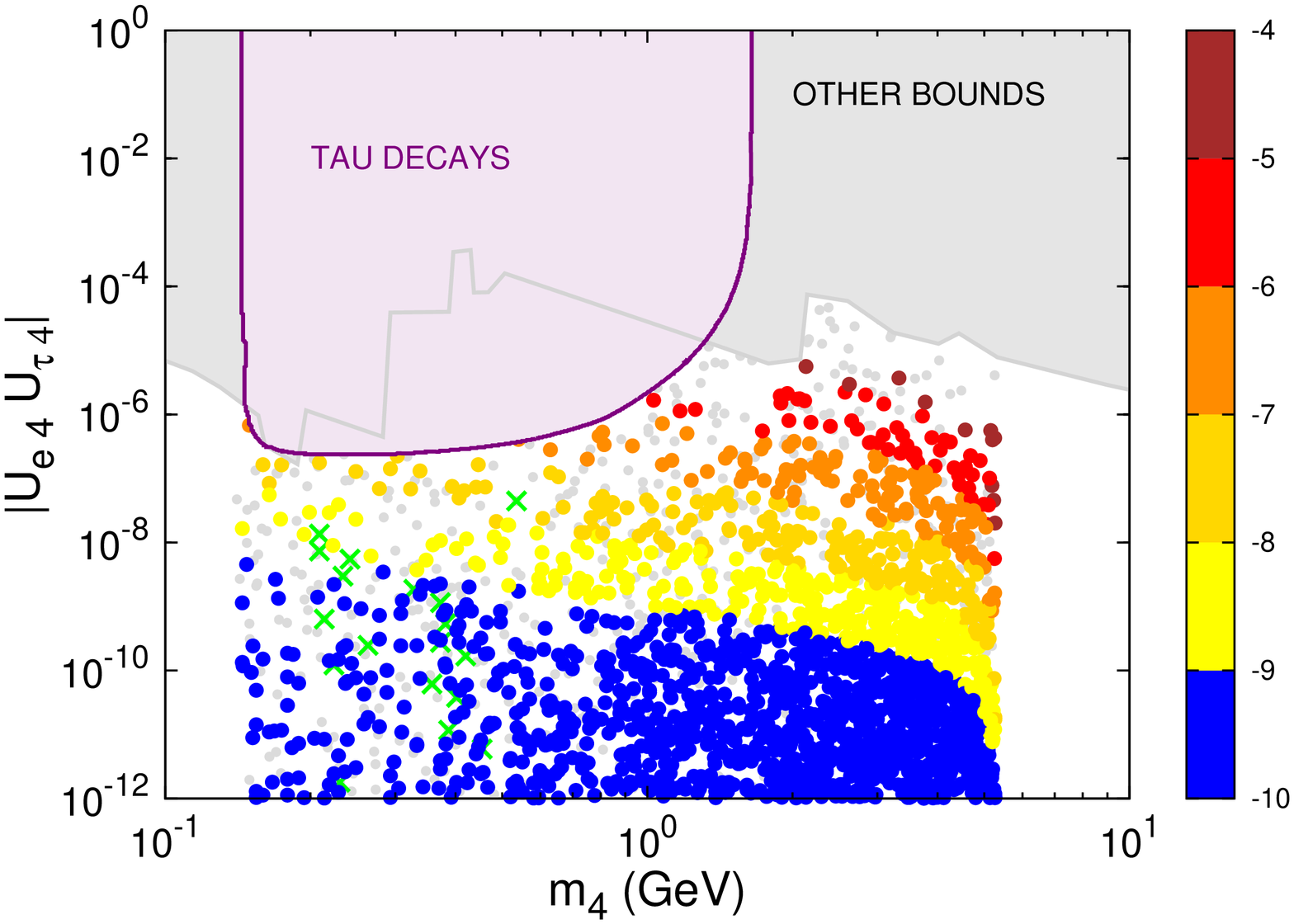,
width=53mm, angle =270}
&
\epsfig{file=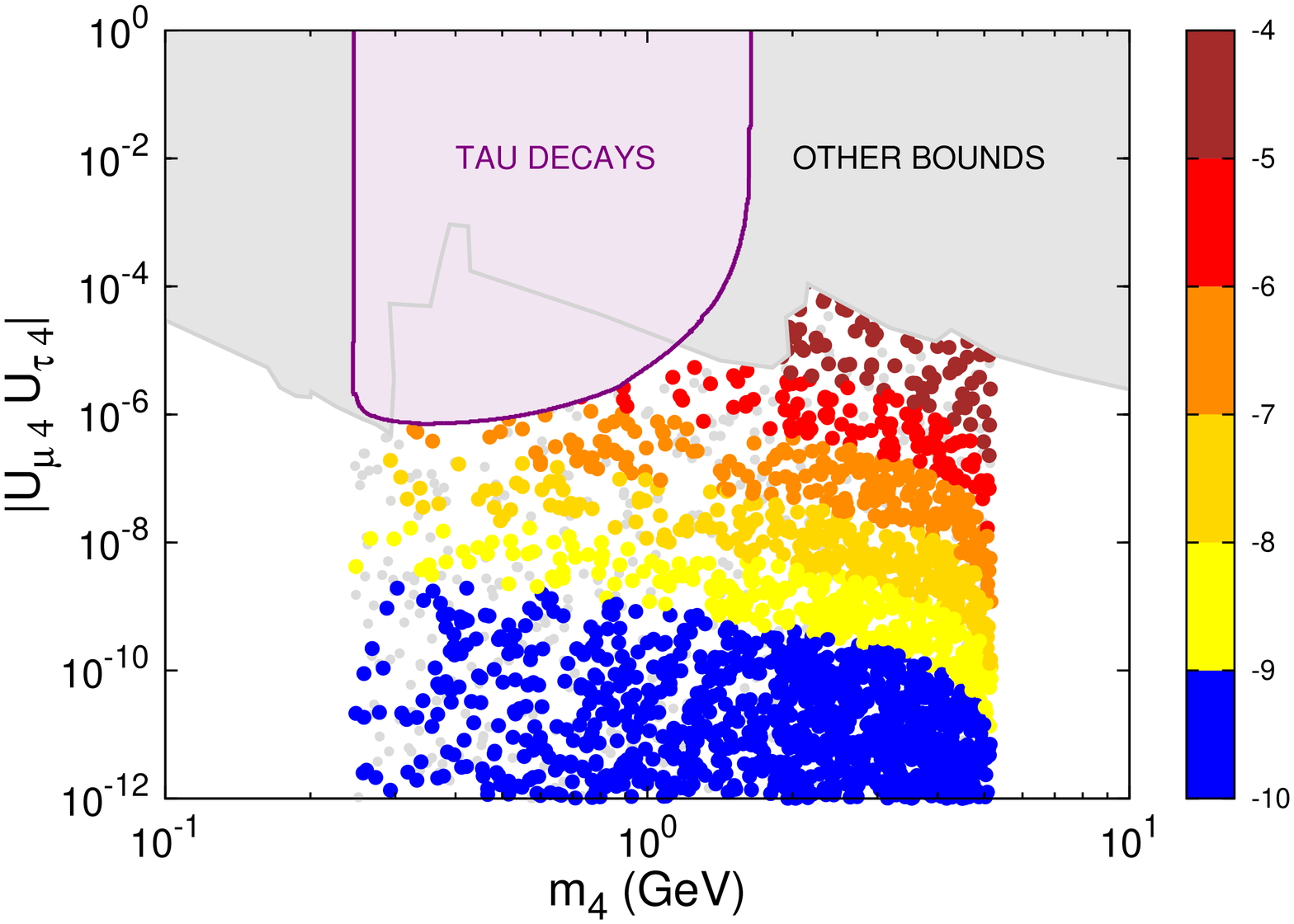,
width=53mm, angle =270}
\end{tabular}
\end{center}
\caption{Same as Fig.~\ref{fig:res:m_eff.Bpi} but without excluding
  the excessively long-lived sterile
  states. }\label{fig:res:m_eff.Bpi_noL10m} 
\end{figure}

As can be seen from the different panels, the constraints on the
mixings of the heavy neutrino to the
active states can be sufficiently weak to allow sizeable
contributions to the effective mass. In agreement with the discussion
of Fig.~\ref{fig:res:LNV:taudecays}, it is manifest that bounds arising
from LNV tau decays already contribute to exclude small regions of the 
sterile neutrino parameter space (pink surfaces on bottom panels).
The same analysis, but without excluding the contribution to LNV
decays of long-lived heavy neutrinos is presented in
Fig.~\ref{fig:res:m_eff.Bpi_noL10m}. Since regimes associated with
smaller mixings are now allowed, the surfaces corresponding to the
regimes experimentally excluded due to conflict with meson and tau
semileptonic bounds are also more important.  

It is also interesting to notice that regimes associated with large
effective masses (e.g. $m_\nu^{\mu \mu}$, as inferred from 
$B \to \mu \mu \pi$) are potentially within reach of future
experiments, such as NA62, SHiP~\cite{Alekhin:2015byh} and FCC-ee. 

\hspace*{-8mm}

\hspace*{-8mm}

\section{Discussion and concluding remarks}\label{sec:disc.concs}

In this work we have revisited the prospects of minimal SM extensions via 
sterile fermions in what concerns the three-body LNV semileptonic decays of 
mesons and of  tau leptons. Motivated by the recent progress in both
experimental searches (new bounds on LNV decays as well as other
relevant observables) and theoretical inputs (improved determination of
decay constants, among others), we have evaluated the
prospects of a large array of lepton number violating decay
modes,
leading to charged leptons of same or different flavours in the
final state. 
Working in the framework of a toy-like extension of the SM via one
sterile Majorana fermion, we have focused on the interesting (and
promising) regime in which the latter state is
on-shell. Our analysis further encompasses a full computation of the
heavy fermion decay width, as well as the additional constraints of
having the propagating neutrino decaying within a realistic (finite)
detector. 

After having carried an up-to-date evaluation of the sterile neutrino
parameter space, including all available observational and
experimental bounds, we have discussed to which extent the bounds on 
three-body LNV decays can further contribute to constrain the sterile
neutrino mass and active-sterile mixing angles. Our findings reveal that
current data on semileptonic tau decays already allow to exclude
certain regimes - in particular for heavy (mostly sterile) neutrino
masses below 1~GeV. Lepton number violating $K$ meson decay modes offer
a remarkably rich laboratory: in
addition to already excluding part of the sterile parameter space
(based on existing bounds), these processes are expected to play a major 
constraining r\^ole with the future sensitivity of NA62. 
The results obtained for other decay
modes (as for example $D_s \to \mu \mu \pi$) further suggest that in the near
future these decays could be within experimental sensitivity, and thus
offer additional means to constrain the active-sterile mixing angles
in the relevant sterile mass interval.  

\bigskip
The study of the distinct LNV decay channels (leading to charged
leptons of same or distinct flavour in the final state) has
allowed to infer bounds on the different independent entries of the
effective neutrino mass matrix. In our work
we have used a general definition of $m_\nu^{\alpha \beta}$ 
(allowing to encompass the standard $0\nu2 \beta$ 
effective mass definition).
The bounds obtained vary depending on
the actual decaying particle and on the final states (strongly
dependent on the mass and mixings of the Majorana mediator); although
for the $ee$ entry, neutrinoless double beta decay remains the most
stringent bound, one can already derive bounds typically below (or
around) $\mathcal{O}(10^{-3}\text{ GeV})$. With the exception of the 
$\tau \tau$ entry of the effective Majorana mass matrix 
- for which the derived bounds are weaker, 
$m_\nu^{\tau \tau} \lesssim \mathcal{O}(10^{-2}\text{ GeV})$ -, the
processes here considered constrain the remaining 
entries to lie below $\mathcal{O}(10^{-3}\text{ GeV})$. 

\begin{figure}[h!]
\begin{center}
\begin{tabular}{cc}
\epsfig{file=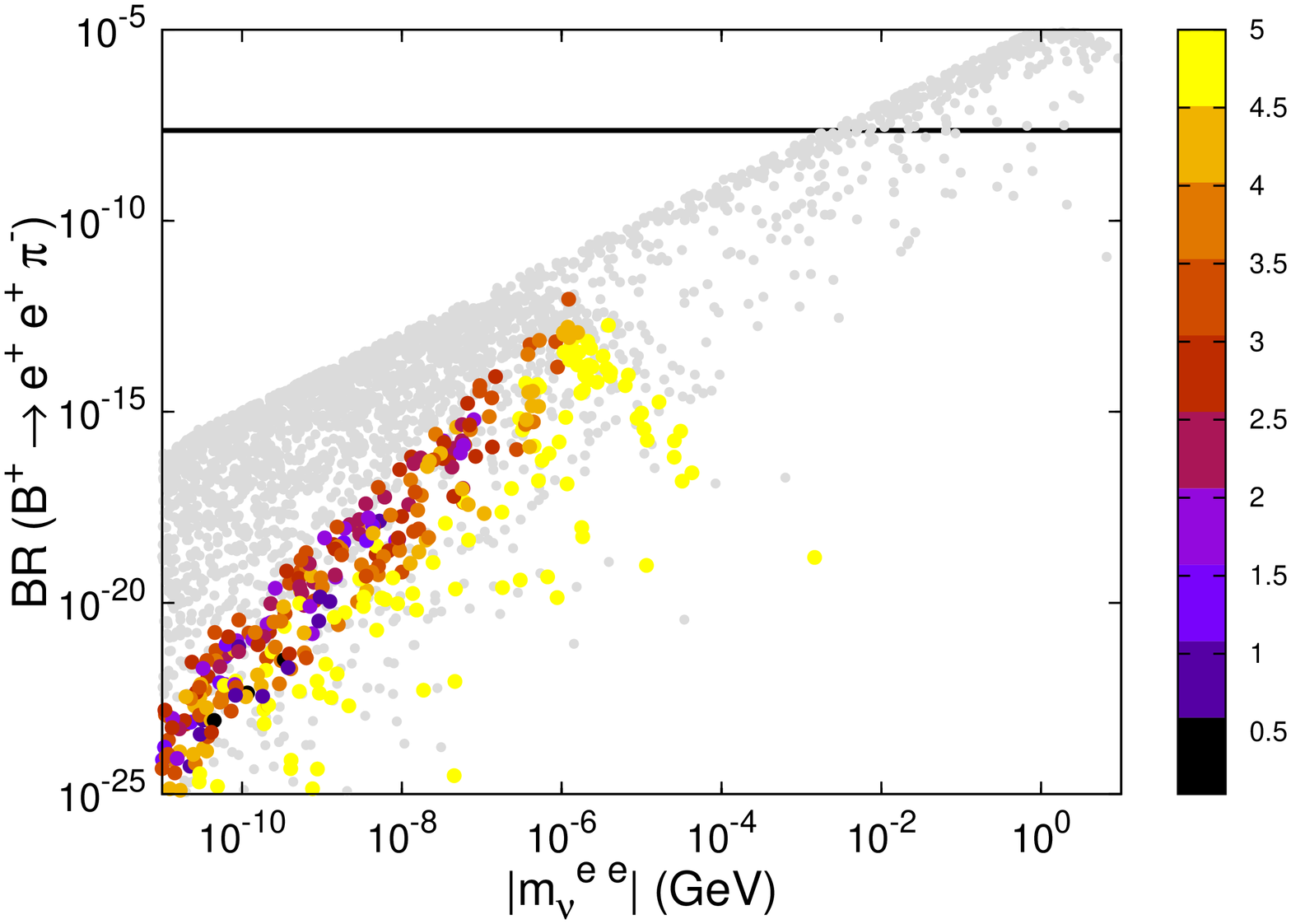,
width=53mm, angle =270}
 & 
\epsfig{file=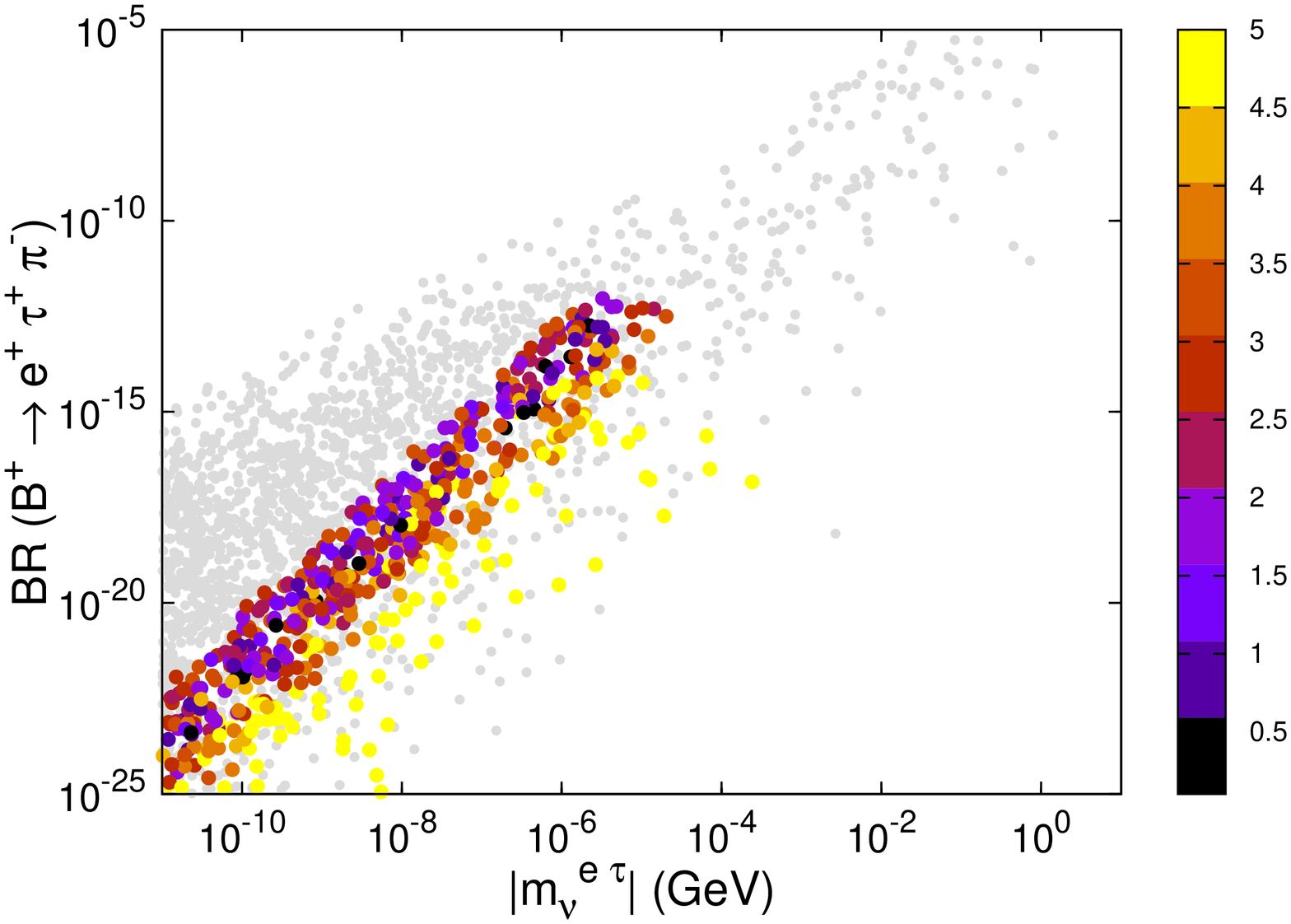,
width=53mm, angle =270} \\
\epsfig{file=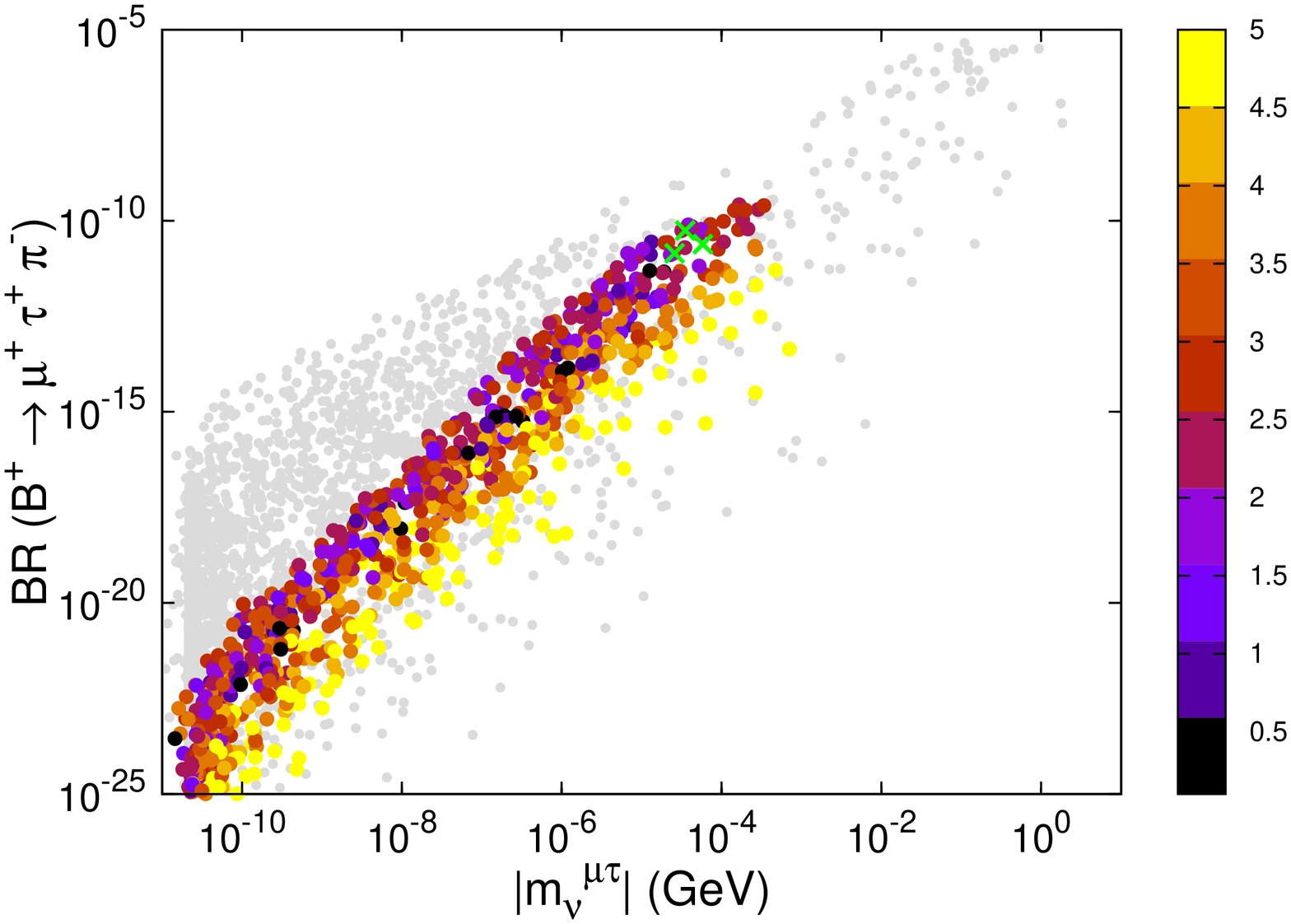,
width=53mm, angle =270}
 & 
 \epsfig{file=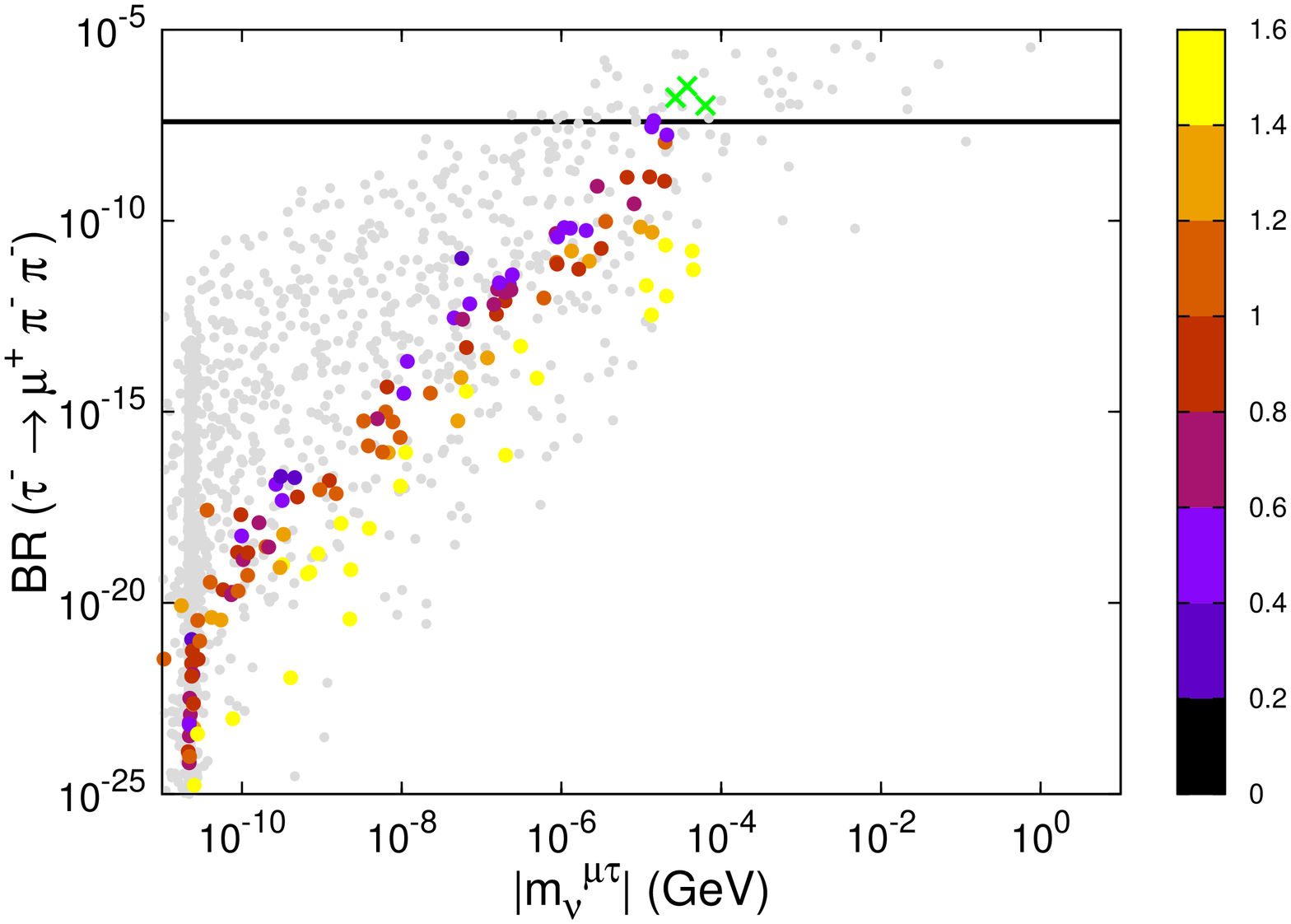,
 width=53mm, angle =270}
\end{tabular}
\end{center}
\caption{Branching ratios for LNV semileptonic meson decays 
(into same - and different - flavour(s) of leptons) 
as a function of the relevant entry of the 
effective mass $m_\nu^{\alpha \beta}$ (in GeV). 
The colour scheme of the 
in-laid points denotes regimes for the mass of the heavy (mostly
sterile) neutrino, $m_4$ in GeV; grey points denote exclusions
due to the violation of experimental/observational
bounds, as well as 
having an excessively long-lived sterile state. Green crosses  denote points 
which are excluded due to excessive contributions to LNV decays.
}\label{fig:BRLNV.meffective}
\end{figure}

We conclude by presenting a global overview of  
the most relevant findings of our work, in particular the overall
prospects of such a minimal SM extension regarding the LNV observables
here addressed and the constraints on the Majorana effective neutrino
mass matrix. 
In Fig.~\ref{fig:BRLNV.meffective}, we thus depict the semileptonic 
LNV branching
ratios for different channels, as a function of the (corresponding) 
relevant entry of
the effective neutrino mass matrix. 
If present, the horizontal lines denote the expected sensitivity of
the corresponding experiment. 

The near future experiments dedicated to neutrinoless double beta
decays are expected to bring down the sensitivity to about 
$|m_\nu^{ee}| \lesssim 0.01$~eV~\cite{DellOro:2016tmg}. Although the future
sensitivity of LHCb, Belle II or NA62 might allow to improve upon 
the current constraints, lepton number  violating 
semileptonic meson decays are unlikely to improve upon the $0\nu2 \beta$
bound for $m_\nu^{ee}$. The two lower panels of 
Fig.~\ref{fig:BRLNV.meffective} offer a critical comparison of the
prospects of very distinct semileptonic decays regarding constraints
on $m_\nu^{\alpha \beta}$, more precisely on $m_\nu^{\mu
  \tau}$: despite the comparatively smaller kinematically allowed
phase space for the $\tau$ decays\footnote{Notice that while the first
three panels of Fig.~\ref{fig:BRLNV.meffective} are associated with the
same colour scheme for the $m_4$ regimes, the fourth - concerning tau
lepton decays - only goes up to 1.6~GeV.} (which also accounts for the
lower density of coloured points), the branching ratios are already
directly testable by current experiments. As already emphasised when
discussing Figs.~\ref{fig:updated.UU.m4.tau.2}, \ref{fig:updated.UU.m4.tau.1},  \ref{fig:res:LNV:taudecays} and~\ref{fig:res:m_eff.m4.D.tau}, tau decays emerge as very promising
channels to effectively constrain  $m_\nu^{\mu \tau}$ to values as low as 
$\mathcal{O}(10^{-5})$~GeV. 

Other than allowing to infer bounds on the electron neutrino
effective mass (thus contributing to address the neutrino absolute
mass scale issue), the panels of Fig.~\ref{fig:BRLNV.meffective}
suggest the importance of constraining the different entries of $m_\nu^{\alpha
  \beta}$: notice the strong underlying correlation between the 
LNV branching ratios and the associated  $m_\nu^{\alpha \beta}$
entries, independently of the sterile fermion mass regime. 
Even if expected from the general definition of
$m_\nu^{\alpha \beta}$ here used (see
Section~\ref{sec:effective.mass.th}), the clear correlation can be
used to rapidly infer information on the BRs from a simple 
estimation of $m_\nu^{\alpha \beta}$.

\hspace*{-8mm}

\bigskip
In this study we have considered a simple toy-like model, in which 
the SM is minimally extended by a massive Majorana sterile fermion,
which is assumed to have non-negligible mixings with the active states
(no hypothesis made on the mechanism of neutrino mass generation). 
The results here obtained - both regarding the updated constraints on
the parameter space and the predictions regarding the LNV branching
fractions and reconstructed effective mass - 
can nevertheless be interpreted as ``benchmark'' results for complete
models of neutrino mass generation in which the SM is extended via
$n_s$ sterile neutrinos (as low-scale type I seesaw
models, and variants).
Notice however that constructions in which the smallness of neutrino
masses is explained via the smallness of a lepton number violating
parameter and/or in which the heavy states form pseudo-Dirac pairs, lead to
a suppression of the expected LNV rates. For instance, this is the case 
of the Inverse Seesaw~\cite{ref:ISS}, Linear Seesaw~\cite{ref:LSS}, 
specific realisations of the $\nu$MSM~\cite{Shaposhnikov:2006nn} 
and low-scale realisations of the type-I seesaw 
mechanism~\cite{Ibarra:2010xw}. 
Further observables, and means to constrain other relevant parameters,
might also be explored in association with the three-body semileptonic LNV
decays here addressed, but this lies beyond the scope of the current work.

As shown in our study, such a minimal SM extension already leads to
contributions to LNV observables (involving leptons of same or
different flavour) which are close - if not in
conflict! - with current data. With the improvements of the
experimental sensitivities, it is possible that one of these decays
will be 
measured; in addition to confirming the Majorana nature of the
mediator, such an observation would allow to test simple SM extensions via
sterile neutrinos.

\section*{Acknowledgements}
We thank X. Marcano for having reported to us an error in the implementation of the experimental constraints that was present in a previous version of the paper. We are grateful to E. Cortina Gil and E. Minucci 
for valuable contributions regarding NA62.
A.A. and A.M.T. acknowledge support within the framework of the
European Union's Horizon 2020 research and innovation programme under
the Marie Sklodowska-Curie grant agreements No 690575 and No 674896.  
V.D.R. acknowledges support by the Spanish grant SEV-2014-0398 (MINECO)
and partial support by the Spanish grants FPA2014-58183-P,  PROMETEOII/2014/084 
(Generalitat Valenciana) and Iberoam\'erica Santander Investigaci\'on 2017.
V.D.R. is also grateful for the kind hospitality received at the
Universidad T\'ecnica Federico Santa Mar\'ia (Campus Santiago San
Joaqu\'in, Chile), for the final stage of this work.  
M.L. acknowledges support from the European Union's Horizon 2020
research and innovation programme under the Marie Sklodowska-Curie
grant agreement No 750627. 
T.T. acknowledges support from JSPS Fellowships for Research Abroad. 
\newpage
\appendix

\section{Extensions of the SM via $\bm{n_S}$ sterile
  fermions}\label{app:SM.sterile}

\subsection{Formalism}\label{app:SM.sterile:formalism} 
Due to possible mixings with the light (mostly active) neutrinos,
extensions of the SM via sterile states open the door to the  
violation of lepton flavour in both neutral and charged leptonic
currents~\cite{Schechter:1980gr,Gronau:1984ct}. 
After electroweak symmetry breaking, and in the charged lepton's physical
basis, the addition of $n_S$ sterile (Majorana) neutrinos leads to the
following modification of vector and scalar currents:
\begin{align}\label{eq:lagrangian:WGHZ}
& \mathcal{L}_{W^\pm}\, =\, -\frac{g_w}{\sqrt{2}} \, W^-_\mu \,
\sum_{\alpha=1}^{3} \sum_{j=1}^{3 + n_S} {\bf U}_{\alpha j} \bar \ell_\alpha 
\gamma^\mu P_L \nu_j \, + \, \text{H.c.}\,, \nonumber \\
& \mathcal{L}_{Z^0}\, 
= \,-\frac{g_w}{4 \cos \theta_w} \, Z_\mu \,
\sum_{i,j=1}^{3 + n_S} \bar \nu_i \gamma ^\mu \left(
P_L {\bf C}_{ij} - P_R {\bf C}_{ij}^* \right) \nu_j\,
-\frac{g_w}{4 \cos \theta_w} \, Z_\mu \,
\sum_{\alpha=1}^{3}  \bar \ell_\alpha \gamma ^\mu \left(
{\bf C}_{V} - {\bf C}_{A} \gamma_5 \right) \ell_\alpha\,, \nonumber \\
& \mathcal{L}_{H^0}\, = \, -\frac{g_w}{2 M_W} \, H^0  \,
\sum_{i,j=1}^{3 + n_S}  {\bf C}_{ij}  \bar \nu_i\left(
P_R m_i + P_L m_j \right) \nu_j + \, \text{H.c.}\,. 
\end{align}
In the above, $g_w$ denotes the weak coupling constant, 
$\cos^2 \theta_w = M_W^2 /M_Z^2$, 
$P_{L,R} = (1 \mp \gamma_5)/2$, 
and $m_i$ are the physical neutrino masses (light and heavy); 
the indices 
$\alpha $ denote the flavour of the charged leptons,
while $i, j = 1, \dots, 3+n_S$ correspond to the physical (massive) 
neutrino states. The (SM) vector and axial-vector 
$Z$-couplings of charged leptons are parametrised by the 
${\bf C}_{V}$ and ${\bf C}_{A}$ coefficients, defined as 
${\bf C}_{V} = -1+4 \sin^2 \theta_w$ and 
${\bf C}_{A} = -1$.  
The rectangular $3 \times (3 +n_S)$ mixing matrix, 
${\bf U}_{\alpha j}$, encodes 
the mixing in charged current interactions (corresponding 
to the (unitary) PMNS matrix, $U_\text{PMNS}$ in the
case of $n_S=0$); its upper $3 \times 3$ 
block, usually denoted $\tilde U_\text{PMNS}$, denotes 
the mixing between the left-handed leptons.
Finally, the matrix ${\bf C}$ parametrises 
lepton flavour violation in neutral currents,
\begin{equation}
\label{eq:Cmatrix:def}
{\bf C}_{ij} \,=\,\sum_{\alpha=1}^{3} {\bf U}_{\alpha i}^*\,{\bf U}_{\alpha j}\,. 
\end{equation} 
We notice that in addition to the vector and scalar currents above
referred to, the interactions with
  neutral and charged Goldstone bosons are also modified:
\begin{align}\label{eq:modified.currents.G}
& \mathcal{L}_{G^0}\, =\,\frac{i g_w}{2 M_W} \, G^0 \,
\sum_{i,j=1}^{3 + n_S} {\bf C}_{ij}  \bar \nu_i  
\left(P_R m_j  - P_L m_i  \right) \nu_j\,+ \, \text{H.c.}\nonumber \\
& \mathcal{L}_{G^\pm}\, =\, -\frac{g_w}{\sqrt{2} M_W} \, G^- \,
\sum_{\alpha=1}^{3}\sum_{j=1}^{3 + n_S} {\bf U}_{\alpha j}   
\bar \ell_\alpha\left(
m_\alpha P_L - m_j P_R \right) \nu_j\, + \, \text{H.c.}\,.
\end{align}

The modification of neutral and charged lepton currents as a
consequence of the addition of sterile neutrinos to the SM content,
opens the door to new contributions to a
vast array of observables, possibly in conflict with current
data. These are summarised in Appendix~\ref{app:constraints}, and will
be applied throughout our phenomenological analysis.

\subsection{Theoretical frameworks - the simple ``3+1 model''}\label{sec:3+1}

A number of theoretical frameworks - ranging from minimal extensions
to complete theoretical constructions - call upon sterile
fermions: the latter are present in  
several mechanisms of neutrino mass generation, which in addition to
accommodating neutrino data, also address in the baryon asymmetry of
the Universe and/or put forward a viable dark matter candidate;
likewise, sterile fermions can be minimally added to the SM in a
simplified, toy-model approach to evaluate their impact regarding
flavour and/or lepton number violation. In our analysis we will
consider the latter approach, which we proceed to
describe.

The SM field content is enlarged via 
the addition of a single Majorana sterile neutral fermion which, as
noticed before, can also be 
interpreted as encoding the effects of a larger number of states
possibly present in the underlying complete new physics model;
this first phenomenological approach - minimal ``toy model'' - 
relies in an ad-hoc construction, which makes no assumption on the mechanism
of neutrino mass generation, thus allowing to decouple the generation
of neutrino masses (possibly occurring at higher scales, or even arise
from interactions not calling upon the lighter sterile
state) from the mechanism responsible for flavour violation and lepton
number violation at low-energies. 

As done in other works 
(see, e.g.~\cite{Abada:2014cca,Abada:2015oba,Abada:2015zea,Abada:2016vzu}),
in the present study we will rely on a simple toy model which is
built under the single hypothesis that
interaction and physical neutrino eigenstates are related via a
$4\times 4$ unitary mixing matrix, ${\bf U}_{ij}$. Other than the
masses of the three light (mostly active) neutrinos, and their mixing
parameters, the simple 
``3+1 model'' is parametrised by the heavier (mostly sterile)
neutrino mass $m_4$, three active-sterile mixing angles as well as
three new CP violating phases (two Dirac and one Majorana).

\subsection{Constraints on sterile fermions}\label{app:constraints}

Due  to the presence of the additional sterile states, 
the modified neutral and charged lepton currents 
might lead to new contributions to a
vast array of observables, possibly in conflict with current
data. These SM extensions via sterile fermions 
must be then confronted to all available constraints
arising from high-intensity, high-energy and cosmological
observations. 

Sterile states, with a mass above the electroweak  scale, can have
sizeable decay widths, a consequence of being sufficiently heavy to 
decay into a $W^\pm$ boson and a charged lepton, or into a 
light (active) neutrino and either a $Z$ or a Higgs boson.
One thus imposes the perturbative
unitarity
condition~\cite{Chanowitz:1978mv,Durand:1989zs,Korner:1992an,
Bernabeu:1993up,Fajfer:1998px,Ilakovac:1999md}, 
$\frac{\Gamma_{\nu_i}}{m_{\nu_i}}\, < \, \frac{1}{2}\, (i \geq 4)$. 
Noticing that the leading contribution to ${\Gamma_{\nu_i}}$
is due to the charged current term, one obtains the following 
bounds~\cite{Chanowitz:1978mv,Durand:1989zs,Korner:1992an,
Bernabeu:1993up,Fajfer:1998px,  
Ilakovac:1999md}:
\begin{equation}\label{eq:sterile:bounds:Ciimi}
m_{\nu_i}^2\,{\bf C}_{ii} \, < 2 \, \frac{M^2_W}{\alpha_w}\, \quad
\quad (i \geq 4)\,,
\end{equation}
where $\alpha_w=g^2_w/4 \pi$, and ${\bf C}_{ii}$ is given in
Eq.~(\ref{eq:Cmatrix:def}). 
However, this constraint is not very relevant in the present analysis,  
as we focus on mass ranges for which 
the sterile neutrino is produced on-shell from meson or tau decays.

Observational constraints on the sterile masses and their mixings with
the active states arise from an extensive number of sources.
If kinematically accessible, sterile neutrinos can be produced in
laboratory experiments via the interactions
in~(\ref{eq:lagrangian:WGHZ}): the negative observation of these
processes constraints the mixings with the
electron~\cite{Ue_exclusions,UeUmu_exclusions,UeUmuUtau_exclusions}, muon~\cite{Umu_exclusions,UeUmu_exclusions,UeUmuUtau_exclusions} and
tau~\cite{Utau_exclusions,UeUmuUtau_exclusions} flavours. 
Moreover, and other than requiring compatibility between the left-handed lepton
mixing matrix $\tilde U_\text{PMNS}$ and the corresponding best-fit
intervals\footnote{We do not impose any constraints on the
(yet undetermined) value of the CP violating Dirac phase $\delta$.}
 defined from {\it $\nu$-oscillation
  data}~\cite{Tortola:2012te,Fogli:2012ua,GonzalezGarcia:2012sz,Forero:2014bxa,nufit,Gonzalez-Garcia:2014bfa,Esteban:2016qun,deSalas:2017kay},
we also impose, when relevant, {\it unitarity bounds} 
as arising from non-standard neutrino interactions with matter,
on the deviation 
of $\tilde U_\text{PMNS}$ from
unitarity~\cite{Antusch:2008tz,Antusch:2014woa,Blennow:2016jkn}.
Further constraints on the active-sterile
mixings (and on the mass regime of new states) arise from 
{\it electroweak precision observables}; these include 
new contributions to the invisible $Z$-decay width 
(addressed in~\cite{Akhmedov:2013hec,Basso:2013jka,
Fernandez-Martinez:2015hxa,Abada:2013aba}), which must comply 
with LEP results
on $\Gamma(Z \to \nu \nu)$~\cite{Patrignani:2016xqp}; 
moreover, any contribution to cLFV
$Z$ decay modes should not exceed the present uncertainty on the 
total $Z$ width~\cite{Patrignani:2016xqp}, $\Gamma (Z 
\to \ell_\alpha^\mp \ell_\beta^\pm) < \delta \Gamma_{\rm  tot}$. In our study
we also take into account current limits on {\it invisible Higgs} decays 
(relevant for $m_{\nu_s} < M_H$), following the approach 
derived in~\cite{BhupalDev:2012zg,Cely:2012bz,Bandyopadhyay:2012px}. 
Likewise, negative results from 
{\it laboratory searches} for monochromatic lines in the
spectrum of muons from  $\pi^\pm \to \mu^\pm \nu$
decays are also taken into account~\cite{Kusenko:2009up,Atre:2009rg}.
The new states (through the modified
currents) induce potentially large contributions to {\it cLFV observables}; 
we evaluate the latter~\cite{Ma:1979px,Gronau:1984ct,Ilakovac:1994kj,Deppisch:2004fa,Deppisch:2005zm,Dinh:2012bp,Alonso:2012ji,Abada:2014kba,Abada:2015oba} 
imposing available limits on a wide variety of observables (some
of them collected in Tables~\ref{tab:LFV:meson:exp}
and~\ref{tab:LFV:Bmeson:exp}). %Table~\ref{table:cLFV:bounds}).   
In addition to the cLFV decays and transitions, which can
prove instrumental to test and 
disentangle these extensions of the SM, important constraints arise from
rare {\it leptonic and semileptonic decays of
pseudoscalar mesons} decays
(including lepton universality violating, cLFV and lepton number
violating modes); we include constraints from numerous 
$\ K,\ D$, $\ D_s$, $B$ modes (see~\cite{Goudzovski:2011tc,Lazzeroni:2012cx} for
kaon decays,~\cite{Naik:2009tk,Li:2011nij} for $D$ and $D_s$ decay
rates, and~\cite{Aubert:2007xj,Adachi:2012mm} for $B$-meson
observations), stressing that in the framework of the SM extended by
sterile neutrinos particularly severe constraints 
arise from the violation of lepton universality
in leptonic meson decays (parametrised by the observables 
$\Delta r_K$, $\Delta r_\pi$ and $\Delta
  r_\tau$)~\cite{Abada:2012mc,Abada:2013aba}. 
(Due to being associated with less robust experimental bounds, we do
  not include in our constraints the observables $R^{e \tau}_\pi$ and
  $R^{\mu \tau}_\pi$.)
Finally, we also take 
into account the recent constraints on 
{\it neutrinoless double beta decay}~\cite{Albert:2014awa}: should the sterile
states be Majorana fermions, they can potentially contribute to 
the effective mass $m_{ee}$~\cite{Benes:2005hn}, which we evaluate 
following~\cite{Blennow:2010th,Abada:2014nwa}.

A number of {\it cosmological observations}~\cite{Smirnov:2006bu,Kusenko:2009up,Hernandez:2014fha,Vincent:2014rja}
put severe constraints on sterile neutrinos with a mass below the GeV
(in particular below 200~MeV). In our study we will in general explore
regimes associated with heavier sterile states ($m_{\nu_s} \gtrsim
0.5$~GeV) so that these constraints are not expected to play a relevant r\^ole.

\section{Computation of lepton number violating decay
  widths}\label{app:sec:widths} 
In this Appendix we describe the theoretical computation of the LNV
decay widths addressed in our work, in particular semileptonic tau
decays and semileptonic decays of mesons into pseudoscalar and vector
meson three-body final states. 

\subsection{Semileptonic tau-lepton LNV decay widths}
Consider the LNV tau decay into two mesons and a charged lepton, 
\begin{equation}
\tau^{-}(p,m_\tau)\, \to\,
M_1^{-}(k_1,m_{M_1})\, 
M_2^{-}(k_2,m_{M_2})\, \ell^{+}(k_3,m_{\ell})\,,
\end{equation}
mediated by the sterile neutrino $\nu_4$, and where $\ell=e$ or $\mu$. 
The amplitude for this process can be computed as 
\begin{eqnarray}
i\mathcal{M}_{\tau}\,=\,
-2iG_F^2\,V_{M_1}^*V\,_{M_2}^*\,
U_{\tau 4}^*\,U_{\ell 4}^*\,m_4\,f_{M_1}\,f_{M_2}
\left[
\frac{\overline{u}(k_3)k\hspace{-0.18cm}/_1k\hspace{-0.18cm}/_2P_Lu(p)}
{m_{31}^2-m_4^2+im_4\Gamma_4}
+
\frac{\overline{u}(k_3)k\hspace{-0.18cm}/_2k\hspace{-0.18cm}/_1P_Lu(p)}
{m_{23}^2-m_4^2+im_4\Gamma_4}
\right]\,,
\label{eq:LNV_tau}
\end{eqnarray}
in which $V_{M_i}$ and $f_{M_i}$ are the CKM matrix and decay constant
corresponding to the final state meson $M_i$, 
$U_{\ell4}$ is the active-sterile mixing, %mixing matrix,
$m_{ij}^2$ are the momentum variables defined by
$m_{ij}\equiv(k_i+k_j)^2$ and $\Gamma_4$ is the total decay width of the
sterile neutrino $\nu_4$ (the computation of the latter is described in
Appendix~\ref{app:sec:widths:nu4}). 
Defining the first and second terms in Eq.~(\ref{eq:LNV_tau}) as
$i\mathcal{M}_{\tau1}$ and $i\mathcal{M}_{\tau2}$, the squared amplitude,
spin-averaged over the initial state and spin-summed over the final
state, is given by 
\begin{equation}\label{eq:LNV_tau-step}
\overline{|\mathcal{M_\tau}|^2}
\, \equiv\, 
\frac{1}{2}\sum_{\text{spin}}
\left[\left|\mathcal{M}_{\tau1}\right|^2
+\left|\mathcal{M}_{\tau2}\right|^2
+2\mathrm{Re}\left(\mathcal{M}_{\tau1}\mathcal{M}_{\tau2}^*\right)
\right],
\end{equation}
in which each term is given by
\begin{eqnarray}
\hspace{-0.2cm}&&\hspace{-0.2cm}
\frac{1}{2}\sum_\text{spin}|\mathcal{M}_{\tau1}|^2
\equiv
\overline{|\mathcal{M}_{\tau1}|^2}\, =\nonumber\\
\hspace{-0.2cm}&=&\hspace{-0.2cm}
\frac{2G_F^4|V_{M_1}|^2|V_{M_2}|^2
|U_{\tau 4}|^2|U_{\ell 4}|^2
m_4^2f_{M_1}^2f_{M_2}^2}
{\left(m_{31}^2-m_4^2\right)^2+m_4^2\Gamma_4^2}
\Biggl[
M_{\tau31M_2}^2M_{\ell31M_1}^2M_{M_212M_1}^2
+m_{M_1}^2m_{M_2}^2M_{\tau12\ell}^2
\nonumber\\
\hspace{-0.2cm}&&\hspace{-0.2cm}
\hspace{6.9cm}
+m_{M_1}^2M_{\tau31M_2}^2M_{\ell23M_2}^2+m_{M_2}^2M_{\tau23M_1}^2M_{\ell31M_1}^2
\Biggr],\hspace{0.6cm}
\end{eqnarray}
\begin{eqnarray}
\hspace{-0.2cm}&&\hspace{-0.2cm}
\frac{1}{2}\sum_\text{spin}|\mathcal{M}_{\tau2}|^2
\equiv
\overline{|\mathcal{M}_{\tau2}|^2}\, =\nonumber\\
\hspace{-0.2cm}&=&\hspace{-0.2cm}
\frac{2G_F^4|V_{M_1}|^2|V_{M_2}|^2
|U_{\tau 4}|^2|U_{\ell 4}|^2
m_4^2f_{M_1}^2f_{M_2}^2}
{\left(m_{23}^2-m_4^2\right)^2+m_4^2\Gamma_4^2}
\Biggl[
M_{\tau23M_1}^2M_{\ell23M_2}^2M_{M_212M_1}^2
+m_{M_1}^2m_{M_2}^2M_{\tau12\ell}^2
\nonumber\\
\hspace{-0.2cm}&&\hspace{-0.2cm}
\hspace{6.9cm}
+m_{M_2}^2M_{\tau23M_1}^2M_{\ell31M_1}^2+m_{M_1}^2M_{\tau31M_2}^2M_{\ell23M_2}^2
\Biggr],\hspace{0.6cm}
\end{eqnarray}
\begin{eqnarray}
\hspace{-0.2cm}&&\hspace{-0.2cm}
\sum_\text{spin}\mathrm{Re}\left(\mathcal{M}_{\tau1}\mathcal{M}_{\tau2}^*\right)
\, = \nonumber\\
\hspace{-0.2cm}&=&\hspace{-0.2cm}
-\frac{2G_F^4|V_{M_1}|^2|V_{M_2}|^2
|U_{\tau4}|^2|U_{\ell4}|^2
m_4^2f_{M_1}^2f_{M_2}^2}{\left(m_{31}^2-m_4^2\right)^2+m_4^2\Gamma_4^2}
\hspace{0.1cm}
\frac{\left(m_{23}^2-m_4^2\right)\left(m_{31}^2-m_4^2\right)+m_4^2\Gamma_4^2}
{\left(m_{23}^2-m_4^2\right)^2+m_4^2\Gamma_4^2}
\hspace{2.5cm}
\nonumber\\
\hspace{-0.2cm}&&\hspace{-0.2cm}
\times\Biggl[
\biggl\{
M_{\ell31M_1}^2M_{\tau31M_2}^2M_{M_112M_2}^2
+2m_{M_2}^2M_{\tau23M_1}^2M_{\ell31M_1}^2
+\Bigl(1\leftrightarrow2\Bigr)\biggr\}
\nonumber\\
\hspace{-0.2cm}&&\hspace{-0.2cm}
\hspace{0.5cm}
-M_{\tau12\ell}^2
\Bigl\{
m_{M_1}^4+m_{M_2}^4+m_{12}^4-2m_{M_1}^2m_{12}^2-2m_{M_2}^2m_{12}^2
\Bigr\}
\Biggr],
\label{eq:tau_inter}
\end{eqnarray}
with $M_{AijB}^2\equiv m_A^2-m_{ij}+m_B^2$. 
The decay width for $\tau^-\to M_1^-M_2^-\ell^+$ is finally given by
\begin{equation}
\Gamma_{\tau\to M_1M_2\ell}\,=\,
\frac{1}{32m_\tau^3\,(2\pi)^3}\int
\overline{|\mathcal{M}_{\tau}|^2}\,dm_{31}^2\,dm_{23}^2\,,
\label{eq:taudecay}
\end{equation}
in which the momentum variable $m_{12}^2$ in the squared amplitude has
been replaced using the relation
$m_{12}^2+m_{23}^2+m_{31}^2=m_\tau^2+m_{M_1}^2+m_{M_2}^2+m_\ell^2$. 
The intervals of the integral can be found in Ref.~\cite{Patrignani:2016xqp}. 
The branching ratio is obtained dividing the decay width by the total 
decay width of the tau lepton, 
$\Gamma_\tau=2.27\times10^{-12}~\mathrm{GeV}$~\cite{Patrignani:2016xqp}, 
\begin{equation}
\text{BR}(\tau\to M_1M_2\ell)\, =\, \frac{\Gamma_{\tau\to
 M_1M_2\ell}}{\Gamma_\tau}. 
\end{equation}

Since we are interested in regimes close to a resonance
($m_{31}^2\approx m_4^2$ or $m_{23}^2\approx m_4^2$), the narrow width
approximation can be applied as a good approximation. 
In this case, the propagator in the amplitude can be replaced by
a $\delta$-function as 
\begin{equation}\label{delta}
\frac{1}{\left(m_{ij}^2-m_4^2\right)^2+m_4^2\Gamma_4^2}
\quad\to\quad
\frac{\pi}{m_4\Gamma_4}\delta\left(m_{ij}^2-m_4^2\right).
\end{equation}
The interference term of Eq.~(\ref{eq:tau_inter}) contains in fact two
resonances; these can be split into two separate parts 
($f_1$ and $f_2$), each including only one 
resonance\footnote{Single-Diagram-Enhanced multi-channel integration~\cite{Maltoni:2002qb}.}, as
\begin{equation}
\overline{|\mathcal{M}_\tau|^2}\, =\, f_{\tau1}+f_{\tau2},
\end{equation}
with $f_{\tau i}$ defined by
\begin{equation}
f_{\tau1}\equiv
\frac{\overline{|\mathcal{M}_{\tau}|^2}}
{\overline{|\mathcal{M}_{\tau1}|^2}+\overline{|\mathcal{M}_{\tau2}|^2}}
\times
\overline{|\mathcal{M}_{\tau1}|^2},\qquad
f_{\tau2}\equiv
\frac{\overline{|\mathcal{M}_{\tau}|^2}}
{\overline{|\mathcal{M}_{\tau1}|^2}+\overline{|\mathcal{M}_{\tau2}|^2}}
\times
\overline{|\mathcal{M}_{\tau2}|^2}.
\end{equation}
This allows to remove one of the integrals in Eq.~(\ref{eq:taudecay}) 
by the $\delta$-function introduced in Eq.~(\ref{delta}). 
After applying the narrow width approximation, 
the remaining integration intervals are given by
\begin{equation}
\left(m_{23}^2\right)_{\mathrm{max}/\mathrm{min}}
=
\left(E_2^*+E_3^*\right)^2-
\left(\sqrt{{E_2^*}^2-m_{M_2}^2}\mp\sqrt{{E_3^*}^2-m_{\ell}^2}\right)^2,
\end{equation}
with $E_2^*=(m_\tau^2-m_{4}^2-m_{M_2}^2)/(2m_{4})$, and
$E_3^*=(m_{4}^2-m_{M_1}^2+m_{\ell}^2)/(2m_{4})$, and 
\begin{equation}
\left(m_{31}^2\right)_{\mathrm{max}/\mathrm{min}}
=
\left(E_3^*+E_1^*\right)^2-
\left(\sqrt{{E_3^*}^2-m_{\ell_\beta}^2}\mp\sqrt{{E_1^*}^2-m_{M_1}^2}\right)^2,
\end{equation}
with $E_1^*=(m_\tau^2-m_{4}^2-m_{M_1}^2)/(2m_{4})$, and
$E_3^*=(m_{4}^2-m_{M_2}^2+m_{\ell}^2)/(2m_{4})$.
Notice that the mass range of the sterile neutrino $\nu_4$ is limited to 
\begin{eqnarray}
m_\ell+m_{M_1}<m_4<m_\tau-m_{M_2},\\
m_\ell+m_{M_2}<m_4<m_\tau-m_{M_1},
\end{eqnarray}
respectively for the integrals $f_{\tau1}$ and $f_{\tau2}$, as a
consequence of the narrow width
approximation.

\subsection{Widths of semileptonic LNV meson decays}
Let us now consider the decay 
\begin{equation}
M_1(p,m_{M_1})\,\to\,
\ell_\alpha(k_1,m_{\ell_\alpha})\,\ell_\beta(k_2,m_{\ell_\beta})\,M_2(k_3,m_{M_2})\,, 
\end{equation}
where both $M_1$ and $M_2$ 
are pseudoscalar mesons. The corresponding amplitude is computed as
\begin{eqnarray}
i\mathcal{M}_P
\hspace{-0.2cm}&\equiv&\hspace{-0.2cm}
i\mathcal{M}_{P1}+i\mathcal{M}_{P2}\nonumber\\
\hspace{-0.2cm}&=&\hspace{-0.2cm}
2i\,G_F^2\,V_{M_1}\,V_{M_2}\,U_{\ell_\alpha 4}\,U_{\ell_\beta 4}\,m_4\,f_{M_1}\,f_{M_2}
\left[
\frac{\overline{u}(k_1)k\hspace{-0.18cm}/_3p\hspace{-0.18cm}/P_Rv(k_2)}
{m_{31}^2-m_4^2+im_4\,\Gamma_4}
+
\frac{\overline{u}(k_1)p\hspace{-0.18cm}/k\hspace{-0.18cm}/_3P_Rv(k_2)}
{m_{23}^2-m_4^2+im_4\,\Gamma_4}
\right].
\end{eqnarray}
The squared amplitude is given by
\begin{equation}
\overline{|\mathcal{M}_P|^2}\equiv
\sum_{\mathrm{spin}}\Bigl[
|\mathcal{M}_{P1}|^2+|\mathcal{M}_{P2}|^2
+2\mathrm{Re}\left(\mathcal{M}_{P1}\mathcal{M}_{P2}^*\right)
\Bigr]\,;
\end{equation}
the individual terms of the above expression can be cast as 
\begin{eqnarray}
\hspace{-0.2cm}&&\hspace{-0.2cm}
\sum_{\mathrm{spin}}|\mathcal{M}_{P1}|^2\equiv\overline{|\mathcal{M}_{P1}|^2}
\, =
\nonumber\\ 
\hspace{-0.2cm}&=&\hspace{-0.2cm}
-\frac{4G_F^4\,|V_{M_1}|^2\,|V_{M_2}|^2|\,U_{\ell_\alpha 4}|^2\,|U_{\ell_\beta 4}|^2
\,m_4^2\,f_{M_1}^2\,f_{M_2}^2}{\left(m_{31}^2-m_4^2\right)^2+m_4^2\,\Gamma_4^2}
\Biggl[
M_{\ell_\alpha31M_2}^2M_{\ell_\beta31M_1}^2M_{M_212M_1}^2
+m_{M_1}^2m_{M_2}^2M_{\ell_\alpha12\ell_\beta}^2
\hspace{-0.2cm}
\nonumber\\
\hspace{-0.2cm}&&\hspace{-0.2cm}
\hspace{7.1cm}
+m_{M_1}^2M_{\ell_\alpha31M_2}^2 M_{\ell_\beta23M_2}^2+ 
m_{M_2}^2M_{\ell_\alpha23M_1}^2M_{\ell_\beta31M_1}^2 
\Biggr],\hspace{0.7cm}
\end{eqnarray}
\begin{eqnarray}
\hspace{-0.2cm}&&\hspace{-0.2cm}
\sum_{\mathrm{spin}}|\mathcal{M}_{P2}|^2\equiv
\overline{|\mathcal{M}_{P2}|^2}\, =
\nonumber\\
\hspace{-0.2cm}&=&\hspace{-0.2cm}
-\frac{4G_F^4\,|V_{M_1}|^2\,|V_{M_2}|^2\,|U_{\ell_\alpha 4}|^2\,|U_{\ell_\beta 4}|^2
\,m_4^2\,f_{M_1}^2\,f_{M_2}^2}{\left(m_{23}^2-m_4^2\right)^2+m_4^2\,\Gamma_4^2}
\Biggl[
M_{\ell_\alpha23M_1}^2M_{\ell_\beta23M_2}^2M_{M_212M_1}^2
+m_{M_1}^2m_{M_2}^2M_{\ell_\alpha12\ell_\beta}^2
\hspace{-0.2cm}
\nonumber\\
\hspace{-0.2cm}&&\hspace{-0.2cm}
\hspace{7.1cm}
+m_{M_2}^2M_{\ell_\alpha23M_1}^2M_{\ell_\beta31M_1}^2+ 
m_{M_1}^2M_{\ell_\alpha31M_2}^2M_{\ell_\beta23M_2}^2
\Biggr],\hspace{0.7cm}
\end{eqnarray}
\begin{eqnarray}
\hspace{-0.2cm}&&\hspace{-0.2cm}
2\sum_{\mathrm{spin}}\mathrm{Re}\left(\mathcal{M}_{P1}\mathcal{M}_{P2}^*\right)
\, = \nonumber\\
\hspace{-0.2cm}&=&\hspace{-0.2cm}
+\, \frac{4G_F^4\,|V_{M_1}|^2\,|V_{M_2}|^2\,|U_{\ell_\alpha 4}|^2\,|U_{\ell_\beta 4}|^2
\,m_4^2\,f_{M_1}^2\,f_{M_2}^2}{\left(m_{31}^2-m_4^2\right)^2+m_4^2\,\Gamma_4^2}
\frac{\left(m_{23}^2-m_4^2\right)\left(m_{31}^2-m_4^2\right)+m_4^2\Gamma_4^2}
{\left(m_{23}^2-m_4^2\right)^2+m_4^2\,\Gamma_4^2}
\hspace{3.cm}
\nonumber\\
\hspace{-0.2cm}&&\hspace{-0.2cm}
\times\Biggl[
\biggl\{M_{M_112M_2}^2M_{M_131\ell_\beta}^2M_{\ell_\alpha31M_2}^2
+2m_{M_1}^2M_{\ell_\alpha31M_2}^2M_{\ell_\beta23M_2}^2
+(1\leftrightarrow2)\biggr\}\nonumber\\
\hspace{-0.2cm}&&\hspace{-0.2cm}
\hspace{0.5cm}
-
M_{\ell_\alpha12\ell_\beta}^2
\Bigl\{
m_{M_1}^4+m_{12}^4+m_{M_2}^4-2\left(m_{M_1}^2+m_{M_2}^2\right)m_{12}^2
\Bigr\}
\Biggr].
\end{eqnarray}
In the case in which $M_2$ is a vector meson, the amplitude is
computed as follows 
\begin{eqnarray}
i\mathcal{M}_V
\hspace{-0.2cm}&\equiv&\hspace{-0.2cm}
i\mathcal{M}_{V1}+i\mathcal{M}_{V2}\nonumber\\
\hspace{-0.2cm}&=&\hspace{-0.2cm}
2iG_F^2\,V_{M_1}\,V_{M_2}\,U_{\ell_\alpha 4}\,U_{\ell_\beta 4}\,m_4\,
f_{M_1}\,f_{M_2}\,m_{M_2}
\left[
\frac{\overline{u}(k_1)\epsilon\hspace{-0.17cm}/p\hspace{-0.18cm}/P_Rv(k_2)}
{m_{31}^2-m_4^2+im_4\,\Gamma_4}
+
\frac{\overline{u}(k_1)p\hspace{-0.18cm}/\epsilon\hspace{-0.17cm}/P_Rv(k_2)}
{m_{23}^2-m_4^2+im_4\,\Gamma_4}
\right],\hspace{0.7cm}
\end{eqnarray}
with $\epsilon_{\mu}(k_3)$ the polarisation vector of $M_2$.
The squared amplitude is then
\begin{equation}
\overline{|\mathcal{M}_V|^2}\,\equiv\,
\sum_{\mathrm{spin}}\Bigl[
|\mathcal{M}_{V1}|^2+|\mathcal{M}_{V2}|^2
+2\mathrm{Re}\left(\mathcal{M}_{V1}\mathcal{M}_{V2}^*\right)
\Bigr],
\end{equation}
with the contributing terms being given by
\begin{eqnarray}
\hspace{-0.2cm}&&\hspace{-0.2cm}
\sum_\mathrm{spin}|\mathcal{M}_{V1}|^2\equiv\overline{|\mathcal{M}_{V1}|^2}
\, = \nonumber\\
\hspace{-0.2cm}&=&\hspace{-0.2cm}
-\frac{4G_F^4\,|V_{M_1}|^2\,|V_{M_2}|^2\,|U_{\ell_\alpha 4}|^2\,|U_{\ell_\beta
4}|^2\,m_4^2\,f_{M_1}^2\,f_{M_2}^2}{(m_{31}^2-m_4^2)^2+m_4^2\,\Gamma_4^2}
\Biggl[
M_{\ell_\beta31M_1}^2M_{M_112M_2}^2M_{\ell_\alpha31M_2}^2
-m_{M_1}^2m_{M_2}^2M_{\ell_\alpha12\ell_\beta}^2
\hspace{-0.2cm}
\nonumber
\\
\hspace{-0.2cm}&&\hspace{-0.2cm}
\hspace{7.2cm}
+m_{M_1}^2M_{\ell_\beta23M_2}^2M_{\ell_\alpha31M_2}^2
-m_{M_2}^2M_{\ell_\beta31M_1}^2M_{\ell_\alpha23M_1}^2
\Biggr],\hspace{0.65cm}
\end{eqnarray}
\begin{eqnarray}
\hspace{-0.2cm}&&\hspace{-0.2cm}
\sum_\mathrm{spin}|\mathcal{M}_{V2}|^2\equiv\overline{|\mathcal{M}_{V2}|^2}
\, =\nonumber\\
\hspace{-0.2cm}&=&\hspace{-0.2cm}
-\frac{4G_F^4\,|V_{M_1}|^2\,|V_{M_2}|^2\,|U_{\ell_\alpha 4}|^2\,|U_{\ell_\beta
4}|^2\,m_4^2\,f_{M_1}^2\,f_{M_2}^2}{(m_{23}^2-m_4^2)^2+m_4^2\,\Gamma_4^2}
\Biggl[
M_{\ell_\alpha23M_1}^2M_{M_112M_2}^2M_{\ell_\beta23M_2}^2
-m_{M_1}^2m_{M_2}^2M_{\ell_\alpha12\ell_\beta}^2
\hspace{-0.2cm}
\nonumber\\
\hspace{-0.2cm}&&\hspace{-0.2cm}
\hspace{7.2cm}
+m_{M_1}^2M_{\ell_\alpha31M_2}^2M_{\ell_\beta23M_2}^2
-m_{M_2}^2M_{\ell_\alpha23M_1}^2M_{\ell_\beta31M_1}^2
\Biggr],
\hspace{0.65cm}
\end{eqnarray}
\begin{eqnarray}
\hspace{-0.2cm}&&\hspace{-0.2cm}
2\sum_\mathrm{spin}\mathrm{Re}\left(\mathcal{M}_1\mathcal{M}_2^*\right)
\, =\nonumber\\
\hspace{-0.2cm}&=&\hspace{-0.2cm}
\frac{4G_F^4\,|V_{M_1}|^2\,|V_{M_2}|^2\,|U_{\ell_\alpha 4}|^2\,|U_{\ell_\beta
4}|^2\,m_4^2\,f_{M_1}^2\,f_{M_2}^2}{(m_{31}^2-m_4^2)^2+m_4^2\,\Gamma_4^2}
\frac{(m_{31}^2-m_4^2)(m_{23}^2-m_4^2)+m_4^2\,\Gamma_4^2}{(m_{23}^2-m_4^2)^2+
m_4^2\,\Gamma_4^2}
\nonumber\\
\hspace{-0.2cm}&&\hspace{-0.2cm}
\times\Biggl[
M_{\ell_\alpha31M_2}^2M_{M_112M_2}^2M_{\ell_\beta31M_1}^2
+M_{\ell_\beta23M_2}^2M_{M_112M_2}^2M_{\ell_\alpha23M_1}^2
\nonumber\\
\hspace{-0.2cm}&&\hspace{-0.2cm}
\hspace{0.6cm}
+2m_{M_1}^2M_{\ell_\alpha31M_2}^2M_{\ell_\beta23M_2}^2
-2m_{M_2}^2M_{\ell_\beta31M_1}^2M_{\ell_\alpha23M_1}^2
-M_{\ell_\alpha12\ell_\beta}^2
\Bigl\{
M_{M_112M_2}^4-2m_{M_1}^2m_{M_2}^2
\Bigr\}\Biggr].
\hspace{1cm}
\end{eqnarray}
After applying the narrow width approximation 
(following the same procedure used in the case of the LNV $\tau$ decays), 
the decay width for the meson decay process is given by 
\begin{equation}
\Gamma_{M_1\to\ell_\alpha \ell_\beta M_2}\,=\,
\frac{1}{32m_{M_1}^3\,(2\pi)^3}
\hspace{-0.1cm}\int\hspace{-0.1cm}
\left(f_{P1/V1}+f_{P2/V2}\right)\,dm_{31}^2\,dm_{23}^2\,.
\end{equation}
The intervals of the integrals and $m_4$ range are obtained by replacing
$\tau\to M_1$, $M_1\to\ell_\alpha$, $M_2\to\ell_\beta$ and $\ell\to M_2$ in
the LNV tau decay width given in the previous subsection.

\section{On-shell sterile neutrino decay width}
\label{app:sec:widths:nu4}
We summarise here the relevant expressions allowing to derive the 
decay width of the sterile neutrino $\nu_4$~\cite{Atre:2009rg}. 
For the two body decay processes $\nu_4\to\ell
P^+,\nu_\ell P^0$ where $P^{+},P^{0}$ denotes a (charged/neutral) pseudoscalar
meson, the decay widths are given by 
\begin{eqnarray}
\Gamma_{\ell P^+}
\hspace{-0.2cm}&=&\hspace{-0.2cm}
\frac{G_F^2}{16\pi}\,f_{P^+}^2\,|U_{\ell4}|^2\,|V_{\overline{q}q^\prime}|^2\,m_4^3
\Bigl[
\left(1+x_\ell-x_{P^+}\right)\left(1+x_\ell\right)-4x_\ell
\Bigr]
\sqrt{\lambda(1,x_\ell,x_{P^+})},\\ 
\Gamma_{\nu_\ell P^0}
\hspace{-0.2cm}&=&\hspace{-0.2cm}
\frac{G_F^2}{64\pi}\,f_{P^0}^2\,|U_{\ell4}|^2\,m_4^3\,(1-x_{P^0})^2,
\end{eqnarray}
with $x_i=m_i^2/m_4^2$ and 
$\lambda(x,y,z)$ is the kinematical
function defined by 
\bea\label{lambda.function}
\lambda(x,y,z) = x^2+y^2+z^2-2xy-2yz-2zx\, , 
\eea
Similarly, the decay widths into vector mesons and a lepton, 
$\nu_4\to\ell V^+$ and
$\nu_\ell V^0$, are given by
\begin{eqnarray}
\Gamma_{\ell V^+}
\hspace{-0.2cm}&=&\hspace{-0.2cm}
\frac{G_F^2}{16\pi}\,f_{V^+}^2\,|U_{\ell4}|^2\,|V_{\overline{q}q^\prime}|^2\,m_4^3\,
\Bigl[
(1+x_{\ell}-x_{V^+})(1+x_\ell+2x_{V^+})-4x_\ell
\Bigr]
\sqrt{\lambda(1,x_\ell,x_{V^+})},\quad\\
\Gamma_{\nu_{\ell}V^0}
\hspace{-0.2cm}&=&\hspace{-0.2cm}
\frac{G_F^2}{2\pi}\,|U_{\ell4}|^2\,f_{V^0}^2\,\kappa_q^2\,m_4^3\,
(1-x_{V^0})^2(1+ 2 x_{V^0}),
\end{eqnarray}
in which $\kappa_q$ is defined by
\renewcommand{\arraystretch}{2}
\begin{equation}
\kappa_q\equiv\left\{
\begin{array}{rl}
\displaystyle
\frac{1}{4}-\frac{2}{3}\sin^2\theta_W &\mbox{ for up-type quarks}\\
\displaystyle
-\frac{1}{4}+\frac{1}{3}\sin^2\theta_W &\mbox{ for down-type quarks}
\end{array}
\right.,
\end{equation}
and $\sin\theta_W$ is the Weinberg angle.
For the three-body decay processes,
$\nu_4\to\ell_\alpha\,\overline{\ell_\beta}\,
\nu_{\ell_\beta}~(\ell_\alpha\neq\ell_\beta)$, 
$\nu_{\ell_\alpha}\,\ell_\beta\,\overline{\ell_\beta}$ and
$\nu_{\ell_\alpha}\,\nu_{\ell_\beta}\,\nu_{\ell_\beta}$,
the decay widths are respectively given by 
\begin{eqnarray}
\Gamma_{\ell_\alpha\overline{\ell_\beta}\nu_{\ell_\beta}}
\hspace{-0.2cm}&=&\hspace{-0.2cm}
\frac{G_F^2\,|U_{\ell_\alpha4}|^2}{192\pi^3}\,m_4^5\,
I_{1}\left(x_{\ell_\alpha},x_{\ell_\beta},x_{\nu_{\ell_\beta}}\right),\\
\Gamma_{\nu_{\ell_\alpha} \ell_\beta \overline{\ell_\beta}}
\hspace{-0.2cm}&=&\hspace{-0.2cm}
\frac{G_F^2\,|U_{\ell_\alpha4}|^2}{96\pi^3}\,m_4^5\Bigl[
\Bigl(
{\kappa_L^{\ell_\alpha}}^2+{\kappa_R^{\ell_\alpha}}^2+\delta_{\ell_\alpha \ell_\beta}
\left(1-2\kappa_L^{\ell_\alpha}\right)\Bigr)
I_1(x_{\nu_{\ell_\alpha}},x_{\ell_\beta},x_{\ell_\beta})
\nonumber\\
\hspace{-0.2cm}&&\hspace{-0.2cm}
\hspace{2.3cm}
+\kappa_L^{\ell_\alpha}\left(\kappa_R^{\ell_\alpha}-\delta_{\ell_\alpha \ell_\beta}\right)
I_2(x_{\nu_{\ell_\alpha}},x_{\ell_\beta},x_{\ell_\beta})
\Bigr],\\
\Gamma_{\nu_{\ell_\alpha}\nu_{\ell_\beta}\nu_{\ell_\beta}}
\hspace{-0.2cm}&=&\hspace{-0.2cm}
\frac{\left(1+\delta_{\ell_\alpha \ell_\beta}\right)}{4}\,
\frac{G_F^2\,|U_{\ell_\alpha4}|^2}{96\pi^3}\,m_4^5\,;
\end{eqnarray}
the quantities $\kappa_L^{\ell}$ and $\kappa_R^{\ell}$ are defined by
\begin{equation}
\kappa_L^{\ell}\equiv\frac{1}{2}-\sin^2\theta_W,\qquad
\kappa_R^{\ell}\equiv-\sin^2\theta_W,
\end{equation}
and the functions $I_1(x,y,z)$ and $I_2(x,y,z)$ given by 
\begin{eqnarray}
I_1(x,y,z)
\hspace{-0.2cm}&=&\hspace{-0.2cm}
12\int_{\left(\sqrt{x}+\sqrt{z}\right)^2}
^{\left(1-\sqrt{y}\right)^2}\frac{ds}{s}
\left(s-x-z\right)\left(1+y-s\right)
\sqrt{\lambda(s,x,z)}\sqrt{\lambda(s,1,y)},\\
I_2(x,y,z)
\hspace{-0.2cm}&=&\hspace{-0.2cm}
24\sqrt{yz}\int_{(\sqrt{y}+\sqrt{z})^2}^{(1-\sqrt{x})^2}\frac{ds}{s}
(1+x-s)\sqrt{\lambda(s,y,z)}\sqrt{\lambda(s,1,x)},
\end{eqnarray}
with the normalisations $I_1(0,0,0)=1$ and 
$\lim_{x,y,z\to0}\frac{I_2(x,y,z)}{\sqrt{yz}}=8$.

The total decay width of the sterile neutrino is then obtained by
summing over the contributions of the above listed possible final states
$\ell_\alpha$, $\ell_\beta$, $\nu_{\ell_\alpha}$, $\nu_{\ell_\beta}$, $P^+$, $P^0$,
$V^+$ and $V^0$. 
Notice that the conjugate processes $\nu_4\to \overline{\ell} P^-$, 
$\nu_4 \to \overline{\ell}
V^-$ and $\nu_4 \to \overline{\ell_\alpha}\ell_\beta
\nu_{\ell_\beta}~(\ell_\alpha\neq\ell_\beta)$
give the same contributions as $\nu_4\to \ell P^+$, $\ell V^+$
and $\ell_\alpha\overline{\ell_\beta}\nu_{\ell_\beta}~(\ell_\alpha\neq\ell_\beta)$, 
and must also be taken into account. 
Also notice that, concerning neutral meson decay channels, 
we only consider flavourless meson final states ($P^0 =
\pi^0,\eta,\eta',\eta_c,\eta_b$, $V^0 =\rho^0,\omega,\phi,J/\psi,\Upsilon$) 
since the decays into flavoured neutral mesons involve penguin-loop 
diagrams and are thus suppressed.

\section{Computation of widths for charged
  meson cLFV semileptonic decays}\label{app:sec:widths:lfv} 

While in SM extensions via additional sterile fermions
the cLFV semileptonic decay of neutral mesons, i.e.
$M_1^0\, \to \ell^\pm_\alpha\, \ell^\mp_\beta\, M_2^0$, necessarily occurs at
loop-level, the analogous decays of charged mesons can proceed via the
so-called ``double $W$'' diagrams, already at the tree-level. 

Should their mass allow for the production and propagation of real,
on-shell neutrinos on the s-channel (as is the case for the regimes we
are interested in our study) then one can write
\begin{equation}\label{eq:gamma:clfv:mesondecay}
\Gamma(M_1^+ \to \ell^\pm_\alpha \ell^\mp_\beta M_2^+) \, \approx \, 
\Gamma(M_1^+ \to \ell^\pm_\alpha \nu_k) \,\times \,
\frac{\Gamma(\nu_k \to \ell^\mp_\beta M_2^+)}{\Gamma^\text{tot}_{\nu_k}}\,,
\end{equation}
which is valid in the limit of narrow-width approximation for the
neutral fermion;  the computation of the different heavy neutrino
decay widths leading 
to $\Gamma^\text{tot}_{\nu_k}$ has been detailed 
in Appendix \ref{app:sec:widths:nu4}.
The width of a leptonic charged meson
decay can be cast as (see, for example~\cite{Abada:2013aba})
\begin{equation}\label{eq:gamma:clfv:mesondecay:Mlnu}
\Gamma(M_1^+ \to \ell^\pm \nu_k) \, =\, 
\frac{G_F^2}{8\, \pi}\, f_{M_1}^2\, m_{M_1}^3\, |U_{\ell k}|^2\,
|V_{M_1}|^2 \, \lambda^{1/2}(1, x_{\ell}, x_{\nu_k}) \,
\left[
x_{\nu_k} + x_{\ell} -(x_{\ell}-x_{\nu_k})^2 \right]\,,
\end{equation}
in which the kinematical
function $\lambda(1, x_{\ell}, x_{\nu_k})$ is given in Eq.~(\ref{lambda.function}), 
and all other quantities have been previously defined.
For completeness, we also include here the full charged meson decay
width, for the generic propagation of virtual, massive neutrinos~\cite{Helo:2010cw}:
\begin{align}\label{eq:gamma:clfv:mesondecay:full}
\Gamma(M_1^+ \to \ell^\pm_\alpha \ell^\mp_\beta M_2^+) \, = &\, 
\frac{G_F^4}{128\, \pi^3}\, f_{M_1}^2\, f_{M_2}^2\,
m_{M_1}^9\, |V_{M_1}|^2 \,|V_{M_2}|^2 \, \times \nonumber\\
& \int^{(1-\sqrt{x_{\ell_\beta}})^2}_{(\sqrt{x_{M_2}}
 +\sqrt{x_{\ell_\alpha}})^2} \,
ds\,
\sum_k \left|
\frac{ U_{\ell_\alpha k}\, U_{\ell_\beta k}^*}{sm_{M_1}^2-
 m_{\nu_k}^2+im_{\nu_k}\Gamma_{\nu_k}}
\right|^2\, 
\mathcal{F}(s,x_{M_2},x_{\ell_\alpha},x_{\ell_\beta},x_{\nu_k})\, ,
\end{align}
where 
\begin{align}\label{eq:gamma:clfv:mesondecay:FF}
\mathcal{F}(s,x_{M_2},x_{\ell_\alpha},x_{\ell_\beta},x_{\nu_k})\,= &\,
\frac{1}{s}\, \lambda^{1/2}(s, x_{M_2},x_{\ell_\alpha})\,
\lambda^{1/2}(s, 1,x_{\ell_\beta})\, \times \nonumber\\
& 
\left[ (s-x_{\ell_\alpha})^2 - x_{M_2} (s+x_{\ell_\alpha})
\right] \,
\left[ (s+x_{\ell_\beta}) - (s-x_{\ell_\beta})^2
\right] \,.
\end{align}


\newpage
{\small
\begin{thebibliography}{99}

 \bibitem{Tortola:2012te} 
D.~V.~Forero, M.~Tortola and J.~W.~F.~Valle,
%``Global status of neutrino oscillation parameters after Neutrino-2012,''
Phys.\ Rev.\ D {\bf 86} (2012) 073012 
[arXiv:1205.4018 [hep-ph]].
  
\bibitem{Fogli:2012ua}
G.~L.~Fogli, E.~Lisi, A.~Marrone, D.~Montanino, A.~Palazzo and A.~M.~Rotunno,
  %``Global analysis of neutrino masses, mixings and phases: entering
  %the era of leptonic CP violation searches,'' 
Phys.\ Rev.\ D {\bf 86} (2012) 013012
[arXiv:1205.5254 [hep-ph]].
 %%CITATION = ARXIV:1205.5254;%%

\bibitem{GonzalezGarcia:2012sz}
M.~C.~Gonzalez-Garcia, M.~Maltoni, J.~Salvado and T.~Schwetz,
%``Global fit to three neutrino mixing: critical look at present precision,''
JHEP {\bf 1212} (2012) 123
[arXiv:1209.3023 [hep-ph]].
%%CITATION = ARXIV:1209.3023;%%

\bibitem{Forero:2014bxa}
  D.~V.~Forero, M.~Tortola and J.~W.~F.~Valle,
  %``Neutrino oscillations refitted,''
  Phys.\ Rev.\ D {\bf 90} (2014) 093006
  [arXiv:1405.7540 [hep-ph]].

\bibitem{nufit}
See also http://www.nu-fit.org/

\bibitem{Gonzalez-Garcia:2014bfa}
  M.~C.~Gonzalez-Garcia, M.~Maltoni and T.~Schwetz,
  %``Updated fit to three neutrino mixing: status of leptonic CP violation,''
  JHEP {\bf 1411} (2014) 052
  [arXiv:1409.5439 [hep-ph]].
 
\bibitem{Esteban:2016qun}
  I.~Esteban, M.~C.~Gonzalez-Garcia, M.~Maltoni, I.~Martinez-Soler and
  T.~Schwetz, 
  ``Updated fit to three neutrino mixing: exploring the
  accelerator-reactor complementarity,'' 
  arXiv:1611.01514 [hep-ph].

\bibitem{deSalas:2017kay}
  P.~F.~de Salas, D.~V.~Forero, C.~A.~Ternes, M.~Tortola and J.~W.~F.~Valle,
  %``Status of neutrino oscillations 2017,''
  arXiv:1708.01186 [hep-ph].
  %%CITATION = ARXIV:1708.01186;%%

 \bibitem{Ali:2001gsa}
  A.~Ali, A.~V.~Borisov and N.~B.~Zamorin,
  %``Majorana neutrinos and same sign dilepton production at LHC and
  %in rare meson decays,'' 
  Eur.\ Phys.\ J.\ C {\bf 21} (2001) 123
  [hep-ph/0104123].
  %%CITATION = doi:10.1007/s100520100702;%%

\bibitem{Atre:2005eb}
  A.~Atre, V.~Barger and T.~Han,
  %``Upper bounds on lepton-number violating processes,''
  Phys.\ Rev.\ D {\bf 71} (2005) 113014
  [hep-ph/0502163].
  %%CITATION = doi:10.1103/PhysRevD.71.113014;%%

\bibitem{Atre:2009rg} 
  A.~Atre, T.~Han, S.~Pascoli and B.~Zhang,
  %``The Search for Heavy Majorana Neutrinos,''
  JHEP {\bf 0905}, 030 (2009)
  %doi:10.1088/1126-6708/2009/05/030
  [arXiv:0901.3589 [hep-ph]].
  %%CITATION = doi:10.1088/1126-6708/2009/05/030;%%
  
\bibitem{Chrzaszcz:2013uz}
  M.~Chrzaszcz,
  ``Searches for LFV and LNV Decays at LHCb,''
  arXiv:1301.2088 [hep-ex].
  
  \bibitem{Deppisch:2015qwa}
  F.~F.~Deppisch, P.~S.~Bhupal Dev and A.~Pilaftsis,
  %``Neutrinos and Collider Physics,''
  New J.\ Phys.\  {\bf 17} (2015) no.7,  075019
  doi:10.1088/1367-2630/17/7/075019
  [arXiv:1502.06541 [hep-ph]].

\bibitem{Cai:2017mow}
  Y.~Cai, T.~Han, T.~Li and R.~Ruiz,
  ``Lepton-Number Violation: Seesaw Models and Their Collider Tests,''
  arXiv:1711.02180 [hep-ph].
  %%CITATION = ARXIV:1711.02180;%%

\bibitem{Helo:2010cw}
  J.~C.~Helo, S.~Kovalenko and I.~Schmidt,
  %``Sterile neutrinos in lepton number and lepton flavor violating decays,''
  Nucl.\ Phys.\ B {\bf 853} (2011) 80
  %doi:10.1016/j.nuclphysb.2011.07.020
  [arXiv:1005.1607 [hep-ph]].
  %%CITATION = doi:10.1016/j.nuclphysb.2011.07.020;%%

\bibitem{Zhang:2010um}
  J.~M.~Zhang and G.~L.~Wang,
  %``Lepton-Number Violating Decays of Heavy Mesons,''
  Eur.\ Phys.\ J.\ C {\bf 71} (2011) 1715
  [arXiv:1003.5570 [hep-ph]].

\bibitem{Cvetic:2010rw}
  G.~Cvetic, C.~Dib, S.~K.~Kang and C.~S.~Kim,
  %``Probing Majorana neutrinos in rare K and D, ~D_s, B, B_c meson
	%decays,''
  Phys.\ Rev.\ D {\bf 82} (2010) 053010
  %doi:10.1103/PhysRevD.82.053010
  [arXiv:1005.4282 [hep-ph]].
  %%CITATION = doi:10.1103/PhysRevD.82.053010;%%

\bibitem{Quintero:2011yh}
  N.~Quintero, G.~Lopez Castro and D.~Delepine,
  %``Lepton number violation in top quark and neutral B meson decays,''
  Phys.\ Rev.\ D {\bf 84} (2011) 096011
   Erratum: [Phys.\ Rev.\ D {\bf 86} (2012) 079905]
  [arXiv:1108.6009 [hep-ph]].

\bibitem{Castro:2012gi}
G.~Lopez Castro and N.~Quintero,
%``Lepton number violating four-body tau lepton decays,''
Phys.\ Rev.\ D {\bf 85} (2012) 076006
 Erratum: [Phys.\ Rev.\ D {\bf 86} (2012) 079904]
%%doi:10.1103/PhysRevD.86.079904, 10.1103/PhysRevD.85.076006
[arXiv:1203.0537 [hep-ph]].

\bibitem{Castro:2012ma}
  G.~Lopez Castro and N.~Quintero,
  %``Lepton number violation in tau lepton decays,''
  Nucl.\ Phys.\ Proc.\ Suppl.\  {\bf 253-255} (2014) 12
  [arXiv:1212.0037 [hep-ph]].

\bibitem{Castro:2013jsn}
  G.~L.~Castro and N.~Quintero,
  %``Bounding resonant Majorana neutrinos from four-body B and D decays,''
  Phys.\ Rev.\ D {\bf 87} (2013) 077901
  [arXiv:1302.1504 [hep-ph]].
  %%CITATION = doi:10.1103/PhysRevD.87.077901;%%

\bibitem{Yuan:2013yba}
H.~Yuan, T.~Wang, G.~L.~Wang, W.~L.~Ju and J.~M.~Zhang,
%``Lepton-number violating four-body decays of heavy mesons,''
JHEP {\bf 1308} (2013) 066
%%doi:10.1007/JHEP08(2013)066
[arXiv:1304.3810 [hep-ph]].
%%CITATION = doi:10.1007/JHEP08(2013)066;%%

\bibitem{Wang:2014lda}
  Y.~Wang, S.~S.~Bao, Z.~H.~Li, N.~Zhu and Z.~G.~Si,
  %``Study Majorana Neutrino Contribution to B-meson Semi-leptonic
  %Rare Decays,'' 
  Phys.\ Lett.\ B {\bf 736} (2014) 428
  [arXiv:1407.2468 [hep-ph]].

\bibitem{Cvetic:2014nla}
G.~Cvetic, C.~S.~Kim and J.~Zamora-Saa,  
%``CP violation in lepton number violating semihadronic decays of
%$K,D,D_s,B,B_c$,''  
  Phys.\ Rev.\ D {\bf 89} (2014) no.9,  093012
  %doi:10.1103/PhysRevD.89.093012
  [arXiv:1403.2555 [hep-ph]].
  %%CITATION = doi:10.1103/PhysRevD.89.093012;%%

\bibitem{Cvetic:2015naa}
  G.~Cvetic, C.~Dib, C.~S.~Kim and J.~Zamora-Saa,
  %``Probing the Majorana neutrinos and their CP violation in decays
  %of charged scalar mesons $\pi, K, D, D_s, B, B_c$,'' 
  Symmetry {\bf 7} (2015) 726
  %doi:10.3390/sym7020726
  [arXiv:1503.01358 [hep-ph]].
  %%CITATION = doi:10.3390/sym7020726;%%

\bibitem{Mandal:2016hpr}
  S.~Mandal and N.~Sinha,
  ``Favoured $B_c$ Decay modes to search for a Majorana neutrino,''
  arXiv:1602.09112 [hep-ph].
  %%CITATION = ARXIV:1602.09112;%%

\bibitem{Milanes:2016rzr}
  D.~Milanes, N.~Quintero and C.~E.~Vera,
  %``Sensitivity to Majorana neutrinos in $\Delta L=2$ decays of $B_c$
	%meson at LHCb,''
  Phys.\ Rev.\ D {\bf 93} (2016) no.9,  094026
  %doi:10.1103/PhysRevD.93.094026
  [arXiv:1604.03177 [hep-ph]].
  %%CITATION = doi:10.1103/PhysRevD.93.094026;%%

\bibitem{Quintero:2016iwi}
  N.~Quintero,
  %``Constraints on lepton number violating short-range interactions
  %from $|\Delta L|=2$ processes,'' 
  Phys.\ Lett.\ B {\bf 764} (2017) 60
  [arXiv:1606.03477 [hep-ph]].
  %%CITATION = doi:10.1016/j.physletb.2016.10.056;%%

\bibitem{Cvetic:2016fbv}
  G.~Cvetic and C.~S.~Kim,
  %``Rare decays of B mesons via on-shell sterile neutrinos,''
  Phys.\ Rev.\ D {\bf 94} (2016) no.5,  053001
   Erratum: [Phys.\ Rev.\ D {\bf 95} (2017) no.3,  039901]
  [arXiv:1606.04140 [hep-ph]].
  %%CITATION = doi:10.1103/PhysRevD.95.039901, 10.1103/PhysRevD.94.053001;%%

\bibitem{Liu:2016oph}
  J.~H.~Liu, J.~Zhang and S.~Zhou,
  %``Majorana Neutrino Masses from Neutrinoless Double-Beta Decays and
  %Lepton-Number-Violating Meson Decays,'' 
  Phys.\ Lett.\ B {\bf 760} (2016) 571
  [arXiv:1606.04886 [hep-ph]].
  %%CITATION = doi:10.1016/j.physletb.2016.07.043;%%

\bibitem{Asaka:2016rwd} 
  T.~Asaka and H.~Ishida,
  %``Lepton number violation by heavy Majorana neutrino in $B$ decays,''
  Phys.\ Lett.\ B {\bf 763}, 393 (2016)
  %doi:10.1016/j.physletb.2016.10.070
  [arXiv:1609.06113 [hep-ph]].
  
\bibitem{Yuan:2017xdp}
  H.~Yuan, Y.~Jiang, T.~h.~Wang, Q.~Li and G.~L.~Wang,
  %``Testing the nature of neutrinos from four-body ? decays,''
  J.\ Phys.\ G {\bf 44} (2017) no.11,  115002
  [arXiv:1702.04555 [hep-ph]].
  %%CITATION = doi:10.1088/1361-6471/aa8a1e;%%

\bibitem{Cvetic:2017vwl}
  G.~Cvetic and C.~S.~Kim,
  %``Sensitivity limits on heavy-light mixing $|U_{\mu N}|^2$ from
  %lepton number violating $B$ meson decays,'' 
  Phys.\ Rev.\ D {\bf 96} (2017) no.3,  035025
  [arXiv:1705.09403 [hep-ph]].

\bibitem{Mejia-Guisao:2017gqp}
  J.~Mejia-Guisao, D.~Milanes, N.~Quintero and J.~D.~Ruiz-Alvarez,
  ``Lepton number violation in $B_s$ meson decays induced by an
  on-shell Majorana neutrino,'' 
  arXiv:1708.01516 [hep-ph].
  %%CITATION = ARXIV:1708.01516;%%

\bibitem{Yuan:2017uyq}
  H.~Yuan, T.~Wang, Y.~Jiang, Q.~Li and G.~L.~Wang,
  ``Four-body decays of $B$ meson with lepton number violation,''
  arXiv:1710.03886 [hep-ph].
  
\bibitem{GomezCadenas:2011it}
  J.~J.~Gomez-Cadenas, J.~Martin-Albo, M.~Mezzetto, F.~Monrabal and M.~Sorel,
  %``The Search for neutrinoless double beta decay,''
  Riv.\ Nuovo Cim.\  {\bf 35} (2012) 29
  [arXiv:1109.5515 [hep-ex]].
  %%CITATION = doi:10.1393/ncr/i2012-10074-9;%%

\bibitem{Flanz:1999ah}
  M.~Flanz, W.~Rodejohann and K.~Zuber,
  %``Bounds on effective Majorana neutrino masses at HERA,''
  Phys.\ Lett.\ B {\bf 473} (2000) 324
   Erratum: [Phys.\ Lett.\ B {\bf 480} (2000) 418]
  [hep-ph/9911298].

\bibitem{Zuber:2000vy}
K.~Zuber,
%``New limits on effective Majorana neutrino masses from rare kaon decays,''
Phys.\ Lett.\ B {\bf 479} (2000) 33
%%doi:10.1016/S0370-2693(00)00333-6
[hep-ph/0003160].

\bibitem{Zuber:2000ca}
  K.~Zuber,
  ``Effective Majorana neutrino masses,''
  hep-ph/0008080.

\bibitem{Rodejohann:2011mu}
W.~Rodejohann,
%``Neutrino-less Double Beta Decay and Particle Physics,''
Int.\ J.\ Mod.\ Phys.\ E {\bf 20} (2011) 1833
%%doi:10.1142/S0218301311020186
[arXiv:1106.1334 [hep-ph]].
  

\bibitem{Patrignani:2016xqp}
  C.~Patrignani {\it et al.} [Particle Data Group],
  %``Review of Particle Physics,''
  Chin.\ Phys.\ C {\bf 40} (2016) no.10,  100001.
  %doi:10.1088/1674-1137/40/10/100001
  %%CITATION = doi:10.1088/1674-1137/40/10/100001;%%
  
\bibitem{Amhis:2016xyh}
  Y.~Amhis {\it et al.},
  ``Averages of $b$-hadron, $c$-hadron, and $\tau$-lepton properties
  %as of summer 2016,'' 
  arXiv:1612.07233 [hep-ex].
  %%CITATION = ARXIV:1612.07233;%%
     
\bibitem{TheMEG:2016wtm}
  A.~M.~Baldini {\it et al.} [MEG Collaboration],
  %``Search for the lepton flavour violating decay $\mu ^+ \rightarrow
  %\mathrm {e}^+ \gamma $ with the full dataset of the MEG
  %experiment,'' 
  Eur.\ Phys.\ J.\ C {\bf 76} (2016) no.8,  434
  %doi:10.1140/epjc/s10052-016-4271-x
  [arXiv:1605.05081 [hep-ex]].
  %%CITATION = doi:10.1140/epjc/s10052-016-4271-x;%%
  
\bibitem{Baldini:2013ke}
  A.~M.~Baldini {\it et al.},
  ``MEG Upgrade Proposal,''
  arXiv:1301.7225 [physics.ins-det].
  %%CITATION = ARXIV:1301.7225;%%
  
\bibitem{Bellgardt:1987du}
  U.~Bellgardt {\it et al.} [SINDRUM Collaboration],
  %``Search for the Decay mu+ ---> e+ e+ e-,''
  Nucl.\ Phys.\ B {\bf 299} (1988) 1.
  %%CITATION = doi:10.1016/0550-3213(88)90462-2;%%
  
\bibitem{Blondel:2013ia}
  A.~Blondel {\it et al.},
  ``Research Proposal for an Experiment to Search for the Decay $\mu \to eee$,''
  arXiv:1301.6113 [physics.ins-det].
  %%CITATION = ARXIV:1301.6113;%%
  
\bibitem{Bertl:2006up}
  W.~H.~Bertl {\it et al.} [SINDRUM II Collaboration],
  %``A Search for muon to electron conversion in muonic gold,''
  Eur.\ Phys.\ J.\ C {\bf 47} (2006) 337.
  %%CITATION = doi:10.1140/epjc/s2006-02582-x;%%
  
\bibitem{Carey:2008zz}
  R.~M.~Carey {\it et al.} [Mu2e Collaboration],
  ``Proposal to search for $\mu^- N \to e^- N$ with a single event
  sensitivity below $10^{-16}$,'' 
  FERMILAB-PROPOSAL-0973.
  %%CITATION = FERMILAB-PROPOSAL-0973;%%
  
\bibitem{Cui:2009zz}
  Y.~G.~Cui {\it et al.} [COMET Collaboration],
  ``Conceptual design report for experimental search for lepton flavor
  violating mu- - e- conversion at sensitivity of 10**(-16) with a
  slow-extracted bunched proton beam (COMET),'' 
  KEK-2009-10.
  %%CITATION = KEK-2009-10;%%
  
\bibitem{Kuno:2013mha}
  Y.~Kuno [COMET Collaboration],
  %%``A search for muon-to-electron conversion at J-PARC: The COMET
  %%experiment,'' 
  PTEP {\bf 2013} (2013) 022C01.
  %%CITATION = doi:10.1093/ptep/pts089;%%
  
  \bibitem{Helo:2011yg}
  J.~C.~Helo, S.~Kovalenko and I.~Schmidt,
  %``On sterile neutrino mixing with \nu_{\tau},''
  Phys.\ Rev.\ D {\bf 84} (2011) 053008
  %doi:10.1103/PhysRevD.84.053008
  [arXiv:1105.3019 [hep-ph]].
  %%CITATION = doi:10.1103/PhysRevD.84.053008;%%
  
\bibitem{Zamora-Saa:2016qlk}
  G.~Moreno and J.~Zamora-Saa,
  %``Rare meson decays with three pairs of quasi-degenerate heavy neutrinos,''
  Phys.\ Rev.\ D {\bf 94} (2016) no.9,  093005
  doi:10.1103/PhysRevD.94.093005
  [arXiv:1606.08820 [hep-ph]].
  
  \bibitem{Zamora-Saa:2016ito}
  J.~Zamora-Saa,
  %``Resonant $CP$ violation in rare $\tau^{\pm}$ decays,''
  JHEP {\bf 1705} (2017) 110
  doi:10.1007/JHEP05(2017)110
  [arXiv:1612.07656 [hep-ph]].
  
\bibitem{Gando:2012zm}
  A.~Gando {\it et al.} [KamLAND-Zen Collaboration],
  %``Limit on Neutrinoless $\beta\beta$ Decay of $^{136}$Xe from the
  %First Phase of KamLAND-Zen and Comparison with the Positive Claim
  %in $^{76}$Ge,'' 
 Phys.\ Rev.\ Lett.\  {\bf 110} (2013) no.6,  062502
[arXiv:1211.3863 [hep-ex]].

\bibitem{KamLAND-Zen:2016pfg}
  A.~Gando {\it et al.} [KamLAND-Zen Collaboration],
  %``Search for Majorana Neutrinos near the Inverted Mass Hierarchy
  %Region with KamLAND-Zen,'' 
  Phys.\ Rev.\ Lett.\  {\bf 117} (2016) no.8,  082503
  Addendum: [Phys.\ Rev.\ Lett.\  {\bf 117} (2016) no.10,  109903]
%%doi:10.1103/PhysRevLett.117.109903, 10.1103/PhysRevLett.117.082503
[arXiv:1605.02889 [hep-ex]].

\bibitem{Agostini:2013mzu}
  M.~Agostini {\it et al.} [GERDA Collaboration],
  %``Results on Neutrinoless Double-$\beta$ Decay of $^{76}$Ge from
  %Phase I of the GERDA Experiment,'' 
  Phys.\ Rev.\ Lett.\  {\bf 111} (2013) no.12,  122503
  [arXiv:1307.4720 [nucl-ex]].

\bibitem{Auger:2012ar}
  M.~Auger {\it et al.} [EXO-200 Collaboration],
  %``Search for Neutrinoless Double-Beta Decay in $^{136}$Xe with EXO-200,''
  Phys.\ Rev.\ Lett.\  {\bf 109} (2012) 032505
  [arXiv:1205.5608 [hep-ex]].

\bibitem{Albert:2017owj}
  J.~B.~Albert {\it et al.} [EXO Collaboration],
  ``Search for Neutrinoless Double-Beta Decay with the Upgraded
  EXO-200 Detector,''  arXiv:1707.08707 [hep-ex]. 
  
\bibitem{Albert:2014awa}
  J.~B.~Albert {\it et al.} [EXO-200 Collaboration],
  %``Search for Majorana neutrinos with the first two years of EXO-200
  %data,''
  Nature {\bf 510} (2014) 229 [arXiv:1402.6956 [nucl-ex]].

\bibitem{DellOro:2016tmg}
  S.~Dell'Oro, S.~Marcocci, M.~Viel and F.~Vissani,
  %``Neutrinoless double beta decay: 2015 review,''
  Adv.\ High Energy Phys.\  {\bf 2016} (2016) 2162659
  [arXiv:1601.07512 [hep-ph]].
  %%CITATION = doi:10.1155/2016/2162659;%%
  
\bibitem{Maneschg:2017mzu}
  W.~Maneschg,
  ``Present status of neutrinoless double beta decay searches,''
  arXiv:1704.08537 [physics.ins-det].
  %%CITATION = ARXIV:1704.08537;%%
  
\bibitem{Tosi:2014zza}
  D.~Tosi [EXO-200 Collaboration],
  ``The search for neutrino-less double-beta decay: summary of current
  experiments,''  arXiv:1402.1170 [nucl-ex]. 

\bibitem{Licciardi:2017oqg}
  C.~Licciardi [nEXO Collaboration],
  %``The Sensitivity of the nEXO Experiment to Majorana Neutrinos,''
  J.\ Phys.\ Conf.\ Ser.\  {\bf 888} (2017) no.1,  012237.

\bibitem{Agostini:2017iyd}
  M.~Agostini {\it et al.},
  %``Background free search for neutrinoless double beta decay with
  %GERDA Phase II,'' 
  Nature {\bf 544} (2017) 47
  [arXiv:1703.00570 [nucl-ex]].

\bibitem{Gorla:2012gd}
  P.~Gorla [CUORE Collaboration],
  %``The CUORE experiment: Status and prospects,''
  J.\ Phys.\ Conf.\ Ser.\  {\bf 375} (2012) 042013.

\bibitem{Aguirre:2014lua}
  D.~R.~Artusa {\it et al.} [CUORE Collaboration],
  %``Initial performance of the CUORE-0 experiment,''
  Eur.\ Phys.\ J.\ C {\bf 74} (2014) no.8,  2956
  [arXiv:1402.0922 [physics.ins-det]].

\bibitem{Artusa:2014lgv}
  D.~R.~Artusa {\it et al.} [CUORE Collaboration],
  %``Searching for neutrinoless double-beta decay of $^{130}$Te with CUORE,''
  Adv.\ High Energy Phys.\  {\bf 2015} (2015) 879871
  [arXiv:1402.6072 [physics.ins-det]].

\bibitem{Hartnell:2012qd}
  J.~Hartnell [SNO+ Collaboration],
  %``Neutrinoless Double Beta Decay with SNO+,''
  J.\ Phys.\ Conf.\ Ser.\  {\bf 375} (2012) 042015
  [arXiv:1201.6169 [physics.ins-det]].
 
  \bibitem{Lozza:2016ghw}
  V.~Lozza [SNO+ Collaboration],
  %``The SNO+ Experiment for Neutrinoless Double-Beta Decay,''
  Nucl.\ Part.\ Phys.\ Proc.\  {\bf 273-275} (2016) 1836.
  
\bibitem{Barabash:2011aa}
  A.~S.~Barabash,
  %``SeperNEMO double beta decay experiment,''
  J.\ Phys.\ Conf.\ Ser.\  {\bf 375} (2012) 042012
  [arXiv:1112.1784 [nucl-ex]].
  
\bibitem{Granena:2009it}
  F.~Granena {\it et al.} [NEXT Collaboration],
  ``NEXT, a HPGXe TPC for neutrinoless double beta decay searches,''
  arXiv:0907.4054 [hep-ex].
  %%CITATION = ARXIV:0907.4054;%%
  
\bibitem{Gomez-Cadenas:2013lta}
  J.~J.~Gomez-Cadenas {\it et al.} [NEXT Collaboration],
  %``Present status and future perspectives of the NEXT experiment,''
  Adv.\ High Energy Phys.\  {\bf 2014} (2014) 907067
  [arXiv:1307.3914 [physics.ins-det]].
  
  \bibitem{Phillips:2011db}
  D.~G.~Phillips, II {\it et al.},
  %``The Majorana experiment: an ultra-low background search for
  %neutrinoless double-beta decay,'' 
  J.\ Phys.\ Conf.\ Ser.\  {\bf 381} (2012) 012044
  [arXiv:1111.5578 [nucl-ex]].
  
\bibitem{Wilkerson:2012ga}
  J.~F.~Wilkerson {\it et al.},
  %``The MAJORANA demonstrator: A search for neutrinoless double-beta
  %decay of germanium-76,'' 
  J.\ Phys.\ Conf.\ Ser.\  {\bf 375} (2012) 042010.
  
\bibitem{LHCb:2011dta}
  The LHCb Collaboration [LHCb Collaboration],
  ``Letter of Intent for the LHCb Upgrade,''
  CERN-LHCC-2011-001.
  %%CITATION = CERN-LHCC-2011-001;%%

  \bibitem{na62}
NA62 Physics Handbook, {http://na62pb.ph.tum.de/}
  
\bibitem{HerediadelaCruz:2016yqi}
  I.~Heredia de la Cruz,
  %``The Belle II experiment: fundamental physics at the flavor frontier,''
  J.\ Phys.\ Conf.\ Ser.\  {\bf 761} (2016) no.1,  012017
  [arXiv:1609.01806 [hep-ex]].

\bibitem{Lees:2011hb}
  J.~P.~Lees {\it et al.} [BaBar Collaboration],
  %``Searches for Rare or Forbidden Semileptonic Charm Decays,''
  Phys.\ Rev.\ D {\bf 84} (2011) 072006
  %doi:10.1103/PhysRevD.84.072006
  [arXiv:1107.4465 [hep-ex]].
  %%CITATION = doi:10.1103/PhysRevD.84.072006;%%

\bibitem{Aaij:2013sua}
  R.~Aaij {\it et al.} [LHCb Collaboration],
  %``Search for $D^+_s \to \pi^+ \mu^+ \mu^-$ and $D^+_s \to \pi^- \mu^+
	%\mu^+$ decays,''
  Phys.\ Lett.\ B {\bf 724} (2013) 203
  %doi:10.1016/j.physletb.2013.06.010
  [arXiv:1304.6365 [hep-ex]].
  %%CITATION = doi:10.1016/j.physletb.2013.06.010;%%

\bibitem{BABAR:2012aa}
  J.~P.~Lees {\it et al.} [BaBar Collaboration],
  %``Search for lepton-number violating processes in $B^+ \to h^- l^+
	%l^+$ decays,''
  Phys.\ Rev.\ D {\bf 85} (2012) 071103
  %doi:10.1103/PhysRevD.85.071103
  [arXiv:1202.3650 [hep-ex]].
  %%CITATION = doi:10.1103/PhysRevD.85.071103;%%

\bibitem{Lees:2013gdj}
  J.~P.~Lees {\it et al.} [BaBar Collaboration],
  %``Search for lepton-number violating decays,''
  Phys.\ Rev.\ D {\bf 89} (2014) no.1,  011102
  %doi:10.1103/PhysRevD.89.011102
  [arXiv:1310.8238 [hep-ex]].
  %%CITATION = doi:10.1103/PhysRevD.89.011102;%%
  
\bibitem{Aaij:2014aba}
  R.~Aaij {\it et al.} [LHCb Collaboration],
  %``Search for Majorana neutrinos in $B^- \to \pi^+\mu^-\mu^-$
	%decays,''
  Phys.\ Rev.\ Lett.\  {\bf 112} (2014) no.13,  131802
  %doi:10.1103/PhysRevLett.112.131802
  [arXiv:1401.5361 [hep-ex]].
  %%CITATION = doi:10.1103/PhysRevLett.112.131802;%%
  
\bibitem{Aaij:2011ex}
  R.~Aaij {\it et al.} [LHCb Collaboration],
  %``Search for the lepton number violating decays $B^{+}\to \pi^- \mu^+
	%\mu^+$ and $B^{+}\to K^- \mu^+ \mu^+$,''
  Phys.\ Rev.\ Lett.\  {\bf 108} (2012) 101601
  %doi:10.1103/PhysRevLett.108.101601
  [arXiv:1110.0730 [hep-ex]].
  %%CITATION = doi:10.1103/PhysRevLett.108.101601;%%

\bibitem{Seon:2011ni}
  O.~Seon {\it et al.} [BELLE Collaboration],
  %``Search for Lepton-number-violating $B^+ \to D^- l^+ l^{\prime +}$
	%Decays,''
  Phys.\ Rev.\ D {\bf 84} (2011) 071106
  %doi:10.1103/PhysRevD.84.071106
  [arXiv:1107.0642 [hep-ex]].
  %%CITATION = doi:10.1103/PhysRevD.84.071106;%%

\bibitem{Aaij:2012zr}
  R.~Aaij {\it et al.} [LHCb Collaboration],
  %``Searches for Majorana neutrinos in $B^-$ decays,''
  Phys.\ Rev.\ D {\bf 85} (2012) 112004
  %doi:10.1103/PhysRevD.85.112004
  [arXiv:1201.5600 [hep-ex]].
  %%CITATION = doi:10.1103/PhysRevD.85.112004;%%

\bibitem{Aoki:2016frl}
  S.~Aoki {\it et al.},
  %``Review of lattice results concerning low-energy particle physics,''
  Eur.\ Phys.\ J.\ C {\bf 77} (2017) no.2,  112
  %doi:10.1140/epjc/s10052-016-4509-7
  [arXiv:1607.00299 [hep-lat]].
  %%CITATION = doi:10.1140/epjc/s10052-016-4509-7;%%

\bibitem{Miyazaki:2012mx}
  Y.~Miyazaki {\it et al.} [Belle Collaboration],
  %``Search for Lepton-Flavor-Violating and Lepton-Number-Violating
	%$\tau \to \ell h h^\prime$ Decay Modes,''
  Phys.\ Lett.\ B {\bf 719} (2013) 346
  %doi:10.1016/j.physletb.2013.01.032
  [arXiv:1206.5595 [hep-ex]].
  %%CITATION = doi:10.1016/j.physletb.2013.01.032;%%

\bibitem{Duplancic:2015zna}
  G.~Duplancic and B.~Melic,
  %``Form factors of B, B$_{s}$ ? ?$^{(}$?$^{)}$ and D, D$_{s}$ ?
  %?$^{(}$?$^{)}$ transitions from QCD light-cone sum rules,'' 
  JHEP {\bf 1511} (2015) 138
  %doi:10.1007/JHEP11(2015)138
  [arXiv:1508.05287 [hep-ph]].
  %%CITATION = doi:10.1007/JHEP11(2015)138;%%
  
\bibitem{Becirevic:2013bsa}
  D.~Becirevic, G.~Duplancic, B.~Klajn, B.~Melic and F.~Sanfilippo,
  %``Lattice QCD and QCD sum rule determination of the decay constants
  %of $\eta_c$, J/$\psi$ and $h_c$ states,'' 
  Nucl.\ Phys.\ B {\bf 883} (2014) 306
  %doi:10.1016/j.nuclphysb.2014.03.024
  [arXiv:1312.2858 [hep-ph]].
  %%CITATION = doi:10.1016/j.nuclphysb.2014.03.024;%%
  
\bibitem{McNeile:2012qf}
  C.~McNeile, C.~T.~H.~Davies, E.~Follana, K.~Hornbostel and G.~P.~Lepage,
  %``Heavy meson masses and decay constants from relativistic heavy
  %quarks in full lattice QCD,'' 
  Phys.\ Rev.\ D {\bf 86} (2012) 074503
  %doi:10.1103/PhysRevD.86.074503
  [arXiv:1207.0994 [hep-lat]].
  %%CITATION = doi:10.1103/PhysRevD.86.074503;%%

\bibitem{Colquhoun:2015oha}
  B.~Colquhoun {\it et al.} [HPQCD Collaboration],
  %``B-meson decay constants: a more complete picture from full lattice QCD,''
  Phys.\ Rev.\ D {\bf 91} (2015) no.11,  114509
  %doi:10.1103/PhysRevD.91.114509
  [arXiv:1503.05762 [hep-lat]].
  %%CITATION = doi:10.1103/PhysRevD.91.114509;%%
  
\bibitem{Gregory:2009hq}
  E.~B.~Gregory {\it et al.},
  %``A Prediction of the B*(c) mass in full lattice QCD,''
  Phys.\ Rev.\ Lett.\  {\bf 104} (2010) 022001
  %doi:10.1103/PhysRevLett.104.022001
  [arXiv:0909.4462 [hep-lat]].
  %%CITATION = doi:10.1103/PhysRevLett.104.022001;%%

\bibitem{Rosner:2015wva}
  J.~L.~Rosner, S.~Stone and R.~S.~Van de Water,
  ``Leptonic Decays of Charged Pseudoscalar Mesons - 2015,''
  arXiv:1509.02220 [hep-ph].
  %%CITATION = ARXIV:1509.02220;%%
  
  
\bibitem{Straub:2015ica}
  A.~Bharucha, D.~M.~Straub and R.~Zwicky,
  %``$B\to V\ell^+\ell^-$ in the Standard Model from light-cone sum rules,''
  JHEP {\bf 1608} (2016) 098
  %doi:10.1007/JHEP08(2016)098
  [arXiv:1503.05534 [hep-ph]].
  %%CITATION = doi:10.1007/JHEP08(2016)098;%%

\bibitem{Lucha:2014xla}
  W.~Lucha, D.~Melikhov and S.~Simula,
  %``Decay constants of the charmed vector mesons $D^*$ and $D^*_s$
  %from QCD sum rules,'' 
  Phys.\ Lett.\ B {\bf 735} (2014) 12
  %doi:10.1016/j.physletb.2014.06.007
  [arXiv:1404.0293 [hep-ph]].
  %%CITATION = doi:10.1016/j.physletb.2014.06.007;%%

\bibitem{Ue_exclusions}
%\bibitem{Britton:1992xv}
  D.~I.~Britton {\it et al.},
  %``Improved search for massive neutrinos in pi+ ---> e+ neutrino decay,''
  Phys.\ Rev.\ D {\bf 46} (1992) R885;\\
  %doi:10.1103/PhysRevD.46.R885
  %%CITATION = doi:10.1103/PhysRevD.46.R885;%%
%\bibitem{Britton:1992pg}
  D.~I.~Britton {\it et al.},
  %``Measurement of the pi+ ---> e+ neutrino branching ratio,''
  Phys.\ Rev.\ Lett.\  {\bf 68} (1992) 3000;\\
  %doi:10.1103/PhysRevLett.68.3000
  %%CITATION = doi:10.1103/PhysRevLett.68.3000;%%
%\bibitem{Baranov:1992vq}
  S.~A.~Baranov {\it et al.},
  %``Search for heavy neutrinos at the IHEP-JINR neutrino detector,''
  Phys.\ Lett.\ B {\bf 302} (1993) 336;\\
  %doi:10.1016/0370-2693(93)90405-7
  %%CITATION = doi:10.1016/0370-2693(93)90405-7;%%
%\bibitem{Back:2003ae}
  H.~O.~Back {\it et al.},
  %``New experimental limits on heavy neutrino mixing in B-8 decay obtained with the Borexino Counting Test Facility,''
  JETP Lett.\  {\bf 78} (2003) 261
   [Pisma Zh.\ Eksp.\ Teor.\ Fiz.\  {\bf 78} (2003) 707];\\
  %doi:10.1134/1.1625721
  %%CITATION = doi:10.1134/1.1625721;%%
%\bibitem{Aad:2015xaa}
  G.~Aad {\it et al.} [ATLAS Collaboration],
  %``Search for heavy Majorana neutrinos with the ATLAS detector in pp collisions at $ \sqrt{s}=8 $ TeV,''
  JHEP {\bf 1507} (2015) 162
  %doi:10.1007/JHEP07(2015)162
  [arXiv:1506.06020 [hep-ex]].
  %%CITATION = doi:10.1007/JHEP07(2015)162;%%

\bibitem{Umu_exclusions}
%\bibitem{Abela:1981nf}
  R.~Abela, M.~Daum, G.~H.~Eaton, R.~Frosch, B.~Jost, P.~R.~Kettle and E.~Steiner,
  %``Search for an Admixture of Heavy Neutrino in Pion Decay,''
  Phys.\ Lett.\  {\bf 105B} (1981) 263
   Erratum: [Phys.\ Lett.\  {\bf 106B} (1981) 513];\\
  %doi:10.1016/0370-2693(81)90884-4
  %%CITATION = doi:10.1016/0370-2693(81)90884-4;%%
%\bibitem{Hayano:1982wu}
  R.~S.~Hayano {\it et al.},
  %``HEAVY NEUTRINO SEARCH USING K(mu2) DECAY,''
  Phys.\ Rev.\ Lett.\  {\bf 49} (1982) 1305;\\
  %doi:10.1103/PhysRevLett.49.1305
  %%CITATION = doi:10.1103/PhysRevLett.49.1305;%%
%\bibitem{Vilain:1994vg}
  P.~Vilain {\it et al.} [CHARM II Collaboration],
  %``Search for heavy isosinglet neutrinos,''
  Phys.\ Lett.\ B {\bf 343} (1995) 453
   [Phys.\ Lett.\ B {\bf 351} (1995) 387];\\
  %doi:10.1016/0370-2693(94)00440-I, 10.1016/0370-2693(94)01422-9
  %%CITATION = doi:10.1016/0370-2693(94)00440-I, 10.1016/0370-2693(94)01422-9;%%
%\bibitem{Gallas:1994xp}
  E.~Gallas {\it et al.} [FMMF Collaboration],
  %``Search for neutral weakly interacting massive particles in the Fermilab Tevatron wide band neutrino beam,''
  Phys.\ Rev.\ D {\bf 52} (1995) 6;\\
  %doi:10.1103/PhysRevD.52.6
  %%CITATION = doi:10.1103/PhysRevD.52.6;%%
%\bibitem{Vaitaitis:1999wq}
  A.~Vaitaitis {\it et al.} [NuTeV and E815 Collaborations],
  %``Search for neutral heavy leptons in a high-energy neutrino beam,''
  Phys.\ Rev.\ Lett.\  {\bf 83} (1999) 4943
  %doi:10.1103/PhysRevLett.83.4943
  [hep-ex/9908011];\\
  %%CITATION = doi:10.1103/PhysRevLett.83.4943;%%
%\bibitem{Kusenko:2004qc}
  A.~Kusenko, S.~Pascoli and D.~Semikoz,
  %``New bounds on MeV sterile neutrinos based on the accelerator and Super-Kamiokande results,''
  JHEP {\bf 0511} (2005) 028
  %doi:10.1088/1126-6708/2005/11/028
  [hep-ph/0405198];\\
  %%CITATION = doi:10.1088/1126-6708/2005/11/028;%%
%\bibitem{Artamonov:2014urb}
  A.~V.~Artamonov {\it et al.} [E949 Collaboration],
  %``Search for heavy neutrinos in $K^+\to\mu^+\nu_H$ decays,''
  Phys.\ Rev.\ D {\bf 91} (2015) no.5,  052001
   Erratum: [Phys.\ Rev.\ D {\bf 91} (2015) no.5,  059903]
  %doi:10.1103/PhysRevD.91.059903, 10.1103/PhysRevD.91.052001
  [arXiv:1411.3963 [hep-ex]].
  %%CITATION = doi:10.1103/PhysRevD.91.059903, 10.1103/PhysRevD.91.052001;%%

\bibitem{Utau_exclusions}
%\bibitem{Astier:2001ck}
  P.~Astier {\it et al.} [NOMAD Collaboration],
  %``Search for heavy neutrinos mixing with tau neutrinos,''
  Phys.\ Lett.\ B {\bf 506} (2001) 27
  %doi:10.1016/S0370-2693(01)00362-8
  [hep-ex/0101041];\\
  %%CITATION = doi:10.1016/S0370-2693(01)00362-8;%%
%\bibitem{Orloff:2002de}
  J.~Orloff, A.~N.~Rozanov and C.~Santoni,
  %``Limits on the mixing of tau neutrino to heavy neutrinos,''
  Phys.\ Lett.\ B {\bf 550} (2002) 8
  %doi:10.1016/S0370-2693(02)02769-7
  [hep-ph/0208075].
  %%CITATION = doi:10.1016/S0370-2693(02)02769-7;%%
  
\bibitem{UeUmu_exclusions}
%\bibitem{Yamazaki:1984sj}
  T.~Yamazaki {\it et al.},
  %``Search for Heavy Neutrinos in Kaon Decay,''
  Conf.\ Proc.\ C {\bf 840719} (1984) 262;\\
%\bibitem{Bergsma:1985is}
  F.~Bergsma {\it et al.} [CHARM Collaboration],
  %``A Search for Decays of Heavy Neutrinos in the Mass Range 0.5-{GeV} to 2.8-{GeV},''
  Phys.\ Lett.\  {\bf 166B} (1986) 473;\\
  %doi:10.1016/0370-2693(86)91601-1
  %%CITATION = doi:10.1016/0370-2693(86)91601-1;%%
%\bibitem{CooperSarkar:1985nh}
  A.~M.~Cooper-Sarkar {\it et al.} [WA66 Collaboration],
  %``Search for Heavy Neutrino Decays in the {BEBC} Beam Dump Experiment,''
  Phys.\ Lett.\  {\bf 160B} (1985) 207;\\
  %doi:10.1016/0370-2693(85)91493-5
  %%CITATION = doi:10.1016/0370-2693(85)91493-5;%%
%\bibitem{Badier:1986xz}
  J.~Badier {\it et al.} [NA3 Collaboration],
  %``Mass and Lifetime Limits on New Longlived Particles in 300-{GeV}/$c \pi^-$ Interactions,''
  Z.\ Phys.\ C {\bf 31} (1986) 21;\\
  %doi:10.1007/BF01559588
  %%CITATION = doi:10.1007/BF01559588;%%
 %\bibitem{Bernardi:1987ek}
  G.~Bernardi {\it et al.},
  %``Further Limits On Heavy Neutrino Couplings,''
  Phys.\ Lett.\ B {\bf 203} (1988) 332;\\
  %doi:10.1016/0370-2693(88)90563-1
  %%CITATION = doi:10.1016/0370-2693(88)90563-1;%%
%\bibitem{Liventsev:2013zz}
  D.~Liventsev {\it et al.} [Belle Collaboration],
  %``Search for heavy neutrinos at Belle,''
  Phys.\ Rev.\ D {\bf 87} (2013) no.7,  071102
   Erratum: [Phys.\ Rev.\ D {\bf 95} (2017) no.9,  099903]
  %doi:10.1103/PhysRevD.95.099903, 10.1103/PhysRevD.87.071102
  [arXiv:1301.1105 [hep-ex]].
  %%CITATION = doi:10.1103/PhysRevD.95.099903, 10.1103/PhysRevD.87.071102;%%
  
\bibitem{UeUmuUtau_exclusions}
%\bibitem{Akrawy:1990zq}
  M.~Z.~Akrawy {\it et al.} [OPAL Collaboration],
  %``Limits on neutral heavy lepton production from Z0 decay,''
  Phys.\ Lett.\ B {\bf 247} (1990) 448;\\
  %doi:10.1016/0370-2693(90)90924-U
  %%CITATION = doi:10.1016/0370-2693(90)90924-U;%%
%\bibitem{Adriani:1992pq}
  O.~Adriani {\it et al.} [L3 Collaboration],
  %``Search for isosinglet neutral heavy leptons in Z0 decays,''
  Phys.\ Lett.\ B {\bf 295} (1992) 371;\\
  %doi:10.1016/0370-2693(92)91579-X
  %%CITATION = doi:10.1016/0370-2693(92)91579-X;%%
%\bibitem{Adriani:1993gk}
  O.~Adriani {\it et al.} [L3 Collaboration],
  %``Results from the L3 experiment at LEP,''
  Phys.\ Rept.\  {\bf 236} (1993) 1;\\
  %doi:10.1016/0370-1573(93)90027-B
  %%CITATION = doi:10.1016/0370-1573(93)90027-B;%%
%\bibitem{Abreu:1996pa}
  P.~Abreu {\it et al.} [DELPHI Collaboration],
  %``Search for neutral heavy leptons produced in Z decays,''
  Z.\ Phys.\ C {\bf 74} (1997) 57
   Erratum: [Z.\ Phys.\ C {\bf 75} (1997) 580];\\
  %doi:10.1007/s002880050370
  %%CITATION = doi:10.1007/s002880050370;%%
%\bibitem{Das:2014jxa}
  A.~Das, P.~S.~Bhupal Dev and N.~Okada,
  %``Direct bounds on electroweak scale pseudo-Dirac neutrinos from $\sqrt s=8$ TeV LHC data,''
  Phys.\ Lett.\ B {\bf 735} (2014) 364
  %doi:10.1016/j.physletb.2014.06.058
  [arXiv:1405.0177 [hep-ph]];\\
  %%CITATION = doi:10.1016/j.physletb.2014.06.058;%%
%\bibitem{Antusch:2015mia}
  S.~Antusch and O.~Fischer,
  %``Testing sterile neutrino extensions of the Standard Model at future lepton colliders,''
  JHEP {\bf 1505} (2015) 053
  %doi:10.1007/JHEP05(2015)053
  [arXiv:1502.05915 [hep-ph]].
  %%CITATION = doi:10.1007/JHEP05(2015)053;%%

 \bibitem{Das:2016hof} 
  A.~Das, P.~Konar and S.~Majhi,
  %``Production of Heavy neutrino in next-to-leading order QCD at the LHC and beyond,''
  JHEP {\bf 1606}, 019 (2016)
  doi:10.1007/JHEP06(2016)019
  [arXiv:1604.00608 [hep-ph]].

\bibitem{Das:2017zjc} 
  A.~Das, P.~S.~B.~Dev and C.~S.~Kim,
  %``Constraining Sterile Neutrinos from Precision Higgs Data,''
  Phys.\ Rev.\ D {\bf 95}, no. 11, 115013 (2017)
  doi:10.1103/PhysRevD.95.115013
  [arXiv:1704.00880 [hep-ph]].
  
\bibitem{Das:2017rsu} 
  A.~Das, Y.~Gao and T.~Kamon,
  %``Heavy Neutrino Search via the Higgs boson at the LHC,''
  arXiv:1704.00881 [hep-ph].

\bibitem{Antusch:2008tz} 
S.~Antusch, J.~P.~Baumann and E.~Fernandez-Martinez,
%``Non-Standard Neutrino Interactions with Matter from Physics Beyond
%the Standard Model,'' 
Nucl.\ Phys.\ B {\bf 810} (2009) 369 
[arXiv:0807.1003 [hep-ph]].

\bibitem{Blennow:2016jkn}
  M.~Blennow, P.~Coloma, E.~Fernandez-Martinez, J.~Hernandez-Garcia
  and J.~Lopez-Pavon, 
  ``Non-Unitarity, sterile neutrinos, and Non-Standard neutrino Interactions,''
  arXiv:1609.08637 [hep-ph].
  
\bibitem{Antusch:2014woa}
  S.~Antusch and O.~Fischer,
  %``Non-unitarity of the leptonic mixing matrix: Present bounds and
  %future sensitivities,'' 
  JHEP {\bf 1410} (2014) 94
  [arXiv:1407.6607 [hep-ph]].

\bibitem{Blennow:2010th} 
  M.~Blennow, E.~Fernandez-Martinez, J.~Lopez-Pavon and J.~Menendez,
  %``Neutrinoless double beta decay in seesaw models,''
  JHEP {\bf 1007} (2010) 096 
  [arXiv:1005.3240 [hep-ph]].
  
\bibitem{Aaij:2017qml}
  R.~Aaij {\it et al.} [LHCb Collaboration],
  ``Measurement of the $B^{\pm}$ production cross-section in pp
  collisions at $\sqrt{s} =$ 7 and 13 TeV,'' 
  arXiv:1710.04921 [hep-ex].
  %%CITATION = ARXIV:1710.04921;%%


%% \bibitem{Aaij:2013noa}
%%   R.~Aaij {\it et al.} [LHCb Collaboration],
%%   %``Measurement of B meson production cross-sections in proton-proton
%%   %collisions at $\sqrt{s}$ = 7 TeV,'' 
%%   JHEP {\bf 1308} (2013) 117
%%   [arXiv:1306.3663 [hep-ex]].

%% \bibitem{Chatrchyan:2011pw}
%%   S.~Chatrchyan {\it et al.} [CMS Collaboration],
%%   %``Measurement of the $B^0$ production cross section in $pp$
%%   %Collisions at $\sqrt{s}=7$ TeV,'' 
%%   Phys.\ Rev.\ Lett.\  {\bf 106} (2011) 252001
%%   [arXiv:1104.2892 [hep-ex]].

%% \bibitem{Aaij:2016avz}
%%   R.~Aaij {\it et al.} [LHCb Collaboration],
%%   %``Measurement of the $b$-quark production cross-section in 7 and 13
%%   %TeV $pp$ collisions,'' 
%%   Phys.\ Rev.\ Lett.\  {\bf 118} (2017) no.5,  052002
%%   [arXiv:1612.05140 [hep-ex]].
%%   %%CITATION = doi:10.1103/PhysRevLett.118.052002;%%

%% \bibitem{Khachatryan:2016csy}
%%   V.~Khachatryan {\it et al.} [CMS Collaboration],
%%   %``Measurement of the total and differential inclusive $B^+$ hadron
%%   %cross sections in pp collisions at $\sqrt{s}$ = 13 TeV,'' 
%%   Phys.\ Lett.\ B {\bf 771} (2017) 435
%%   [arXiv:1609.00873 [hep-ex]].
%%   %%CITATION = doi:10.1016/j.physletb.2017.05.074;%%

%% \bibitem{Kniehl:2011bk}
%%   B.~A.~Kniehl, G.~Kramer, I.~Schienbein and H.~Spiesberger,
%%   %``Inclusive B-Meson Production at the LHC in the GM-VFN Scheme,''
%%   Phys.\ Rev.\ D {\bf 84} (2011) 094026
%%   [arXiv:1109.2472 [hep-ph]].
%%   %%CITATION = doi:10.1103/PhysRevD.84.094026;%%

%% \bibitem{Aad:2015zix}
%%   G.~Aad {\it et al.} [ATLAS Collaboration],
%%   %``Measurement of $D^{*\pm}$, $D^\pm$ and $D_s^\pm$ meson production
%%   %cross sections in $pp$ collisions at $\sqrt{s}=7$ TeV with the
%%   %ATLAS detector,'' 
%%   Nucl.\ Phys.\ B {\bf 907} (2016) 717
%%   [arXiv:1512.02913 [hep-ex]].
%%   %%CITATION = doi:10.1016/j.nuclphysb.2016.04.032;%%

\bibitem{Alonso:2012ji}
 R.~Alonso, M.~Dhen, M.~B.~Gavela and T.~Hambye,
 %``Muon conversion to electron in nuclei in type-I seesaw models,''
 JHEP {\bf 1301} (2013) 118
 [arXiv:1209.2679 [hep-ph]].

\bibitem{Abada:2014kba}
  A.~Abada, M.~E.~Krauss, W.~Porod, F.~Staub, A.~Vicente and C.~Weiland,
  %``Lepton flavor violation in low-scale seesaw models: SUSY and
  %non-SUSY contributions,'' 
  JHEP {\bf 1411} (2014) 048
  [arXiv:1408.0138 [hep-ph]].

\bibitem{Abada:2014cca}
  A.~Abada, V.~De Romeri, S.~Monteil, J.~Orloff and A.~M.~Teixeira,
  %``Indirect searches for sterile neutrinos at a high-luminosity Z-factory,''
  JHEP {\bf 1504} (2015) 051
  [arXiv:1412.6322 [hep-ph]].
  %%CITATION = doi:10.1007/JHEP04(2015)051;%%

%\cite{Abada:2015zea}
\bibitem{Abada:2015zea}
  A.~Abada, D.~Becirevic, M.~Lucente and O.~Sumensari,
  %``Lepton flavor violating decays of vector quarkonia and of the $Z$ boson,''
  Phys.\ Rev.\ D {\bf 91} (2015) no.11,  113013
  doi:10.1103/PhysRevD.91.113013
  [arXiv:1503.04159 [hep-ph]].
  %%CITATION = doi:10.1103/PhysRevD.91.113013;%%
  %26 citations counted in INSPIRE as of 19 Dec 2017

\bibitem{Abada:2015oba}
  A.~Abada, V.~De Romeri and A.~M.~Teixeira,
  %``Impact of sterile neutrinos on nuclear-assisted cLFV processes,''
  JHEP {\bf 1602} (2016) 083
  [arXiv:1510.06657 [hep-ph]].
  %20 citations counted in INSPIRE as of 10 Oct 2017

\bibitem{Abada:2016vzu}
  A.~Abada, V.~De Romeri, J.~Orloff and A.~M.~Teixeira,
  %``In-flight cLFV conversion: ${e-\mu }$ , ${e-\tau }$ and ${\mu
  %-\tau }$ in minimal extensions of the standard model with sterile
  %fermions,'' 
  Eur.\ Phys.\ J.\ C {\bf 77} (2017) no.5,  304
  [arXiv:1612.05548 [hep-ph]].
  %%CITATION = doi:10.1140/epjc/s10052-017-4864-z;%%
      
      
      
\bibitem{Alekhin:2015byh}
  S.~Alekhin {\it et al.},
  %``A facility to Search for Hidden Particles at the CERN SPS: the
  %SHiP physics case,'' 
  Rept.\ Prog.\ Phys.\  {\bf 79} (2016) no.12,  124201
%%  doi:10.1088/0034-4885/79/12/124201
  [arXiv:1504.04855 [hep-ph]].
  %%CITATION = doi:10.1088/0034-4885/79/12/124201;%%
  
\bibitem{ref:ISS}
%\bibitem{Mohapatra:1986bd}
  R.~N.~Mohapatra and J.~W.~F.~Valle,
  %``Neutrino Mass and Baryon Number Nonconservation in Superstring Models,''
  Phys.\ Rev.\ D {\bf 34} (1986) 1642.
  %doi:10.1103/PhysRevD.34.1642;
  %%CITATION = doi:10.1103/PhysRevD.34.1642;%%  
  %\bibitem{GonzalezGarcia:1988rw}
  M.~C.~Gonzalez-Garcia and J.~W.~F.~Valle,
  %``Fast Decaying Neutrinos and Observable Flavor Violation in a New Class of Majoron Models,''
  Phys.\ Lett.\ B {\bf 216} (1989) 360.
  %doi:10.1016/0370-2693(89)91131-3;
  %%CITATION = doi:10.1016/0370-2693(89)91131-3;%%
  %\bibitem{Deppisch:2004fa}
  F.~Deppisch and J.~W.~F.~Valle,
  %``Enhanced lepton flavor violation in the supersymmetric inverse seesaw model,''
  Phys.\ Rev.\ D {\bf 72} (2005) 036001
  %doi:10.1103/PhysRevD.72.036001
  [hep-ph/0406040].
  %%CITATION = doi:10.1103/PhysRevD.72.036001;%%
  %\bibitem{Abada:2014vea}
  A.~Abada and M.~Lucente,
  %``Looking for the minimal inverse seesaw realisation,''
  Nucl.\ Phys.\ B {\bf 885} (2014) 651
  %doi:10.1016/j.nuclphysb.2014.06.003
  [arXiv:1401.1507 [hep-ph]].
  %%CITATION = doi:10.1016/j.nuclphysb.2014.06.003;%%
  
\bibitem{ref:LSS}
%\bibitem{Barr:2003nn}
  S.~M.~Barr,
  %``A Different seesaw formula for neutrino masses,''
  Phys.\ Rev.\ Lett.\  {\bf 92} (2004) 101601
  %doi:10.1103/PhysRevLett.92.101601
  [hep-ph/0309152];
  %%CITATION = doi:10.1103/PhysRevLett.92.101601;%%  
  %\bibitem{Malinsky:2005bi}
  M.~Malinsky, J.~C.~Romao and J.~W.~F.~Valle,
  %``Novel supersymmetric SO(10) seesaw mechanism,''
  Phys.\ Rev.\ Lett.\  {\bf 95} (2005) 161801
  %doi:10.1103/PhysRevLett.95.161801
  [hep-ph/0506296];
  %%CITATION = doi:10.1103/PhysRevLett.95.161801;%%
  %\bibitem{Gavela:2009cd}
  M.~B.~Gavela, T.~Hambye, D.~Hernandez and P.~Hernandez,
  %``Minimal Flavour Seesaw Models,''
  JHEP {\bf 0909} (2009) 038
  %doi:10.1088/1126-6708/2009/09/038
  [arXiv:0906.1461 [hep-ph]].
  %%CITATION = doi:10.1088/1126-6708/2009/09/038;%%

  
\bibitem{Shaposhnikov:2006nn}
  M.~Shaposhnikov,
  %``A Possible symmetry of the nuMSM,''
  Nucl.\ Phys.\ B {\bf 763} (2007) 49
  %doi:10.1016/j.nuclphysb.2006.11.003
  [hep-ph/0605047].
  %%CITATION = doi:10.1016/j.nuclphysb.2006.11.003;%%  
  
  \bibitem{Ibarra:2010xw}
  A.~Ibarra, E.~Molinaro and S.~T.~Petcov,
  %``TeV Scale See-Saw Mechanisms of Neutrino Mass Generation, the Majorana Nature of the Heavy Singlet Neutrinos and $(\beta\beta)_{0\nu}$-Decay,''
  JHEP {\bf 1009} (2010) 108
  %doi:10.1007/JHEP09(2010)108
  [arXiv:1007.2378 [hep-ph]].
  %%CITATION = doi:10.1007/JHEP09(2010)108;%%

\bibitem{Schechter:1980gr}
  J.~Schechter and J.~W.~F.~Valle,
  %``Neutrino Masses In SU(2) X U(1) Theories,''
  Phys.\ Rev.\ D {\bf 22} (1980) 2227.
  %%CITATION = PHRVA,D22,2227;%%

\bibitem{Gronau:1984ct}
  M.~Gronau, C.~N.~Leung and J.~L.~Rosner,
  %``Extending Limits on Neutral Heavy Leptons,''
  Phys.\ Rev.\ D {\bf 29} (1984) 2539.
  %%CITATION = PHRVA,D29,2539;%% 

\bibitem{Chanowitz:1978mv}
  M.~S.~Chanowitz, M.~A.~Furman and I.~Hinchliffe,
  %``Weak Interactions of Ultraheavy Fermions. 2.,''
  Nucl.\ Phys.\ B {\bf 153} (1979) 402.

\bibitem{Durand:1989zs}
  L.~Durand, J.~M.~Johnson and J.~L.~Lopez,
  %``Perturbative Unitarity Revisited: A New Upper Bound on the Higgs
  %Boson Mass,'' 
  Phys.\ Rev.\ Lett.\  {\bf 64} (1990) 1215.

\bibitem{Korner:1992an}
  J.~G.~Korner, A.~Pilaftsis and K.~Schilcher,
  %``Leptonic flavor changing Z0 decays in SU(2) x U(1) theories with
  %right-handed neutrinos,'' 
  Phys.\ Lett.\ B {\bf 300} (1993) 381
  [hep-ph/9301290].

\bibitem{Bernabeu:1993up}
  J.~Bernabeu, J.~G.~Korner, A.~Pilaftsis and K.~Schilcher,
  %``Universality breaking effects in leptonic Z decays,''
  Phys.\ Rev.\ Lett.\  {\bf 71} (1993) 2695
  [hep-ph/9307295].
    
\bibitem{Fajfer:1998px}
  S.~Fajfer and A.~Ilakovac,
  %``Lepton flavor violation in light hadron decays,''
  Phys.\ Rev.\ D {\bf 57} (1998) 4219.

\bibitem{Ilakovac:1999md}
  A.~Ilakovac,
  %``Lepton flavor violation in the standard model extended by heavy
  %singlet Dirac neutrinos,''  
  Phys.\ Rev.\ D {\bf 62} (2000) 036010
  [hep-ph/9910213].

  
  \bibitem{Akhmedov:2013hec}
  E.~Akhmedov, A.~Kartavtsev, M.~Lindner, L.~Michaels and J.~Smirnov,
  %``Improving Electro-Weak Fits with TeV-scale Sterile Neutrinos,''
  JHEP {\bf 1305} (2013) 081
  [arXiv:1302.1872 [hep-ph]].
  %%CITATION = ARXIV:1302.1872;%%
  
\bibitem{Fernandez-Martinez:2015hxa}
  E.~Fernandez-Martinez, J.~Hernandez-Garcia, J.~Lopez-Pavon and M.~Lucente,
%  ``Loop level constraints on Seesaw neutrino mixing,''
   JHEP {\bf 1510} (2015) 130 [arXiv:1508.03051 [hep-ph]].

\bibitem{Basso:2013jka}
L.~Basso, O.~Fischer and J.~J.~van der Bij,
%``Precision tests of unitarity in leptonic mixing,''
Europhys.\ Lett.\  {\bf 105} (2014) 11001
[arXiv:1310.2057 [hep-ph]].
%%CITATION = ARXIV:1310.2057;%%

\bibitem{Abada:2013aba}
  A.~Abada, A.~M.~Teixeira, A.~Vicente and C.~Weiland,
  %``Sterile neutrinos in leptonic and semileptonic decays,''
  JHEP {\bf 1402} (2014) 091
  [arXiv:1311.2830 [hep-ph]].
  
\bibitem{BhupalDev:2012zg}
P.~S.~Bhupal Dev, R.~Franceschini and R.~N.~Mohapatra,
%``Bounds on TeV Seesaw Models from LHC Higgs Data,''
Phys.\ Rev.\ D {\bf 86} (2012) 093010
[arXiv:1207.2756 [hep-ph]].
%%CITATION = ARXIV:1207.2756;%%
 
\bibitem{Cely:2012bz}
C.~G.~Cely, A.~Ibarra, E.~Molinaro and S.~T.~Petcov,
%``Higgs Decays in the Low Scale Type I See-Saw Model,''
Phys.\ Lett.\ B {\bf 718} (2013) 957
[arXiv:1208.3654 [hep-ph]].
%%CITATION = ARXIV:1208.3654;%%
  
\bibitem{Bandyopadhyay:2012px}
  P.~Bandyopadhyay, E.~J.~Chun, H.~Okada and J.~-C.~Park,
  %``Higgs Signatures in Inverse Seesaw Model at the LHC,''
  JHEP {\bf 1301} (2013) 079
  [arXiv:1209.4803 [hep-ph]].
  %%CITATION = ARXIV:1209.4803;%%
  
 \bibitem{Kusenko:2009up}
  A.~Kusenko,
  %``Sterile neutrinos: The Dark side of the light fermions,''
  Phys.\ Rept.\  {\bf 481} (2009) 1
  [arXiv:0906.2968 [hep-ph]].
  %%CITATION = ARXIV:0906.2968;%%

\bibitem{Ma:1979px}
 E.~Ma and A.~Pramudita,
 %``Flavor Changing Effective Neutral Current Couplings in the
%{Weinberg-Salam} Model,''
 Phys.\ Rev.\ D {\bf 22} (1980) 214.

\bibitem{Ilakovac:1994kj}
  A.~Ilakovac and A.~Pilaftsis,
  %``Flavor violating charged lepton decays in seesaw-type models,''
  Nucl.\ Phys.\ B {\bf 437} (1995) 491
  [hep-ph/9403398].

\bibitem{Deppisch:2004fa}
  F.~Deppisch and J.~W.~F.~Valle,
  %``Enhanced lepton flavor violation in the supersymmetric inverse
  %seesaw model,'' 
  Phys.\ Rev.\ D {\bf 72} (2005) 036001
  [hep-ph/0406040].
  %%CITATION = HEP-PH/0406040;%%

\bibitem{Deppisch:2005zm}
 F.~Deppisch, T.~S.~Kosmas and J.~W.~F.~Valle,
 %``Enhanced mu- - e- conversion in nuclei in the inverse seesaw model,''
 Nucl.\ Phys.\ B {\bf 752} (2006) 80
 [hep-ph/0512360].

\bibitem{Dinh:2012bp}
 D.~N.~Dinh, A.~Ibarra, E.~Molinaro and S.~T.~Petcov,
 %``The $\mu - e$ Conversion in Nuclei, $\mu \to e \gamma, \mu \to 3e$
Decays and TeV Scale See-Saw Scenarios of Neutrino Mass Generation,''
 JHEP {\bf 1208} (2012) 125
  [Erratum-ibid.\  {\bf 1309} (2013) 023]
 [arXiv:1205.4671 [hep-ph]].


\bibitem{Goudzovski:2011tc} 
  E.~Goudzovski [NA48/2 and NA62 numbers Collaborations],
  %``Kaon programme at CERN: recent results,''
  PoS EPS {\bf -HEP2011} (2011) 181 
  [arXiv:1111.2818 [hep-ex]].
  
\bibitem{Lazzeroni:2012cx} 
  C.~Lazzeroni {\it et al.}  [NA62 Collaboration],
  %``Precision Measurement of the Ratio of the Charged Kaon Leptonic
  %Decay Rates,'' 
  Phys.\ Lett.\ B {\bf 719} (2013) 326 
  [arXiv:1212.4012 [hep-ex]].
  
\bibitem{Naik:2009tk} 
  P.~Naik {\it et al.}  [CLEO Collaboration],
  %``Measurement of the Pseudoscalar Decay Constant f(D(s)) Using
  %D(s)+ ---> tau+ nu, tau+ ---> rho+ anti-nu Decays,'' 
  Phys.\ Rev.\ D {\bf 80} (2009) 112004 
  [arXiv:0910.3602 [hep-ex]].
  
\bibitem{Li:2011nij} 
  H.~-B.~Li,
  ``Proceedings, 4th International Workshop on Charm Physics (Charm
  2010) : Beijing, China, October 21-24, 2010,'' 
  Int.\ J.\ Mod.\ Phys.\ Conf.\ Ser.\  {\bf 02} (2011).
  
 \bibitem{Aubert:2007xj} 
  B.~Aubert {\it et al.}  [BaBar Collaboration],
  %``A Search for $B^{+} \to \tau^{+} \nu$ with Hadronic $B$ tags,''
  Phys.\ Rev.\ D {\bf 77} (2008) 011107.
  
\bibitem{Adachi:2012mm} 
  I.~Adachi {\it et al.}  [Belle Collaboration],
  %``Evidence for $B^- \to \tau^- \bar{\nu}_\tau$  with a Hadronic
  %Tagging Method Using the Full Data Sample of Belle,'' 
  Phys.\ Rev.\ Lett.\  {\bf 110}  (2013) 131801
  [arXiv:1208.4678 [hep-ex]].
 
 \bibitem{Abada:2012mc}
  A.~Abada, D.~Das, A.~M.~Teixeira, A.~Vicente and C.~Weiland,
  %``Tree-level lepton universality violation in the presence of
  %sterile neutrinos: impact for $R_K$ and $R_\pi$,'' 
  JHEP {\bf 1302} (2013) 048
  [arXiv:1211.3052 [hep-ph]].

 \bibitem{Benes:2005hn}
  P.~Benes, A.~Faessler, F.~Simkovic and S.~Kovalenko,
  %``Sterile neutrinos in neutrinoless double beta decay,''
  Phys.\ Rev.\ D {\bf 71} (2005) 077901
  [hep-ph/0501295].  
  
\bibitem{Abada:2014nwa}
  A.~Abada, V.~De Romeri and A.~M.~Teixeira,
  %``Effect of steriles states on lepton magnetic moments and
  %neutrinoless double beta decay,'' 
  JHEP {\bf 1409} (2014) 074
  [arXiv:1406.6978 [hep-ph]].
  %%CITATION = doi:10.1007/JHEP09(2014)074;%%

\bibitem{Smirnov:2006bu}
  A.~Y.~Smirnov and R.~Zukanovich Funchal,
  %``Sterile neutrinos: Direct mixing effects versus induced mass
  %matrix of active neutrinos,'' 
  Phys.\ Rev.\ D {\bf 74} (2006) 013001
  [hep-ph/0603009].

\bibitem{Hernandez:2014fha}
  P.~Hernandez, M.~Kekic and J.~Lopez-Pavon,
  %``$N_{\rm eff}$ in low-scale seesaw models versus the lightest
  %neutrino mass,'' 
  Phys.\ Rev.\ D {\bf 90} (2014) no.6,  065033
  [arXiv:1406.2961 [hep-ph]].

\bibitem{Vincent:2014rja}
  A.~C.~Vincent, E.~F.~Martinez, P.~Hernandez, M.~Lattanzi and O.~Mena,
  %``Revisiting cosmological bounds on sterile neutrinos,''
  JCAP {\bf 1504} (2015) no.04,  006
  [arXiv:1408.1956 [astro-ph.CO]].

\bibitem{Maltoni:2002qb}
  F.~Maltoni and T.~Stelzer,
  %``MadEvent: Automatic event generation with MadGraph,''
  JHEP {\bf 0302} (2003) 027
  doi:10.1088/1126-6708/2003/02/027
  [hep-ph/0208156].


\end{thebibliography}
}
\end{document}